\documentclass[pdftex,twocolumn,epjc3_preprint,runningheads]{svjour3}

\usepackage[T1]{fontenc}
\usepackage{lmodern}
\usepackage{calc}
\usepackage{graphicx}
\usepackage{booktabs}
\usepackage{textcomp}
\usepackage{xspace}
\usepackage{relsize}
\usepackage{amssymb}
\usepackage{amsmath}
\usepackage{listings}
\usepackage{microtype}
\usepackage{multirow}
\usepackage{tabularx}
\usepackage{array}
\usepackage{placeins}
\usepackage{cuted}
\usepackage{soul} % only for \st; delete if this causes you problems.
\usepackage{fixltx2e}
\usepackage{slashed}
\usepackage{bm}
\usepackage[numbers,sort&compress]{natbib}
\usepackage[labelfont=bf,font=small]{caption}
\usepackage[skip=-2pt]{subcaption}
\usepackage[colorlinks,citecolor=blue,urlcolor=blue,linkcolor=blue,breaklinks=breakall]{hyperref}
\usepackage{breakurl}
\usepackage[dvipsnames]{xcolor}
\usepackage[clockwise,figuresright]{rotating}
\usepackage{siunitx}
\usepackage{tikz}
\usepackage[normalem]{ulem}
\usepackage[utf8]{inputenc}

\usepackage{todonotes}
\usepackage{etoolbox}
\AfterEndEnvironment{strip}{\leavevmode}

\allowdisplaybreaks

\newcolumntype{L}{>{\raggedright\let\newline\\\arraybackslash\hspace{0pt}}X}
\newcolumntype{R}{>{\raggedleft\let\newline\\\arraybackslash\hspace{0pt}}X}
\newcolumntype{C}{>{\centering\let\newline\\\arraybackslash\hspace{0pt}}X}

\newcommand{\gambitinstitute}[1]{\expandafter\csname #1\endcsname \label{#1}}

\newcommand{\monash}{School of Physics and Astronomy, Monash University, Melbourne, VIC 3800, Australia}

\newcommand{\grappa}{GRAPPA, Institute of Physics, University of Amsterdam, Science Park 904, 1098 XH Amsterdam, Netherlands}

\newcommand{\cern}{European Organization for Nuclear Research (CERN), Geneva, Switzerland}
\newcommand{\krakow}{H.~Niewodnicza\'nski Institute of Nuclear Physics, Polish Academy of Sciences, 31-342  Krak\'ow, Poland}

\newcommand{\munich}{Physik Department T70, Technische Universit\"at M\"unchen, James-Franck-Straße 1, D-85748 Garching, Germany}

\makeatletter

\newcommand{\preprintnumber}[1]{\gdef\@preprintnumber{\begin{flushright}{#1}\end{flushright}}}

\g@addto@macro\bfseries{\boldmath}
\makeatother

\newcommand{\subparagraph}{} %< workaround for svjour not defining subparagraph
\usepackage{titlesec}
\titleformat*{\paragraph}{\bfseries}
\journalname{Eur. Phys. J. C}
\smartqed
\let\underscore\_
\renewcommand{\_}{\discretionary{\underscore}{}{\underscore}}

\makeatletter
\let\orgdescriptionlabel\descriptionlabel
\renewcommand*{\descriptionlabel}[1]{%
  \let\orglabel\label
  \let\label\@gobble
  \phantomsection
  \protected@edef\@currentlabel{#1}%
  \let\label\orglabel
  \orgdescriptionlabel{#1}%
}
\makeatother

\newcommand\postnewlinemarker{\hbox{\ensuremath{\hookrightarrow}}}
\newcommand\cpp[1]{{\lstinline!#1!}}  % Apparently curly braces are only "experimental"

\newcommand\yaml[1]{{\lstset{style=yaml}\lstinline!#1!\lstset{style=cpp}}}

\newcommand\term[1]{{\lstset{style=terminal}\lstinline!#1!\lstset{style=cpp}}}
\newcommand\fortran[1]{{\lstset{style=fortran}\lstinline!#1!\lstset{style=cpp}}}
\newcommand\py[1]{{\lstset{style=python}\lstinline!#1!\lstset{style=cpp}}}
\newcommand\customtilde{{\raisebox{0.2ex}{\scalebox{0.6}{\boldmath$\sim$}}}}
\newcommand\mathematica[1]{{\lstset{style=Mathematica}\lstinline!#1!\lstset{style=cpp}}}

\lstnewenvironment{lstlistingyaml}{\lstset{style=yaml}}{\lstset{style=cpp}}
\lstnewenvironment{lstlistingterm}{\lstset{style=terminal}}{\lstset{style=cpp}}
\lstnewenvironment{lstlistingfortran}{\lstset{style=fortran}}{\lstset{style=cpp}}
\lstnewenvironment{lstcpp}{\lstset{style=cpp}}{\lstset{style=cpp}}
\lstnewenvironment{lstcppalt}{\lstset{style=cppalt}}{\lstset{style=cpp}}
\lstnewenvironment{lstcppnum}{\lstset{style=cppnum}}{\lstset{style=cpp}}
\lstnewenvironment{lstyaml}{\lstset{style=yaml}}{\lstset{style=cpp}}
\lstnewenvironment{lstterm}{\lstset{style=terminal}}{\lstset{style=cpp}}
\lstnewenvironment{lsttermalt}{\lstset{style=terminalalt}}{\lstset{style=cpp}}
\lstnewenvironment{lsttext}{\lstset{style=text}}{\lstset{style=cpp}}
\lstnewenvironment{lstfortran}{\lstset{style=fortran}}{\lstset{style=cpp}}
\lstnewenvironment{lstpy}{\lstset{style=python}}{\lstset{style=cpp}}
\lstnewenvironment{lstmathematica}{\lstset{style=mathematica}}{\lstset{style=cpp}}

\newcommand{\tmpname}{}
\newcommand{\tmplistingname}{}
\makeatletter
\newif\ifATOlabelname
\lst@Key{labelname}{Listing}{\def\ATOlabelname{#1}\global\ATOlabelnametrue}
\makeatother
\lstnewenvironment{lstcpplabel}[1][]{
  \lstset{style=cpp,#1} % #1 allows to add new options with [] same as for normal lstlistings environment
  \ifATOlabelname
    \renewcommand{\tmpname}{\lstlistingname}
    \renewcommand{\tmplistingname}{\lstlistlistingname}
    \renewcommand{\lstlistingname}{\ATOlabelname}% Listing -> labelname
    \renewcommand{\lstlistlistingname}{List of \lstlistingname s}% List of Listings -> List of labelname
  \fi
}{
  \renewcommand{\lstlistingname}{\tmpname}
  \renewcommand{\lstlistlistingname}{\tmplistingname}
  \lstset{style=cpp}
}
\definecolor{solarized@base03}{HTML}{002B36}
\definecolor{solarized@base02}{HTML}{073642}
\definecolor{solarized@base01}{HTML}{586e75}
\definecolor{solarized@base00}{HTML}{657b83}
\definecolor{solarized@base0}{HTML}{839496}
\definecolor{solarized@base1}{HTML}{93a1a1}
\definecolor{solarized@base2}{HTML}{EEE8D5}
\definecolor{solarized@base3}{HTML}{FDF6E3}
\definecolor{solarized@yellow}{HTML}{B58900}
\definecolor{solarized@orange}{HTML}{CB4B16}
\definecolor{solarized@red}{HTML}{DC322F}
\definecolor{solarized@magenta}{HTML}{D33682}
\definecolor{solarized@violet}{HTML}{6C71C4}
\definecolor{solarized@blue}{HTML}{268BD2}
\definecolor{solarized@cyan}{HTML}{2AA198}
\definecolor{solarized@green}{HTML}{859900}
\definecolor{darkred}{HTML}{550003}
\definecolor{darkgreen}{HTML}{00AA00}

\newcommand\YAMLstringstyle{\footnotesize\color{solarized@green}\mdseries}
\newcommand\YAMLkeystyle{\footnotesize\color{solarized@blue}\ttfamily}
\newcommand\YAMLvaluestyle{\footnotesize\color{blue}\mdseries}
\newcommand\ProcessThreeDashes{\llap{\color{cyan}\mdseries-{-}-}}
\newcommand\CPPcommentstyle{\color{solarized@violet}\footnotesize\ttfamily}
\newcommand\CPPdirectivestyle{\color{solarized@magenta}\footnotesize\ttfamily}
\newcommand\termplainstyle{\footnotesize\ttfamily}

\newcommand\processLongMacroDelimiter
{%
\CPPdirectivestyle%
\#define%
}

\lstdefinestyle{cpp}
{
  language=C++,
  basicstyle=\footnotesize\ttfamily,
  basewidth={0.53em,0.44em}, %Ben: experimenting a bit with the fixed-width width (first argument); feels a bit more readable to me with the slightly smaller width (was 0.6em by default)
  numbers=none,
  tabsize=2,
  breaklines=true,
  escapeinside={@}{@},
  showstringspaces=false,
  numberstyle=\tiny\color{solarized@base01},
  keywordstyle=\color{solarized@orange},
  stringstyle=\color{solarized@red}\ttfamily,
  identifierstyle=\color{solarized@blue},
  commentstyle=\CPPcommentstyle,
  directivestyle=\CPPdirectivestyle,
  emphstyle=\color{solarized@green},
  frame=single,
  rulecolor=\color{solarized@base2},
  rulesepcolor=\color{solarized@base2},
  literate={~} {\customtilde}1,
  moredelim=*[directive]\ \ \#,
  moredelim=*[directive]\ \ \ \ \#
}

\lstdefinestyle{cppalt}
{
  language=C++,
  basicstyle=\footnotesize\ttfamily,
  basewidth={0.53em,0.44em}, %Ben: experimenting a bit with the fixed-width width (first argument); feels a bit more readable to me with the slightly smaller width (was 0.6em by default)
  numbers=none,
  tabsize=2,
  breaklines=true,
  escapeinside={*@}{@*},
  showstringspaces=false,
  numberstyle=\tiny\color{solarized@base01},
  keywordstyle=\color{solarized@orange},
  stringstyle=\color{solarized@red}\ttfamily,
  identifierstyle=\color{solarized@blue},
  commentstyle=\CPPcommentstyle,
  directivestyle=\CPPdirectivestyle,
  emphstyle=\color{solarized@green},
  frame=single,
  rulecolor=\color{solarized@base2},
  rulesepcolor=\color{solarized@base2},
  literate={~}{\customtilde}1,
  moredelim=**[is][\processLongMacroDelimiter]{BeginLongMacro}{EndLongMacro} %special delimiter for long macros that go over several lines
}

\lstdefinestyle{cppnum}
{
  language=C++,
  basicstyle=\footnotesize\ttfamily,
  basewidth={0.53em,0.44em}, %Ben: experimenting a bit with the fixed-width width (first argument); feels a bit more readable to me with the slightly smaller width (was 0.6em by default)
  numbers=none,
  tabsize=2,
  breaklines=true,
  escapeinside={@}{@},
  numberstyle=\tiny\color{solarized@base01},
  showstringspaces=false,
  numberstyle=\tiny\color{solarized@base01},
  keywordstyle=\color{solarized@orange},
  stringstyle=\color{solarized@red}\ttfamily,
  identifierstyle=\color{solarized@blue},
  commentstyle=\CPPcommentstyle,
  directivestyle=\CPPdirectivestyle,
  emphstyle=\color{solarized@green},
  frame=single,
  rulecolor=\color{solarized@base2},
  rulesepcolor=\color{solarized@base2},
  literate={~} {\customtilde}1,
  moredelim=*[directive]\ \ \#,
  moredelim=*[directive]\ \ \ \ \#
}

\lstdefinestyle{python}
{
  language=Python,
  basicstyle=\footnotesize\ttfamily,
  basewidth={0.53em,0.44em},
  numbers=none,
  tabsize=2,
  breaklines=true,
  escapeinside={@}{@},
  showstringspaces=false,
  numberstyle=\tiny\color{solarized@base01},
  keywordstyle=\color{blue},
  stringstyle=\color{orange}\ttfamily,
  identifierstyle=\color{darkred},
  commentstyle=\color{purple},
  emphstyle=\color{green},
  frame=single,
  rulecolor=\color{solarized@base2},
  rulesepcolor=\color{solarized@base2},
  literate = {~}{\customtilde}1
             {\ as\ }{{\color{blue}\ as\ \color{black}}}3
}

\lstdefinestyle{fortran}
{
  language=Fortran,
  basicstyle=\footnotesize\ttfamily,
  basewidth={0.53em,0.44em},
  numbers=none,
  tabsize=2,
  breaklines=true,
  escapeinside={@}{@},
  showstringspaces=false,
  numberstyle=\tiny\color{solarized@base01},
  keywordstyle=\color{blue},
  stringstyle=\color{orange}\ttfamily,
  identifierstyle=\color{Periwinkle},
  commentstyle=\color{purple},
  emphstyle=\color{green},
  morekeywords={and, or, true, false},
  frame=single,
  rulecolor=\color{solarized@base2},
  rulesepcolor=\color{solarized@base2},
  literate={~}{\customtilde}1
}

\lstdefinestyle{terminal}
{
  language=bash,
  basicstyle=\termplainstyle,
  numbers=none,
  tabsize=2,
  breaklines=true,
  escapeinside={@}{@},
  frame=single,
  showstringspaces=false,
  numberstyle=\tiny\color{solarized@base01},
  keywordstyle=\color{solarized@orange},
  stringstyle=\color{solarized@red}\ttfamily,
  identifierstyle=\color{black},
  commentstyle=\color{solarized@violet},
  emphstyle=\color{solarized@green},
  frame=single,
  rulecolor=\color{solarized@base2},
  rulesepcolor=\color{solarized@base2},
  morekeywords={gambit, cmake, make, mkdir},
  deletekeywords={test},
  literate = {\ gambit}{{\ }{\color{black}}gambit}7
             {/gambit}{{/}{\color{black}}gambit}6
             {gambit/}{{\color{black}}gambit{/}}6
             {/include}{{/}{\color{black}}include}8
             {cmake/}{{\color{black}}cmake/}6
             {.cmake}{{.}{\color{black}}cmake}6
             {~}{\customtilde}1
}

\lstdefinestyle{terminalalt}
{
  language=bash,
  basicstyle=\footnotesize\ttfamily,
  numbers=none,
  tabsize=2,
  breaklines=true,
  escapeinside={*@}{@*},
  frame=single,
  showstringspaces=false,
  numberstyle=\tiny\color{solarized@base01},
  keywordstyle=\color{solarized@orange},
  stringstyle=\color{solarized@red}\ttfamily,
  identifierstyle=\color{black},
  commentstyle=\color{solarized@violet},
  emphstyle=\color{solarized@green},
  frame=single,
  rulecolor=\color{solarized@base2},
  rulesepcolor=\color{solarized@base2},
  morekeywords={gambit, cmake, make, mkdir},
  deletekeywords={test},
  literate = {\ gambit}{{\ }{\color{black}}gambit}7
             {/gambit}{{/}{\color{black}}gambit}6
             {gambit/}{{\color{black}}gambit{/}}6
             {/include}{{/}{\color{black}}include}8
             {cmake/}{{\color{black}}cmake/}6
             {.cmake}{{.}{\color{black}}cmake}6
             {~}{\customtilde}1
}

\lstdefinestyle{text}
{
  language={},
  basicstyle=\footnotesize\ttfamily,
  identifierstyle=\color{black},
  numbers=none,
  tabsize=2,
  breaklines=true,
  escapeinside={*@}{@*},
  showstringspaces=false,
  frame=single,
  rulecolor=\color{solarized@base2},
  rulesepcolor=\color{solarized@base2},
  literate={~}{\customtilde}1
}

\lstdefinestyle{yaml}
{
  language=bash,
  escapeinside={@}{@},
  keywords={true,false,null},
  otherkeywords={},
  keywordstyle=\color{solarized@base0}\bfseries,
  basicstyle=\footnotesize\color{black}\ttfamily,
  identifierstyle=\YAMLkeystyle,
  sensitive=false,
  commentstyle=\color{solarized@orange}\ttfamily,
  morecomment=[l]{\#},
  morecomment=[s]{/*}{*/},
  stringstyle=\YAMLstringstyle\ttfamily,
  moredelim=**[s][\YAMLkeystyle]{,}{:},   % switch to value style at : but back to key style at,
  moredelim=**[l][\YAMLvaluestyle]{:},    % switch to value style at :
  morestring=[b]',
  morestring=[b]",
  literate =    {---}{{\ProcessThreeDashes}}3
                {>}{{\textcolor{solarized@red}\textgreater}}1
                {|}{{\textcolor{solarized@red}\textbar}}1
                {\ -\ }{{\mdseries\color{black}\ -\ \negmedspace}}3
                {\}}{{{\color{black} \}}}}1
                {\{}{{{\color{black} \{}}}1
                {[}{{{\color{black} [}}}1
                {]}{{{\color{black} ]}}}1
                {~}{\customtilde}1,
  breakindent=0pt,
  breakatwhitespace,
  columns=fullflexible
}

\lstdefinestyle{mathematica}
{
  language={Mathematica},
  basicstyle=\footnotesize\ttfamily,
  basewidth={0.53em,0.44em},
  numbers=none,
  tabsize=2,
  breaklines=true,
  escapeinside={@}{@},
  numberstyle=\tiny\color{black},
  showstringspaces=false,
  numberstyle=\tiny\color{solarized@base01},
  keywordstyle=\color{solarized@orange},
  stringstyle=\color{solarized@red}\ttfamily,
  identifierstyle=\color{solarized@orange}\ttfamily,
  commentstyle=\color{solarized@gray}\ttfamily,
  directivestyle=\color{solarized@orange}\ttfamily,
  emphstyle=\color{solarized@green},
  frame=single,
  rulecolor=\color{solarized@base2},
  rulesepcolor=\color{solarized@base2},
  literate={~} {\customtilde}1,
  moredelim=*[directive]\ \ \#,
  moredelim=*[directive]\ \ \ \ \#,
  mathescape=true
}

\newcommand{\doublecross}[2]{\hyperref[#2]{\textbf{#1}}}
\newcommand{\doublecrosssf}[2]{\hyperref[#2]{\textbf{\textsf{#1}}}}

\newcommand{\startglossary}{\section{Glossary}\label{glossary}Here we explain some terms that have specific technical definitions in \GB.\begin{description}}
\newcommand{\finishglossary}{\end{description}}

\newcommand{\bcode}{\begin{lstlisting}}
\newcommand{\ecode}{\end{lstlisting}}

\newcommand{\gambit}{\textsf{GAMBIT}\xspace}

\newcommand{\darkbit}{\textsf{DarkBit}\xspace}

\newcommand{\flavbit}{\textsf{FlavBit}\xspace}

\newcommand{\decaybit}{\textsf{DecayBit}\xspace}
\newcommand{\precisionbit}{\textsf{PrecisionBit}\xspace}

\newcommand{\neutrinobit}{\textsf{NeutrinoBit}\xspace}

\newcommand{\GB}{\gambit}

\newcommand\pippi{\textsf{pippi}\xspace}

\newcommand\diver{\textsf{Diver}\xspace}

\newcommand\YAML{\textsf{YAML}\xspace}

\newcommand\beq{\begin{equation}}
\newcommand\eeq{\end{equation}}

\renewcommand{\url}[1]{\href{#1}{#1}}

\usepackage{slashed}
\usepackage{todonotes}
\usepackage{upgreek}
\newcommand{\M}{\bar{M}}
\newcommand{\Muu}{\upmu}
\newcommand{\Mu}{\mu}
\newcommand{\Epsilon}{\upepsilon}

\newcommand{\louvainlaneuve}{Centre for Cosmology, Particle Physics and Phenomenology, Université catholique de Louvain, Louvain-la-Neuve B-1348, Belgium}

\begin{document}

\preprintnumber{}

\title{A Frequentist Analysis of Three Right-Handed Neutrinos with GAMBIT}

\author
{
Marcin Chrzaszcz\thanksref{inst:a,inst:b} \and
Marco Drewes\thanksref{inst:c} \and
Tom\'as E. Gonzalo\thanksref{inst:h,e1} \and 
Julia Harz\thanksref{inst:e} \and
Suraj Krishnamurthy\thanksref{inst:f,e2} \and
Christoph Weniger\thanksref{inst:f} 
%\and
%Aaron Vincent\thanksref{inst:g}
}

\institute{%
  \cern\label{inst:a} \and
  \krakow\label{inst:b} \and
  \louvainlaneuve\label{inst:c} \and
  \monash\label{inst:h} \and
  \munich\label{inst:e} \and
  \grappa\label{inst:f}
}

\thankstext{e2}{s.krishnamurthy@uva.nl}
\thankstext{e1}{tomas.gonzalo@monash.edu}

\titlerunning{Right-handed neutrinos}
\authorrunning{Authors}

\date{Received: date / Accepted: date}

\maketitle

\begin{abstract}
The extension of the Standard Model by right-handed neutrinos can not only explain the active neutrino masses via the seesaw mechanism, it is also able solve a number of long standing problems in cosmology. Especially, masses below the TeV scale are of particular interest as they can lead to a plethora of signatures in experimental searches. We present the first full frequentist analysis of the extension of the Standard Model by three right-handed neutrinos, with masses between 60 MeV and 500 GeV, using the Global and Modular BSM (beyond the Standard Model) Inference Tool GAMBIT. Our analysis is based on the Casas-Ibarra parametrisation and includes a large range of experimental constraints: active neutrino mixing, indirect constraints from, e.g., electroweak precision observables and lepton universality, and numerous direct searches for right-handed neutrinos. To study their overall effect, we derive combined profile likelihood results for the phenomenologically most relevant parameter projections. Furthermore, we discuss the role of (marginally) statistically preferred regions in the parameter space. Finally, we explore the flavour mixing pattern of the three right-handed neutrinos for different values of the lightest neutrino mass. Our results comprise the most comprehensive assessment of the model with three right-handed neutrinos model below the TeV scale so far, and provide a robust ground for exploring the impact of future constraints or detections.
  
\end{abstract}

\newpage

\tableofcontents

%%%%%%%%%%%%%%%%%%%%%%%%%%%%%%%%%%%%%%%%%%%%%%%%%%%%%%%
%%%%%%%%%%%%%%%%%%%%%%%%%%%%%%%%%%%%%%%%%%%%%%%%%%%%%%%
\section{Introduction} 
\label{sec:intro}

\subsection{Motivation}
The observation of neutrino flavour oscillations is one of the strongest hints for the existence of particle physics beyond the Standard Model (SM).
The oscillations imply that neutrinos have small masses, while the minimal SM predicts that they are massless. At the same time neutrinos are the only elementary fermions that are only known to exist with left handed chirality $\nu_L$.
If right handed neutrinos $\nu_R$ exist, one could immediately add a Dirac mass term $\bar{\nu_L}M_D\nu_R$ to the SM Lagrangian in analogy to all other known fermions. The fact that the $\nu_R$ have not been seen yet could easily be explained because they are "sterile", i.e., not charged under any known gauge interactions.
The same property also makes it possible for them to have a Majorana mass term $\bar{\nu_R}M_M\nu_R^c$ in addition to the Dirac mass. 
For eigenvalues of $M_M$ that are much larger than the observed light neutrino masses,
the smallness of the neutrino masses can be explained via the seesaw mechanism~\cite{Minkowski:1977sc, Mohapatra:1979ia, Mohapatra:1980yp, GellMann:1980vs, Yanagida:1980xy}. Neutrino oscillation data is, however, not sufficient to pin down the value of $M_M$, known as \emph{seesaw scale}, because it is primarily sensitive to the combination $M_D M_M^{-1} M_D^T$. The range of allowed values spans from a few eV \cite{deGouvea:2005er} up to the scale of Grand Unification \cite{Asaka:2015eda}.
For specific choices of their Majorana mass the $\nu_R$ could in addition solve a number of long standing problems in cosmology. For instance, they could explain the baryon asymmetry of our Universe via leptogenesis during the decay~\cite{Fukugita:1986hr} or production \cite{Akhmedov:1998qx,Asaka:2005pn} of the heavy neutrinos 
or provide a viable dark matter candidate~\cite{Dodelson:1993je, Shi:1998km}.
An overview of the cosmological implications of different choices of $M_M$ can e.g.~be found in Ref.~\cite{Drewes:2013gca}.

Experiments can directly search for heavy neutrinos if $M_M$ is below the TeV scale. Such searches have been performed in various different facilities, including high energy colliders and fixed target experiments.
This is the mass range we consider in the present article.
In addition, the $\nu_R$ would indirectly affect precision observables or searches for rare processes.
A summary of different existing constraints can be found in the reviews~\cite{Atre:2009rg,Boyarsky:2009ix,Drewes:2013gca,Deppisch:2015qwa,Cai:2017mow}.
For the future a wide range of different searches have been proposed, an overview can be found in Refs.~\cite{Antusch:2016ejd,Beacham:2019nyx,Alimena:2019zri}.
In order to decide about the best possible search strategy is it important to understand which parameter region is already ruled out by past experiments. This is in fact a non-trivial question because different observables are correlated in the seesaw model, and the requirement to \emph{simultaneously} respect all known experimental results imposes stronger constrains on the model parameter space than superimposing individual bounds. 
Such \emph{global constraints} can only be derived within a given model.
An important quantity in this context is the unknown number $n$ of right handed neutrino flavours. 
The minimal number that is required to explain the light neutrino oscillation data is $n=2$, which would necessarily require the lightest SM neutrino to be massless. The minimal number that is required to generate masses for all three SM neutrinos is $n=3$. This choice is also somewhat appealing in view of the fact that there are three fermion generations in the SM, and it is mandatory for anomaly freedom in many gauge extensions of the SM.
The goal of the present work is to impose global constraints on the parameter space of the model with $n=3$, based on the combination of direct, indirect and cosmological constraints summarised in section \ref{sec:obslike}.

Several authors have previously imposed global constraints on the properties of right handed neutrinos. Here we exclusively focus on models in which the right handed neutrinos can explain the light neutrino oscillation data.\footnote{The authors of Ref.~\cite{deGouvea:2015euy} considered a single heavy neutrino, but made the conservative assumption that this particle may predominantly decay into a dark sector via new interactions.}
This e.g.~excludes most sterile neutrino Dark Matter models because the feeble coupling of such particles that is required to ensure their longevity implies that its contribution to the light neutrino mass generation can be neglected  \cite{Boyarsky:2006jm}.\footnote{We refer the reader to Refs.~\cite{Adhikari:2016bei,Boyarsky:2018tvu} for recent reviews on sterile neutrino Dark Matter.}
One of the most complete studies of indirect constraints on the parameter space for $n=2$ in the last few years was presented in Ref.~\cite{Antusch:2014woa}, where multiple electroweak precision observables and flavour-violating decays were included, along with tests of lepton universality and the unitarity of the CKM matrix. Loop corrections to some of these relations were considered in Ref.~\cite{Fernandez-Martinez:2015hxa}.
The authors of \cite{Drewes:2016jae} included direct search constraints and those from big bang nucleosynthesis.
The model with $n=3$ is much less studied.
Recent analyses of indirect constraints include Refs.~\cite{Escrihuela:2015wra,Fernandez-Martinez:2016lgt},
direct search constraints and BBN have been added to this in Ref.~\cite{Drewes:2015iva}.

%\medskip
\subsection{Main improvements compared to previous studies}\label{subsec:MainImprovements}

In this paper, we present the first full frequentist analysis of the $n=3$ right-handed neutrino (RHN) extension of the SM, for a wide range of RHN masses from about 60 MeV to 500 GeV. We opted for a frequentist analysis rather than a Bayesian analysis since this is best suited to fully explore the valid parameter space while avoiding prior dependence and volume effects of the parameter space (however, we emphasize that we do not perform a full sampling-based goodness-of-fit analysis and instead resort for practical reasons to an approximate treatment of likelihood and their sampling statistics). We improve on different aspects of earlier analyses by combining all the strongest
limits exerted by experiments as well as indirect signatures in a statistically
consistent manner. Previous studies that examined the parameter space for $n=3$ either used a subset of the constraints included here~\cite{Escrihuela:2015wra,Fernandez-Martinez:2016lgt} or
used less rigorous statistical methods~\cite{Drewes:2015iva}
and focused on specific regions of the parameter space~\cite{Gorbunov:2014ypa}.
\begin{itemize}
  \item While most previous studies fixed the mixing angles and mass differences in the active neutrino sector to the best fit values as presented in \cite{Esteban:2016qun}, we take into account likelihoods for the active neutrino observables.
  \item Electroweak observables require precise calculations for its comparison with the extremely accurate measurements. We therefore use the calculation of the SM prediction for $\sin \theta_w^{eff}$ up to two-loop order \cite{Ferroglia:2012ir}.
  \item Most studies of lepton flavour violation in neutrino models focus exclusively on the most constraining processes, such as $\mu \to e \gamma$ and $\mu \to eee$~\cite{Antusch:2014woa,Drewes:2015iva}. In this work we include all lepton flavour violating processes, in particular all leptonic $\tau$ decays, for which we use the most recent average of experimental results provided by HFLAV~\cite{Amhis:2016xyh}, as well as $\mu -e$ conversion in nuclei (Pb, Au and Ti).
  \item For neutrinoless double-beta decay, in comparison with~\cite{Drewes:2015iva}, we opt to carry out our analysis conservatively; in addition, the upper limit on the effective Majorana mass and hence the mixing is encoded in the form of a (one-sided) Gaussian likelihood, not as a strict cut.
  \item Lepton universality tests are often centered on leptonic decays of mesons, $K$ and $\pi$, $\tau$-leptons and $W$-bosons~\cite{Antusch:2014woa}. We supplement these tests of universality with the recently observed semileptonic decays of B-mesons~\cite{Aaij:2014ora, Aaij:2015yra,Aaij:2017vbb}.
  \item We improve the treatment of CKM unitarity with respect to the discussion in Ref.~\cite{Drewes:2015iva}.
  \item Concerning direct searches, previous studies have used only a subset of the experiments considered here~\cite{Ruchayskiy:2011aa,deGouvea:2015euy}, or chose to place a hard cut at the upper limits presented in the individual papers~\cite{Drewes:2015iva,Drewes:2016jae}. We implement the strongest constraints over the mass range as likelihoods. The statistical combination of these likelihoods also leads to more accurate profile likelihood contours in comparison to simply overlaying individual limits.
  \item We study in detail the flavour mixing pattern of the three RHN, for different values of the lightest neutrino mass. We discuss the limit where the lightest neutrino is massless and the connection to the $n=2$ case.
\end{itemize}

We use here the open-source software package
\GB~\cite{gambit}.  It includes an interface to Diver~\cite{ScannerBit}, a differential evolution-based scanner that provides
efficient sampling performance for frequentist scans.

\medskip

This paper is organised as follows. In section~\ref{sec:RHN}, the model, parametrisation used and essential quantities are defined. All the observables and experiments that are considered are subsequently discussed in detail in section~\ref{sec:obslike}. Our scanning strategy, parameter ranges and applied priors are mentioned in section~\ref{sec:scanning}. The results are presented in section~\ref{sec:results} and we discuss the implications of the combined constraints for future searches in section~\ref{sec:concoutlook}. In Appendix~\ref{app:gambit} we comment on the details of the implementation in \GB, 
in Appendix~\ref{app:equations} we explicitly give the expressions for the different observables,
in Appendix~\ref{FakeSymmetryPoints} we provide details on how we interpret our results in view of the criterion of technical naturalness, and in Appendix~\ref{app:partial} we show the different partial likelihoods.

\section{Right-handed neutrino physics}
\label{sec:RHN}

\subsection{Basic definitions}

The addition of three RHNs to the particle content of the Standard Model introduces in total 18 new parameters. In this section we summarise basic relations in the seesaw model and define our notation, following
Ref.~\cite{Drewes:2015iva}.

The most general renormalisable Lagrangian that can be constructed from SM fields and the $\nu_R$ has the following form:
\begin{align}\label{Lagrangian}
\notag\mathcal{L} &= \mathcal{L}_{SM}+i\overline{\nu}_{R}\slashed{\partial}\nu_R-\bar{\ell_L}F\nu_R\tilde{\Phi}-\tilde{\Phi}^{\dagger}\bar{\nu_R}F^{\dagger}\ell_L\\
&-\frac{1}{2}\left(\bar{\nu_R^c}M_M\nu_R+\bar{\nu_R}M_M^{\dagger}\nu_R^c\right)\,.
\end{align}
Hereby, $\ell_L=(\nu_L,e_L)^T$ indicate the left-handed leptons\footnote{Throughout this article we use four component spinor notation, where the chiral spinors $\nu_R$ and $\ell_L$ have only two non-zero components ($P_R\nu_R=\nu_R$ and $P_L\ell_L=\ell_L$). As a result, no explicit chiral projectors are necessary in the weak interaction term (\ref{WeakWW}).} 
of the SM and $\Phi$ is the Higgs doublet with $\tilde{\Phi}=\epsilon\Phi^{\ast}$ and $\epsilon$ being the Levi-Civita tensor. $M_M$ is the Majorana mass matrix for $\nu_R$ and $F$ is the Yukawa coupling matrix. We work in a flavour basis where $M_M=\text{diag}(M_1,M_2,M_3)$.

After electroweak symmetry breaking (EWSB), the complete neutrino mass term reads
\begin{equation}
\frac{1}{2}(\bar{\nu_L} \bar{\nu_R^c})\mathcal{M}
\begin{pmatrix}
\nu_L^c
\\
\nu_R
\end{pmatrix}\;,
\end{equation}
with
\begin{equation}
\mathcal{M}=\begin{pmatrix}{\delta}m_{\nu}^{1loop} & M_D \\ M_D^T & M_M + {\delta}M_N^{1loop} \end{pmatrix}\;,\label{FullNeutrinoMass}
\end{equation}
where $M_D=Fv$, $v$ being the Higgs vacuum expectation value ($v=174$ GeV in the ground state).
We include the one loop corrections
$\delta m_{\nu}^{1loop}$ 
and 
${\delta}M_N^{1loop}$
as we aim for performing an analysis to be consistent at second order in the Yukawa couplings $F$.
The mass matrix (\ref{FullNeutrinoMass}) can be diagonalised by a matrix of the form \cite{Fernandez-Martinez:2015hxa}
\begin{equation}
\mathcal{U}= \begin{pmatrix} \cos(\theta) & \sin(\theta) \\ -\sin(\theta^\dagger) & \cos(\theta^\dagger)  \end{pmatrix}
\begin{pmatrix} U_{\nu} & \\ & U_N^{\ast} \end{pmatrix}
\end{equation}
with 
\begin{align}
\cos(\theta)&=\sum_{n=0}^\infty \frac{(-\theta\theta^\dagger)^n}{(2n)!}\\
 \sin(\theta)&=\sum_{n=0}^\infty \frac{(-\theta\theta^\dagger)^n\theta}{(2n+1)!}.
 \end{align}
Hereby, $\theta$ indicates the matrix that mediates the mixing between the active neutrinos $\nu_L$ and the sterile neutrinos $\nu_R$.
We can generally write 
\begin{equation}\label{DiagonaliseTheMass}
\mathcal{U}^{\dagger}\mathcal{M}\mathcal{U}^{\ast}=\begin{pmatrix}m_{\nu}^{\rm diag} & \\ & M_N^{\rm diag} \end{pmatrix}
\end{equation}
with 
\begin{align}  
M_N^{\rm diag}&=U_N^T M_N U_N=\text{diag}(M_1,M_2,M_3)\label{MNdiagDef}\\
m_{\nu}^{\rm diag}&= U_{\nu}^{\dagger}m_{\nu}U_{\nu}^{\ast}=\text{diag}(m_1,m_2,m_3). 
\end{align}
The additional complex conjugation of $U_N$ ensures that the relation among mass and flavour eigenstates will be analogous for left-handed neutrinos (LHNs) and RHNs within the notation.
In the second relation in eq.~(\ref{MNdiagDef}) we have neglected the difference between the eigenvalues of $M_M$ and $M_N$, which is of second order in $\theta$. This is justified for the present purpose because of the experimental constraints on the magnitude of the elements $\theta_{\alpha I}$, which we discuss further below. 

\subsection{The seesaw limit}

The limit of small $\theta_{\alpha I}$ is usually referred to as the \emph{seesaw limit}, it corresponds to $M_D\ll M_M$ (in terms of eigenvalues).
It allows the approximation
\begin{equation}
\theta = M_D M_M^{-1} = v F M_M^{-1}
\end{equation}
and
\begin{equation}
\mathcal{U}=\Bigg[ \begin{pmatrix} \mathbb{I}-\frac{1}{2}\theta\theta^{\dagger} & \theta \\ -\theta^{\dagger} & \mathbb{I}-\frac{1}{2}\theta^{\dagger}\theta \end{pmatrix} + \mathcal{O}(\theta^3) \Bigg]\begin{pmatrix} U_{\nu} & \\ & U_N^{\ast} \end{pmatrix},
\end{equation}
leading to
\begin{equation}\label{mnutreeandloop}
m_{\nu}=m_{\nu}^{\rm tree}+\delta m_{\nu}^{1loop}
\end{equation}
with
\begin{align}
\notag m_{\nu}^{\rm tree}&=-M_D M_M^{-1}M_D^T=-{\theta}M_M\theta^T=-v^2FM_M^{-1}F^T
\end{align}
and
\begin{equation}
M_N = M_M+\frac{1}{2}\left( \theta^{\dagger}\theta M_M+M_M^T\theta^T\theta^{\ast} \right) + \delta M_{N}^{1loop}\,.\label{MNDefinition}
\end{equation}
The loop correction to the light neutrino mixing matrix is given by \cite{Pilaftsis:1991ug}:
\begin{equation}
\left( \delta m_{\nu}^{1loop} \right)_{\alpha\beta}=\sum_I F_{\alpha I}M_IF_{I\beta}^Tl(M_I)\,,
\end{equation}
where $l(M_I)$ is a loop function given by
\begin{multline}
  l(M_I) = \frac{1}{{(4\pi})^2}\left[\left(\frac{3\text{ln}[(M_I/m_Z)^2]}{(M_I/m_Z)^2 - 1}\right) \right.\\
  \left.+ \left(\frac{\text{ln}[(M_I/m_H)^2]}{(M_I/m_H)^2 - 1}\right)\right]\,.
\end{multline}
The light and heavy neutrino mass eigenstates are described by the flavour vectors
\begin{equation}  \nu=V_{\nu}^{\dagger}\nu_L-U_{\nu}^{\dagger}\theta \nu_R^c+V_{\nu}^T\nu_L^c-U_{\nu}^T\theta^{\ast} \nu_R
\end{equation}
and
\begin{equation}
  N=V_N^{\dagger}\nu_R+\Theta^T \nu_L^c+V_N^T\nu_R^c+\Theta^{\dagger} \nu_L,
\end{equation}
respectively.
We can define the matrices $V_{\nu}$ and $V_N$ that represent the mixing between mass and interaction eigenstates in the respective sectors as
\begin{align}
\label{defV}
  V_{\nu}&\equiv \left( \mathbb{I}-\frac{1}{2}\theta\theta^{\dagger} \right) U_{\nu}\\
  V_N&\equiv \left( \mathbb{I}-\frac{1}{2}\theta^T\theta^{\ast} \right) U_N,
\end{align}
while mixing between the two sectors is encoded in the matrix
\begin{equation}
\Theta=\theta U_N^{\ast}\,.
\label{Thetadef}
\end{equation}
This quantity is of primary interest 
because it controls the interactions of the heavy neutrinos
with the physical Higgs field $h$ and the gauge bosons $W$ and $Z$,
\begin{align}
  &- \frac{g}{\sqrt{2}}\overline{N}_I \Theta^\dagger_{I \alpha}\gamma^\mu e_{L \alpha} W^+_\mu
  - \frac{g}{\sqrt{2}}\overline{e_{L \alpha}}\gamma^\mu \Theta_{\alpha I} N_I W^-_\mu\notag\\
 &- \frac{g}{2\cos\theta_W}\overline{N_I} \Theta^\dagger_{I \alpha}\gamma^\mu \nu_{L \alpha} Z_\mu
  - \frac{g}{2\cos\theta_W}\overline{\nu_{L \alpha}}\gamma^\mu \Theta_{\alpha I} N_i Z_\mu\notag\\
 &- \frac{g}{\sqrt{2}}\frac{M_I}{m_W}\Theta_{\alpha i} h \overline{\nu_{L \alpha}}N_I
  - \frac{g}{\sqrt{2}}\frac{M_I}{m_W}\Theta^\dagger_{I \alpha} h \overline{N_I}\nu_{L \alpha}
\ \label{WeakWW}
\end{align}
Here $g$ is the weak gauge coupling constant and $\theta_W$ the Weinberg angle.
For convenience, we introduce the notation
\begin{align}
U_{\alpha I}^2 &\equiv |\Theta_{\alpha I}|^2
\label{UaI}\\
U_I^2 &\equiv U_{eI}^2+U_{\mu I}^2+U_{\tau I}^2
\label{UI}\\
U_{\alpha}^2 &\equiv \sum_I U_{\alpha I}^2\,.
\label{Ua}
\end{align}
From the relations \eqref{FullNeutrinoMass} and \eqref{DiagonaliseTheMass} it is straightforward to derive the relation 
\begin{eqnarray}
({\delta}m_{\nu}^{1loop})_{\alpha\alpha}&=&\sum_i m_i (V_\nu)_{\alpha i}^2 + \sum_I M_I \Theta_{\alpha I}^2.
\end{eqnarray}

\subsection{The role of the matrix $U_N$}

In our numerical scan we approximate $U_N$ by unity.\footnote{Note that the approximation $U_N=\mathbb{I}$ also allows to neglect $\delta M_{N}^{1loop}$ because it only amounts to a change in the matrix $U_N$ \cite{Drewes:2019mhg}.}
For generic parameter choices this can be justified because we work in a basis where $M_N$ is diagonal, and the physical mass matrix \eqref{MNDefinition} is also diagonal up to corrections of second order in $\theta$. 
These corrections can lead to a large deviation of $U_N$ from unity only if the eigenvalues of $M_M$ are quasi-degenerate, so that the $\mathcal{O}[\theta^2]$ terms in the matrix \eqref{MNDefinition} are relevant.

If a degeneracy between only two of the RHNs is caused by a symmetry, cf. sec.~\ref{sec:symmetries}, then it can be shown that the effect of  $U_N$ on the $U_{\alpha I}^2$ is small even if individual entries of $U_N$ are larger than the $U_{\alpha I}^2$ \cite{Drewes:2019mhg}. 
This means that the production cross sections for heavy neutrinos are not affected. 
However, the branching ratio between lepton number violating (LNV) and lepton number conserving heavy neutrino decays is affected by $U_N$ \cite{Drewes:2019byd}.
This has no big effect on our scan because constraints from searches for LNV are sub-dominant in almost the entire mass range that we consider, but it may have important implications for future searches. 

$U_N$ can have a big impact on the individual mixings $U_{\alpha I}^2$ of each heavy neutrino if all three Majorana masses are degenerate. 
This can be accommodated in technically natural scenarios discussed in the following section \ref{sec:symmetries}, cf. in particular footnote \ref{MassCommunism}.
The practical impact on experimental searches is, however, limited because most experiments are not able to kinematically resolve small mass splittings and therefore only probe $U_\alpha^2$ in this regime (rather than the couplings $U_{\alpha I}^2$ of individual heavy neutrino flavours).
Also in this case observables that are sensitive to LNV are the only ones that are likely to be affected.

Finally, if the degeneracy between the heavy neutrino masses is accidental, then the proof in Ref.~\cite{Drewes:2019mhg} does not apply, and $U_N$ can have a significant effect on the $U_{\alpha I}^2$ even if only two heavy neutrinos have degenerate masses. Our results contain a significant number of points of this kind because we performed several scans with "agnostic" parameter ranges that do not suppress fine-tuned points, cf.~ table \ref{tab:scanpars}. 
However, the fact that experiments are unlikely to resolve the individual resonances in this case implies that they are only sensitive to the quantities $U_a^2$, where the summation is to be taken over the mass degenerate heavy neutrino flavours only. 
As in the previous two cases, the effect of $U_N$ on the total production rate is minor because the matrix mainly re-distributes coupling between the mass degenerate states. 
The main affect would again be on LNV observables. 

In summary, if any heavy neutrinos are discovered in the future, a comparison between the branching ratios of lepton number violating and lepton number conserving decays will give important insight into the mechanism of neutrino mass generation and will be crucial to identify any underlying symmetries.

\subsection{Casas-Ibarra parametrisation}
\label{sec:CI}

In the current work, we  use the Casas-Ibarra (C-I) parametrisation \cite{Casas:2001sr}, generalised
to include the 1-loop correction to the left-handed neutrino mass matrix \cite{Lopez-Pavon:2015cga}.
This provides a simple way to impose constraints from light neutrino oscillation data in our scan.
This parametrisation is based on the observation that $m_\nu$ in eq.~(\ref{mnutreeandloop}) can be expressed as
\begin{equation}\label{SeesawFull}
m_\nu=-\theta\tilde{M}\theta^T
\end{equation}
with
\begin{equation}
\tilde{M} = \big[ 1 - \frac{1}{v^2} M_M M_N^{\rm diag} l(M_N^{\rm diag})
\big] M_M\,.  
\end{equation}
Since the loop function is smooth we can neglect the difference in the eigenvalues of $M_M$ and $M_N$, 
\begin{equation}
\label{eq:Mtildediag}
\tilde{M}_{IJ}\simeq
\tilde{M}^{\rm diag}_{IJ} =
M_I\delta_{IJ}\left(1 - \frac{M_I^2}{v^2}l(M_I)\right).
\end{equation}
In this scheme the sterile
neutrino mixing matrix, i.e. the matrix encoding the mixing among LHNs and RHNs~\eqref{Thetadef} can be written as 
\begin{align}
\Theta = iU_{\nu}\sqrt{m_{\nu}^\text{diag}}\mathcal{R}\sqrt{\tilde{M}^\text{diag}}^{-1}
\;,
\label{CItheta}
\end{align}
where $U_{\nu}$ is the PMNS matrix introduced above, $m_{\nu}^\text{diag}$ is the diagonalised, one-loop-corrected LHN mass matrix and $\tilde{M}^\text{diag}$ is the analogous RHN mass matrix, given by \eqref{eq:Mtildediag}. 
Furthermore, $\mathcal{R}$ is a complex, orthogonal matrix that is parametrised by complex angles $\omega_{ij}$
\begin{align}
\mathcal{R} = \mathcal{R}^{23}\mathcal{R}^{13}\mathcal{R}^{12}\;,
\label{Rorder}
\end{align}
where $\mathcal{R}^{ij}$ has the non-zero elements
\begin{align}
\mathcal{R}^{ij}_{ii} &= \mathcal{R}^{ij}_{jj} = \cos\omega_{ij}, \\
\mathcal{R}^{ij}_{ij} &= -\mathcal{R}^{ij}_{ji} = \sin\omega_{ij}, \\
\mathcal{R}^{ij}_{kk} &= 1; k \neq i,j\;.
\end{align}

Since we work in the flavour basis in which the Yukawa couplings of the charged
leptons are diagonal, $U_{\nu}$ can be parametrised as
\begin{align}\label{UnuParameterisation}
  U_{\nu} = V^{23}U_{\delta}V^{13}U_{-\delta}V^{12}\mathrm{diag}(e^{i\alpha_1/2},e^{i\alpha_2/2},1)\;,
\end{align}
where $U_{\pm\delta} = \mathrm{diag}(e^{{\mp}i\delta/2},1,e^{{\pm}i\delta/2})$
and $V^{ij}$, parametrised by the LHN mixing angles $\theta_{ij}$, has non-zero
elements analogous to $\mathcal{R}$.  Furthermore, $\alpha_1$, $\alpha_2$ and $\delta$ are
CP-violating phases.

The C-I parametrisation scheme generates by construction Yukawa couplings and mixing angles $\Theta$ that are consistent with light neutrino oscillation data up to second order in $\theta$. 
This has two disadvantages.
First, one may find it unsatisfactory that we treat light neutrino oscillation data differently from other constraints.
Second, the C-I is a ``bottom up'' parametrisation. There is usually no simple relation between the C-I parameters and parameters that may be well-motivated from a model building viewpoint, 
and any theory-motivated prior on the RHNs' mixings and masses
would acquire a rather convoluted form in the C-I parametrisation. 
In particular, there is no simple way to distinguish ``natural'' from ``fine tuned'' parameter choices.
Hence, we refrain from performing Bayesian scans in the current work, and instead concentrate on a likelihood-based frequentist treatment. In view of the high dimensionality of the parameter space and the complicated functional form of the different constraints, the disadvantages of the C-I parametrisation are, however, compensated for by the numerical advantage that one gains.

\subsection{The symmetry protected scenario}
\label{sec:symmetries}

The smallness of the light neutrino masses $m_i$ can be explained in different ways by the \emph{seesaw relation}~\eqref{SeesawFull}. One possibility is that the $N_I$ are very heavy, i.e., $M_I\gg v$, in which case the smallness of $m_i$ is due to the smallness of the ratio $v/M_I$. This choice for the mass scale(s) $M_I$ is well-motivated by Grand Unified Theories,\footnote{See \cite{Fukugita:2003en,Croon:2019kpe} for a review on neutrino masses in the context of Grand Unified Theories.} but raises the question of radiative corrections to the Higgs potential from the Yukawa couplings of the RHNs \cite{Vissani:1997ys}.

This ``hierarchy problem'' can be avoided in low scale seesaw scenarios. Low values of $M_I$ are natural because in the limit $M_I \to 0$ the $B-L$ symmetry in the SM is restored. In this case, however, the smallness of $m_i$ can no longer be explained efficiently by the suppression of $v/M_I$, as it typically requires couplings
\begin{equation}\label{NaiveSeesaw}
\Theta_{\alpha I} \simeq i (U_\nu)_{\alpha I}\sqrt{\frac{m_i}{M_I}} \ , \ 
F_{\alpha I} \simeq i (U_\nu)_{\alpha I}\frac{\sqrt{m_i M_I}}{v}
\end{equation}
that are very small, in particular for \textit{seesaw scales} as low as 100 MeV.

Such small values for fundamental parameters are considered 'unnatural' by many theorists~\cite{Giudice:2008bi}, though some possible explanations have been proposed~\cite{Froggatt:1978nt}. However, this estimate relies on the underlying assumption that there are no cancellations (accidental or otherwise) in the seesaw relation~\eqref{SeesawFull}, which would allow for much larger $U_{\alpha I}^2=|\Theta_{\alpha I}|^2$ than the naive estimate \eqref{NaiveSeesaw} suggests while keeping the eigenvalues $m_i^2$ of $m_\nu^\dagger m_\nu$ small.

Hence, a \textit{technically natural} \cite{tHooft:1979rat} way to obtain small neutrino masses $m_i$ can be realised if the Lagrangian (\ref{Lagrangian}) approximately respects a $B-\bar{L}$ symmetry~\cite{Shaposhnikov:2006nn, Kersten:2007vk} (cf.~also \cite{Gluza:2002vs}), where $\bar{L}$ is a generalised lepton number under which combinations of the $\nu_{Ri}$ are charged. Such $B - \bar{L}$ symmetry is exact if the Yukawa coupling and mass matrix take the form~\cite{Moffat:2017feq}
\begin{eqnarray}\label{SymmExact}
 M_M^{B-\bar{L}} =\begin{pmatrix} \bar M & 0 & 0 \\ 0 & \bar M   & 0  \\ 0 &0 & M'    \end{pmatrix} \, \ 
 F^{B-\bar{L}} =\begin{pmatrix}\ F_e \ & \ iF_e & 0 \\ \ F_\mu \ & \ iF_\mu &  0  \\  F_\tau \ & \ iF_\tau  & 0 
\end{pmatrix}\;,
\end{eqnarray}
in which case the light neutrinos are exactly massless $m_i = 0$. In order to generate non-zero light neutrino masses this symmetry has to be slightly broken, i.e.,
\begin{eqnarray}\label{GeneralParam}
M_M =  M_M^{B-\bar{L}} (1 + \Muu)
\ , \ 
F =  F^{B-\bar{L}}(1 + \Epsilon),
\end{eqnarray}
where the entries of the matrices $\Muu$ and $\Epsilon$ are small symmetry breaking parameters.

If the symmetry is not exact 
$M_M$ can have off-diagonal elements, see for example Ref.~\cite{Abada:2007ux}.
Throughout this work we use a basis in which $M_M$ is diagonal.
The diagonalisation affects the form of the Yukawa matrix $F$, but as long as the off diagonal elements of $\Muu$ are small, this only leads to a small modification of the flavour structure.
For the following discussion we will therefore adapt the simpler form \cite{Abada:2018oly}\footnote{\label{MassCommunism}
An important exception is the case $\Mu\ll1$, $\M'\simeq \M$. In that situation even small off-diagonal elements $\Muu_{ij}$ can lead to a comparably large misalignment between the basis in which $F$ has the form \eqref{SymmProtectParam} and the heavy neutrino mass basis, which means which that all heavy neutrinos have unsuppressed Yukawa couplings $\sim F_a$ in spite of the fact that $\epsilon'_a\ll 1$, cf. ref.~\cite{Drewes:2019byd} for a discussion.
However, in this case all three mass eigenstate $N_I$ have approximately the same mass $\M$ and cannot be distinguished kinematically.
In this case  the experimentally relevant mixing is $U_a^2$, the magnitude of which is controlled by the large entries $F_a$. 
Heavy neutrino oscillations in the detector \cite{Boyanovsky:2014una,Cvetic:2015ura,Anamiati:2016uxp,Dib:2016wge,Das:2017hmg, Antusch:2017ebe,Antusch:2017pkq, Cvetic:2018elt,Hernandez:2018cgc,Cvetic:2019rms} could provide an indirect way to access the  small mass splitting and phenomenologically study this specific case.
}

\begin{eqnarray}\label{SymmProtectParam}
\notag M_M&=&\begin{pmatrix} \bar M(1 - \mu) & 0 & 0 \\ 0 & \bar M (1 + \mu)  &0  \\ 0 &0 & M'    \end{pmatrix}  , \\
F&=&\begin{pmatrix}\ F_e(1 + \epsilon_e) \ & \ iF_e(1 - \epsilon_e) & F_e \epsilon'_e \\ \ F_\mu(1 + \epsilon_\mu) \ & \ iF_\mu(1 - \epsilon_\mu) & F_\mu \epsilon'_\mu  \\  F_\tau(1 + \epsilon_\tau) \ & \ iF_\tau (1 - \epsilon_\tau) & F_\tau \epsilon'_\tau \end{pmatrix},
\end{eqnarray}
with $\epsilon_\alpha', \epsilon_\alpha, \mu,\ll1$ being small symmetry breaking parameters and $F_\alpha$ being of the order of one. This means that one heavy neutrino practically decouples while the other two approximately form a Dirac spinor with mass $\bar{M}$.

In this \emph{symmetry protected scenario} there is no upper limit on $U_{\alpha I}^2$ from neutrino oscillation data. In the mass range considered here the upper limit comes from the experimental constraints, while for larger masses there is a theoretical bound $U_{\alpha I}^2<4\pi(n-1)(v/\bar{M})^2$ from the requirement that the Yukawa couplings remain perturbative \cite{Asaka:2015eda}. This provides a theoretical motivation for a low scale seesaw with experimentally accessible mixings $U_{\alpha I}^2$. Specific examples that motivate this limit include ``inverse seesaw'' ~\cite{Wyler:1982dd,Mohapatra:1986aw,Mohapatra:1986bd,Bernabeu:1987gr}, ``linear seesaw''~\cite{Akhmedov:1995ip,Akhmedov:1995vm}, scale invariant~\cite{Khoze:2013oga} and some technicolour-type models~\cite{Appelquist:2002me,Appelquist:2003uu} and also the  $\nu$MSM~\cite{Asaka:2005pn,Shaposhnikov:2006nn}.

\subsection{Connection to the model with $n=2$}\label{sec:n2model}

The parametrisation \eqref{SymmProtectParam} suggests that the $B-\bar{L}$ symmetric limit for the model with $n=3$ should contain the model with $n=2$, as the third heavy neutrino decouples for $\epsilon'_a\to 0$. 
This is, for example, observed in the $\nu$MSM.
However, some care is required when taking this limit if one wants to be consistent with neutrino oscillation data. 

First, it is clear that not all seven symmetry breaking parameters $\epsilon_a, \epsilon'_a, \mu$ can be set to zero because this would give exactly massless light neutrinos. Which of these parameters are non-zero and how small they are with respect to each other depends on the way how the symmetry is broken and thus on the particle physics model in which the Lagrangian \eqref{Lagrangian} is embedded. It is not possible to make a model independent statement about the relative size of the $\epsilon'_a$ in relation to other model parameters. 

Second, the parametrisation \eqref{SymmProtectParam} is not the most general one: If we allow for small off diagonal elements in the general form \eqref{GeneralParam}, then all three heavy neutrinos can have unsuppressed interactions if $\M'\simeq\M$, cf. footnote \ref{MassCommunism}.
Hence, if $\M'$ is degenerate with $\M$, one cannot expect to recover the $n=2$ model even if $\epsilon_\alpha \ll 1$. 

Finally, as discussed in more detail in Appendix \ref{FakeSymmetryPoints}, there are Casas-Ibarra parameter choices that yield small values of $m_{\nu_0}$, but correspond to highly fine-tuned scenarios where this smallness is due to accidental cancellations. These solutions can imitate the symmetry protected scenario and can also circumvent the seesaw upper limit and thus reach high values of $U_{\alpha I}^2$.

\section{Observables, experiments and likelihoods}
\label{sec:obslike}

Models with heavy right-handed neutrinos, as described above, will alter the SM predictions for different observables that are already significantly constrained by experimental results. In this analysis, we implemented all relevant constraints such as active neutrino likelihoods (\ref{sec:activeneutrino})  and direct detection experiments which currently exert the strongest bounds over the considered
mass range (\ref{sec:direct}); these include beam dump and peak search experiments, which looked for RHNs in meson, tau and gauge boson decays. Besides, we similarly include the most relevant indirect constraints: electroweak precision observables (\ref{sec:ewpo}), lepton flavour violating processes (\ref{sec:lfv}), lepton universality constraints (\ref{sec:lepuniv}), BBN (\ref{sec:bbn}), neutrinoless double-beta decay
(\ref{sec:0nubb}) and CKM unitarity (\ref{sec:ckm}).

In this section, we will focus on the physics and statistics aspects of our likelihood
functions.  The corresponding implementation of \GB capabilities and module functions associated
with the various observables are discussed in detail in Appendix~\ref{app:gambit}.

\subsection{Active neutrino mixing}
\label{sec:activeneutrino}
In contrast to previous studies, we include likelihoods for the active neutrino mixing observables in our analysis: the three mixing angles $\theta_{12},\theta_{13},\theta_{23}$, the mass splittings $\Delta m^2_{21}$ and $\Delta m^2_{3\ell}$ with $\ell = 1$ for normal ordering and $\ell = 2$ for inverted ordering, as well as the CP-phase $\delta_{\mathrm{CP}}$. We use the most recent publically available results of the global analysis of solar, atmospheric, reactor and accelerator neutrino data in the framework of three neutrino oscillations provided by the NuFIT collaboration (as of January 2018)~\cite{Esteban:2016qun,NuFit}, including
\begin{itemize}
\item the solar neutrino experiments Homestake chlorine~\cite{Cleveland:1998nv}, Gallex/GNO~\cite{Kaether:2010ag} and SAGE~\cite{Abdurashitov:2009tn}, SNO~\cite{Aharmim:2011vm}, the four phases of Super-Kamiokande~\cite{Hosaka:2005um,Cravens:2008aa,Abe:2010hy} and two phases of Borexino~\cite{Bellini:2011rx,Bellini:2008mr,Bellini:2014uqa},
\item the atmospheric experiments IceCube/DeepCore~\cite{Aartsen:2014yll},
\item the reactor experiments KamLAND~\cite{Gando:2013nba}, Double-Chooz~\cite{An:2016srz}, Daya-Bay~\cite{An:2016ses} and Reno~\cite{reno},
\item the accelerator experiments MINOS~\cite{Adamson:2013whj,Adamson:2013ue}, T2K~\cite{t2k} and NO$\nu$A~\cite{nova},
\item the cosmic microwave background measurement Planck~\cite{Ade:2015xua}
\end{itemize}
For our global fit, we take the provided one-dimensional $\Delta \chi^2$ tables for both orderings of the NuFIT collaboration~\cite{NuFit}. For more detailed information, we refer to \cite{Esteban:2016qun} and references therein.
We emphasize that using higher dimensional tables that account for correlations would in general lead to (slightly) more stringent results on the RHN parameter space, hence our treatment can be considered as conservative.

\subsection{Indirect constraints}

\subsubsection{Electroweak precision observables}
\label{sec:ewpo}

The leptonic charge currents are modified by the RHNs, and hence the value of
$G_\mu$ that is measured via the muon decay will differ from the actual Fermi constant $G_F$
which is defined in terms of the fine structure constant and mass of the $Z$ boson.
The correction can be written as~\cite{Drewes:2015iva}
\begin{equation}
  G_\mu^2 = G_F^2 (1 - (\theta \theta^\dagger)_{\mu\mu} - (\theta
  \theta^\dagger)_{ee})\
  \label{gmu}
\end{equation}
and is caused by the non-unitarity of the flavour mixing matrix $V_\nu$, see
Eq.~\eqref{defV}, which leads to a slight suppression of the muon decay.

Both the weak mixing angle $\theta_w$ and the mass of the $W$ boson $m_W$
depend on $G_\mu$ at one loop, which means they also get a correction from
the active-sterile mixing matrix $\Theta$, which is given by~\cite{Antusch:2014woa}
\begin{align}
\notag s^2_w &= [s^2_w]_{SM} \sqrt{1 -  (\theta \theta^\dagger)_{\mu\mu} - (\theta \theta^\dagger)_{ee}}, \\
  \frac{m_W^2}{[m_W^2]_{SM}} &= \frac{[s^2_w]_{SM}}{s^2_w}  \sqrt{1 -  (\theta \theta^\dagger)_{\mu\mu} - (\theta \theta^\dagger)_{ee}}\;,
\end{align}
where $s_w^2 = \sin^2\theta_w$. Since experiments typically measure the effective Weinberg angle $s^2_{eff}$, and assuming the QCD corrections factorize from the leptonic corrections~\cite{Antusch:2015mia}, we use for the SM prediction the highly accurate calculation, including corrections up to two-loops, from \cite{Ferroglia:2012ir} 
\begin{align}
\notag[s^2_{eff}]_{SM} &= 0.23152 \pm 0.00010, \\
[m_W]_{SM} &= 80.361 \pm 0.010 \text{ GeV}.
\end{align}
Other electroweak precision observables affected by the presence of the heavy
neutrinos are the decays of the $Z$ and $W$ bosons, in particular the invisible
decay width of the $Z$ boson, $\Gamma_{\rm{inv}}$, and the leptonic decays of
$W$. Under the assumption that the radiative corrections factorize from the heavy neutrino contribution, at least up to order $\theta^2$~\cite{Antusch:2015mia,Fernandez-Martinez:2015hxa}, one can write the invisible decay width of the Z as~\cite{Abada:2013aba}
\begin{align}
 \notag \Gamma_{\rm{inv}} &= \sum_{i,j} |\Gamma_{Z\to\nu_i\nu_j}|^{\text{SM}} \Big(|V_\nu^\dagger V_\nu|^2 _{ij} \\ 
 &+ |V_\nu^\dagger \Theta|^2_{ij} (1 - \frac{m_{N_j}^2}{m_Z^2})^2(1 + \frac{1}{2}\frac{m_{N_j}^2}{m_Z^2}) \Big)\;,
\end{align}
where we have neglected the contribution from $Z \to N_i N_j$ due to being of order $\theta^4$, and for the SM decay $Z \to \nu_i \nu_j$ we use the 2-loop calculation from \cite{Dubovyk:2018rlg}.

The contribution of heavy neutrinos to the $W$ decay widths to leptons can be written as~\cite{Antusch:2014woa}
\begin{equation}\Gamma_{W\to l_\alpha \bar\nu} = \frac{G_\mu m_W^3}{6\sqrt{2}\pi}\frac{(1-\tfrac{1}{2} \theta\theta^\dagger)_{\alpha\alpha})(1-x_\alpha)^2(1 + x_\alpha)}{\sqrt{1 - (\theta \theta^\dagger)_{\mu\mu} -(\theta \theta^\dagger)_{ee})}}\;,
\end{equation}
where we defined $x_\alpha \equiv m_{l_\alpha}^2 / m_W^2$.

\medskip

\begin{table}[h]
  \centering
  \begin{tabular}{lr}
    \toprule
    \textbf{Observable} & \textbf{Value} \\
    \midrule
    \textit{Input parameters}\\
    $G_\mu$ [GeV$^{-2}$] & $1.1663787(6) \times 10^{-5}$ \\
    $m_Z$ [GeV] & $91.1875(21)$ \\\\
    \textit{Constraints}\\
    $m_W$ [GeV] & $80.385(15)$  \\
    $s_{eff}^2$ & $0.23155 \pm 0.00005$ \\
    $\Gamma_{\rm{inv}}$ [MeV] & $499.0\pm1.6$\\
    $\Gamma_{W\to e \bar\nu_e}$ [MeV] & $223\pm 6$ \\
    $\Gamma_{W\to \mu \bar\nu_\mu}$ [MeV] & $222\pm5$ \\
    $\Gamma_{W\to \tau \bar\nu_\tau}$ [MeV] & $237\pm6$ \\
    \bottomrule
  \end{tabular}
  \caption{Electroweak precision observables measurements and uncertainties, taken from Ref.~\cite{PDG17}.}
  \label{tab:ewpo}
\end{table}

We construct Gaussian likelihoods for these observables using the experimental
measurements and uncertainties displayed in Table~\ref{tab:ewpo}. All these observables depend on $G_\mu$ (eq.\eqref{gmu}) either directly or through another observable ($s_w$ or $m_W$). Since the experimental measurements of these quantities are independent of each other, we assume them to be uncorrelated.

\subsubsection{Lepton flavour violation}
\label{sec:lfv}

Flavour changing neutral processes, such as lepton flavour violation (LFV), are
strongly suppressed in the Standard Model at one loop due to the GIM
mechanism~\cite{Glashow:1970gm}. Hence, any non-trivial contribution to these
processes from physics beyond the Standard Model would dominate over the SM
contribution, which in turn makes the experimental determination of these
observables a smoking gun of new physics. Several experiments have attempted to
measure LFV processes with outstanding precision and they have imposed a set of
upper limits on their branching fractions. In Table~\ref{tab:lfvexp} we list
the most significant of these observables, along with the experimental upper
bound on their branching ratios and the experiment that provided it.

\begin{table}[h]
 \centering
 \begin{tabular}{lrr}
 \toprule
 \textbf{Process} & \textbf{Branch.~Frac.} & \textbf{Reference} \\
 \midrule
 \textit{LFV decay}\\
  $\mu^- \to e^- \gamma$ & $4.2\times 10^{-13}$ & MEG~\cite{TheMEG:2016wtm}\\
  $\tau^- \to e^- \gamma$ & $5.4\times 10^{-8}$ & BaBar~\cite{Aubert:2009ag},Belle~\cite{Hayasaka:2007vc}\\
  $\tau^- \to \mu^- \gamma$ & $5.0\times 10^{-8}$ & BaBar~\cite{Aubert:2009ag},Belle~\cite{Hayasaka:2007vc}\\
  $\mu^- \to e^-e^-e^+$ & $1.0\times 10^{-12}$ & SINDRUM~\cite{Bellgardt:1987du}\\
  $\tau^- \to e^-e^-e^+$ & $1.4\times 10^{-8}$ & BaBar~\cite{Lees:2010ez},Belle~\cite{Hayasaka:2010np}\\
  $\tau^- \to \mu^- \mu^- \mu^+$ & $1.2\times 10^{-8}$ & ATLAS~\cite{Aad:2016wce},BaBar~\cite{Lees:2010ez}\\
  & & Belle~\cite{Hayasaka:2010np},LHCb~\cite{Aaij:2014azz}\\
  $\tau^- \to \mu^-e^-e^+$ & $ 1.1\times 10^{-8}$ & BaBar~\cite{Lees:2010ez},Belle~\cite{Hayasaka:2010np}\\
  $\tau^- \to e^-e^-\mu^+$ & $0.84\times 10^{-8}$& BaBar~\cite{Lees:2010ez},Belle~\cite{Hayasaka:2010np}\\
  $\tau^- \to e^-\mu^-\mu^+$ & $1.6\times 10^{-8}$& BaBar~\cite{Lees:2010ez},Belle~\cite{Hayasaka:2010np}\\
  $\tau^- \to \mu^-\mu^-e^+$ & $0.98\times 10^{-8}$& BaBar~\cite{Lees:2010ez},Belle~\cite{Hayasaka:2010np}\\\\
 \textit{LFV conversion}\\
  $\mu - e$ (Ti) & $4.3\times 10^{-12}$& SINDRUM II~\cite{Dohmen:1993mp}\\
  $\mu - e$ (Au) & $7 \times 10^{-13}$& SINDRUM II~\cite{Bertl:2006up}\\
  $\mu - e$ (Pb) & $4.6\times 10^{-11}$& SINDRUM II~\cite{Honecker:1996zf}\\
  \bottomrule
 \end{tabular}
 \caption{Experimental upper bounds on LFV processes, along with the experiments
 that provided that bound. When more than one experiment is cited, the HFLAV average is used~\cite{Amhis:2016xyh}. All upper bounds are given at the $90\%$ C.L..}
 \label{tab:lfvexp}
\end{table}

The experimental upper bounds for LFV $\mu$ and $\tau$ decays in Table
\ref{tab:lfvexp} are given as branching fractions with respect to the total
decay width of the respective lepton~\cite{PDG17,Ilakovac:1994kj},
\begin{align}
  \notag \Gamma_{\mu} &=  (2.995984 \pm 0.000003) \times 10^{-19}\;,\\
  \Gamma_{\tau} &= (2.2670 \pm 0.0039) \times 10^{-12}\;.
\end{align}

In the model with three heavy neutrinos the leading
contributions to these observables arise from dipole and box diagrams with
mixing between the active and sterile neutrinos, given by the active-sterile mixing matrix $\Theta$.
The relevant LFV processes containing these diagrams are of the form
$l_\alpha^- \to l_\beta^- \gamma$, $l_\alpha^- \to l_\beta^- l_\beta^-
l_\beta^+$, $l_\alpha^- \to l_\beta^- l_\gamma^- l_\gamma^+$ and $l_\alpha^-
\to l_\gamma^- l_\gamma^- l_\beta^+$.  The associated decay widths can be found
in Appendix~\ref{app:lfv}.

\medskip

Lastly, LFV processes can result in a neutrinoless $\mu - e$ conversion inside a nucleus. Muons captured by a nucleus typically decay in orbit providing a continuous spectrum of energy for the electron in the final state. In coherent flavour violating conversion, $\mu^- N \to e^- N$, final state electrons have a discrete energy spectrum, corresponding to the mass of the decaying muon. Consequently experiments measure the rate at which this conversion happens, with respect to the rate of capture by the nucleus, 
\begin{equation}
R_{\mu -e} = \Gamma_{\text{conv}}/\Gamma_{\text{capt}}. 
\end{equation}
The corresponding expressions for the conversion ratio, as well as the nuclear parameters for the two nuclei studied, $\text{Ti}_{22}^{48}$, $\text{Au}_{79}^{197}$ and $\text{Pb}_{82}^{208}$, can be found in Appendix~\ref{app:lfv}.
\medskip

The likelihoods for these LFV observables are all Gaussian upper limit
likelihoods. They are computed as
\begin{equation}
  \ln\mathcal{L} = \left\{
    \begin{array}{lr}
      -\frac{1}{2}\log(2\pi\sigma^2), & \quad x < x_0 \\
      -\frac{1}{2}\log(2\pi\sigma^2) - \frac{1}{2} \frac{(x - x_0)^2}{\sigma^2}, & \quad  x > x_0
  \end{array} \right.\;,
  \label{upperlimitlnL}
\end{equation}
using the experimental data from Table \ref{tab:lfvexp}.  More specifically, we
assume a measured value of $x_0$ for all observables\footnote{In the cases where the experiments do not provide a measured value we take $x_0 = 0$.}, and set $\sigma = v/1.64$ for full Gaussians and $\sigma = v/1.28$ for one-sided Gaussians, where $v$ is the quoted upper 90\% C.L. limit. 

\bigskip

\subsubsection{Lepton universality}
\label{sec:lepuniv}

Recent measurements of meson decays~\cite{Aaij:2014ora, Aaij:2015yra,
Aaij:2017vbb} have put into question the flavour-independence of leptonic
charged currents, as predicted by the SM.  Previous tests of lepton
universality performed by LEP and SLC, using lifetime measurements of the tau
and muon as well as the partial decay widths of the Z boson, showed no such
deviation. This has lead to the formulation of many BSM theories attempting to
explain the deviation shown in meson decays with sterile neutrinos~\cite{Abada:2013aba,
Boucenna:2015raa,Bryman:2019ssi}.

The presence of right-handed neutrinos modifies the leptonic currents and thus
triggers a contribution to processes testing lepton universality such as in the fully leptonic
decays of charged mesons, $X^+ \to l^+ \nu$, or the semileptonic decays of $B$
mesons $B^{0/\pm} \to X^{0/\pm} l^+ l^-$.

\medskip

In order to cancel the considerable hadronic uncertainties present in the
decays of pseudoscalar mesons, lepton universality tests are best formulated
using ratios between lepton species. For fully leptonic and semileptonic decays
of mesons, these ratios are expressed as
\begin{align}
R_{\alpha\beta}^X &= \frac{\Gamma(X^+ \to l_{\alpha}^+\nu_{\alpha})}{\Gamma(X^+ \to l_{\beta}^+\nu_{\beta})}\;, \\
R_{X} &= \frac{\Gamma(B^{0/\pm} \to X^{0/\pm} l^+_\alpha l^-_\alpha)}{\Gamma(B^{0/\pm} \to X^{0/\pm} l^+_\beta l^-_\beta)}\;,
\end{align}
respectively.

In case of fully leptonic decays, one can express the test of lepton
universality in terms of deviations from the SM prediction as
\begin{equation}
  R_{\alpha\beta}^X = R^X_{\alpha\beta,SM}(1 + {\Delta}r_{\alpha\beta}^X)\;,
  \label{lepuniv}
\end{equation}
where the sterile neutrino contribution can be calculated from the active-sterile mixing matrix
$\Theta$ as~\cite{Shrock:1980ct,Drewes:2015iva}
\begin{equation}
  {\Delta}r_{\alpha\beta}^X = \frac{1+\sum_I{|\Theta_{{\alpha}I}|^2[G_{{\alpha}I} - 1]}}{1+\sum_I{|\Theta_{{\beta}I}|^2[G_{{\beta}I} - 1]}} - 1\;,
  \label{rXab}
\end{equation}
where we used
\begin{align}
  \notag G_{{\alpha}I} &= \vartheta(m_X - m_{l_{\alpha}} - M_I)\frac{r_{\alpha} + r_I + (r_{\alpha} - r_I)^2}{r_{\alpha}(1 - r_{\alpha})^2}\\
                       &\cdot \sqrt{1 - 2(r_{\alpha} + r_I) + (r_{\alpha} - r_I)^2}\;,
\end{align}
with $\vartheta$ being the Heaviside step function, $r_{\alpha} \equiv
m_{l_{\alpha}}^2/m_X^2$ and $r_I \equiv M_I^2/m_X^2$. The SM predictions used in eq. \eqref{lepuniv} for the tests of lepton universality for pions and kaons are $R^\pi_{e\mu,SM} = 1.2354 \times 10^{-4}$ and $R^K_{e\mu,SM} = 2.477 \times 10^{-5}$, respectively~\cite{Cirigliano:2007xi}.

The contribution from heavy right-handed neutrinos to the semileptonic decays
of $B$ mesons is much less significant than to the leptonic decays. As argued in Ref.~\cite{Abada:2013aba}, the effect on B decays to charmed mesons, $B^\pm \to D l \nu$, is
completely negligible. Semileptonic decays
to $K$ mesons are more affected, particularly the decays $B^+ \to K^+ l^+ l^-$
and $B^0 \to K^{*0} l^+ l^-$. Assuming that $m_l \ll m_{K^{(*)}}$ and that the
Wilson coefficient $C_7 \ll C_9, C_{10}$, one can approximate the ratios $R_K$
and $R_{K^*}$ as~\cite{Ghosh:2014awa}
\begin{align}
  \notag R_{K^{(*)}} &= \frac{\Gamma(B^{\pm/0} \to K^{\pm/*0} \mu^+ \mu^-)}{\Gamma(B^{\pm/0} \to K^{\pm/*0} e^+ e^-)} \\
                     &\approx \frac{|C_{10}^{SM} + \Delta C_{10}^\mu|^2 + |C_{9}^{SM} + \Delta C_{9}^\mu|^2}{|C_{10}^{SM} + \Delta C_{10}^e|^2 + |C_{9}^{SM} + \Delta C_{9}^e|^2}\;,
  \label{RKstar}
\end{align}
and the BSM contributions to the Wilson coefficients $\Delta C_9^\alpha$ and $\Delta C_{10}^\alpha$ can be expressed as~\cite{He:2017osj}
\begin{equation}
  \Delta C_9^\alpha = - \Delta C_{10}^\alpha = -\frac{1}{4 s_w^2} \sum_I |\Theta_{\alpha I}|^2 E(x_t, x_I)\;,
\end{equation}
with $x_t = m_t^2/m_W^2$, $x_I = M_I^2/m_W^2$ and the loop function 
\begin{multline}
  E(x,y) = xy \Bigg\{-\frac{3}{4}\frac{1}{(1-x)(1-y)} \\
  + \Big(\frac{1}{4} - \frac{3}{2(x-1)} - \frac{3}{4(x-1)^2}\Big)\frac{\log x}{x-y} \\ 
         + \Big(\frac{1}{4} - \frac{3}{2(y-1)} - \frac{3}{4(y-1)^2}\Big)\frac{\log y}{y-x} \Bigg\}\;.
\end{multline}
NNL calculations for the Standard Model contribution to the Wilson
coefficients $C_9$ and $C_{10}$ used in Eq.~\eqref{RKstar} gives $C_9^{SM} =
4.211$ and $C_{10}^{SM} =
-4.103$~\cite{Altmannshofer:2008dz,Choudhury:2017ijp}.

\medskip

In addition to meson decays, other common tests of lepton universality include
the decays of the $W$ boson to leptons as well as $\tau$ decays. The ratio of
decay widths of $W$ to charged leptons $l_\alpha$ and $l_\beta$ can be written
as~\cite{Antusch:2014woa}
\begin{equation}
  R^W_{\alpha\beta} = \frac{\Gamma(W^+ \to l^+_\alpha\nu_\alpha)}{\Gamma(W^+
  \to l_\beta^+ \nu_\beta)}= \sqrt{\frac{1 - (\theta
  \theta^\dagger)_{\alpha\alpha}}{1 - (\theta \theta^\dagger)_{\beta\beta}}}\;.
\end{equation}
Deviations from the SM for the lepton universality test in $\tau$ decays follow the same form as in Eq.~\eqref{rXab} and the SM prediction is $R^\tau_{\mu e, SM} = 0.973$~\cite{Pich:2009zza}.

\begin{table}[h]
  \centering
  \begin{tabular}{lr}
    \toprule
    \textbf{Obs.}  & \textbf{Measured} \\
    \midrule
    $R^\pi_{e\mu}$ & $(1.2327 \pm 0.0023) \times 10^{-4}$ ~\cite{Tanabashi:2018oca} \\
    $R^K_{e\mu}$ &  $(2.488 \pm 0.010) \times 10^{-5}$ ~\cite{Lazzeroni:2012cx} \\
    $R^\tau_{\mu e}$ & $0.9762 \pm 0.0028$ ~\cite{Amhis:2016xyh}\\
    $R^W_{\mu e}$ & $0.980 \pm 0.018$ ~\cite{Aaij:2016qqz} \\
    $R^W_{\tau e}$ & $1.063 \pm 0.027$ ~\cite{Schael:2013ita} \\
    $R^W_{\tau\mu}$ & $1.070 \pm 0.026$ ~\cite{Schael:2013ita} \\
    $R^B_K$ & $0.745 \pm 0.089$ ~\cite{Aaij:2014ora}\\
    $R^B_{K^*}$ (1)\hspace{-1cm} & $0.66 \pm 0.09 $ ~\cite{Aaij:2017vbb} \\
    $R^B_{K^*}$ (2)\hspace{-1cm} & $0.69 \pm 0.10$ ~\cite{Aaij:2017vbb}\\
    \bottomrule
  \end{tabular}
  \caption{Experimental measurements for all tests of lepton universality.}
  \label{tab:luv}
\end{table}

These tests of lepton universality are implemented as Gaussian likelihoods
centered on the experimentally measured value. The experimental measurements, with their
corresponding uncertainties\footnote{The experimental uncertainties for
$R^B_{K^{(*)}}$ are obtained as the sum in quadrature of the statistical and
systematic uncertainties provided by~\cite{Aaij:2014ora} and~\cite{Aaij:2017vbb}.}, are shown in Table~\ref{tab:luv}. The measurements of $R^\pi_{e\mu}$ include subleading decays with $\gamma$'s, hence the upper limit shown is the PDG average of the ratios of $\Gamma(\pi^+ \to l_\alpha^+ \nu_\alpha) + \Gamma(\pi^+ \to l_\alpha^+ \nu_\alpha\gamma)$, based on the measurements in ~\cite{Britton:1992pg,Czapek:1993kc,Aguilar-Arevalo:2015cdf}. Two experimental
measurements are shown for $R_{K^*}$ corresponding to two regions of the
dilepton invariant mass $0.045 < q^2 < 1.1 (\text{GeV}^2/c^4)$ for (1) and $1.1
< q^2 < 6.0 (\text{GeV}^2/c^4)$ for (2).

\subsubsection{CKM unitarity}
\label{sec:ckm}

The determination of the CKM matrix elements $(V_{CKM}^{exp})_{ab}^i$ is
usually done under the implicit assumption of a zero active-sterile mixing matrix, $\Theta = 0$. 
The measurements of the  $(V_{CKM}^{exp})_{ab}^i$ therefore need to be adjusted to take into account effects of RHNs.

Firstly, the smallest element of the CKM matrix, $(V_{CKM})_{ub}$, can be neglected in our study as its absolute value $|(V_{CKM})_{ub}|^2 \sim 10^{-5}$ is much smaller than our sensitivity to the  $\Theta $ parameter. Hence, under the assumption of the unitary of the CKM matrix, one can derive the following relation:
\begin{equation}
  |(V_{CKM})_{ud}|^2 + |(V_{CKM})_{us}|^2 = 1\;.
  \label{eq:CKMunitarity}
\end{equation}

Thus, we use the various experimental measurements of
$(V_{CKM}^{exp})_{us}$~\cite{Antonelli:2010yf,Follana:2007uv,Amhis:2012bh} and $(V_{CKM}^{exp})_{ud}$~\cite{Patrignani:2016xqp} to 
simultaneously constrain the true value of $|(V_{CKM})_{us}|$ and active-sterile mixing matrix $\Theta$.

Following Refs.~\cite{Antusch:2014woa, Drewes:2015iva},
the experimental measurements and true value of CKM matrix element $(V_{CKM})_{us,ud}$ are 
related via
\begin{equation}
  |(V_{CKM}^{exp})_{us,ud}^i|^2 = |(V_{CKM})_{us,ud}|^2[1+f^i(\Theta)]\;,
\end{equation}
where we defined the functions $f^i$ to encode the contribution of RHNs to the
process considered in each experiment. The decay processes considered to extract the value of $|(V_{CKM}^{exp})_{us}|$, and the $f(\Theta)$ functions, are given by~\cite{Antusch:2014woa}
\begin{align}
  K_L \to \pi^+e^-\bar{\nu}_e: 1+f^1(\Theta) = \frac{G_F^2}{G_{\mu}^2}[1-(\theta\theta^{\dagger})_{ee}],
\end{align}
\begin{align}
  K_S \to \pi^+e^-\bar{\nu}_e: f^2(\Theta) = f^1(\Theta),
\end{align}
\begin{align}
  K^- \to \pi^0e^-\bar{\nu}_e: f^3(\Theta) = f^1(\Theta),
\end{align}
\begin{align}
  K_L \to \pi^+\mu^-\bar{\nu}_{mu}: 1+f^4(\Theta) = \frac{G_F^2}{G_{\mu}^2}[1-(\theta\theta^{\dagger})_{\mu\mu}],
\end{align}
\begin{align}
  K^- \to \pi^0\mu^-\bar{\nu}_{mu}: f^5(\Theta) = f^4(\Theta),
\end{align}
\begin{align}
  \frac{\tau^- \to K^-\nu_{\tau}}{\tau^- \to \pi^-\nu_{\tau}}: 1+f^6(\Theta) = 1+(\theta\theta^{\dagger})_{\mu\mu},
\end{align}
\begin{multline}
  \tau^- \to \pi^-\bar{\nu}_{\tau}:\\ 1+f^7(\Theta) = 1+(\theta\theta^{\dagger})_{ee}+(\theta\theta^{\dagger})_{\mu\mu}
                                                         -(\theta\theta^{\dagger})_{\tau\tau},
\end{multline}
\begin{multline}
  \tau \to s : 1 + f^8(\Theta) = \\ 1 + 0.2(\theta\theta^\dagger)_{ee} - 0.9(\theta\theta^\dagger)_{\mu\mu} - 0.2(\theta\theta^\dagger)_{\tau\tau}. 
\end{multline}
The situation is simpler in the determination of the $|(V_{CKM}^{exp})_{ud}|$ element as the uncertainty is dominated by the superallowed $0^+ \to 0^+$ nuclear beta transitions measurements, which need to be modified accordingly to:
\begin{align}
0^+ \to 0^+: 1+f^1(\Theta) = \frac{G_F^2}{G_{\mu}^2}[1-(\theta\theta^{\dagger})_{ee}]
\end{align}

The experimentally measured values of $|(V_{CKM}^{exp})^i_{us}|$ in each of the decay processes above are listed in Tab.~\ref{tab:ckm}, and the value of $|(V_{CKM}^{exp})_{ud}|=0.97417\pm0.00021$ is taken from the world average~\cite{Patrignani:2016xqp}.

\begin{table}[h]
  \centering
  \begin{tabular}{llrr}
    \toprule
    \textbf{Parameter} & \textbf{Process} & \textbf{Value} & \textbf{Ref.}\\
    \midrule
     & $K_L \to \pi e \nu$ & $0.2163(6)$ & \\
     & $K_L \to \pi \mu \nu$ & $0.2166(6)$ & \\
    $|(V^\text{exp}_\text{CKM})_{us}|f_+(0)$ & $K_S \to \pi e \nu$ & $0.2155(13)$ & \cite{Antonelli:2010yf,Aoki:2016frl}\\
    & $K^{\pm} \to \pi^0 e \nu$ & $0.2160(11)$ & \\
    & $K^{\pm} \to \pi^0 \mu \nu$ & $0.2158(14)$ & \\ \midrule
    & $\frac{\text{BR}(\tau \to K \nu)}{\text{BR}(\tau \to \pi \nu)}$ & $0.2262(13)$ & \\
    $|(V^\text{exp}_\text{CKM})_{us}|$ & $\tau \to K \nu$ & $0.2214(22)$ & \cite{Follana:2007uv,Amhis:2012bh}\\
    & $\tau \to l, \tau \to s$ & $0.2173(22)$ & \\ \midrule
    $|(V^\text{exp}_\text{CKM})_{ud}|$ & Average & $0.97417(21)$ & \cite{Patrignani:2016xqp}\\
    \bottomrule
  \end{tabular}
  \caption{Experimental values of $(V_{CKM})_{us}$ and the average value of $(V_{CKM})_{ud}$ used in the calculation of the CKM likelihood. The factor $f_+(0)=0.959\pm0.005$ is taken from \cite{Aoki:2016frl}.}
  \label{tab:ckm}
\end{table}

We thus construct the likelihood for this constraint from a chi-squared function, $2\ln\mathcal{L} = -\chi^2$, where the discriminant measures the deviation of the true value $(V_{CKM})_{us,ud}$ and the experimental measurements $(V_{CKM}^{exp})^i_{us,ud}$, and is given by
\begin{align}
  \chi^2 = \sum_{i=1}^7 \frac{\left((V_{CKM}^{exp})_{us}^i -
  (V_{CKM})_{us} \cdot (1+f^i(\Theta)  \right)^2}{\sigma_i^2} \nonumber  \\
  + \frac{\left((V_{CKM}^{exp})_{ud} -
  (V_{CKM})_{ud} \cdot (1+f^1(\Theta)  \right)^2}{\sigma^2}
  \;.
  \label{eq:chi2CKM}
\end{align}

Due to the unitarity relation in Eq.~\ref{eq:CKMunitarity}, the value $(V_{CKM})_{ud}$ is obtained from $(V_{CKM})_{us}$ for every parameter point, and thus the only free floating parameters are the value of $(V_{CKM})_{us}$ and the active-sterile mixing matrix, $\Theta$. For simplicity, and since this is the only constraint to depend strongly on the value of $(V_{CKM})_{us}$, we optimise on its value for each $\Theta$, which removes the necessity of making $(V_{CKM})_{us}$ part of the scanning model. This approach is similar to the discussion in \cite{Drewes:2015iva}, but we improve upon it by optimising on the true value $(V_{CKM})_{us}$, including the $\Theta$ corrections, for each parameter point, rather than the value measured experimentally.

\subsubsection{Neutrinoless double-beta decay}
\label{sec:0nubb}

Double-beta decay refers to the decay of two neutrons into two protons while emitting two
electrons and two anti-neutrinos. In case of neutrinos having a Majorana nature, lepton number would be violated and neutrinoless double-beta decay ($0\nu\beta\beta$) induced. Besides the exchange the light neutrinos, the exchange of RHNs is similarly possible and would alter the expected effective neutrino mass $m_{\beta\beta}$. The effective mass is constrained by half life measurements of $0\nu\beta\beta$ decay. The most stringent limits are currently set by the  GERDA experiment (Germanium)~\cite{Agostini:2017iyd} with $m_{\beta\beta}<0.15-0.33\;\text{eV}$ (90\% CL), and KamLAND-Zen (Xenon)~\cite{KamLAND-Zen:2016pfg}, $m_{\beta\beta}<0.061-0.165\;\text{eV}$ (90\% CL).
The effective mass $m_{\beta\beta}$, can be theoretically evaluated in term of the mixings and masses of the light and right handed neutrinos~\cite{Drewes:2016lqo}
\begin{equation}
m_{\beta\beta} = |\sum_i{(U_{\nu})_{ei}^2 m_i}+\sum_I{\Theta_{eI}^2 M_I f_A(M_I)}|\,.
\end{equation}
Hereby, the first term denotes the contribution from LHNs, the second the one from RHNs. With a typical momentum exchange of around 100 MeV in $0\nu\beta\beta$ decay, RHNs with a mass above this threshold participate in the process only virtually. This suppression is taken into account by the factor~\cite{Drewes:2016lqo}
\begin{equation}
f_A(M) \approx \frac{p^2}{p^2+M^2}\,.
\end{equation}
The typical momentum exchange $p^2$ depends not only on the specific isotope in consideration but is also subject to the theoretical model in which the constraints are derived and the value of the nucleon axial-vector constant.  An overview is given in~\cite{Faessler:2014kka}: For our analysis, we use the ``Argonne'' model and the lower of the two values for $p^2$ (quenched), which yields the most conservative constraints: 
$\sqrt{\langle p^2\rangle} = 178 \; \text{MeV}$ for xenon, and
$\sqrt{\langle p^2\rangle} = 159 \; \text{MeV}$ for germanium.
A more dedicated analysis of the impact of different limits due to nuclear uncertainties is beyond the scope of this work. Since we are focusing on 
profile likelihood for our results, this approach is largely equivalent to
profiling over systematic uncertainties assuming a flat prior that spans the entire
range of values $\langle p^2\rangle$ in ref.~\cite{Faessler:2014kka}.
For our analysis we use the experimental values, as stated above, as one-sided Gaussian likelihoods, choosing the higher of the two values in order to remain conservative.

\subsubsection{Big Bang Nucleosynthesis}
\label{sec:bbn}

If RHNs decay shortly before or during BBN, the typical energy of decay
products, here $\sim M_I \geq 50\rm\; MeV$, is significantly higher than the
plasma temperature at that time, $\sim 100$ keV.  Therefore, either by
dissociating formed nuclei, or by causing deviations from thermal equilibrium,
they will affect the abundances of primordial elements, which are however
observationally well constrained.  The requirement that the RHN decay happens
sufficiently early enough before BBN implies an upper limit on the
lifetime ($\tau_I$) of RHNs, or equivalently, a \emph{lower} bound on the
mixing $U_I^2$~\cite{Dolgov:2003sg}. However, in the presence of multiple RHN species, BBN cannot constrain individual mixing angles $U_{\alpha I}^2$~\eqref{UaI} but only the total mixing $U_I^2~\eqref{UI}$.

We consider leptonic decay channels for all RHNs masses, when kinematically allowed, as well as hadronic decays to mesons and leptons. As shown in~\cite{Bondarenko:2018ptm}, for low masses the hadronic decay width is dominated by channels with a single meson and a lepton, while for masses above the hadronisation scale, $\Lambda_{\textrm{had}} \sim 1$ GeV, it can be approximated by computing the decay to free quarks. The decay width for each topology is listed in Appendix~\ref{app:bbndw}, with expressions and values for the decay constants taken from ~\cite{Atre:2009rg},~\cite{Gorbunov:2007ak},~\cite{Canetti:2012kh},~\cite{Bondarenko:2018ptm} and~\cite{Ballett:2019bgd}, along with a detailed comparison of the various expressions.

In the current study, we require the lifetime of
each RHN to be less than $0.1 s$~\cite{Ruchayskiy:2012si}, which is implemented
in the likelihood as a step function.  In principle, this limit can
be weakened if the lightest active neutrino has a mass $< \mathcal{O}(10^{-3})$
eV, since the RHNs do not necessarily thermalize in this case~\cite{Hernandez:2014fha}. We leave, however, the implementation of refined BBN constraints in \GB for future work.
Note that a lifetime bound that is stronger by a factor of two would lead to proportionally stronger constraints on the total mixing $U_I^2$.

\subsection{Direct RHN searches}
\label{sec:direct}

Different experiments search with various approaches directly for RHNs. One can distinguish between three types: peak searches (PIENU), searches at beam dump experiments (PS-191, CHARM, E949, NuTeV), and searches at $e^+ e^-$ or $pp$ colliders (DELPHI, ATLAS, CMS).

One possibility to look for RHN, is to search for peaks in the lepton energy spectrum of a meson decay. If, for example, a meson of mass $m_X$ decays into an RHN of mass $M_I$ and an
electron/muon with mass $m_{l_{\alpha}}$, this peak will be approximately at
\begin{align}
  E_{peak} \simeq \frac{m_X^2 + m_{l_{\alpha}}^2 - M_I^2}{2m_X}\;.
\end{align}
Even in situations where backgrounds are sizeable, a peak search can hence be
used to impose constraints on the mixing.

In beam dump experiments, the large background signal that is usually present
near the target hinders the detection of charged particles that are produced
along with the RHNs. On the other hand, RHNs with mass below the D meson scale
can be long-lived enough to travel macroscopic distances.  Looking for their
charged decay products some distance away from the target leads to (almost)
background-free experimental situations.

In collision experiments ($e^+ e^-$ or $pp$), vector bosons or mesons get produced that subsequently can decay leptonically. The bounds on these processes are then able to constrain the corresponding active-sterile mixing angles in a certain mass range.

\medskip

To implement the direct detection constraints as likelihoods, we follow two
different approaches, depending on the information that is provided in each
study.  Firstly, some of the experiments found no signal events and had no
background counts after cuts (DELPHI, CHARM, PS191 and NuTeV). In this case,
since the processes in the experiments are essentially Poissonian, we construct
the likelihood (to observe $n$ events) as a Poisson distribution. The number of
expected counts, $\mu$, is a function of the RHN masses and mixings, i.e. $\mu
= \mu(M_I, U_{{\alpha}I}^4)$ (assuming the experiment does so as well, the fourth power takes both production and decay
of RHNs into account). For expected $\mu$ events and background $b$, the
likelihood is:
\begin{align}
  \mathcal{L}(n | \mu) = (\mu+b)^n\frac{e^{-(\mu+b)}}{n!}\,.
\end{align}
With no reported detections ($n = 0$) and background cuts
reducing $b$ to approximately zero,
\begin{align}
\ln\mathcal{L}(n=0 | \mu) = -\mu\,.
\label{lnLpoisson}
\end{align}
To connect $\mu$ with our model parameters, we use the fact that the expected
signal counts are proportional to the LHN-RHN mixing, $\mu \propto
U_{{\alpha}I}^4$. The factor of proportionality is set to reproduce the results from the experimental papers (assuming that these limits are based on the common Feldman-Cousins procedure~\cite{Feldman:1997qc}, where e.g.~a 95\% CL upper limit would correspond to an expected number of signal counts of $\mu=3.09$).

On the other hand, for the experiments which either quote non-zero signal
events and/or backgrounds, or if this information is ambiguous (CHARM
($\nu_{\tau}$ re-interpretation), PIENU, ATLAS and E949), we model the
constraint likelihood as Gaussian upper limits, \textit{i.e.}~we model them as
half-Gaussians with zero mean and error set according to the confidence level
at which the results are presented. For example, in the case of an experiment that presents limits at
$90\%$ CL, for a half Gaussian, this lies within $1.28\sigma$ of the mean.

It is worth noting that collider experiments often use simplified model assumptions to compute the confidence level intervals presented in their results. Since we use these to construct our likelihoods, we are incorporating these assumptions as well, in spite of the fact that our confidence intervals are computed by profiling over the multidimensional parameter space. Given that a full collider simulation is beyond the scope of this study, we employ the provided simplified model limits as given. We acknowledge, however, that the true limits may be slightly weaker due to, e.g a reduction of the production cross-section, and we defer the exploration of the differences between the collider predictions of simplified and full models to future work.

\subsubsection{PIENU}

The PIENU experiment~\cite{PIENU:2011aa} sought to detect RHNs in the mass range of $68-129$ MeV by searching for peaks in the energy spectrum of the decay process $\pi^+ \rightarrow e^+ \nu$. It was, hence, sensitive to the mixing $|\Theta_{eI}|^2 \equiv U_{eI}^2$ and $\mu$ in eq.~\eqref{lnLpoisson} is also taken to scale as $U_{eI}^2$ in our analysis. Although no peaks were found, exact information on the number of background events is unavailable. Further, production processes in peak searches are, in general, unaffected by the Majorana/Dirac nature of the RHNs; hence, no correction is necessary here.

The constraints on $U_{eI}^2$ are at 90\% CL, so it is implemented in \GB as a half-Gaussian with zero mean and error set at 1.28$\sigma$.

After our analysis was complete we became aware of the slightly stronger updated constraints presented in Ref.~\cite{Aguilar-Arevalo:2017vlf}, which are not included in our scan.

\subsubsection{PS-191}
\label{sec:ps191}

This experiment~\cite{Bernardi:1987ek} was designed for the purpose of detecting neutrino decays. RHNs would be produced via either of the following mechanisms: $\pi^+/K^+ \rightarrow e^+ \nu_e$, or $\pi^+/K^+ \rightarrow \mu^+ \nu_{\mu}$, and would then decay via $\nu_R \rightarrow \mu^- e^+ \nu$, $\nu_R \rightarrow e^- \mu^+ \nu$, $\nu_R \rightarrow e^- \pi^+$, $\nu_R \rightarrow \mu^- \mu^+ \nu$, $\nu_R \rightarrow \mu^- \pi^+$ or $\nu_R \rightarrow e^- \pi^+ \pi^0$. Thus, PS-191 could constrain the quantities $U_{eI}^4$ and $U_{\mu I}^4$ for RHNs with a mass between $20-450$ MeV.

Having found no signal or background events, it placed constraints on these quantities at 90\% CL. We deviate from the original analysis in two ways. The first is necessitated by the fact that in the original analysis, the constraints were derived under the assumption that the RHNs interact only through the charged current. In~\cite{Ruchayskiy:2011aa}, these limits were re-interpreted with the inclusion of neutral current interactions. Thus, instead of the signal count being proportional to the fourth power of the relevant flavour mixing, it is proportional to $U_{e/\mu I}^2 \times \sum_{\alpha} c_{\alpha}U_{\alpha I}^2$, with the coefficients given by
\begin{align}
    c_e &= \frac{1+4\sin^2\theta_W+8\sin^4\theta_W}{4} \notag,\\
    c_\mu,c_\tau &= \frac{1-4\sin^2\theta_W+8\sin^4\theta_W}{4}.
\label{eqn:coeff}
\end{align}
We use these revised bounds here. The limits are encoded in likelihood form as in eqn.~\eqref{lnLpoisson}, with the aforementioned proportionality factor being $2.44$.

\subsubsection{CHARM}
\label{sec:charm}

RHNs were searched for in CHARM~\cite{Bergsma:1985is} using two strategies, one with a neutrino beam from dumping protons on copper (BD) and another using a wide-band neutrino beam (WBB) from primary protons.

In BD, the production of RHNs was assumed to occur through the decay of D mesons. They would then decay via $\nu_R \rightarrow e^+ e^- \nu_e$, $\nu_R \rightarrow \mu^+ \mu^- \nu_{\mu}$ or $\nu_R \rightarrow e^+ \mu^- \nu_e$, $\mu^+ e^- \nu_{\mu}$ (and the anti-particle counterparts) and the decay products were looked for.

In WBB, RHN production was assumed to occur via neutrino-nucleus neutral current scattering $\nu_{\mu} N \rightarrow \nu_R X$. The subsequent decay $\nu_R \rightarrow \mu R$, R representing hadrons, was then searched for. The limits from the WBB analysis are, however, weaker than those exerted by other experiments in the same mass range, and are not considered here.

The BD analysis yielded no candidate events or background and hence placed limits on $U_{eI}$ and $U_{\mu I}$ at 90\% CL. Further, the original analysis assumed the possibility of RHNs interacting solely via the charged current; we use the results re-interpreted after the inclusion of neutral current interactions~\cite{Ruchayskiy:2011aa} as discussed in section~\ref{sec:ps191}, i.e. the signal count is proportional to $U_{e/\mu I}^2 \times \sum_{\alpha} c_{\alpha}U_{\alpha I}^2$ and once again use eqn.~\eqref{lnLpoisson} to represent the likelihood, with the proportionality factor being $2.44$.

In~\cite{Orloff:2002de}, the data from the CHARM experiment was re-analyzed assuming that RHNs mix solely with tau-flavoured leptons, and was able to place limits at 90\% CL on $U_{\tau I}$, which we implement as a half-Gaussian with zero mean and error set at 1.28$\sigma$.

Dirac RHNs were assumed in both the original and tau-specific analyses, so the limits presented are also re-scaled by dividing them by $\sqrt{2}$.

\subsubsection{E949}

In this experiment~\cite{Shaykhiev:2011zz,Artamonov:2014urb}, RHNs were searched for in the decay of kaons produced in a beam dump: $K^+ \rightarrow \mu^+ \nu_R$. Constraints on $U_{\mu I}$ were placed at $90\%$ CL in the mass range $175-300$ MeV; we also divide the limits by a factor of $\sqrt{2}$ to account for the Majorana nature of RHNs in our model.

The likelihood is modeled as a half-Gaussian with zero mean, error set at 1.28$\sigma$ and $\mu \propto U_{\mu I}^2$.

\subsubsection{NuTeV}

The NuTeV experiment~\cite{Vaitaitis:1999wq} searched for RHNs through their decay into the following final states: $\mu e \nu$, $\mu \mu \nu$, $\mu\pi$ and $\mu \rho$. They were assumed to be produced in the decay of mesons. $90\%$ CL limits on $U_{\mu I}$ were placed for RHNs with a mass between $0.25-2$ GeV.

Information about the assumed Dirac or Majorana nature of the RHNs is not present, so we take the conservative route and presume Majorana RHNs were considered in the analysis. No candidate events or background were detected, so the likelihood is modeled as in eq.~\eqref{lnLpoisson}, with a proportionality factor of $2.44$ and $\mu$ scaling as $U_{\alpha I}^4$.

\subsubsection{DELPHI}

At DELPHI~\cite{Abreu:1997uq}, $e^+ e^- \rightarrow Z^0 \rightarrow \nu_R \bar{\nu}$ was the dominant RHN production mechanism; the process $Z^0 \rightarrow \nu_R \bar{\nu_R}$ would be suppressed due to the additional $U^2$ factor. The products of the RHN decaying via the weak and neutral current were then searched for, according to: $\nu_R \rightarrow \nu Z^{\ast}$, $Z^{\ast} \rightarrow \nu \bar{\nu}$, $l \bar{l}$, $q \bar{q}$ or $\nu_R \rightarrow l' W^{\ast}$, $W^{\ast} \rightarrow \nu \bar{l},q \bar{q'}$. DELPHI could constrain $\Theta_{eI}$, $\Theta_{\mu I}$ and $\Theta_{\tau I}$ for RHNs having a mass between $0.5-80$ GeV.

Since the RHNs could have existed long enough to travel macroscopic distances of upto $100$ cm, different signatures had to be considered and the analysis was split to tackle the short- and long-lived cases separately.

In the short-lived RHN case, depending on the particle mass, two signatures were looked for. For masses less than about 30 GeV, due to the large boost received by the RHNs, the signature would be a monojet. Background coming from leptonic Z boson decays or $\gamma\gamma$ processes were accounted for. Higher masses open the decay channel into $q\bar{q}$ (and a lepton, depending on the channel), and the signature in this case would be two acollinear jets which are also acoplanar with respect to the beam axis. Most of the background in this scenario came from hadronic Z decays with missing energy; a neural network was used to remove all of them from the final data.

Longer-lived RHNs were looked for using displaced vertices and calorimeter clusters. The former was useful in tracking RHNs with an intermediate lifetime; however, a cluster finding algorithm along with vertex reconstruction did not find any signals. Calorimeter clusters were used to detect the longest-lived RHNs, whose decay products would interact with the outermost layers/components of the experimental setup: the signature would be a cluster of hits in a small angular region coincident with the beam collision, which could be traced back to the initial interaction point.

The analysis was carried out assuming Majorana RHNs and yielded one candidate event and no background events. In our analysis, this means the proportionality factor is $3.09$ and $\mu$ scales as $U_{\alpha I}^4$.

A caveat must be mentioned here: the DELPHI analysis presented bounds on the mixing in a flavour-independent manner: the limit on $U^2$, as presented in the paper, applies equally to $U_e^2$, $U_\mu^2$ and $U_\tau^2$, as they mention. In the mass range under consideration, the mass of the tauon will, of course, influence the strength of the limit and, as they quote, the presented bounds become weaker for masses below $\sim 4$ GeV. However, the extent of the kinematic suppression due to the tauon mass is not quantitatively discussed; we use the limits as is, noting that it is highly likely that NA62 will subsume these bounds in the near future~\cite{Lurkin:2017tmu}.

\subsubsection{ATLAS}

The process relevant for RHN production in ATLAS~\cite{Aad:2015xaa} is $pp \rightarrow (W^{\pm})^{\ast} \rightarrow l^{\pm} \nu_R$. The RHNs were taken to be heavier than the W boson, allowing it to decay to a lepton a W boson: $\nu_R \rightarrow l^{\pm} W^{\mp}$; the W boson would then decay predominantly into a quark-antiquark pair, and the signature of this decay chain was searched for, with either two electrons or muons in the final state.\footnote{
There is an ongoing dispute in the literature on whether the rate of LNV processes at collider experiments are always suppressed by the small parameters $\epsilon_i$ and $\mu$ in Eq.~(\ref{SymmProtectParam}) and therefore unobservably small (roughly of the order of the "naive seesaw estimate") \cite{Kersten:2007vk,Moffat:2017feq} or whether coherent flavour oscillations can lead to LNV signatures in spite of the smallness of these parameters \cite{Anamiati:2016uxp,Antusch:2017ebe,Antusch:2017pkq}. In the range of $M_I$ below the electroweak scale under consideration here, the strongest direct search constraints do not come from experimental signatures that rely on LNV, and our results are therefore only mildly affected by the outcome of this discussion.} Hence, in our analysis, $\mu \propto U_{\alpha I}^4$, $\alpha = e, \mu$. The original analysis was carried out under the assumption of Majorana RHNs, so no additional correction is necessary.

The analysis placed $95\%$ CL limits on the two mixing angles in the mass range of $100-500$ GeV. Details on the number of observed/expected events and background is available and could be cast into a likelihood function combining Poissonian and Gaussian errors; however, we find that implementing the limits in \GB as a half-Gaussian with zero mean and error set at 1.64$\sigma$ reproduces the experimental limits well enough for the purpose of a global fit.

\subsubsection{CMS}

With the LHC having run with a center-of-mass energy of 13 TeV, the CMS detector searched for different event signatures of the same process as ATLAS. $95\%$ CL limits were calculated for $U_{eI}$ and $U_{\mu I}$ for RHNs with mass between 1 GeV and 1.2 TeV~\cite{Sirunyan:2018mtv}.

As before, Majorana RHNs were assumed in the analysis, and our implementation of the limits mirrors that of ATLAS.

Note that updated bounds from ATLAS~\cite{Aad:2019kiz} and CMS~\cite{CMS:2018szz,Sirunyan:2018xiv} have been released, but are not included, since these papers came out after our scans were completed. However, the new bounds from ATLAS are comparable to those from DELPHI, and the newer dilepton search from CMS only produces stronger bounds for RHN masses above $\sim 500$ GeV, which is beyond our range of study.

\subsubsection{LHCb}
LHCb has performed direct searches for heavy neutrinos.
The most recent results \cite{Aaij:2014aba} were derived with an inconsistent model and have been corrected in Ref.~\cite{Shuve:2016muy}. They are subdominant in the mass range considered here.
In Ref.~\cite{Antusch:2017hhu} the results of a generic long lived particle search \cite{Aaij:2016xmb} has been re-interpreted in the context of heavy neutrinos. We do not include these results here because the conservative interpretation does not yield stronger bounds than the ones we include.

\subsubsection{Other experiments}
Further measurements at 
Borexino \cite{Back:2003ae},
Bugey \cite{Hagner:1995bn},
SIN \cite{Abela:1981nf},
BEBC \cite{CooperSarkar:1985nh}, 
JINR \cite{Baranov:1992vq},
TRIUMF \cite{Britton:1992pg,Britton:1992xv},
OKA \cite{Sadovsky:2017qsr,Aguilar-Arevalo:2019owf},
ISTRA \cite{Duk:2011yv},
NOMAD \cite{Astier:2001ck},
NA62 \cite{CortinaGil:2017mqf}, 
Belle \cite{Liventsev:2013zz},
KEK \cite{Hayano:1982wu,Yamazaki:1984sj} and T2K \cite{Abe:2019kgx} have both published constraints on RHNs. We do not indculde them here because, with the present data, they are subdominant or cover a different mass range.

\section{Scanning strategy and parameter ranges}
\label{sec:scanning}

In this work, we focus on the exploration of the RHN parameter space using frequentist statistics. Our main goal is to establish the ranges of RHN parameters that are not yet explored by experiments, and a frequentist approach delivers a suitable and prior-independent method.  We are dealing with a high dimensional parameter space, which we have to project into two-dimensional plots.  To this end, the central quantity of interest is the \textit{profile likelihood},
\begin{equation}
    \ln\mathcal{L}_\text{prof}(\theta_1, \theta_2) = \max_{\vec\eta} \ln\mathcal{L}(\theta_1, \theta_2, \vec\eta)\;.
\end{equation}
which is, for fixed parameters of interest $\theta_1$ and $\theta_2$, the maximum value of the (log-)likelihood function that can be obtained when maximizing over the remaining parameters $\vec\eta$.  

We emphasize that the main goal of this work is to establish conservative constraints on RHN mixings and masses by profiling over all relevant parameters.  We do \emph{not} perform a proper goodness-of-fit analysis to experimental data, which would require sampling of experimental results; given the large range of included experimental results and the sometimes limited knowledge about individual experiments this is beyond the scope of the current work.
Instead, likelihoods are included in a approximate fashion that allows to reproduce published experimental results, and we use Wilks' theorem~\cite{Wilks:1938dza} to approximate the sampling statistics of log likelihood ratios and estimate confidence contours when necessary.

Our scanning strategy is designed in order to explore the complex parameter space of the RHN model such that we obtain reliable results for the projections shown in this work. To this end, we perform a large set of scans with different settings which we then merge into a single dataset. We study the normal (NH) and inverted (IH) hierarchy independently, in order to avoid artificially favouring one over the other due to the different normalisation of the active neutrino likelihoods (c.f. Section \ref{sec:activeneutrino}). Hence, we make independent scans for each of the neutrino mass hierarchies, normal and inverted, for the full set of scans described below.

\subsection{Parameters and priors}

The parameter ranges and priors for the original scans can be seen in Table \ref{tab:scanpars}. We emphasize that `priors' do here not correspond to priors in the Bayesian sense, but rather determine the efficiency with which different regions of the parameter space are explored.  For convergent scans, the results are prior-independent.  We have chosen to split the complex angles $\omega_{ij}$ into their real and imaginary parts. The active-sterile mixings depend strongly on the imaginary parts of $\omega_{ij}$ $\left(\Theta^2 \sim \frac{\exp(2\text{Im}(\omega))}{M}\right)$ and large values of Im$\omega$ produce mixings that are too large to pass any constraints, so we take a conservative range Im$\omega \in [-15,15]$, and also pre-emptively disallow choices that lead to $|\Theta|_{ij}^2 > 1$. As discussed in~\ref{sec:symmetries}, a condition for an approximate $B - \bar L$ symmetry to be realized is for two RHNs to have almost degenerate masses, which extends the range of the mixings so that they can be probed by experiments. This provides motivation for using a logarithmic prior on the RHN masses, also allowing the scanner to sample better the region close to the limits of the most constraining experiments/observables.

\begin{table}[t]
  \centering
  \begin{tabular}{lrrr}
    \toprule
    \textbf{Parameter} & \textbf{Value/Range} & \textbf{Prior}\\
    \midrule
    \textit{Active neutrino parameters}\\
    $\theta_{12}$ [rad] & $[0.547684,0.628144]$ & flat \\
    $\theta_{23}$ [rad] & $[0.670206, 0.925025]$ & flat\\
    $\theta_{13}$ [rad] & $[0.139452, 0.155509]$ & flat\\ 
    $m_{\nu_0}$ [eV] & $[10^{-7},0.23]$ & log \\
    $\Delta m^2_{21}$ $[10^{-5}\,\text{eV}^2]$ & $[6, 9]$ & flat \\
    $\Delta m^2_{3l}$ $[10^{-3}\,\text{eV}^2]$ & $[\pm 2, \pm 3]$ & flat \\
    $\alpha_1$, $\alpha_2$ [rad] & $[0,2\pi]$ & flat\\ \\
    \textit{Sterile neutrino parameters}\\
    $\delta$ [rad] & $[0,2\pi]$ & flat\\
    Re $\omega_{ij}$ [rad] & $[0,2\pi]$ & flat\\
    Im $\omega_{ij}$ & $[-15,15]$ &  flat\\
    $M_I$ [GeV] & $[0.06,500]$ & log\\
    $R_{\rm{order}}$ & [1,6] & flat\\ \\
    \textit{Nuisance parameters} \\
    $m_H$ [GeV] & $[124.1,127.3]$ & flat \\
    \bottomrule
  \end{tabular}
  \caption{Parameter ranges adopted for the full model scans, with $+$ ($-$) for normal (inverted) hierarchy of the active neutrino masses.}
  \label{tab:scanpars}
\end{table}

The C-I parametrisation, as defined in Section~\ref{sec:CI}, together with the particular parametrisation choice of $R$ in eq.~\ref{Rorder}, was found to not fully cover the entire parameter space. To circumvent this and ensure that all possible couplings are covered by the scans, we introduce an additional parameter to the scan $R_{\rm{order}}$ with discrete values $[1,6]$ corresponding to each of the possible permutations of the definition of $R$ in terms of $R^{ij}$. This allows full coverage of the coupling space and, since the likelihood is conceptually independent of the order in $R$ (and confirmed by the data), it ensures an uniform distribution of values in the parameter $R_{\rm{order}}$.

Out of the active neutrino parameters, only $\alpha_1$ and $\alpha_2$ are unconstrained by oscillation data, hence they are allowed to vary freely from 0 to $2\pi$ with flat priors. The ranges for the other neutrino phases and angles are taken as the widest of the $3\sigma$ ranges, for normal or inverted hierarchy, from the NuFit collaboration~\cite{NuFit}, also with flat priors. The mass of the lightest active neutrino, $m_{\nu_0}$, has a definite impact on the lower bound of $U_I^2$~\eqref{UI}~\cite{Drewes:2015iva}, so we choose a logarithmic prior, which enables us to examine this impact in greater detail than a flat prior would allow and keeps the BBN limits relevant~\cite{Hernandez:2014fha}. The upper limit on $m_{\nu_0}$ is chosen as the broad cosmological bound given by Planck~\cite{Ade:2015xua}, $\sum m_\nu < 0.23$ eV\footnote{This upper limit is not very conservative in light of Planck data, a more conservative bound would be $\sum m_\mu < 0.6$ eV~\cite{Abe:2018emu}. However, there is no effect of this constraint on our data as most high likelihood data points lie in the limit $m_{\nu_0}\to0$. We have, nevertheless, studied a subset of cases with the conservative bound and indeed found them to not be relevant.}. In order to better fit the active neutrino data, the mass splittings $\Delta m^2_{21}$ and $\Delta m^2_{3l}$ are chosen as scan parameters, where $l = 1$ and $\Delta m^2_{3l} > 0$ for normal hierarchy and $l = 2$ and $\Delta m^2_{3l} < 0$ for inverted hierarchy.

Since the construction of the mixing matrix in the C-I parametrisation depends on $m_H$ (1-loop correction), as seen in \ref{sec:CI}, we take $m_H$ as a nuisance parameter with a Gaussian distribution around its averaged measured value~\cite{PDG17} and a flat prior. Other SM parameters are fixed to their PDG values~\cite{PDG17}.

\subsection{Targeted scans}\label{subsec:TargetedScans}

We encountered a number of challenges while sampling the full RHN parameter space. One reason is connected to the behaviour of the likelihood function over the whole parameter range. The adopted scanning algorithm (\diver, see below for details) is designed to find regions of maximum likelihood across the parameter space.  However, as we will discuss later when we study the effect of each individual observable, most constraints have flat contributions to the likelihood in a large portion of the parameter space.  Hence, the scanner often does not fully explore large regions with equal or worse likelihood. This happens especially near the experimental bounds.  Furthermore, although high couplings are possible between active and sterile neutrino sector, they often lie in the symmetry protected regime, as described in Section~\ref{sec:symmetries} and/or require severe fine-tuning of the parameters. Again, exploring these regions turned out to be challenging.

Therefore, we designed and performed a large set of targeted scans to fully saturate the experimental bounds, the list of which can be found in Table~\ref{tab:scans}. The design strategies we adopted for these targeted scans can be summarised as follows.

First, all targeted scans were performed using a differential RHN model, where the parameter $M_2$ is replaced by $\Delta M_{21}$, with a logarithmic prior. This allows the exploration of the symmetry protected region, with near degenerate masses for two right-handed neutrinos.

Most of the experimental bounds occur at high couplings, thus in order to encourage the scanner to explore the high coupling regions, we added an artificial likelihood to the scan to drive the scan to the unexplored boundaries. To saturate the experimental bounds for each coupling $U_{\alpha I}^2$, $\alpha=e,\mu,\tau$, different targeted scans were performed using this \textit{coupling slide} likelihood on each of the couplings, of the form $s\log U_{\alpha I}^2 + m \log M_I$. Table~\ref{tab:scans} shows the parameter that is optimised in each scan, $\alpha$, and the coefficients, $(s,m)$. This contribution was later removed from the data in the postprocessing stage.

The targeted scans were further split along the $M_I$ axis following the limits of the various experimental constraints (mostly from direct searches). This ensures that each coupling (with the selection above) saturates the most relevant experimental upper bound in each mass range. Additionally, some scans used different values of $\Delta M_{21}$ and/or $m_{\nu_0}$ to further force the scan into fine-tuned regions of parameter space. The ranges used for $M_I$, $\Delta M_{21}$ and $m_{\nu_0}$ for each scan are specified in Table~\ref{tab:scans}.

A similar strategy was used to saturate the BBN bound at low couplings. Three scans were performed for each hierarchy, with slide coefficients $(s,m)=(-0.5,-0.5)$ on each coupling  $U_{\alpha I}^2$, $\alpha=e,\mu,\tau$ . To further optimise on low couplings, these scans were performed for fixed $m_{\nu_0} = 10^{-10}$ and a narrow range on Im$\omega$ $\in [-0.5,0.5]$. With these settings the BBN bound was fully saturated in the explored mass range.

We found that some of the experimental likelihoods provide positive contributions to the total likelihood in specific regions of the parameter space. This forced the scan towards those regions, leaving others unexplored. Although this is a rather interesting feature, and will be discussed in detail later, it prevented a thorough exploration of the full parameter space. We thus chose to remove the likelihood contribution of $R^K_{e\mu}$ from the total likelihood that drives the scan, adding it later in postprocessing. Other likelihoods with positive contributions, $\Gamma_{\rm{inv}}$,  CKM and $R^\tau_{e\mu}$, tended to force the scan towards large $U_{\tau I}^2$ couplings. Although desirable to saturate the limits, this also left regions with low $\tau$ coupling undersampled. Thus, a cut on the coupling $U_{\tau I}^2$ was enforced in some scans to fully sample all regions.

\begin{table*}[t]
 \centering
 \begin{tabular}{ccccccc}
  \toprule
  $M_1$ [GeV] & $\Delta M_{21}$ [GeV] & $m_{\nu_0}$ [eV] &$\alpha$ & $(s,m)$  & Hierarchy & Other\\
  \midrule
  $[0.1,0.3162]$ & $[10^{-10},0.1]$ & $[10^{-7},0.23]$ & $(e,\mu,\tau)$ & $(0.5,-0.5)$ & N, I & $U_{\tau I}^2 < 10^{-4}$ \\
  $[0.1,0.4217]$ & $[10^{-10},0.1]$ & $[10^{-7},0.23]$ & $(e,\mu,\tau)$ & $(0.5,-0.5)$ & N, I & $U_{\tau I}^2 < 10^{-4}$ \\
  $[0.3162,2.0]$ & $[10^{-10},0.1]$ & $[10^{-7},0.23]$ & $(e,\mu,\tau)$ & $(0.5,0.5)$ & N, I & $U_{\tau I}^2 < 10^{-4}$ \\
  $[2.0,60]$ & $[10^{-20},10^{-10}]$ & $[10^{-6},0.23]$ & $(e,\mu,\tau)$ & $(0.5,0)$ & N, I & $U_{\tau I}^2 < 10^{-4}$, flat $m_{\nu_0}$ prior \\
  $[2.0,60]$ & $[10^{-20},10^{-10}]$ &  $10^{-4},10^{-5},10^{-6}$ & $(e,\mu,\tau)$ & $(0.5,0)$ & N, I & fixed $m_{\nu_0}$  \\
  $[60,500]$ & $[10^{-20},10^{-10}]$ & $[10^{-6},0.23]$ & $(e,\mu,\tau)$ & $(0.7,0.25)$ & N, I & $U_{\tau I}^2 < 10^{-4}$, flat $m_{\nu_0}$ prior \\
  $[60,500]$ & $[10^{-20},10^{-10}]$ & $10^{-4},10^{-5},10^{-6}$ & $(e,\mu,\tau)$ & $(0.7,0.25)$ & N, I & fixed $m_{\nu_0}$ \\
  $[0.06,0.14]$ & $[10^{-10},0.1]$ & $[10^{-7},0.23]$ & $(e,\mu)$ & $(0.5,-0.5)$ & N, I & $U_{\tau I}^2 < 10^{-4}$, flat $m_{\nu_0}$ prior \\
  $[60,500]$ & $[10^{-20},10^{-10}]$ & $[10^{-6},0.23]$ & $(e,\mu)$ & $(0.7,0.25)$ & I & flat $m_{\nu_0}$ prior \\
  $[0.14,0.2]$ & $[10^{-10},0.1]$ & $[10^{-7},0.23]$ & $(e)$ & $(0.5,-0.5)$ & N,I & $U_{\tau I}^2 < 10^{-4}$, flat $m_{\nu_0}$ prior \\
  $[0.2,0.4217]$ & $[10^{-10},0.1]$ & $[10^{-7},0.23]$ & $(e)$ & $(0.5,-0.5)$ & N,I & $U_{\tau I}^2 < 10^{-4}$, flat $m_{\nu_0}$ prior \\
  $[0.14,0.3162]$ & $[10^{-10},0.1]$ & $[10^{-7},0.23]$ & $(\mu)$ & $(0.5,-0.5)$ & N,I & $U_{\tau I}^2 < 10^{-4}$, flat $m_{\nu_0}$ prior \\
  $[0.1,0.3162]$ & $[10^{-10},0.1]$ & $[10^{-7},0.23]$ & $(\tau)$ & $(0.5,-0.5)$ & N, I & - \\
  $[0.1,0.4217]$ & $[10^{-10},0.1]$ & $[10^{-7},0.23]$ & $(\tau)$ & $(0.5,-0.5)$ & N, I & - \\
  $[0.175,0.3611]$ & $[10^{-20},10^{-10}]$ & $[10^{-2},0.23]$ & $(\tau)$ & $(0.5,0.5)$ & N, I & - \\
  $[0.25,0.3611]$ & $[10^{-20},10^{-10}]$ & $[10^{-2},0.23]$ & $(\tau)$ & $(0.5,0.5)$ & N, I & - \\
  $[0.25,0.4]$ & $[10^{-20},10^{-10}]$ & $[10^{-2},0.23]$ & $(\tau)$ & $(1.0,0)$ & N, I & - \\
  $[0.3611,0.4492]$ & $[10^{-10},0.1]$ & $[10^{-2},0.23]$ & $(\tau)$ & $(0.5,-0.5)$ & N, I & - \\
  $[0.3611,0.4492]$ & $[10^{-20},10^{-10}]$ & $[10^{-2},0.23]$ & $(\tau)$ & $(0.5,0.5)$ & N, I & - \\
  $[0.4,0.5]$ & $[10^{-10},0.1]$ & $[10^{-7},0.23]$ & $(\tau)$ & $(0.5,-0.5)$ & N, I & - \\ 
  $[0.3162,2.0]$ & $[10^{-10},0.1]$ & $[10^{-7},0.23]$ & $(\tau)$ & $(0.5,0.5)$ & N, I & - \\ $[0.3162,1.4]$ & $[10^{-10},0.1]$ & $[0.03,0.23]$ & $(\tau)$ & $(0.5,0.5)$ & I & $U_{\tau I}^2 < 10^{-3}$ \\
  $[1.0,1.5]$ & $[10^{-7},0.01]$ & $[0.03,0.23]$ & $(\tau)$ & $(0.5,0.5)$ & I & $U_{\tau I}^2 < 10^{-3}$ \\
  $[1.25,1.45]$ & $[10^{-20},10^{-10}]$ & $[0.01,0.23]$ & $(\tau)$ & $(0.5,0.5)$ & I & -\\
  $[1.4,1.78]$ & $[10^{-7},0.01]$ & $[0.03,0.23]$ & $(\tau)$ & $(0.5,0.5)$ & I & $U_{\tau I}^2 < 10^{-3}$ \\
  $[1.65,1.85]$ & $[10^{-20},10^{-10}]$ & $[0.01,0.23]$ & $(\tau)$ & $(0.5,0.5)$ & I & -\\
  $[1.25,1.45]$ & $[10^{-20},10^{-10}]$ & $[0.01,0.23]$ & $(\tau)$ & $(0.5,0.5)$ & I & -\\
  $[2.0,60]$ & $[10^{-20},10^{-10}]$ & $[10^{-6},0.23]$ & $(\tau)$ & $(0.5,0)$ & N, I & flat $m_{\nu_0}$ prior \\
  $[60,500]$ & $[10^{-20},10^{-10}]$ & $[10^{-6},0.23]$ & $(\tau)$ & $(0.7,0.25)$ & N, I & flat $m_{\nu_0}$ prior \\
  \bottomrule
 \end{tabular}
 \caption{Set of targeted scans performed for normal (N) and inverted (I) hierarchy in addition to the full parameter scans. Parameters not shown in this table are taken as in Table~\ref{tab:scanpars}.}
 \label{tab:scans}
\end{table*}

The adopted strategy for scanning was driven by the need to fully sample the parameter space. The results from all the diverse scans were combined into a single dataset after some postprocessing (see below).  This does not pose a problem for the statistical interpretation, since we are interested in the profile likelihood, which only becomes more accurately estimated when adding additional chains.

\subsection{Scanning framework}

To perform the detailed scans, we make use of the \GB framework, as described in Appendix \ref{app:gambit}, and the differential evolution scanner \diver, version 1.0.4~\cite{ScannerBit}, which is a self-adaptive sampler, capable of sampling the profile likelihood more efficiently than other scanners. We choose a population size of \fortran{NP} = 19200 and a convergence threshold of \fortran{convthresh} = $10^{-10}$. After some tests, we have concluded that the aggressive $\lambda$jDE setting in \diver provides an improvement on the sampling of the parameter space, since it is more suited for sampling fine-tuned regions.

These scanner settings, including the very low convergence threshold, together with the scanning strategy described above, ensure a thorough exploration of the parameter space, albeit at the price of CPU time. Despite the fact that none of the observables used required heavy computation or simulations, most scans took between 2 and 10 hours of running time on a large number of supercomputer cores varying between 250 and 780. All tests and scans were carried out across several supercomputer facilities, including the MareNostrum supercluster in Barcelona, Marconi in Bologna, LISA/Surfsara through the University of Amsterdam and Prometheus in Krakow.

\subsection{Data postprocessing}

Upon completion of the scans, a number of postprocessing tasks were performed on the data to prepare it for plotting. As previously mentioned, the first of these tasks was to remove the artificial \textit{coupling slide} likelihood used to drive the scans to high couplings.

Due to the large amount of scans performed and the low convergence threshold used, the size of the samples surpassed 1TB for each hierarchy, rendering them unmanageable for most plotting routines. We hence performed a few operations on the scan results prior to combining them. With the target of showing profile likelihood plots in the $M_I$ vs $U_{\alpha I}^2$ planes, we hence extracted a subset of the data points optimised in these planes, with a resolution of $10^{-5}$. Since most scans were targeted to saturate the limits for a particular coupling (see Table~\ref{tab:scans}) we perform this reduction of the data in the respective mass vs coupling two-dimensional planes. The combined set will hence be optimised for all couplings. Additionally, and independent reduction of the data is performed on the planes $m_{\nu_0}$ vs $U_{\alpha I}^2$, since we intent to study the effect of $m_{\nu_0}$ cuts on the coupling limits.

The flavour label of the heavy neutrinos is arbitrary, and the experimental constraints on a heavy neutrino with a given mass cannot depend on the labelling. However, for reasons explained in more detail in appendix \ref{FakeSymmetryPoints}, the scanning strategy outlined in Sec.~\ref{subsec:TargetedScans} introduces a bias that suggests that the constraints differ for $N_1$, $N_2$ and $N_3$.
Hence, to remove this bias in the labels, after combining the reduced datasets for all the scans, we conduct a symmetrization procedure over the combined datasets. We therefore symmetrize over $M_I$ as well as $U_{\alpha I}$, which will increase the size of the datasets six fold.

Lastly, in order to compare with the $n=2$ case, two further datasets were obtained, for normal and inverted ordering, where the data points are required to lie in the symmetry protected region.

Out of the incalculable amount of data points we collected through our scanning procedures, a total of 40.7 million valid data samples were used for plotting. Of which 11M correspond to normal hierarchy and 10M for inverted hierarchy, optimised on $M_I$ vs $U_{\alpha I}^2$ planes, and 9.9M for normal and 9.7M for inverted hierarchy, optimised on $m_{\nu_0}$ vs  $U_{\alpha I}^2$ planes. The datasets with points in the symmetry protected region have over 71k and 20k valid data samples for normal and inverted hierarchy, respectively. These samples can be found in Zenodo~\cite{Zenodo_RHN}.

\subsection{Capped likelihood}
\label{sec:capped} 

The figures in this article show the so-called \textit{capped} profile likelihood (unless stated otherwise), which is defined in each of the scanned point to an \textit{equal or worse} fit than the SM:  $\mathcal{L} = \min[\mathcal{L}_{\rm{SM}}, \mathcal{L}_{\rm{RHN}}]$. It can thus be interpreted as exclusion-only likelihood. Capped likelihoods have been used in previous studies, particularly in the context of collider searches~\cite{ColliderBit,EWMSSM}. The rationale behind the use of this \textit{capped} likelihood is the presence of positive (above SM) contributions to the log likelihood from various observables.  Importantly, these `excesses' would not show up as localized features in the total profile likelihood, as there is enough of freedom to add points in the $M_I-U_{\alpha I}^2$ plane to find $M_J$, $J\neq I$ with values that would saturate the excess likelihood.  Thus a very large fraction of the parameter points would have the maximum allowed likelihood from the combination of all excesses. This effect forces to separate the exclusion studies from the possible signal observation. Thus, in most of the paper, we use the  \textit{capped} likelihood to present parameter constraints.  
The excess likelihoods will be discussed separately in Sec.~\ref{sec:excesses}.

\section{Results and discussion}
\label{sec:results}

\subsection{General constraints on the RHN mass and mixing}

\begin{figure*}[t]
  \centering
  \includegraphics[width=0.45\linewidth]{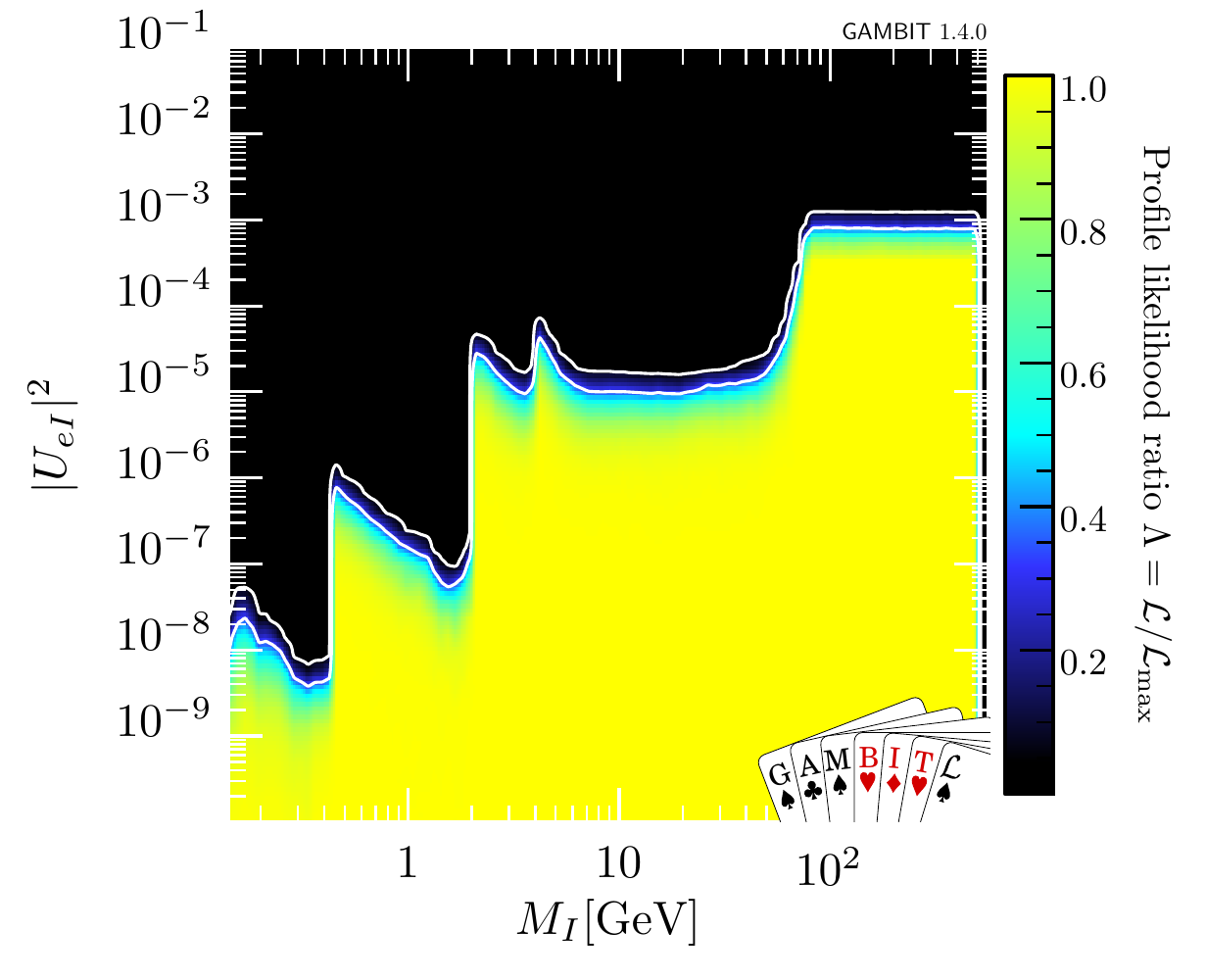}
  \includegraphics[width=0.45\linewidth]{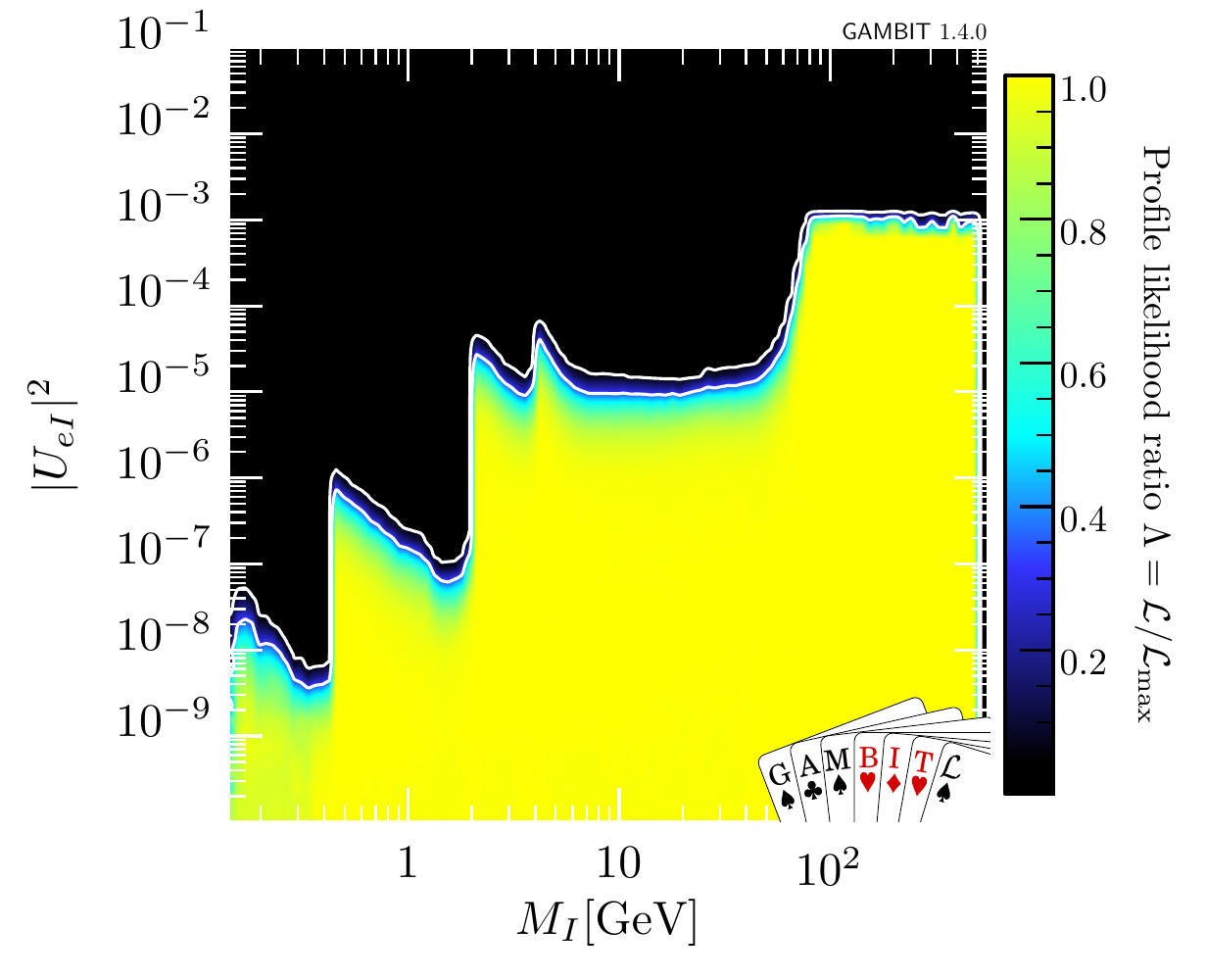}
  \caption{Profile likelihood in $M_I$ vs $U_{eI}^2$ plane for normal (left) and inverted hierarchy (right). Tables with the 90\% and 95\% CLs for both hierarchies can be found in Zenodo~\cite{Zenodo_RHN}.}
  \label{fig:M_Ue_capped}
\end{figure*}

\begin{figure*}[t]
  \centering
  \includegraphics[width=0.45\linewidth]{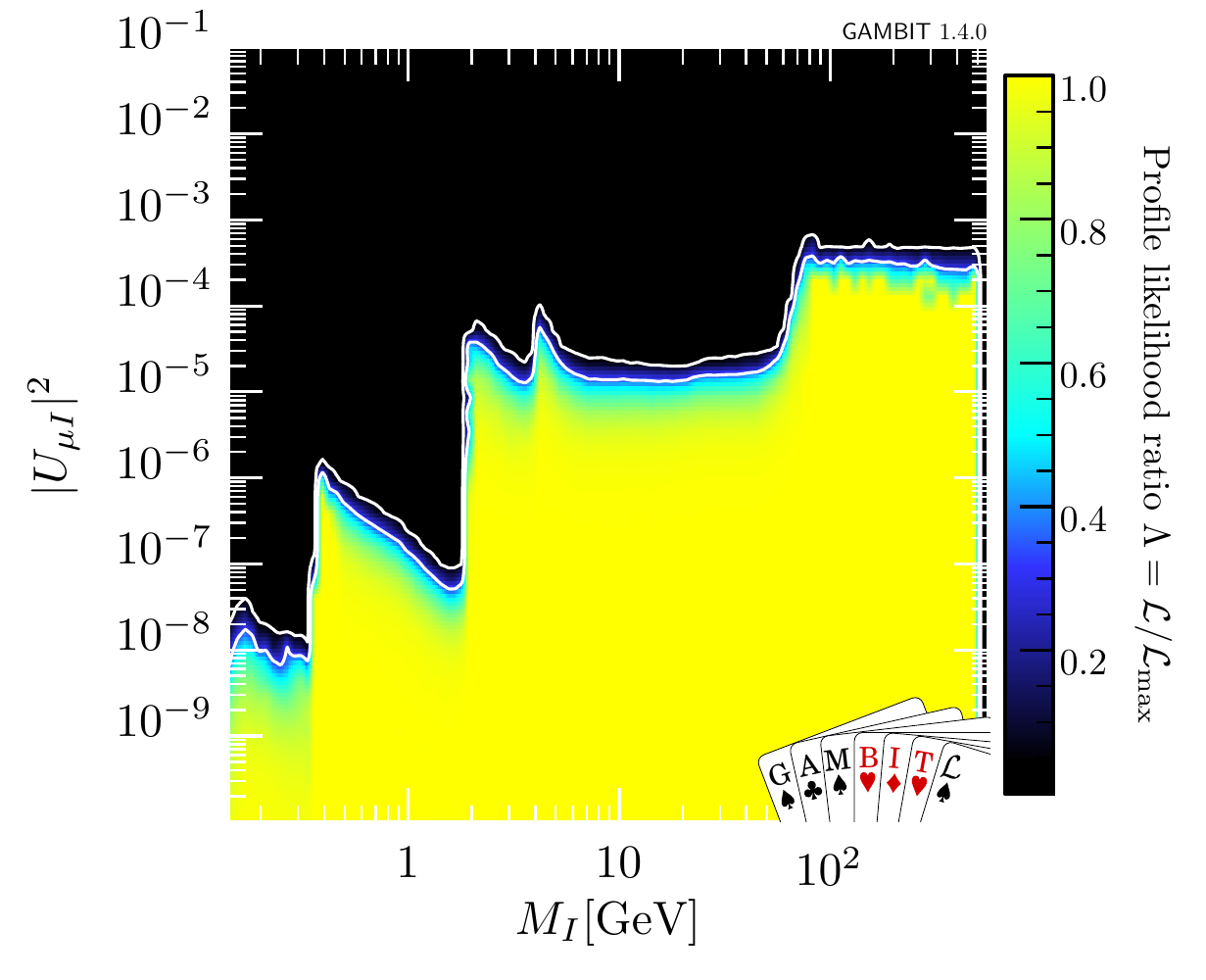}
  \includegraphics[width=0.45\linewidth]{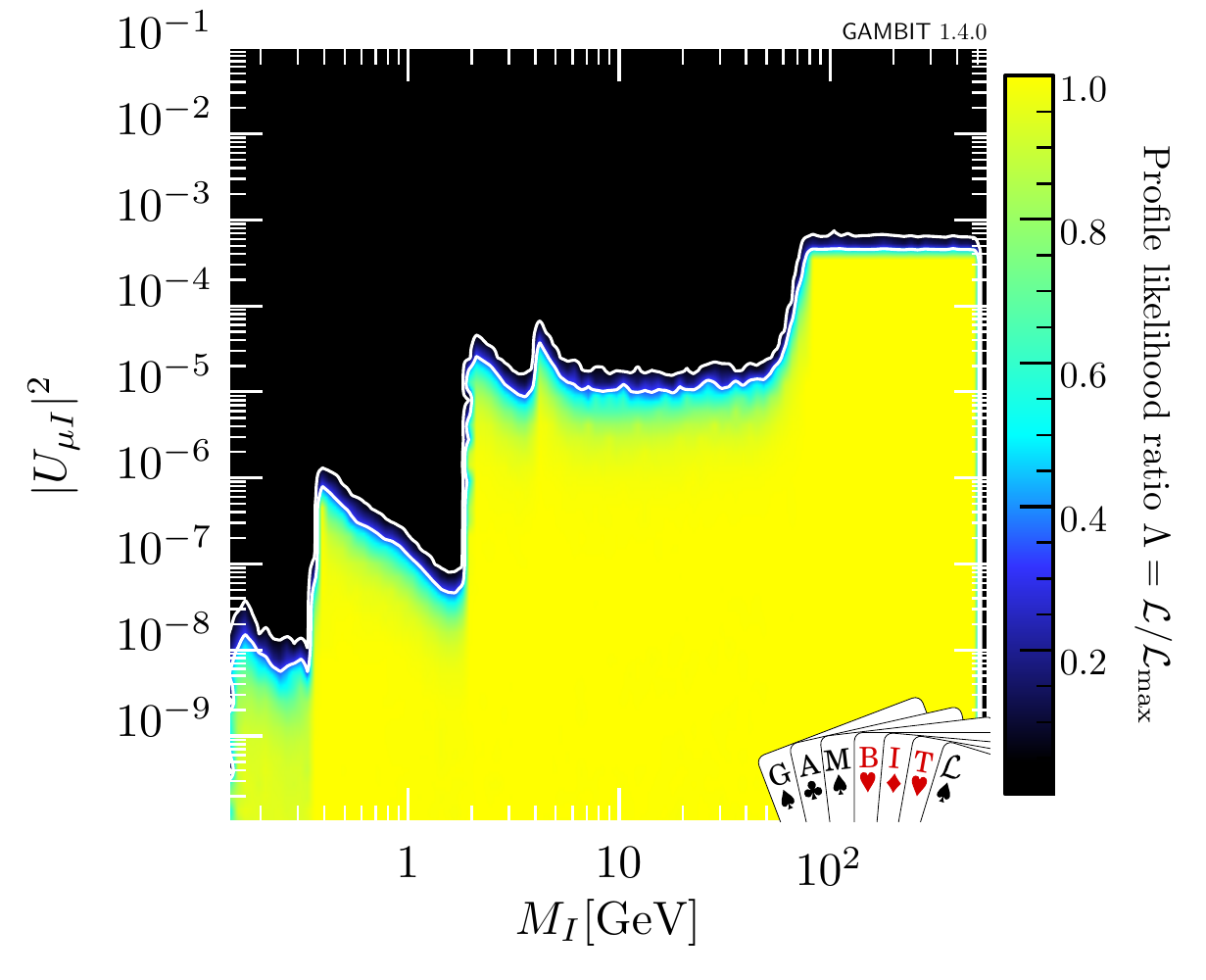}
  \caption{Profile likelihood in $M_I$ vs $U_{\mu I}^2$ plane for normal (left) and inverted hierarchy (right). Tables with the 90\% and 95\% CLs for both hierarchies can be found in Zenodo~\cite{Zenodo_RHN}.}
  \label{fig:M_Umu_capped}
\end{figure*}

\begin{figure*}[t]
  \centering
  \includegraphics[width=0.45\linewidth]{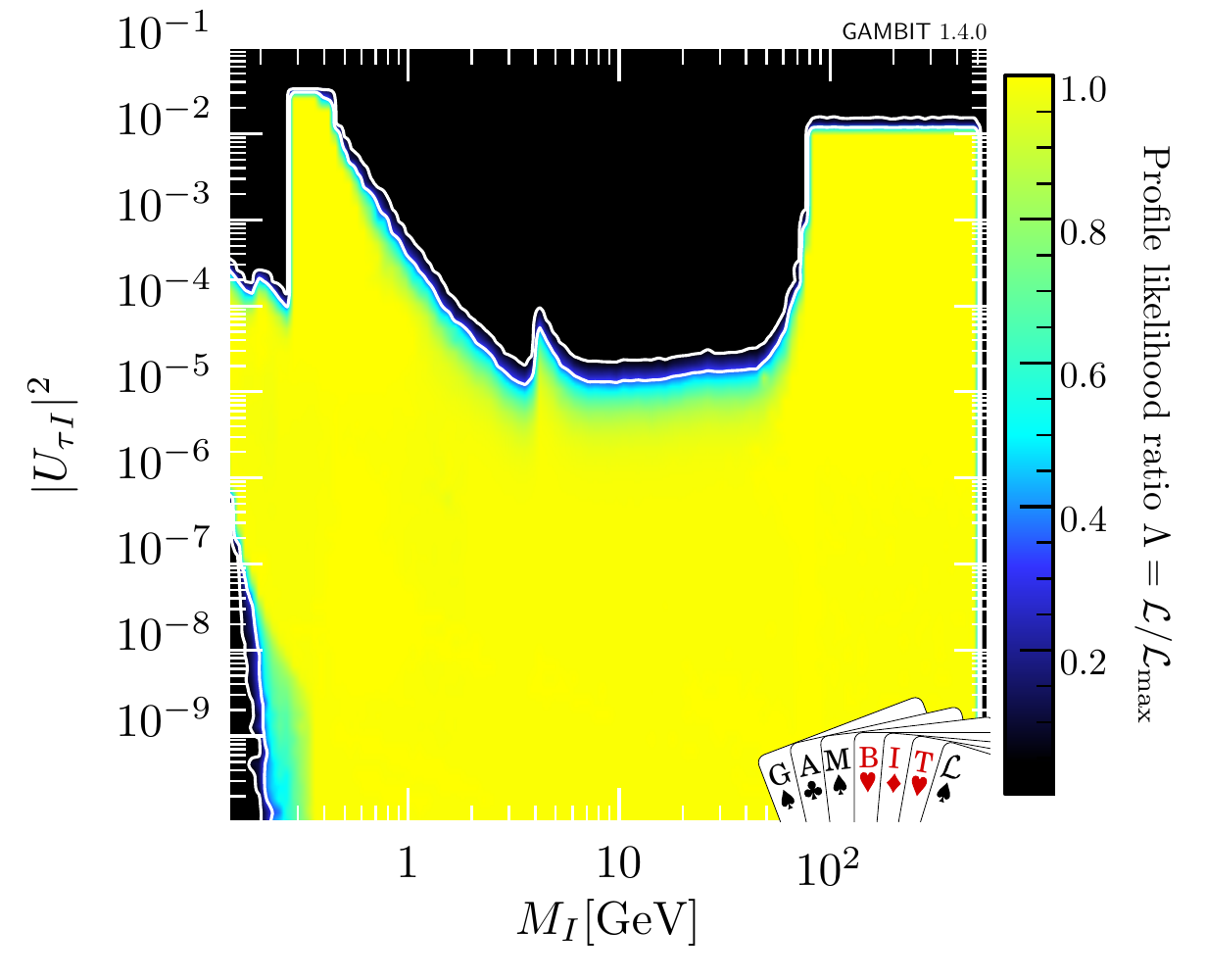}
  \includegraphics[width=0.45\linewidth]{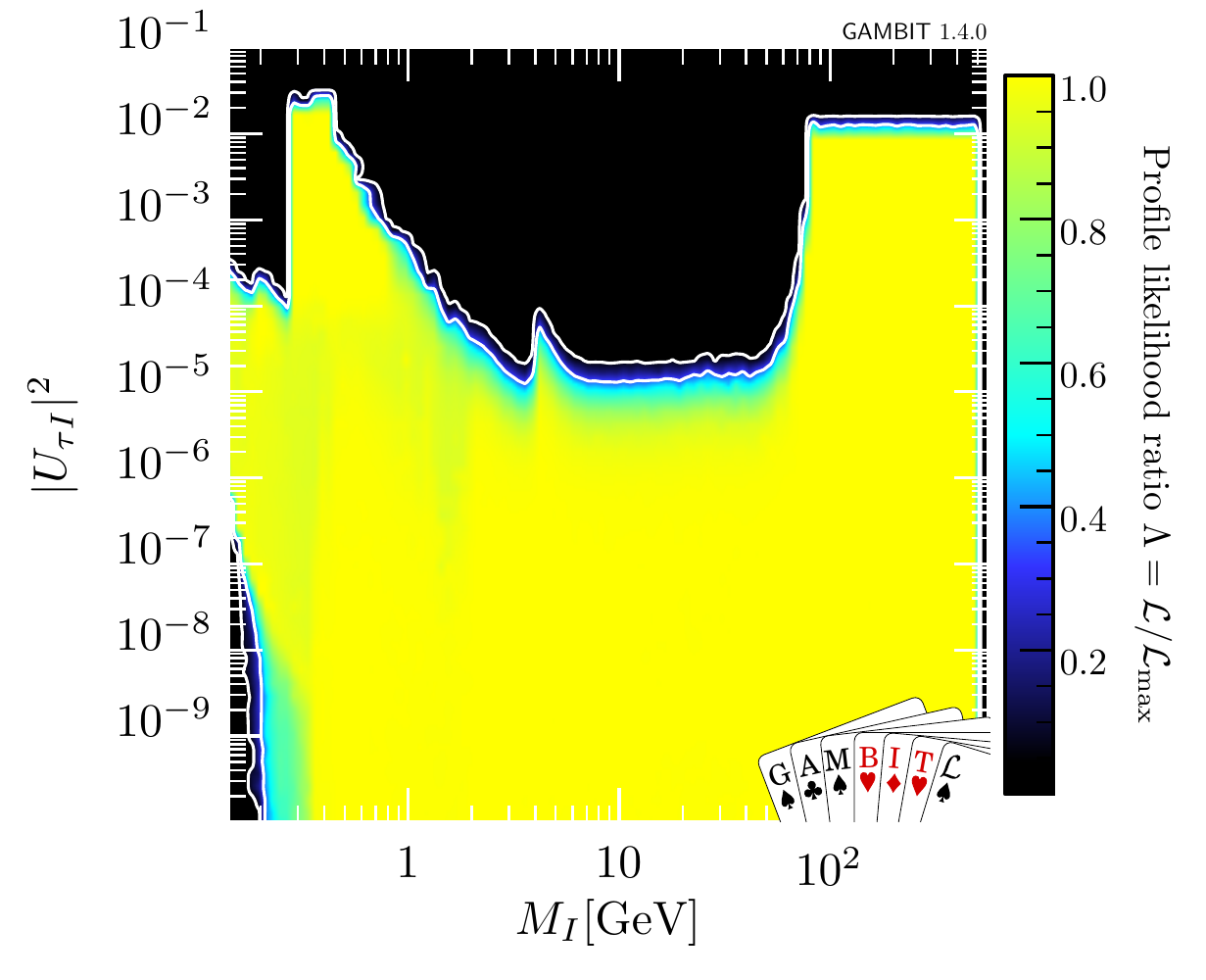}
  \caption{Profile likelihood in $M_I$ vs $U_{\tau I}^2$ plane for normal (left) and inverted hierarchy (right). Tables with the 90\% and 95\% CLs for both hierarchies can be found in Zenodo~\cite{Zenodo_RHN}.}
  \label{fig:M_Utau_capped}
\end{figure*}

\begin{figure*}[t]
  \centering
  \includegraphics[width=0.45\linewidth]{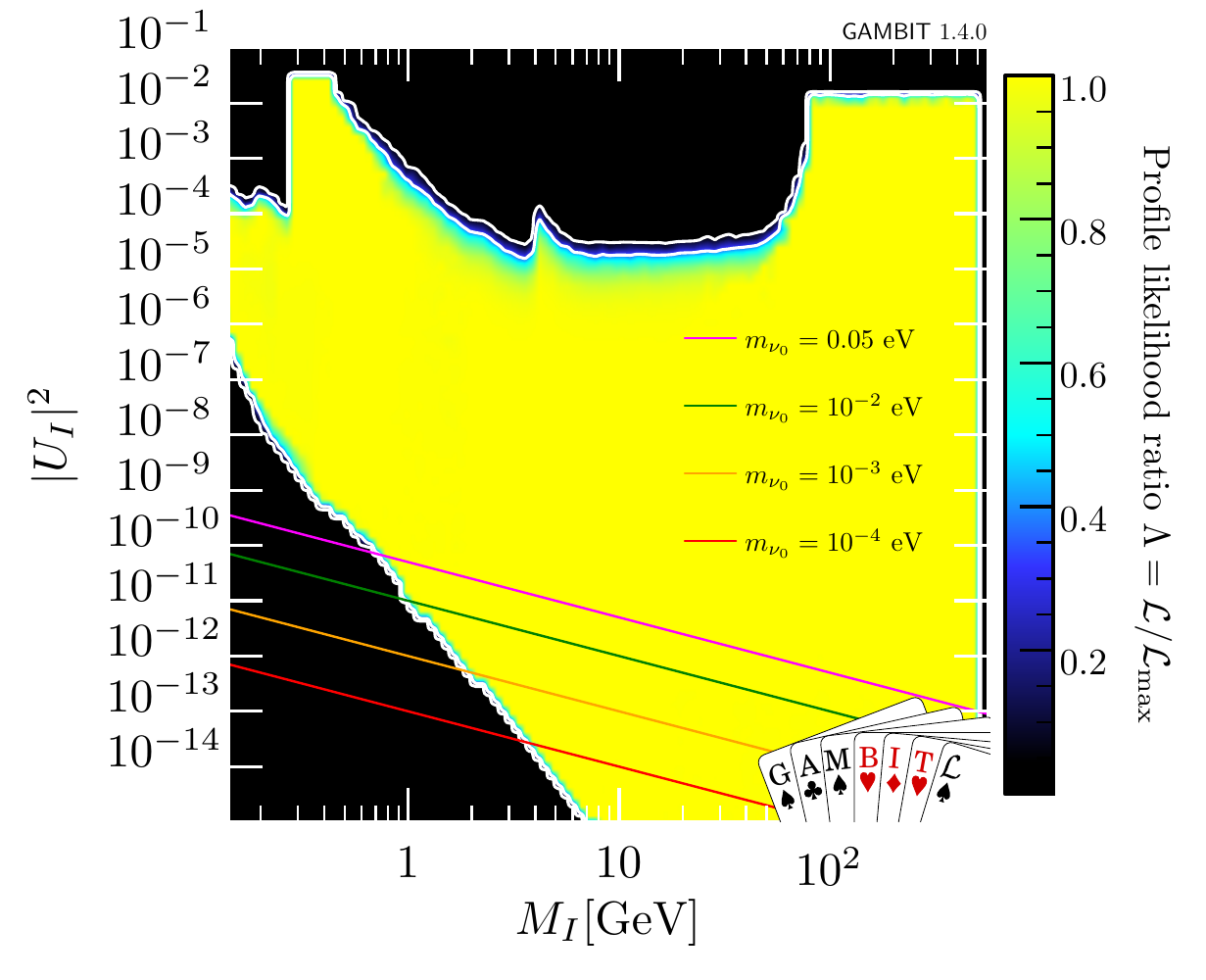}
  \includegraphics[width=0.45\linewidth]{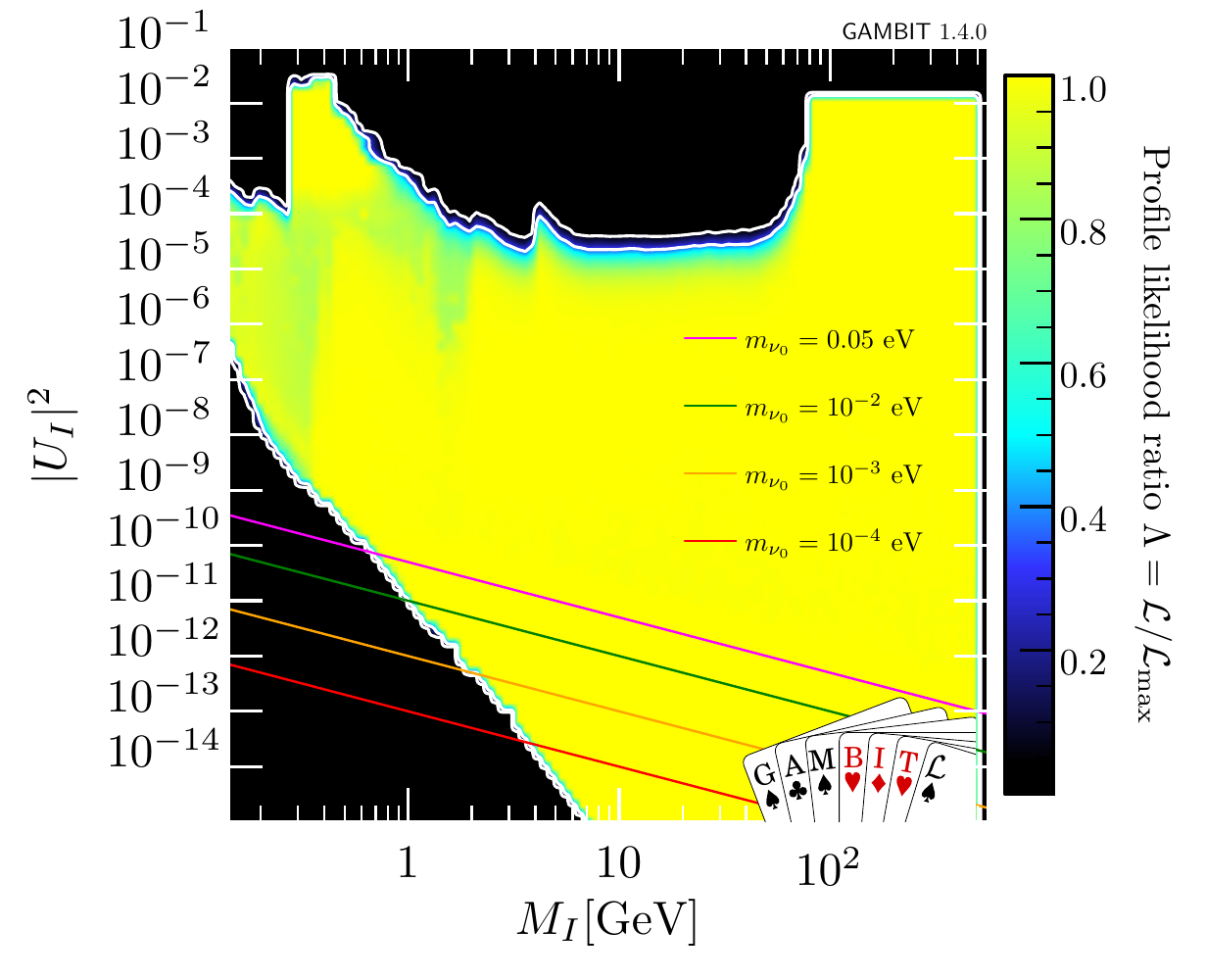}
  \caption{Profile likelihood in $M_I$ vs $U_I^2$ plane for normal (left) and inverted hierarchy (right). Overlaid are the lowest limits for various values of $m_{\nu_0}$ \cite{Drewes:2019mhg}. Tables with the 90\% and 95\% CLs for both hierarchies can be found in Zenodo~\cite{Zenodo_RHN}.}
  \label{fig:M_U_capped}
\end{figure*}

\begin{figure*}[t]
  \centering
  \includegraphics[width=0.45\linewidth]{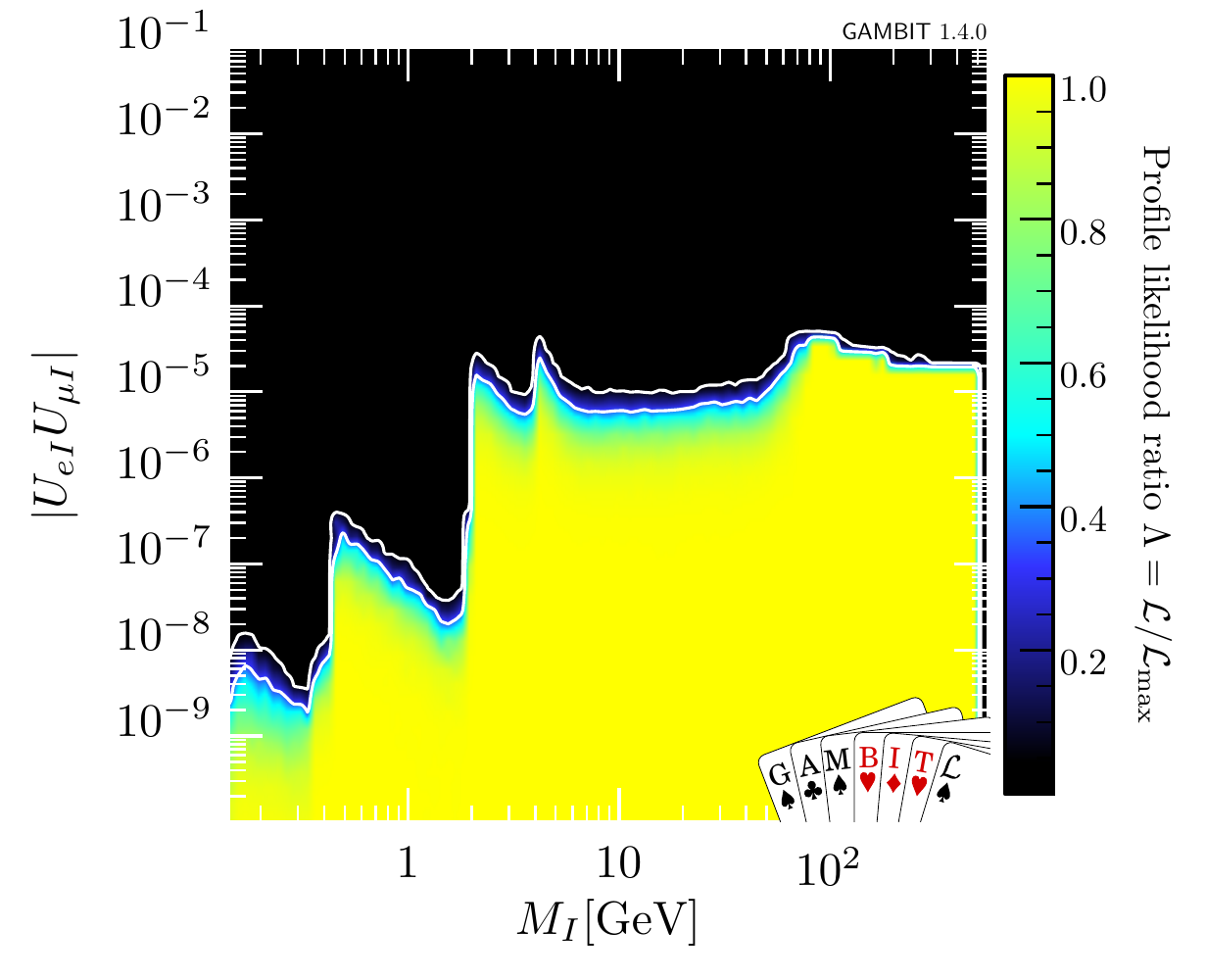}
  \includegraphics[width=0.45\linewidth]{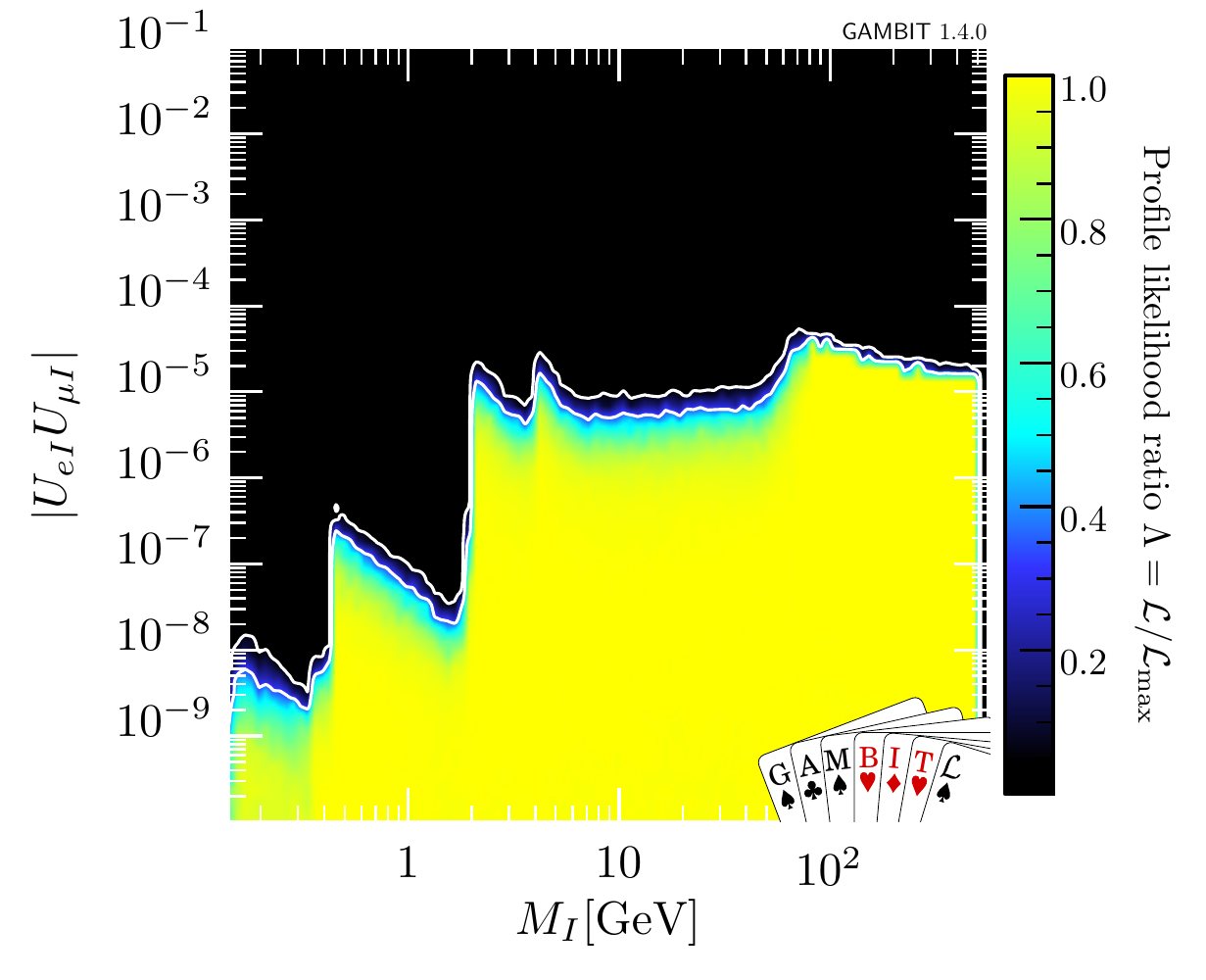}
  \caption{Profile likelihood in $M_I$ vs $|U_{eI}U_{\mu I}|$ plane for normal (left) and inverted hierarchy (right). Tables with the 90\% and 95\% CLs for both hierarchies can be found in Zenodo~\cite{Zenodo_RHN}.}
  \label{fig:M_UeUmu_capped}
\end{figure*}

\begin{figure*}[t]
  \centering
  \includegraphics[width=0.45\linewidth]{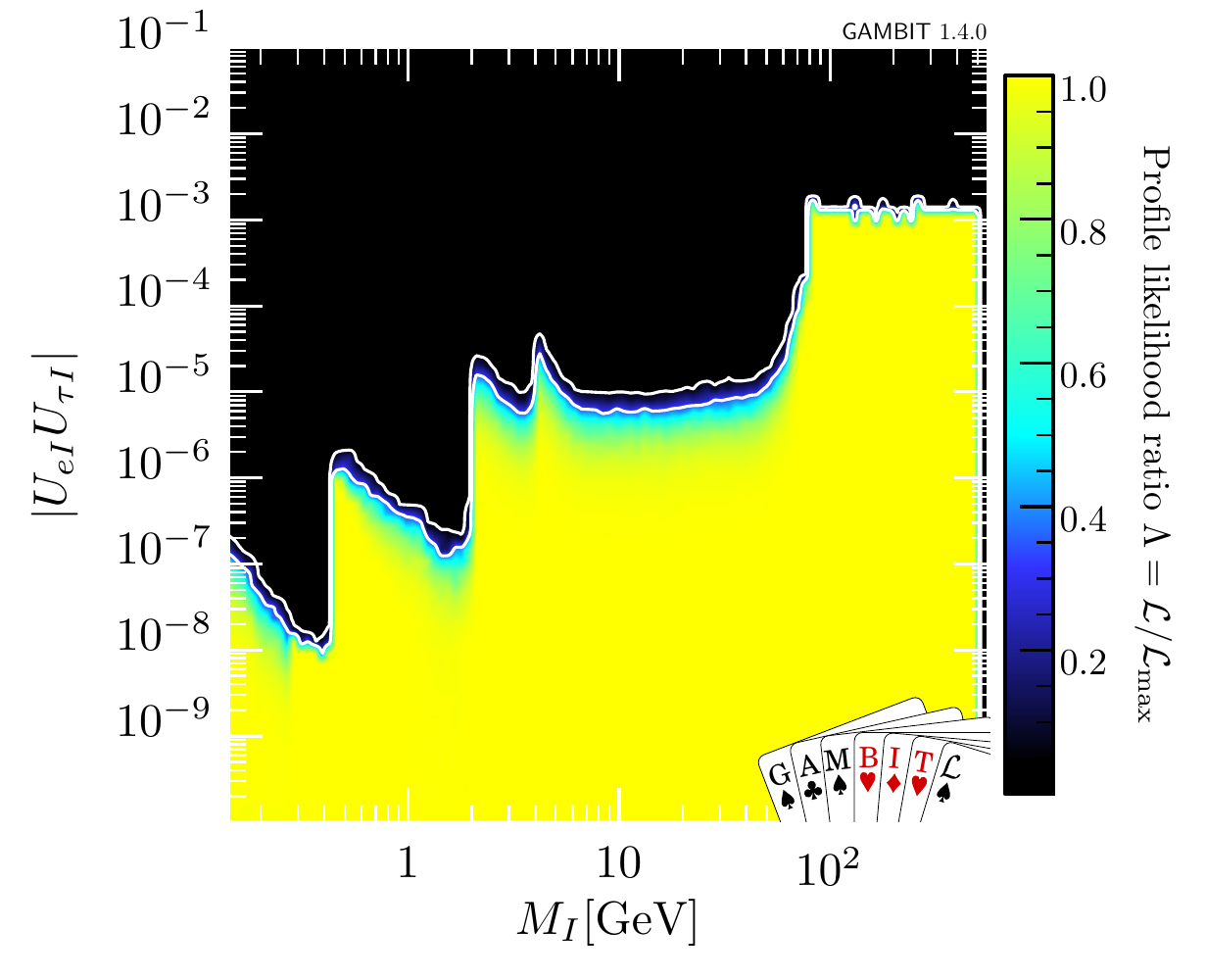}
  \includegraphics[width=0.45\linewidth]{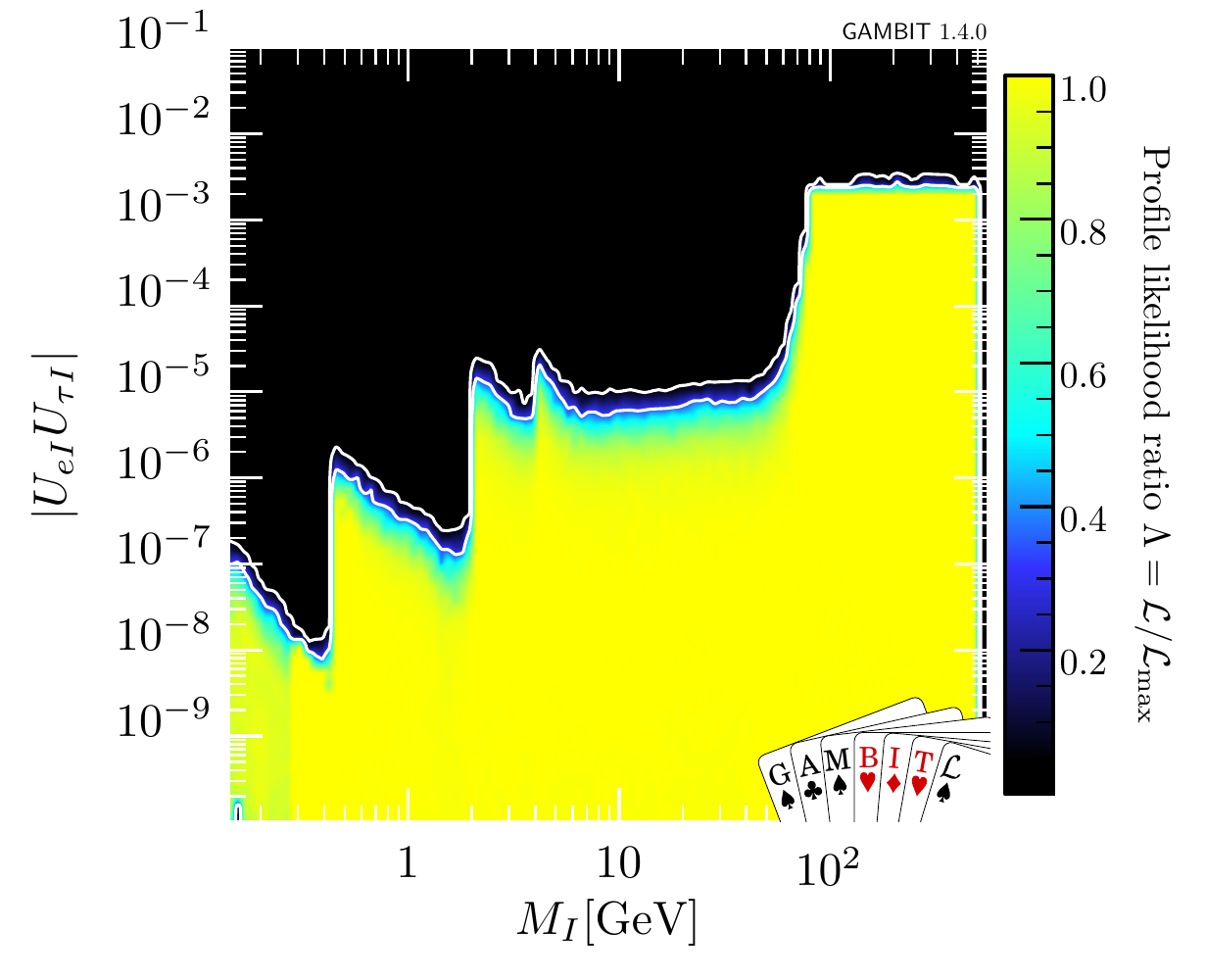}
  \caption{Profile likelihood in $M_I$ vs $|U_{eI}U_{\tau I}|$ plane for normal (left) and inverted hierarchy (right). Tables with the 90\% and 95\% CLs for both hierarchies can be found in Zenodo~\cite{Zenodo_RHN}.}
  \label{fig:M_UeUtau_capped}
\end{figure*}

\begin{figure*}[t]
  \centering
  \includegraphics[width=0.45\linewidth]{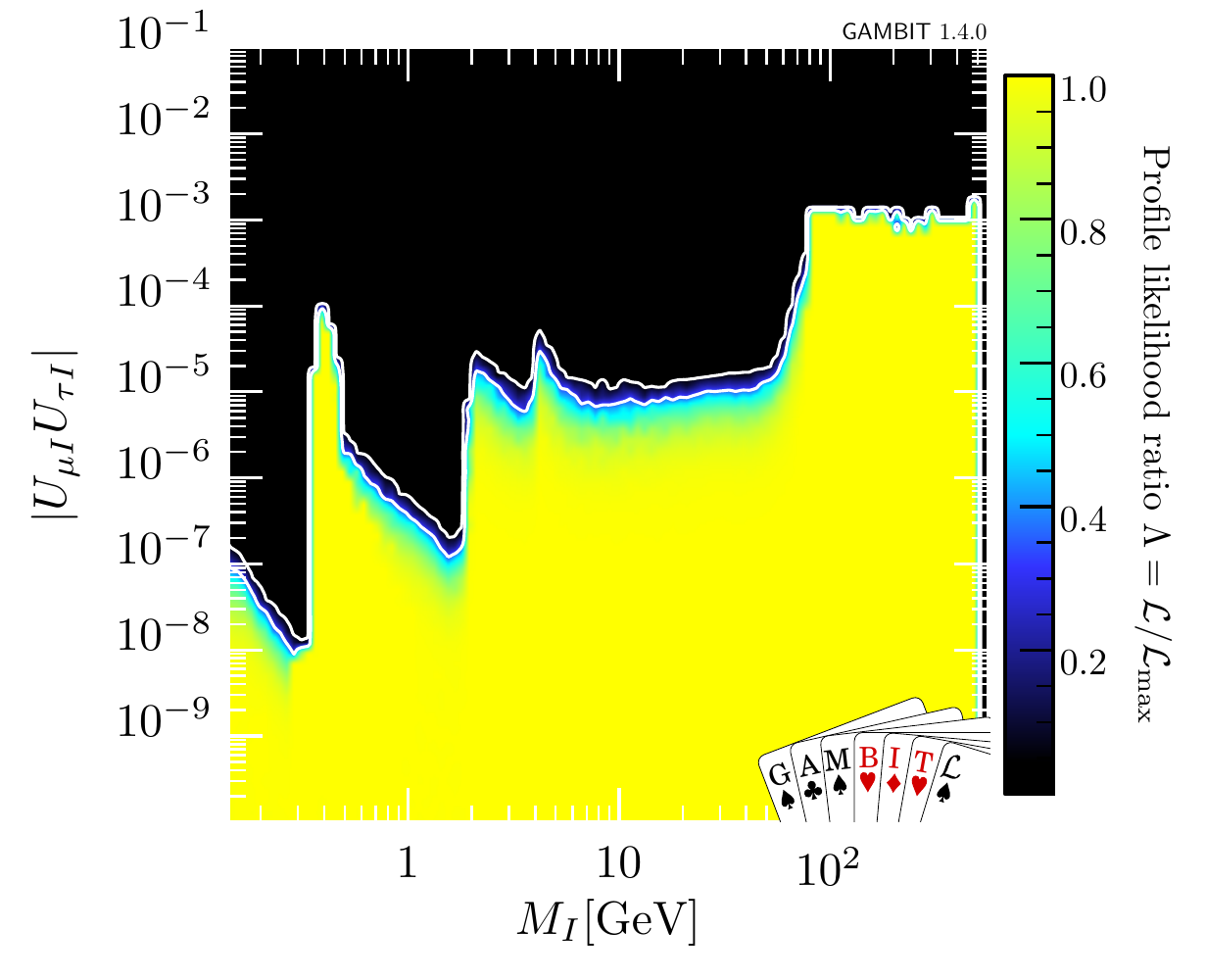}
  \includegraphics[width=0.45\linewidth]{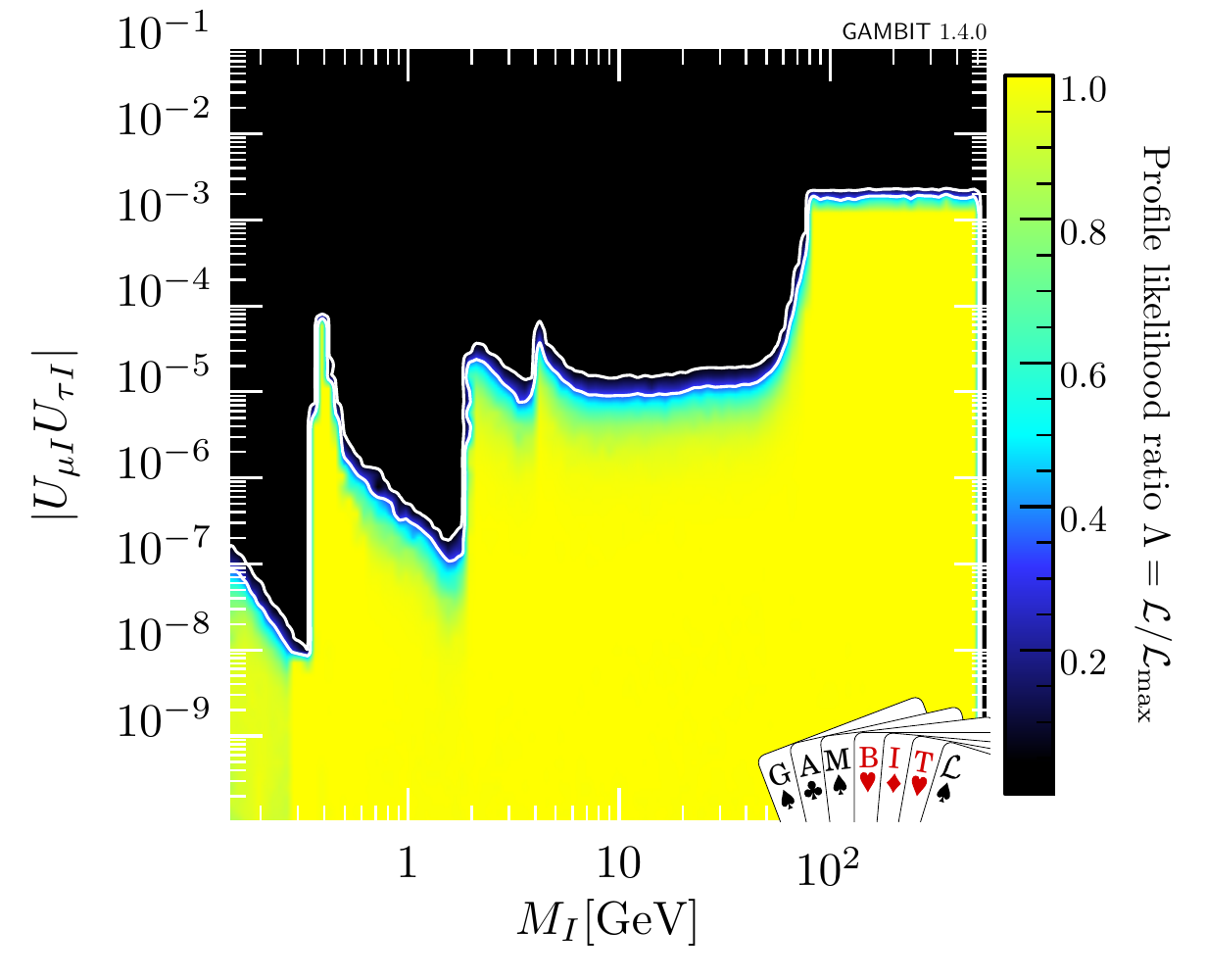}
  \caption{Profile likelihood in $M_I$ vs $|U_{\mu I}U_{\tau I}|$ plane for normal (left) and inverted hierarchy (right). Tables with the 90\% and 95\% CLs for both hierarchies can be found in Zenodo~\cite{Zenodo_RHN}.}
  \label{fig:M_UmuUtau_capped}
\end{figure*}

The constraints are shown in Figs.~\ref{fig:M_Ue_capped}-\ref{fig:M_U_capped} for the couplings $U_{\alpha I}^2$ to the active neutrino flavours $\alpha = (e,\mu,\tau)$, as well as their combination $U_I^2 = \sum_\alpha U_{\alpha I}^2$, as functions of the heavy neutrino masses $M_I$.  Here, the second index can refer to any of the heavy neutrino flavours $I=(1,2,3)$, because their labelling is not physical. Figs.~\ref{fig:M_UeUmu_capped}-\ref{fig:M_UmuUtau_capped} show the combinations of couplings $U_{\alpha I}U_{\beta I}$ with $\alpha \neq \beta$. The allowed profile likelihood regions are flat for most of the parameter space, in particular for small couplings $U_{\alpha I}^2$, and drop smoothly at high couplings following the relevant upper limits. The white lines around the experimental limits mark the 1$\sigma$ and 2$\sigma$ contours, which are estimated assuming Wilks' theorem with 2 degrees of freedom\footnote{All profile likelihood plots were created using \pippi~\cite{pippi}.}. 

The largest values of mixings $U_{\alpha I}^2$ and $U_{\alpha I}U_{\beta I}$ for all flavours are allowed for $M_I$ above the masses of the weak gauge bosons. 
In this regime the direct searches at colliders are sub-dominant, and the heavy neutrino properties are primarily constrained from above due to electroweak precision observables, lepton flavour violation and CKM constraints.
The upper limits on the couplings $U_{\alpha I}^2$ and $U_{\alpha I}U_{\beta I}$ within $2\sigma$ of the highest likelihood for each hiearchy and flavour in the high mass region can be found in Table~\ref{tab:upperlimits}. It can be readily noticed that the upper limits for the $\tau$ couplings is much larger than for the other two flavours, which can be understood because the limits from EWPO and LFV are stronger for $e$ and $\mu$ (see also Sec.~\ref{sec:excesses}). In particular, the combination $U_{eI}U_{\mu I}$ has the smallest of upper limits, as shown as well in Fig.~\ref{fig:M_UeUmu_capped}, due to strong constraints from LFV observables, specifically $\mu \to e\gamma$ and $\mu - e$ conversion (see Figs.~\ref{fig:lnL_U121_lnL_l2lgamma_NH} and \ref{fig:lnL_U121_lnL_mu2e_NH} in Appendix ~\ref{app:partial}).

\begin{table}[h]
  \centering
  \begin{tabular}{ccc}
    \toprule
    \textbf{Hierarchy} & \textbf{Coupling} & \textbf{Upper limit ($2\sigma$)}\\
    \midrule
    N & $U_{eI}^2$ &  $4.92\times 10^{-4}$ \\
    N & $U_{\mu I}^2$ &  $2.42\times 10^{-4}$\\
    N & $U_{\tau I}^2$ &  $9.59\times 10^{-3}$ \\
    N & $U_{eI}U_{\mu I}$ &  $3.49\times 10^{-5}$ \\
    N & $U_{eI}U_{\tau I}$ &  $1.37\times 10^{-3}$ \\
    N & $U_{\mu I}U_{\tau I}$ &  $1.25\times 10^{-3}$ \\
    I & $U_{eI}^2$ &  $8.15\times 10^{-4}$ \\
    I & $U_{\mu I}^2$ &  $3.46\times 10^{-4}$ \\
    I & $U_{\tau I}^2$ &  $9.91\times 10^{-3}$ \\
    I & $U_{eI}U_{\mu I}$ &  $3.34\times 10^{-5}$ \\
    I & $U_{eI}U_{\tau I}$ &  $2.19\times 10^{-3}$ \\
    I & $U_{\mu I}U_{\tau I}$ &  $1.43\times 10^{-3}$ \\
    \bottomrule
  \end{tabular}
  \caption{Upper limits on $U_{\alpha I}^2$ and $U_{\alpha I}U_{\beta I}$ within 2$\sigma$ in the high mass region $M_I\gtrsim 80$ GeV, for normal (N) and inverted (I) hierarchy.}
  \label{tab:upperlimits}
\end{table}

For $M_I$ between the masses of the $D$ mesons and the $W$ boson  the limits from direct searches dominate because the heavy neutrinos can be produced efficiently via the $s$-channel exchange of on-shell $W$ bosons. In the range between the $D$ meson masses and the $W$ boson mass, the limits from the DELPHI~\cite{Abreu:1997uq} and CMS~\cite{Sirunyan:2018mtv} experiments compete to impose the strongest bound.

\begin{figure*}[t]
  \centering
  \includegraphics[width=0.49\linewidth]{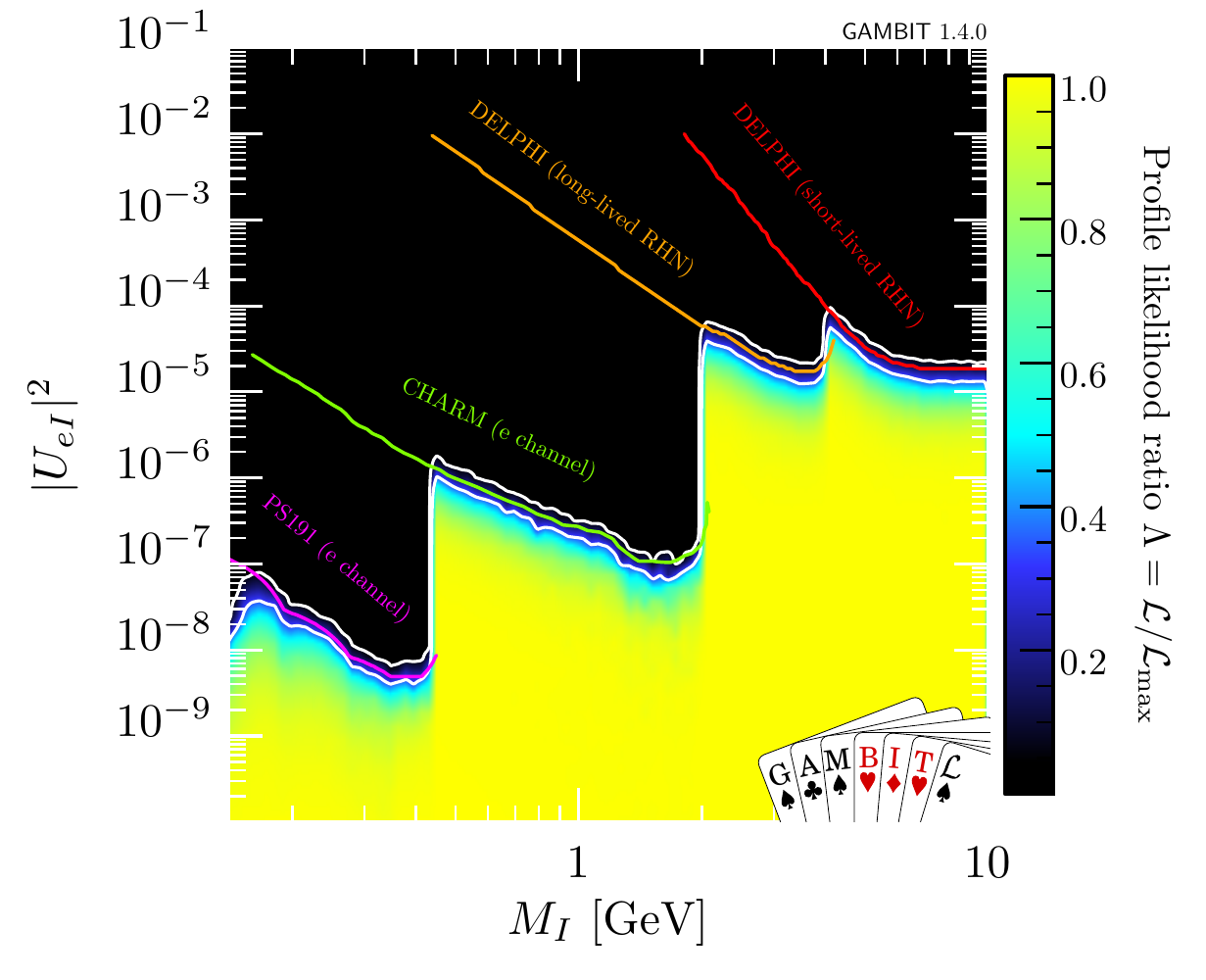}
  \includegraphics[width=0.49\linewidth]{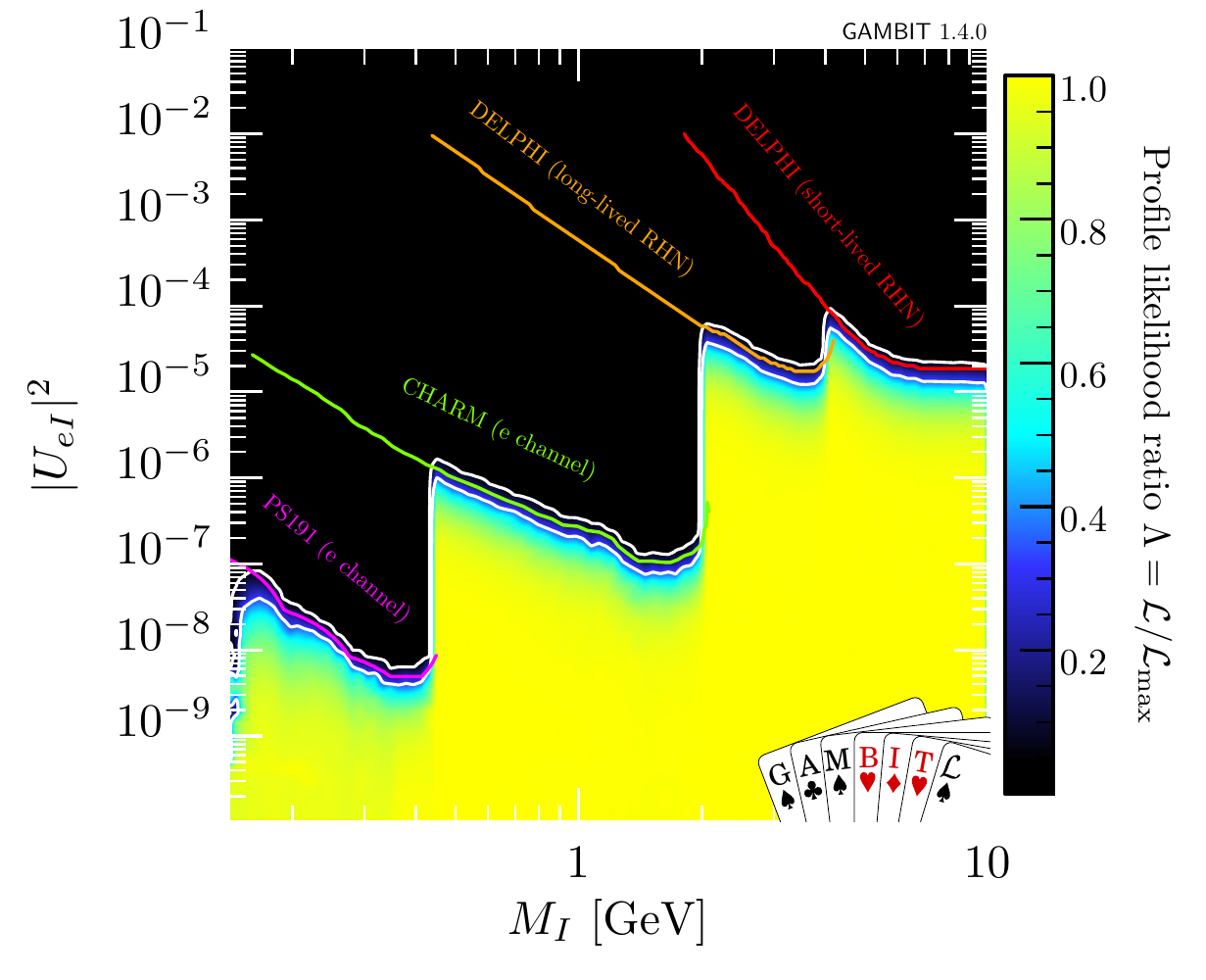}
  \caption{Profile likelihood in $M_I$ vs $U_{eI}^2$ plane with $M_I <10$ GeV and overlaid direct detection limits, for normal (left) and inverted hierarchy (right).}
  \label{fig:M_Ue_limits}
\end{figure*}

\begin{figure*}[t]
  \centering
  \includegraphics[width=0.49\linewidth]{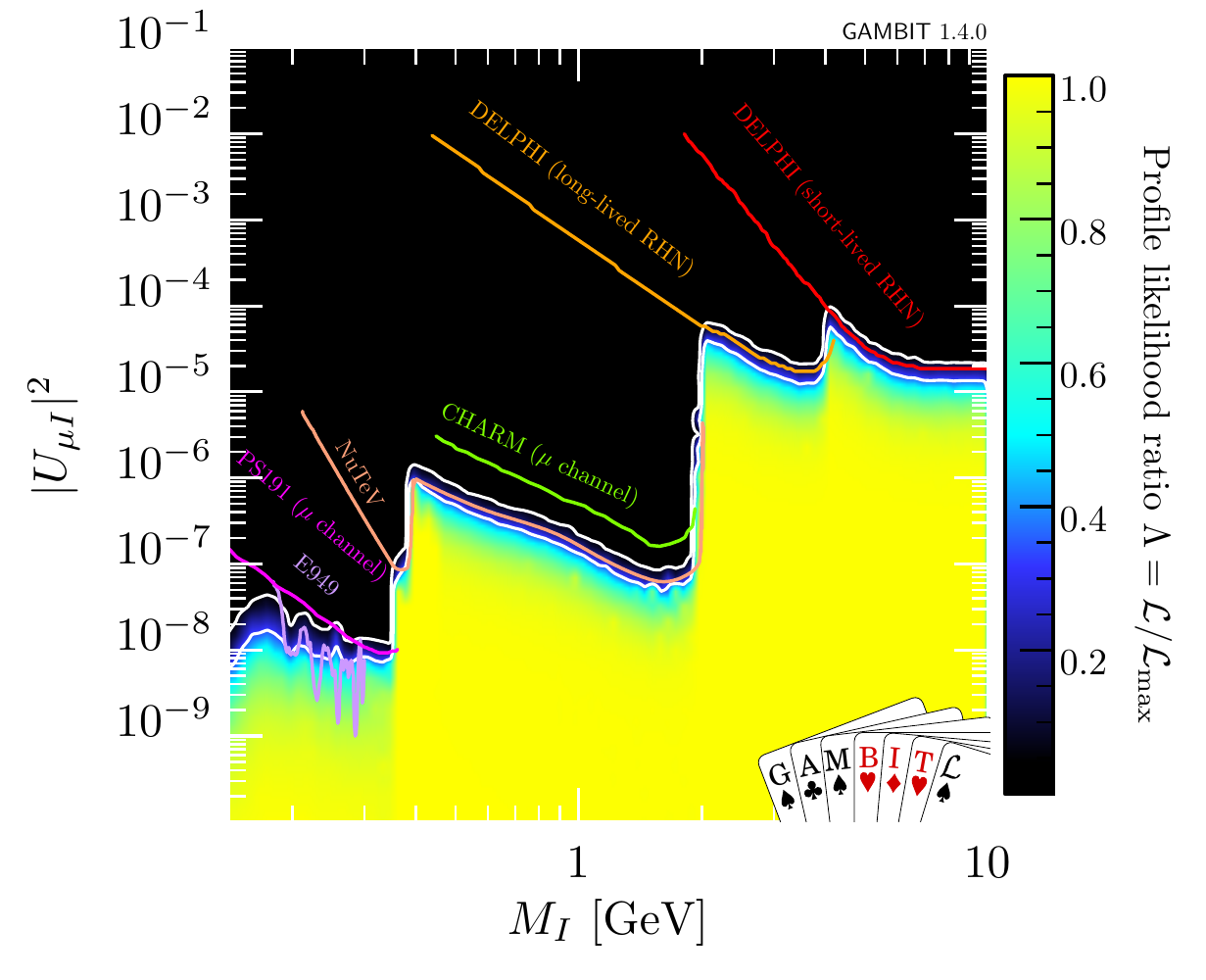}
  \includegraphics[width=0.49\linewidth]{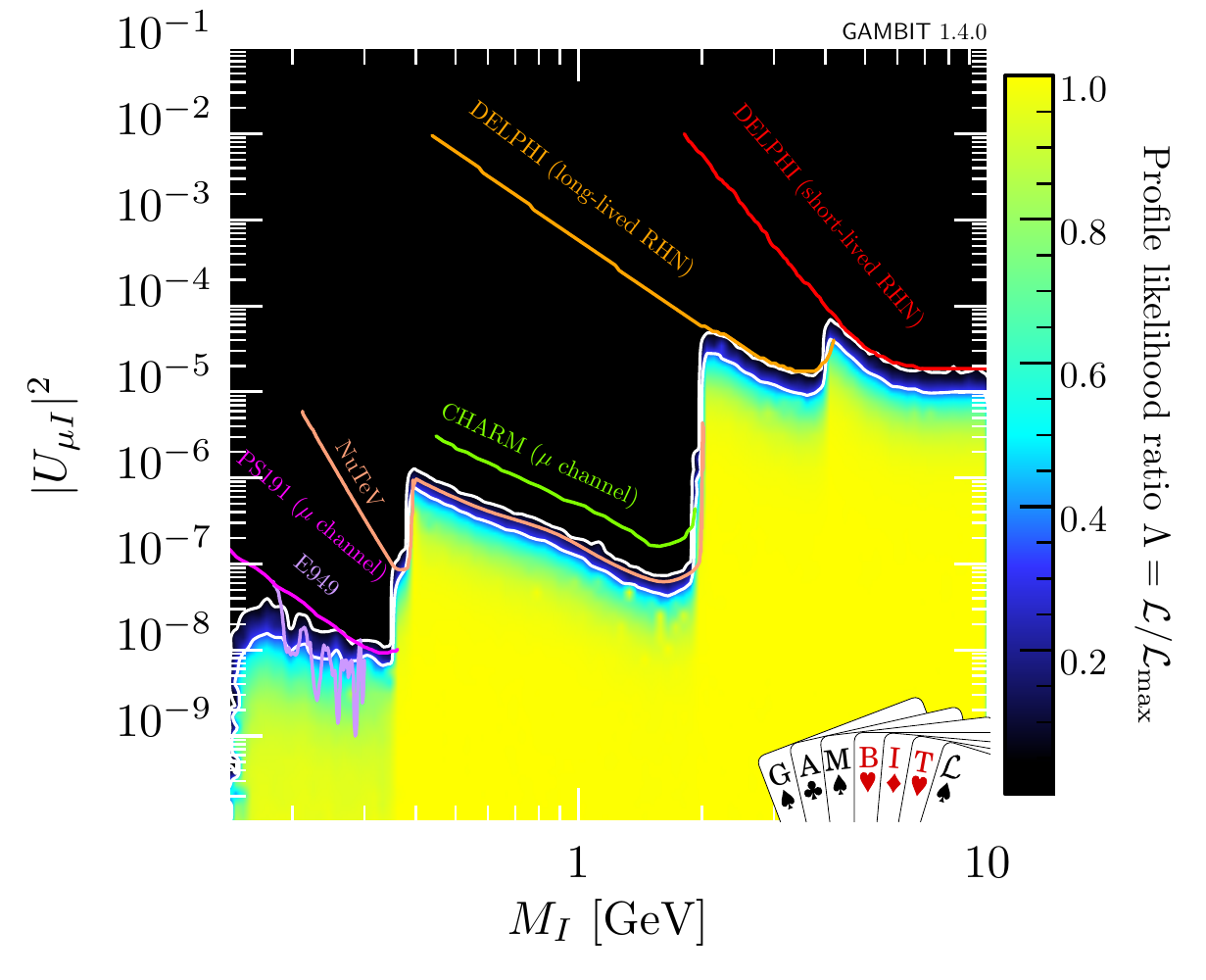}
  \caption{Profile likelihood in $M_I$ vs $U_{\mu I}^2$ plane with $M_I < 10$ GeV and overlaid direct detection limits, for normal (left) and inverted hierarchy (right).}
  \label{fig:M_Umu_limits}
\end{figure*}

\begin{figure*}[t]
  \centering
  \includegraphics[width=0.49\linewidth]{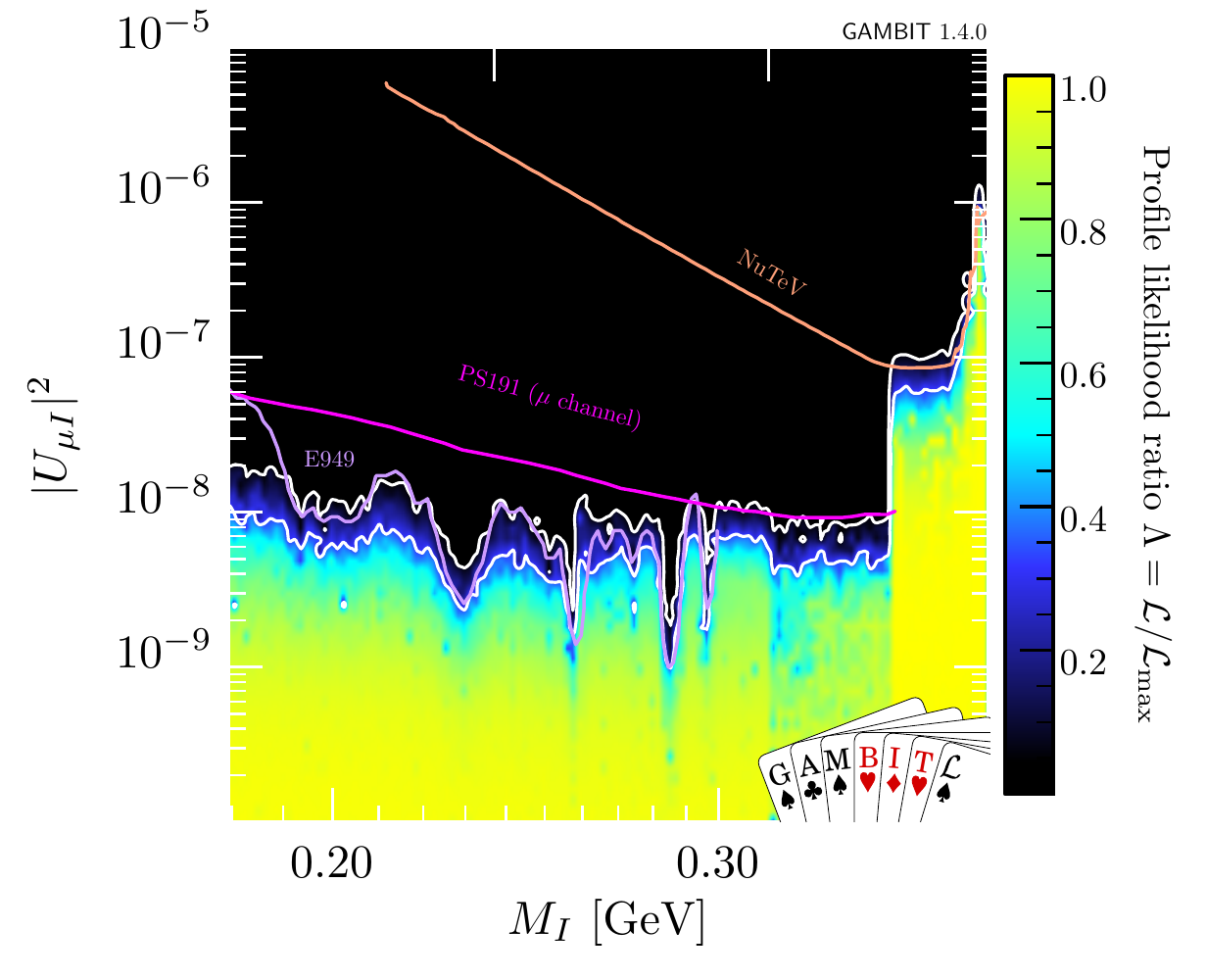}
  \includegraphics[width=0.49\linewidth]{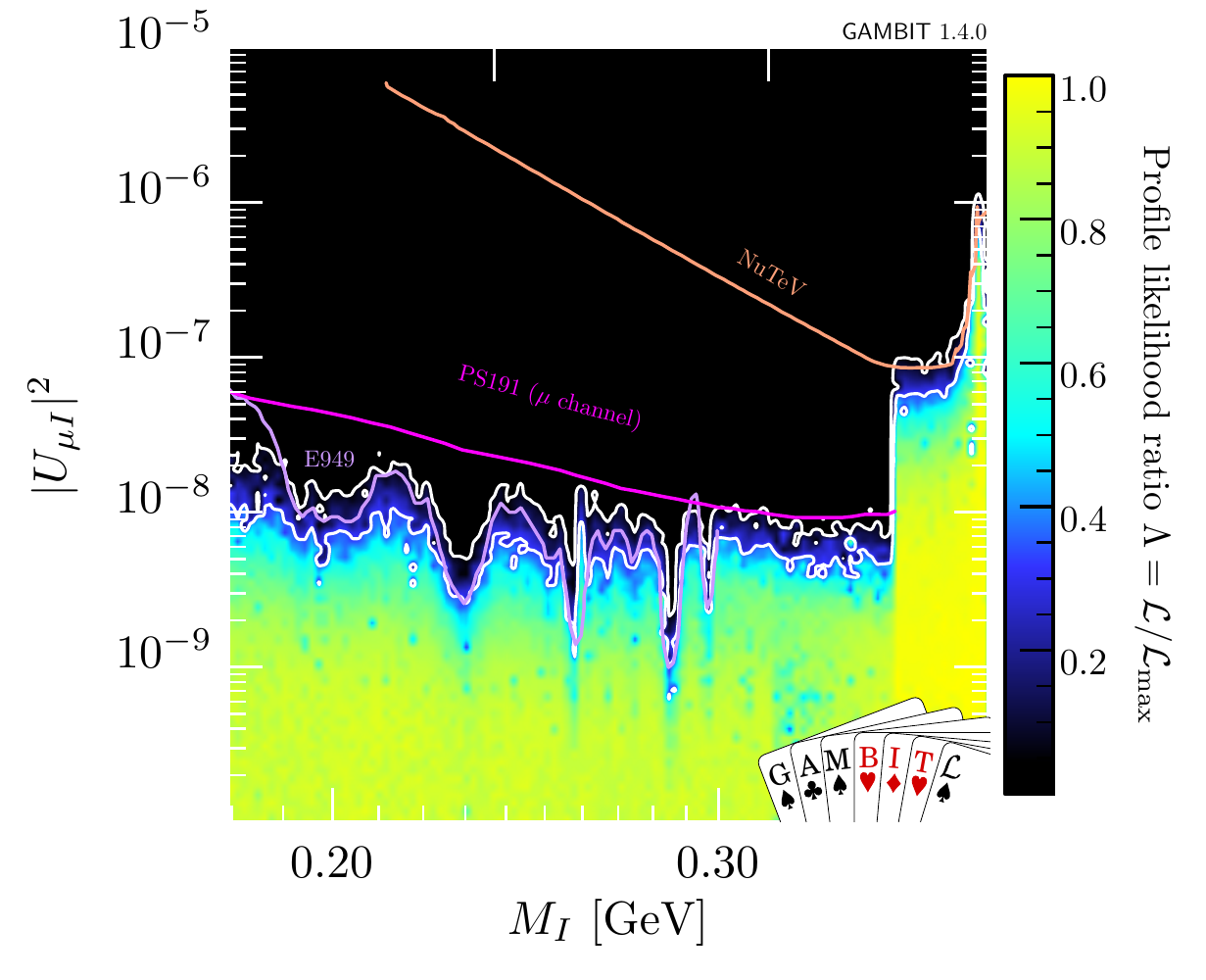}
  \caption{Profile likelihood in $M_I$ vs $U_{\mu I}^2$ plane with $M_I < 0.4$ GeV and overlaid direct detection limits, for normal (left) and inverted hierarchy (right).}
  \label{fig:M_Umu_limits_low}
\end{figure*}

Below the $D$ meson mass the constraints on $U_{e I}^2$ and $U_{\mu I}^2$ are dominated by direct search constraints from fixed target experiments,  in particular CHARM~\cite{Bergsma:1985is} and NuTeV~\cite{Vaitaitis:1999wq} above the kaon mass, PS-191~\cite{Bernardi:1987ek} and E949~\cite{Artamonov:2014urb} between the pion and kaon mass  and pion decay experiments at even lower masses.
In this regime the global constraints  on $U_{e I}^2$ and $U_{\mu I}^2$ are in good approximation given by the direct search constraints, as discussed in Sec.~\ref{Sec:IndividualBounds} and Figs.~\ref{fig:M_Ue_limits}-\ref{fig:M_Umu_limits_low}. 
This is in contrast to the model with $n=2$, where the global fits
rule out a significant mass range below the kaon mass that appears to be allowed if one simply superimposes the direct constraints in the mass-mixing planes \cite{Drewes:2016jae}.
For $U_{\tau I}^2$, the direct search constraints are much weaker, the limit from long-lived particle searches by DELPHI remains the most significant one in our scans.
Figure~\ref{fig:M_Utau_capped} shows that direct searches become subdominant for the $\tau$ coupling and the EWPO limit is saturated for a considerable range of masses below the kaon mass.

For masses below roughly $0.3$ GeV the global constraints are stronger than the sum of their ingredients due to an interplay of the lower bound from BBN on the mixings, the upper bounds on $U_{e I}^2$ and $U_{\mu I}^2$ from direct searches and the constraints on the heavy neutrino flavour mixing pattern from neutrino oscillation data (discussed further below in Sec.~\ref{sec:flavourmixing}). 
The latter disfavours large hierarchies amongst the couplings to individual SM flavours, though these constraints are weaker than in the model with $n=2$ \cite{Drewes:2016jae,Drewes:2018gkc}. This implies that upper bounds on combinations of $U_{eI}^2$ and $U_{\mu I}^2$ indirectly constrain $U_{\tau I}^2$. 
The BBN constraint on the lifetime does not impose a constraint on any individual coupling $U_{\alpha I}^2$, but requires at least some of them to be sizeable and practically translates into a lower bound on $U_I^2$ that is visible in Figure \ref{fig:M_U_capped}.
Both, the BBN constraint and the constraint on the flavour mixing pattern (that will be discussed in more detail in Sec.~\ref{sec:flavourmixing} and is visible in Fig.~\ref{fig:Triangle}) leads to the lower and upper bounds on $U_{\tau I}^2$ that are visible in Figure \ref{fig:M_Utau_capped}.

The upper bound on the total mixing $U_I^2$ from the global constraints can roughly be identified with the bound on $U_{\tau I}$ across the entire mass range as it is constrained the weakest. The lower bound is again given by the lifetime constraint from BBN. In addition, there is a lower bound from the requirement to explain the light neutrino oscillation data that depends on $m_{\nu_0}$ and is therefore only visible if one imposes a cut on this unknown quantity.
Our results agree with the analytic estimates made in Ref.~\cite{Drewes:2019mhg}, as will be discussed in Sec.~\ref{sec:flavourmixing}, and are illustrated in Figure \ref{fig:M_U_capped}.

\subsection{Discussion of individual bounds}\label{Sec:IndividualBounds}

Figures~\ref{fig:M_Ue_limits} and \ref{fig:M_Umu_limits} show explicitly the effect of direct searches on the upper limits of the $e$ and $\mu$ couplings in the mass range $M_I \in [0.1, 10]$ GeV. Most of the limits shown are at 90\%CL, with the exception of DELPHI at 95\%. As expected, they lie between the 1$\sigma$ and $2\sigma$ contours. Some of the experimental limits, PS-191 and CHARM, do not directly constrain an individual coupling, but rather the combination $U_{e/\mu I}^2 \times \sum_{\alpha} c_{\alpha}U_{\alpha I}^2$ (as mentioned in Sections~\ref{sec:ps191} and \ref{sec:charm}), with the coefficients $c_\alpha$ from eq.~\eqref{eqn:coeff}. As we profile over the other two couplings, the strongest limit for the $\alpha$ flavour for these experiments would correspond to $(U_{I}^{\rm{exp}})^2 / c_\alpha$, with $(U_I^{\rm{exp}})^2$ being the reported limit by the experiment. Hence the former ratio is what is shown in the figures as the PS-191 and CHARM limits. As observed, $U_{e I}^2$ is constrained by PS-191 and CHARM in the lowest and next-to-lowest mass regions, whereas they are superseded by the limits from E949 and NuTeV for $U_{\mu I}^2$. In the lowest mass region for the $\mu$ coupling it would appear that the E949 bound is in fact not saturated as the experimental limit falls below the data. This is however just an artifact of binning and interpolation in that region and the fact that the E949 limit is quite jagged. To illustrate this, we show in Figure~\ref{fig:M_Umu_limits_low} a zoom into the lowest mass region from Figure~\ref{fig:M_Umu_limits}, where it can be seen clearly that the profile likelihood follows the limits of E949.

Neutrino oscillation data imposes very strong constraints on the parameter space and disfavours vast volumina in the 18 dimensional model parameter space. In the scenario with $n=2$ this has a visible effect on the projections of the global constraints on the $M_I$-$U_{\alpha I}^2$ planes  \cite{Drewes:2016jae}, in particular for heavy neutrinos lighter than the kaon, where the interplay between neutrino oscillation data, BBN and direct searches rules our most values of $M_I$. This effect strongly depends on the light neutrino mass ordering, and varying the light neutrino oscillation parameters within their experimentally allowed limits leads to visible differences \cite{Drewes:2016jae}. In the present analysis with $n=3$ the impact of neutrino oscillation data on the likelihoods in the $M_I$-$U_{\alpha I}^2$ planes is much smaller. This is primarily visible in the third generation and the total mixing, cf.~Figs.~\ref{fig:M_Utau_capped} and \ref{fig:M_U_capped}, where the dependence on the light neutrino mass ordering is weak. The reason is that the larger dimensionality of the parameter space with $n=3$ makes it easier to avoid conflicts with direct or other indirect bounds. With $n=2$ neutrino oscillation data also imposes strong constraints on the flavour mixing pattern \cite{Hernandez:2016kel,Drewes:2016jae,Drewes:2018gkc}. These are also visible in the present analysis, cf.~Figs.~\ref{fig:Triangle}-\ref{fig:Triangle_symmetric}, but can be avoided by choosing a sufficiently large value for $m_{\nu_0}$. The constraints on the flavour mixing pattern strongly depend on the light neutrino mass ordering, and varying the light neutrino oscillation parameters within their experimentally allowed range has a considerable impact on the predictions.

As mentioned above, EWPO (including $\Gamma_{\rm{inv}}$, $m_W$, $W$-decays and $s_w$), LFV and CKM constraints become relevant for very large couplings and are thus the dominant limit in the high mass region, as well as a small region at small masses for the $\tau$ coupling (Figure~\ref{fig:M_Utau_capped}). Besides providing constraints, in particular $\Gamma_{\rm{inv}}$ and CKM observables are also responsible for the slight excesses in the total likelihood, which we will discuss in Sec.~\ref{sec:excesses}. 
Other constraints included in the analysis have little to no effect on the profile likelihoods as shown above. 

Among the leptonic decays, only $R^K_{e\mu}$ has some impact on the likelihood, with a negative contribution at masses below $0.45$ GeV. Both $R^K_{e\mu}$ and $R^\tau_{e\mu}$ show minor excesses in total likelihood, which again will be discussed later.
Other lepton universality constraints have only little effect on the likelihood.

Neutrinoless double beta decay sets strong upper bounds on $U_{eI}^2$ for generic parameter choices, which strongly disfavours considerable regions of parameter space. However, in the limit where lepton number is approximately preserved the expected signal from $0\nu\beta\beta$ is suppressed. Since many of our parameter points are in this symmetry protected scenario, particularly at high couplings, the impact of this constraint on the likelihoods in the projection on the mass-mixing plane is minimal.
This is consistent with what was found in Refs.~\cite{Drewes:2015iva,Drewes:2016jae,Bolton:2019pcu}.

The effect of BBN can be seen in the lower limits of Figure~\ref{fig:M_Utau_capped} and \ref{fig:M_U_capped}. The lower limit on $U_I^2$ is a direct consequence of BBN, as lower couplings would mean that right-handed neutrinos would not decay before BBN and thus affect the abundance of primordial elements. Although no individual limits are imposed by the BBN constraint on the couplings, the strong upper limit on the $e$ and $\mu$ flavours at low masses has the side effect of setting a lower limits on $U_{\tau I}^2$, as seen in Figure~\ref{fig:M_Utau_capped}. 

For a more detailed explanation of the effect of each partial likelihood, and associated figures, we refer to Appendix~\ref{app:partial}.

\subsection{Lightest neutrino mass and flavour mixing}
\label{sec:flavourmixing}

Oscillation data strongly constrains most of the active neutrino parameters, in particular the mass splittings $\Delta m^2_{21}$ and $\Delta m^2_{3l}$, the mixing angles $\theta_{ij}$ and CP phase $\delta_{CP}$, whereas the lightest neutrino mass $m_{\nu_0}$ remains unknown. 
There are upper bounds from cosmology on the sum $\sum_i m_i$ that depend on the active neutrino mass hierarchy, the data set used and the underlying cosmological model.
The value quoted by the Planck collaboration for a standard cosmological model is $\sum_i m_i < 0.12$ eV \cite{Aghanim:2018eyx}, a discussion of how this value changes with different assumptions can e.g.~be found in the Particle Data Group Report \cite{Tanabashi:2018oca}.
In fact, using the best fit values for the mass splittings from the NuFit data~\cite{NuFit} and the conservative value $\sum_i m_i < 0.23$ eV, we can infer the upper limits of $m_{\nu_0} < 7.12 \times 10^{-2}$ eV for normal and $m_{\nu_0} < 6.55 \times 10^{-2}$ eV for inverted hierarchy. 

The value of $m_{\nu_0}$ strongly impacts on the lower limit on $U_I^2$. 
One can obtain a reliable estimate of the lower bounds on $U_I^2$ by setting $\mathcal{R}=1$ \cite{Drewes:2019mhg}.
This makes the PMNS matrix unitary, and the lower limit one the smallest mixing can be estimated as $U_I^2 \gtrsim m_{\nu_0}/M_I$.  Using this approximation, we show in Figure~\ref{fig:M_U_capped} the lower limits on $U_I^2$ that we obtain in our scans for different values of $m_{\nu_0} = (0.05, 10^{-2}, 10^{-3}, 10^{-4})$ eV. In the case of $m_{\nu_0} = 0$ there is no absolute lower limit on the coupling from the seesaw mechanism, and the residual lower limit on $U_I^2$ is due to the BBN constraint.

\begin{figure*}[t]
  \centering
  \includegraphics[width=0.49\linewidth]{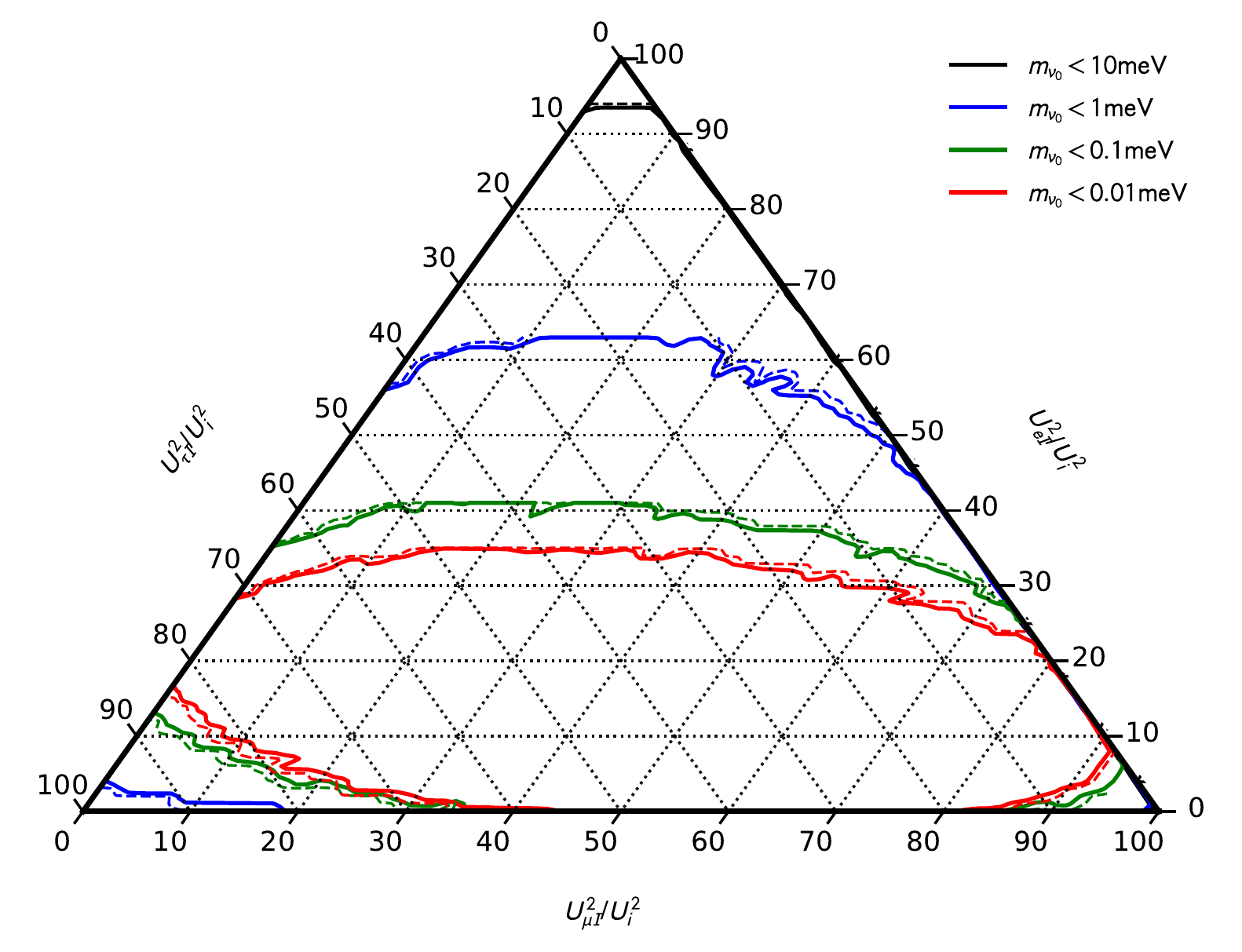}
  \includegraphics[width=0.49\linewidth]{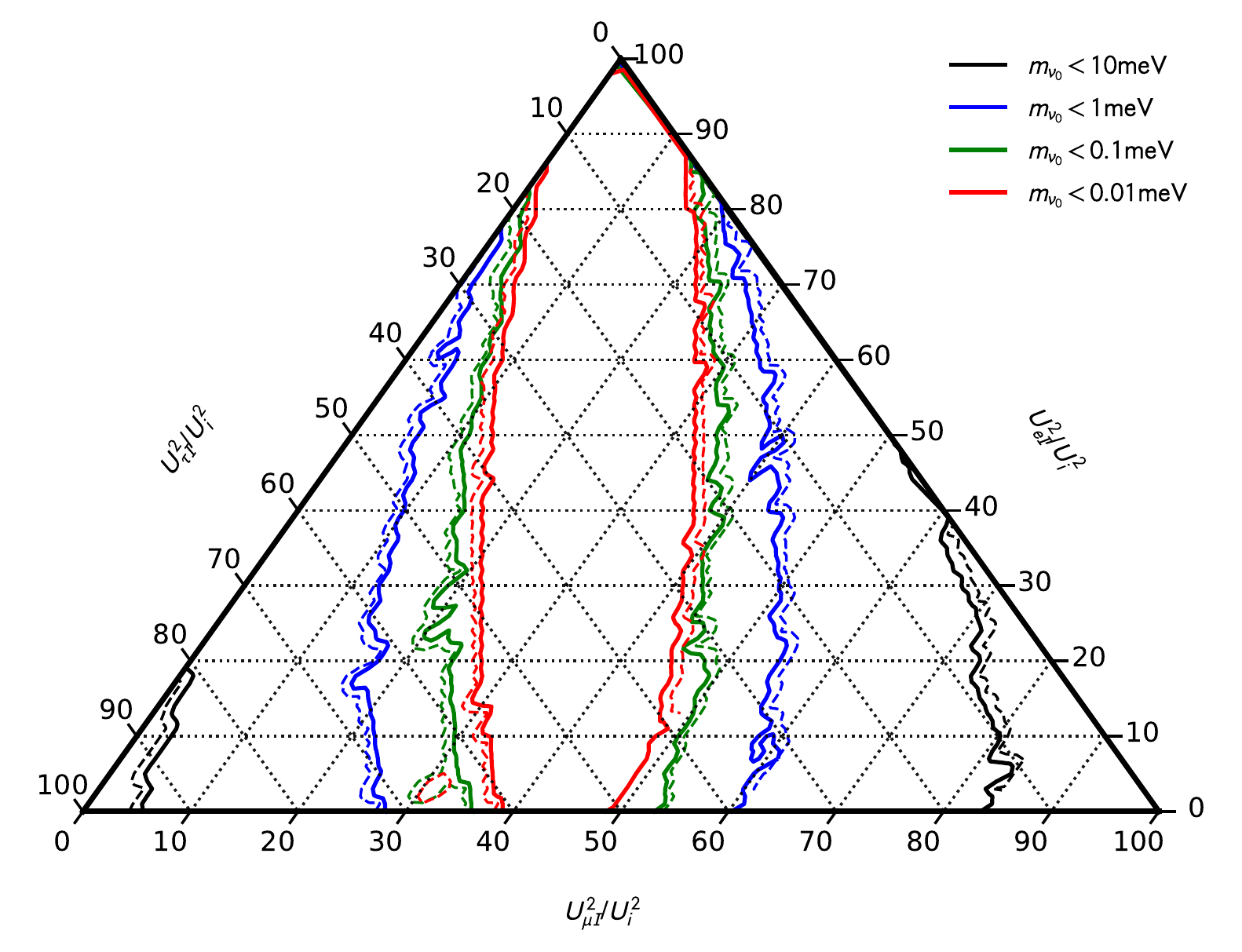}
  \caption{$U_{\alpha I}^2/U_I^2$ (in percent) for different upper limits of $m_{\nu_0}$ (see legend). Solid (dashed) lines delineate the 1$\sigma$ (2$\sigma$) contours, for normal (left) and inverted hierarchy (right).
  As discussed in footnote \ref{FlavourMixingFootnote}, these constraints apply to those heavy neutrinos that can be found experimentally.
  }
  \label{fig:Triangle}
\end{figure*}

\begin{figure*}[t]
  \centering
  \includegraphics[width=0.49\linewidth]{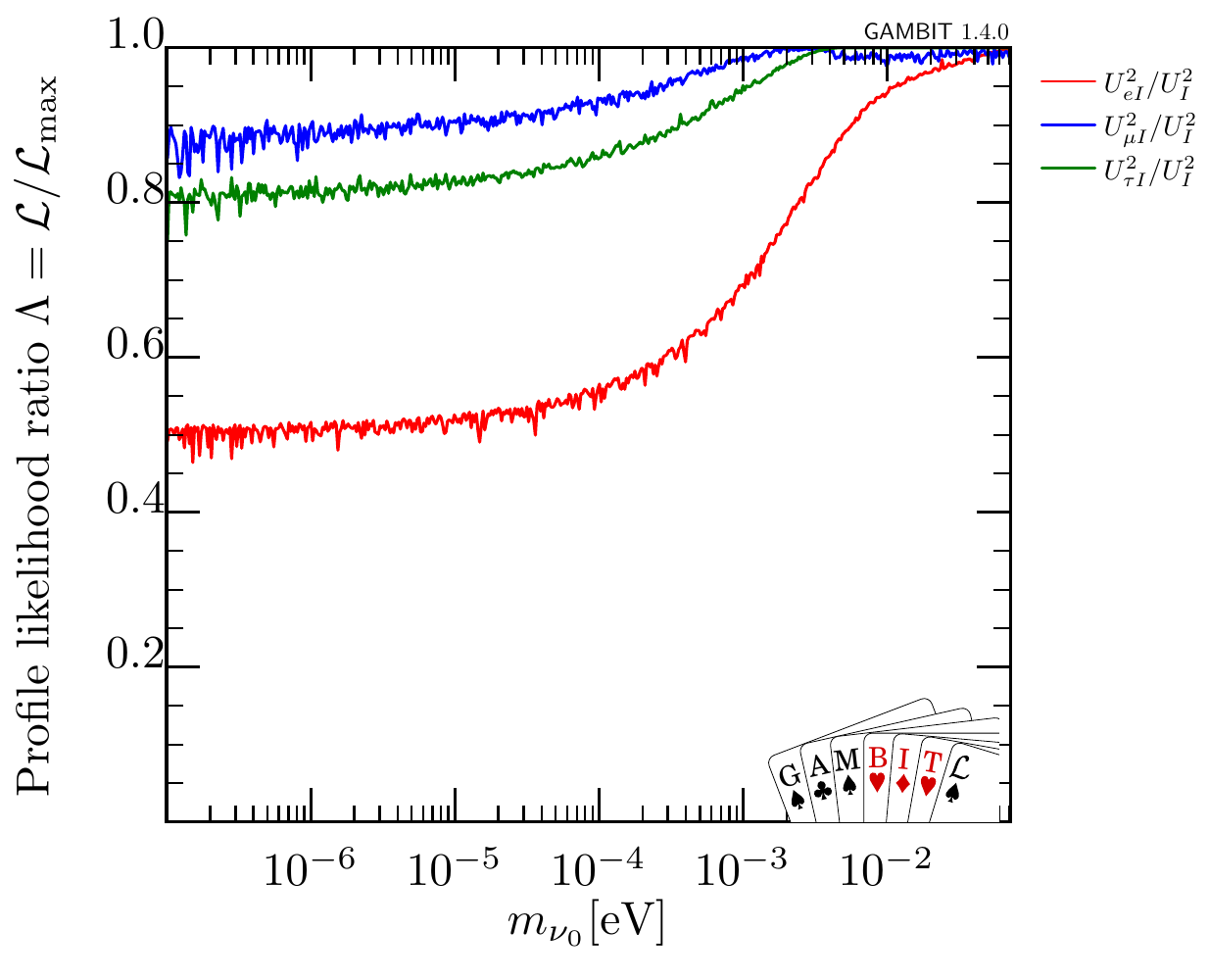}
  \includegraphics[width=0.49\linewidth]{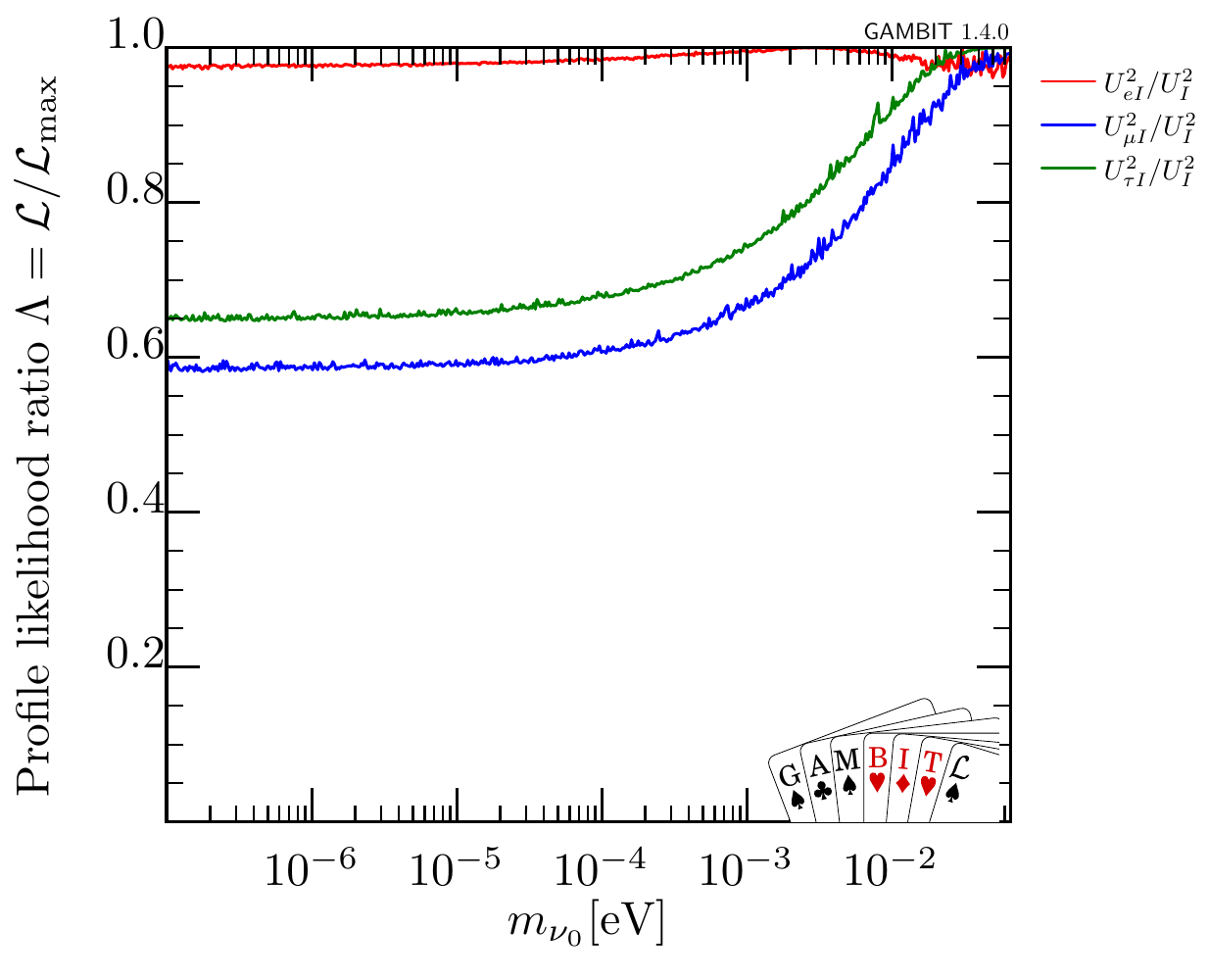}
  \caption{Upper limits on the coupling ratios $U_{\alpha I}^2/U_I^2$ within $2\sigma$ as a function of the lightest active neutrino mass $m_{\nu_0}$, for normal (left) and inverted hierarchy (right).
  As discussed in footnote \ref{FlavourMixingFootnote}, these constraints apply to those heavy neutrinos that can be found experimentally.
  }
  \label{fig:mnu0_U_upperlimits}
\end{figure*}

\begin{figure*}[t]
  \centering
   \includegraphics[width=0.49\linewidth]{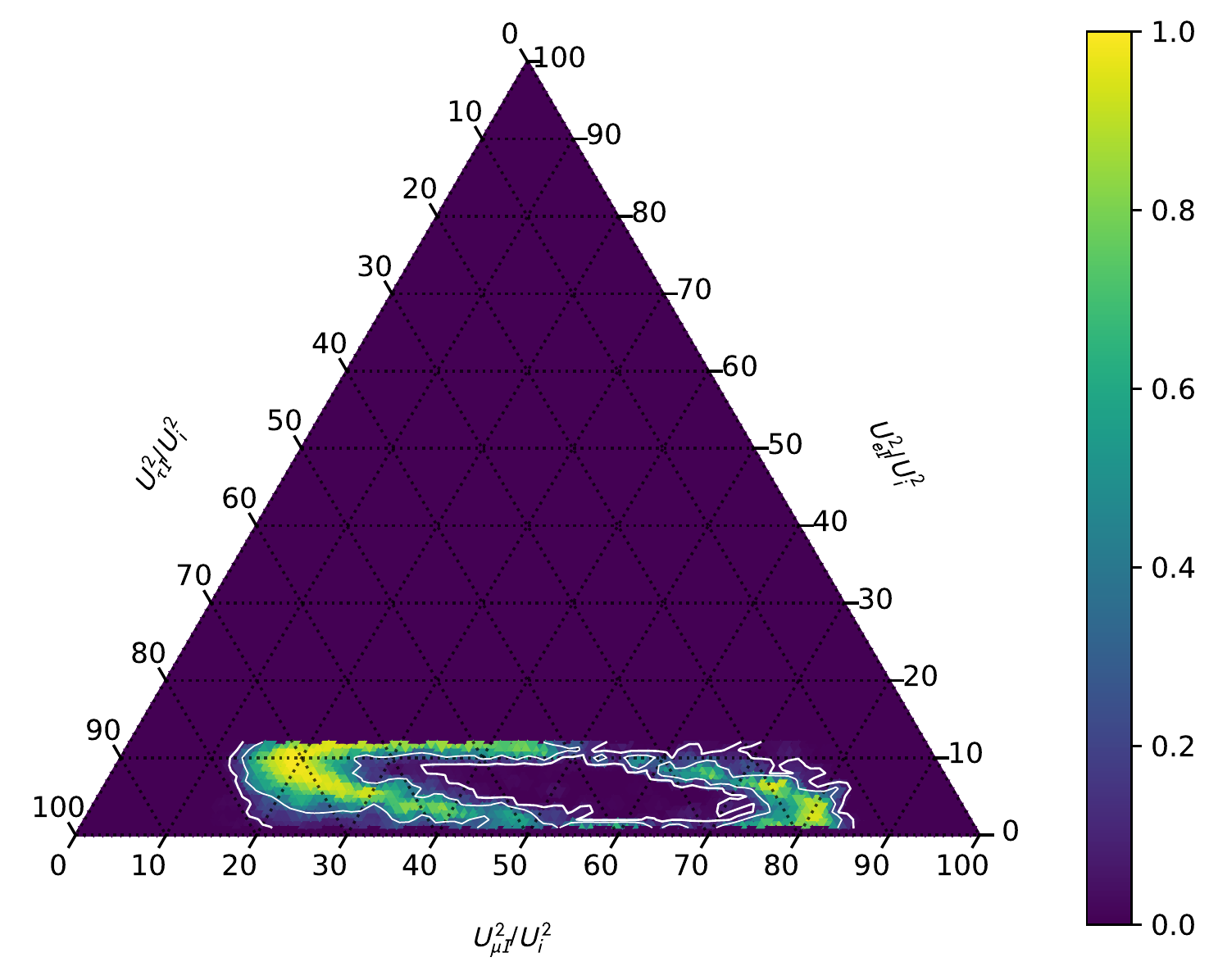}
    \includegraphics[width=0.49\linewidth]{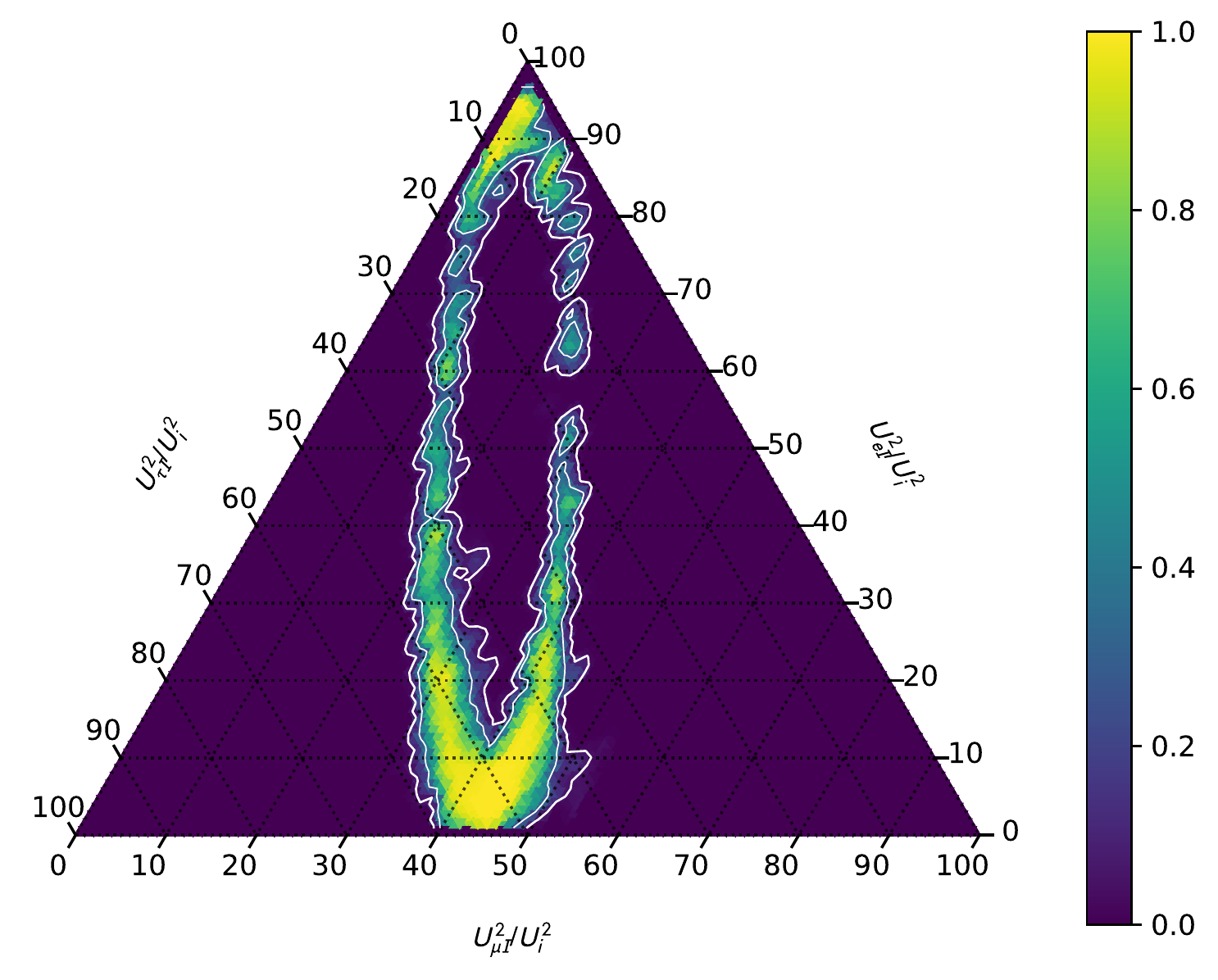}
  \caption{Profile likelihood for $U_{\alpha I}^2/U_I^2$ (in percent) in the limit of $n=2$ in the symmetry protected region for normal (left) and inverted (right) hierarchy. For the detailed cuts we refer to the text.}
  \label{fig:Triangle_symmetric}
\end{figure*}

\smallskip

The lightest neutrino mass has an important effect on the pattern of flavour mixing. In the limit of large $m_{\nu_0}$, there is almost no constraint on the allowed flavour ratios $U_{\alpha I}^2/U_I^2$. This is shown in Figure~\ref{fig:Triangle} by the black solid (dashed) contours, which indicate the allowed region within $1\sigma$ ($2\sigma$) where the lightest neutrino mass is $m_{\nu_0}$ < 10 meV (close to the cosmological bound stated above). In this case, there is no visible upper limit on $U_{\mu I}^2/U_I^2$ or $U_{\tau I}^2/U_I^2$ for normal hierarchy, whereas $U_{eI}^2/U_I^2$ is constrained $\lesssim 0.95$. Conversely, for inverted hierarchy there is an upper limit for the $\mu$ and $\tau$ flavours, but none for the $e$ flavour. However, for smaller values of $m_{\nu_0}$, the allowed range for the flavour mixing pattern becomes significantly constrained.\footnote{\label{FlavourMixingFootnote} When constraining $m_{\nu_0}$ to very small values, we almost decouple one right handed neutrinos. 
The contribution of this feebly coupled state to the generation of light neutrino masses is negligible, which in return implies that its properties are almost unconstrained by neutrino oscillation data, and such is its flavour mixing pattern.  Thus extreme ratios $U_{\alpha I}^2/U_I^2$ can in principle occur for this particular heavy neutrino, although the absolute values of $U_I^2$ remains negligible, and it has no effect on any near future experiment.
Since our focus is primarily on heavy neutrinos that make a measurable contribution to the generation of light neutrino masses and/or may be discovered in experiments, we applied a cut on $M_I U_I^2 > 10^{-10}$ GeV in Figure~\ref{fig:Triangle} and Figure~\ref{fig:mnu0_U_upperlimits} to remove artefacts arising from states that are practically decoupled.  
The value of the cut is motivated by experimental sensitivities as demonstrated. The NA62 experiment, for instance, will only be able to probe up to sensitivities of $M_I U_I^2 \approx \mathcal{O}(10^{-8})$ \cite{Drewes:2018gkc}; under optinistic assumptions the LHC may test $M_I U_I^2 \approx \mathcal{O}(10^{-8})$ \cite{Drewes:2019fou} and the FCC $M_I U_I^2 \approx \mathcal{O}(10^{-11})$ \cite{Antusch:2017pkq}. } 
This is shown by the lines for $m_{\nu_0} < 1$ meV (blue), $m_{\nu_0} < 0.1$ meV (green) and $m_{\nu_0} < 0.01$ meV (red).  For masses lower than 0.01 meV the constraints saturate and the size of the ellipse remains almost constant. This can be also seen in Figure~\ref{fig:mnu0_U_upperlimits}, where the largest coupling ratio is plotted for each flavour as function of neutrino mass. 

It is instructive to compare our results to the constraints on the flavour mixing pattern in the scenario with $n=2$ that were found in Refs.~\cite{Hernandez:2016kel,Drewes:2016jae,Drewes:2018gkc}.
For this purpose it is not sufficient to simply insert very small values for $m_{\nu_0}$ in the parameterisation \eqref{CItheta} because such values can also be achieved due to accidental cancellations in the light neutrino mass matrix (without decoupling of any of the heavy neutrinos), cf.~Section \ref{sec:n2model}. 
To remove such fine tuned points we impose the following cuts %on the data to generateFigure~\ref{fig:Triangle_symmetric}
\begin{align}
\label{eq:tuningcut}
\notag \frac{|M_2-M_1|}{M_2+M_1} < \epsilon, &\quad \frac{m_{\nu_0}}{\mu\textrm{eV}} < 1, \\
|F_{\alpha 3}| < \epsilon, &\quad 
\frac{|F_{\alpha 1} + i F_{\alpha 2}|}{|F_{\alpha 1}| + |F_{\alpha 2}|} <  \epsilon.
\end{align}
Here $\epsilon$ is an arbitrarily small number, which we choose as $\epsilon = 0.01$ for convenience. In addition, we work in the limits as defined by $|\mathrm{Im}\omega_{23}| \gg 1$ and $\mathrm{Re}\omega_{13} \sim \pi/2$ for normal hierarchy, and $|\mathrm{Im}\omega_{12}| \gg 1$ for inverted hierarchy (c.f. Appendix \ref{FakeSymmetryPoints}). Note that we randomised the order of the matrices $\mathcal{R}^{ij}$, 
and hence for normal hierarchy we can only reproduce the true symmetry protected regime for the permutation $\mathcal{R} = \mathcal{R}^{23} \mathcal{R}^{13} \mathcal{R}^{12}$. The inverted hierarchy limit is independent of permutations as two of the $\omega_{ij}$ are zero. In Figure~\ref{fig:Triangle_symmetric}, we show the triangle plots with 1$\sigma$ and 2$\sigma$ contours for NH and IH in the symmetry protected region after applying the aforementioned cuts to remove fine-tuned points. 
The results are consistent with what was found in Ref.~\cite{Drewes:2018gkc} for $n=2$ RHNs. It is worth noting that there is a sharp upper limit on $U_{eI}^2/U_I^2$ where the contours do not show. This is due to the hard upper limit imposed on $m_{\nu_0}$ in order to reach the $n=2$ case and it is, as before, consistent with the results in \cite{Drewes:2018gkc}.

\subsection{Discussion of excesses likelihoods}
\label{sec:excesses}

In the previous subsections, we have made use of an \textit{exclusion-only} `capped' profile likelihood to study the constraining effect of the various observables on the parameter space (for a justification see Sec.~\ref{sec:capped}). The \emph{total} likelihood, however shows a pattern of excesses in some small regions of the parameter space.  As discussed in Sec.~\ref{sec:capped}, experimental results with a preference for specific heavy neutrino masses and mixings would in general not show up as localized excesses in the total profile likelihood.  This is due to the fact that for each value of $M_I$ it would be in general possible to find a value of $M_J$ (with $J\neq I$) and associated couplings that would maximize the excess likelihood, irrespective of the values of $M_I$.  In order to extract the specific masses and couplings preferred by an excess likelihood, we adopt throught this subsection the following strategy.  We only allow \emph{one} of the three RHNs (which we take to be $I=1$) to acquire the required masses and couplings, while disallowing the other two RHNs to enter the preferred region.  This is emphasized in the plots by specifying $M_1$ and $|U_{\alpha 1}|^2$ instead of $M_I$ and $U_{\alpha I}^2$.  Mind that these results would be identical for $M_2$ and $M_3$.

\begin{figure*}[t]
  \centering
  \includegraphics[width=0.45\linewidth]{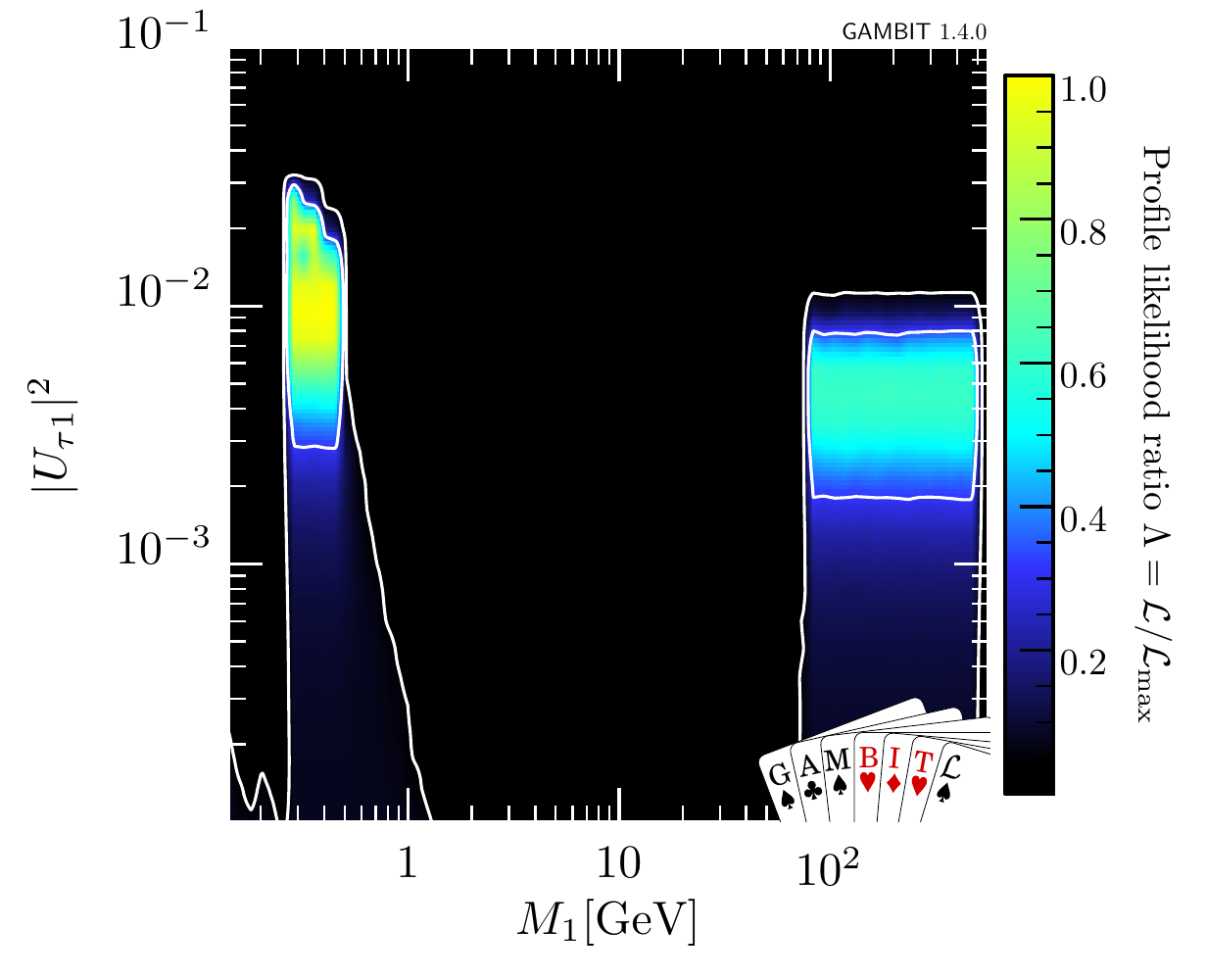}
  \includegraphics[width=0.45\linewidth]{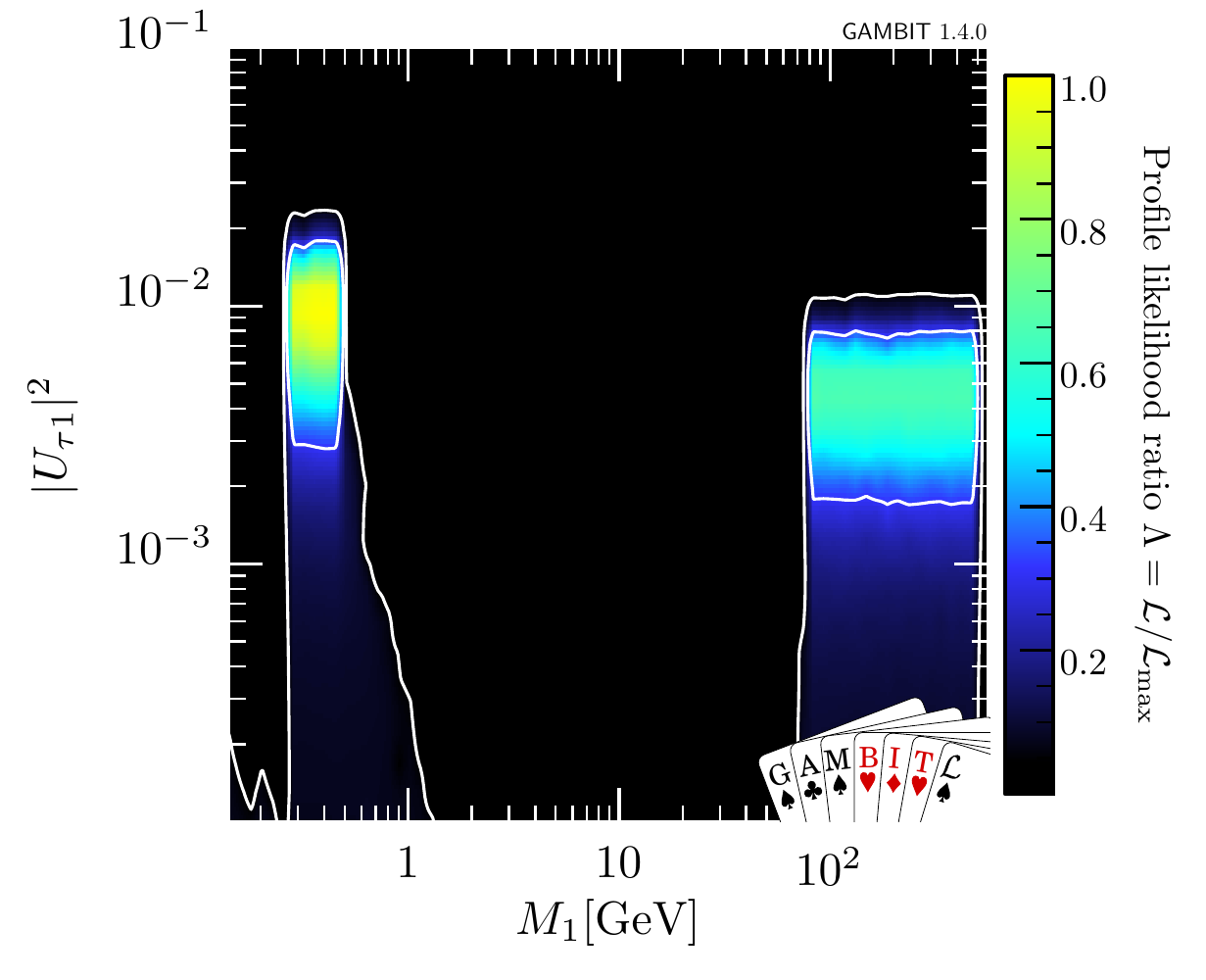}
  \caption{Profile likelihood in $M_I$ vs $U_{\tau I}^2$ plane without likelihood cap showing the excesses due to the $\Gamma_{inv}$, CKM and $R_\tau$ constraints, for normal (left) and inverted hierarchy (right).}
  \label{fig:M_Utau_CKM_Zinv_Rtau}
\end{figure*}

The invisible width of the $Z$ boson is modified by the presence of the right-handed neutrinos through their mixing, as described in Section~\ref{sec:ewpo}. For very high $\tau$ couplings, $U_{\tau I}^2 > 10^{-3}$, the prediction from the RHN model is actually a better fit to the experimental measurement than the SM, and thus there is a slight ($ < 2\sigma$) excess. A similar effect occurs for the CKM and $R^\tau_{e\mu}$ constraints, where the modified contribution on the neutrino mixing in the decay products of $K$-mesons and $\tau$, enhances the prediction with respect to that of the SM. Figure~\ref{fig:M_Utau_CKM_Zinv_Rtau} shows the excesses on the total profile likelihood in the $M_1$ vs $U_{\tau I}^2$ plane, zoomed in at high couplings (as discussed above, we excluded $M_2$ and $M_3$ from entering the excess regions).  Since there are no constraints from direct searches at masses above $M_1 > 80$ GeV or in the range $0.3 < M_1 < 0.5$ GeV, there is a combined excess shown of about $2\sigma$.

\begin{figure*}[t]
  \centering
  \includegraphics[width=0.49\linewidth]{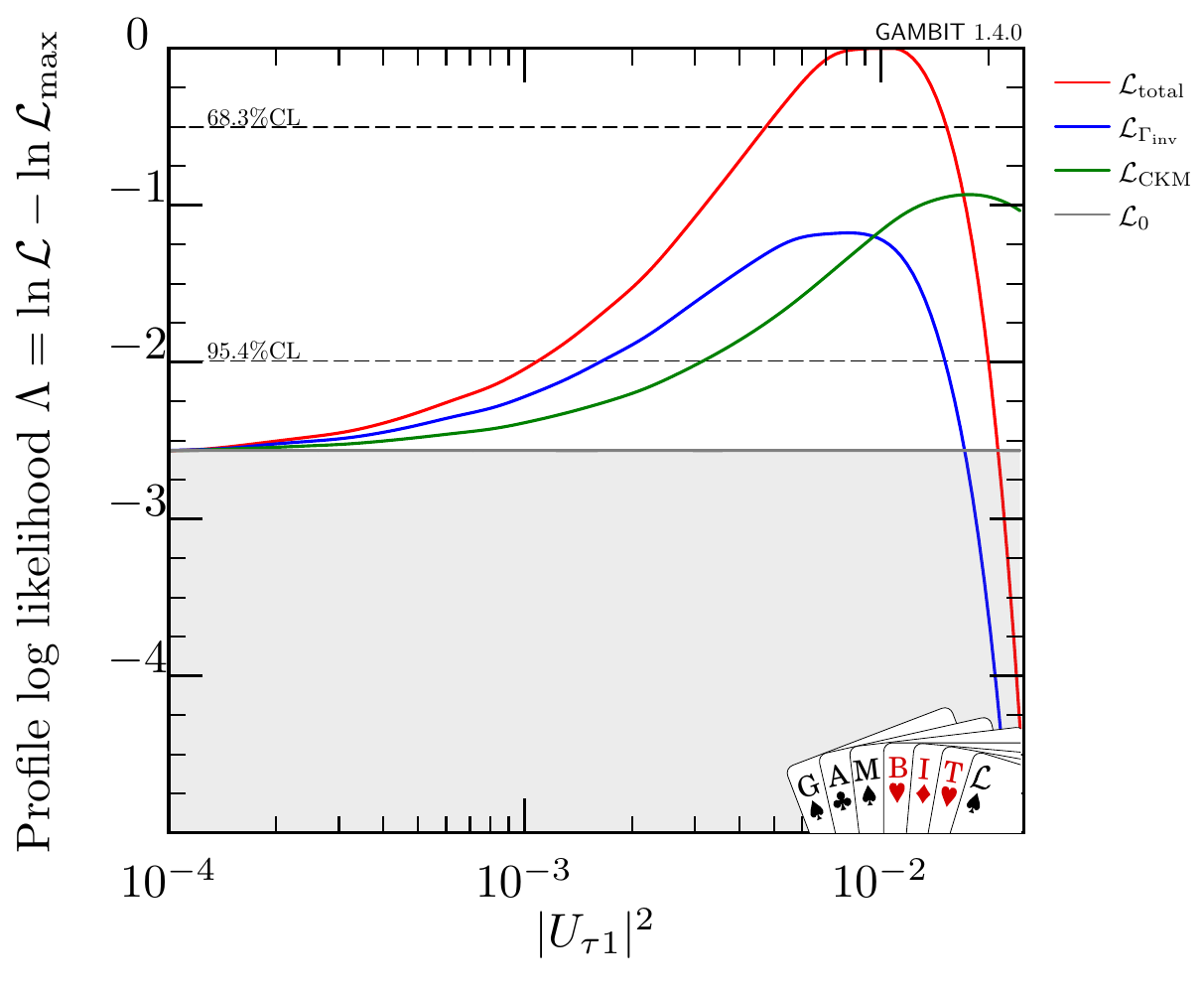}
  \includegraphics[width=0.49\linewidth]{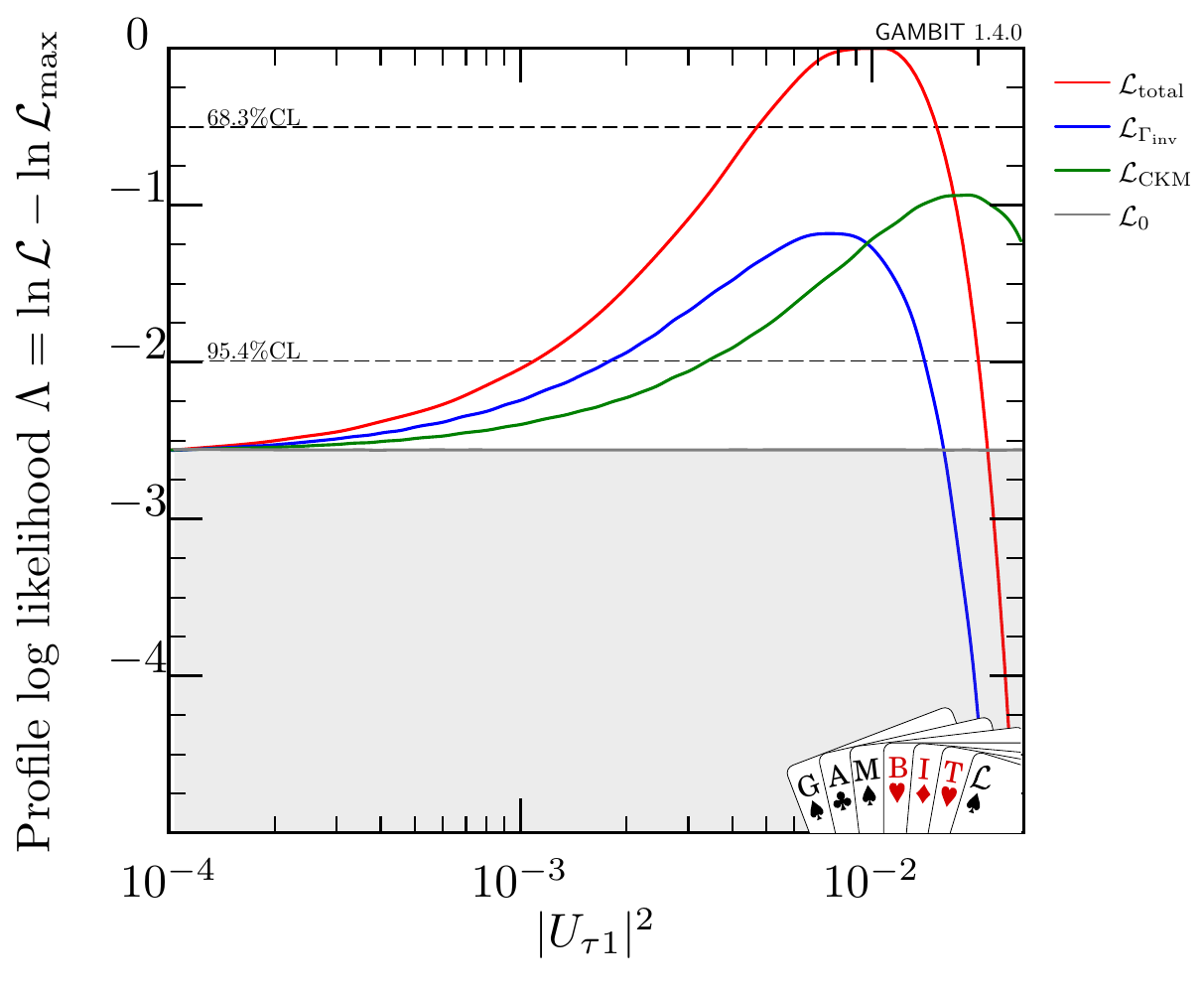}
  \caption{One-dimensional profile likelihood for $U_{\tau 1}^2$, $\mathcal{L}_{total}$, and partial likelihoods for $\Gamma_{Z}$, CKM and combination of the rest of constraints, $\mathcal{L}_0$, in the low mass region, $M_1 < 1$ GeV, for normal (left) and inverted hierarchy (right).}
  \label{fig:Utau_lnL_lowmass}
\end{figure*}
\begin{figure*}[t]
  \centering
  \includegraphics[width=0.49\linewidth]{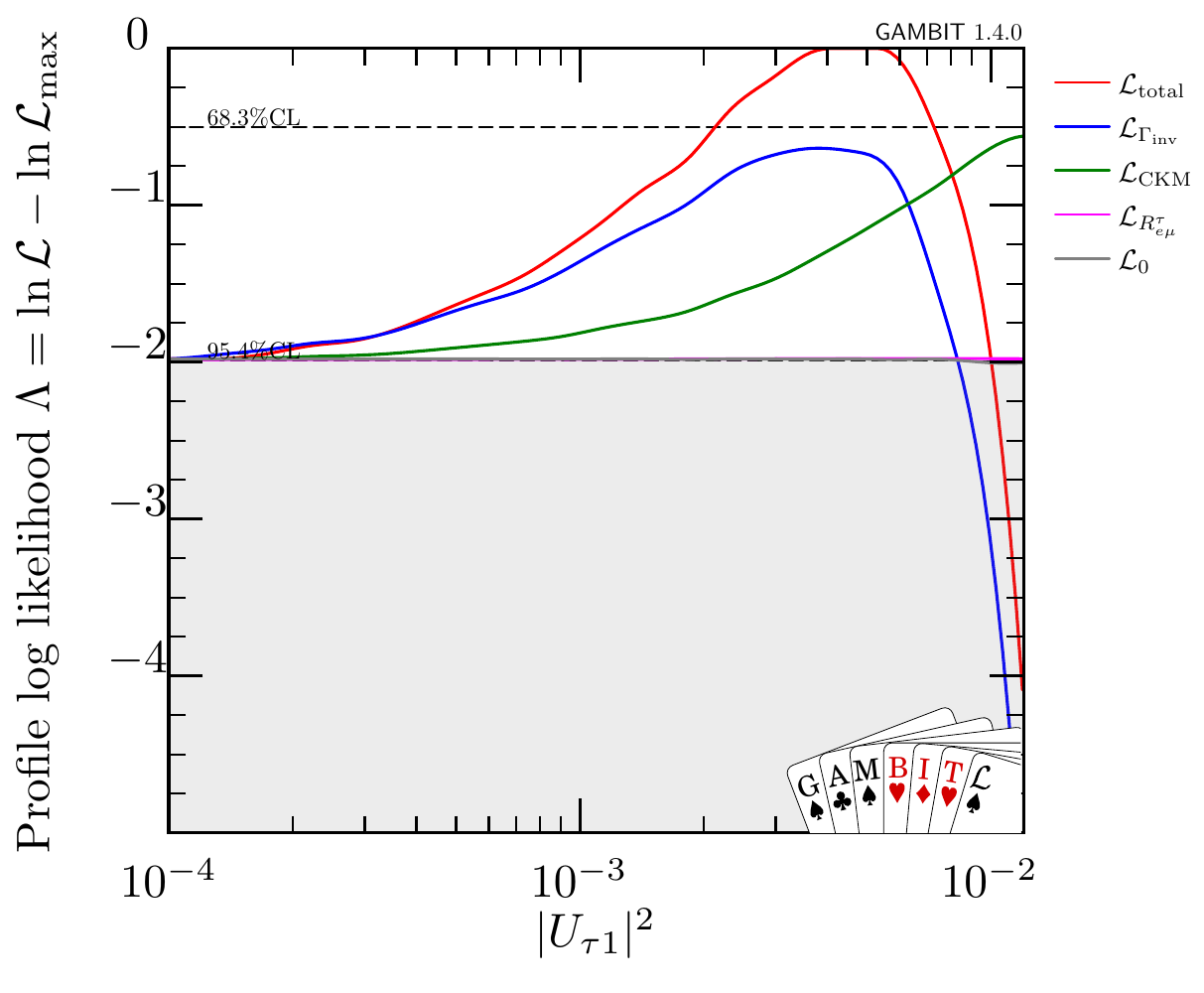}
  \includegraphics[width=0.49\linewidth]{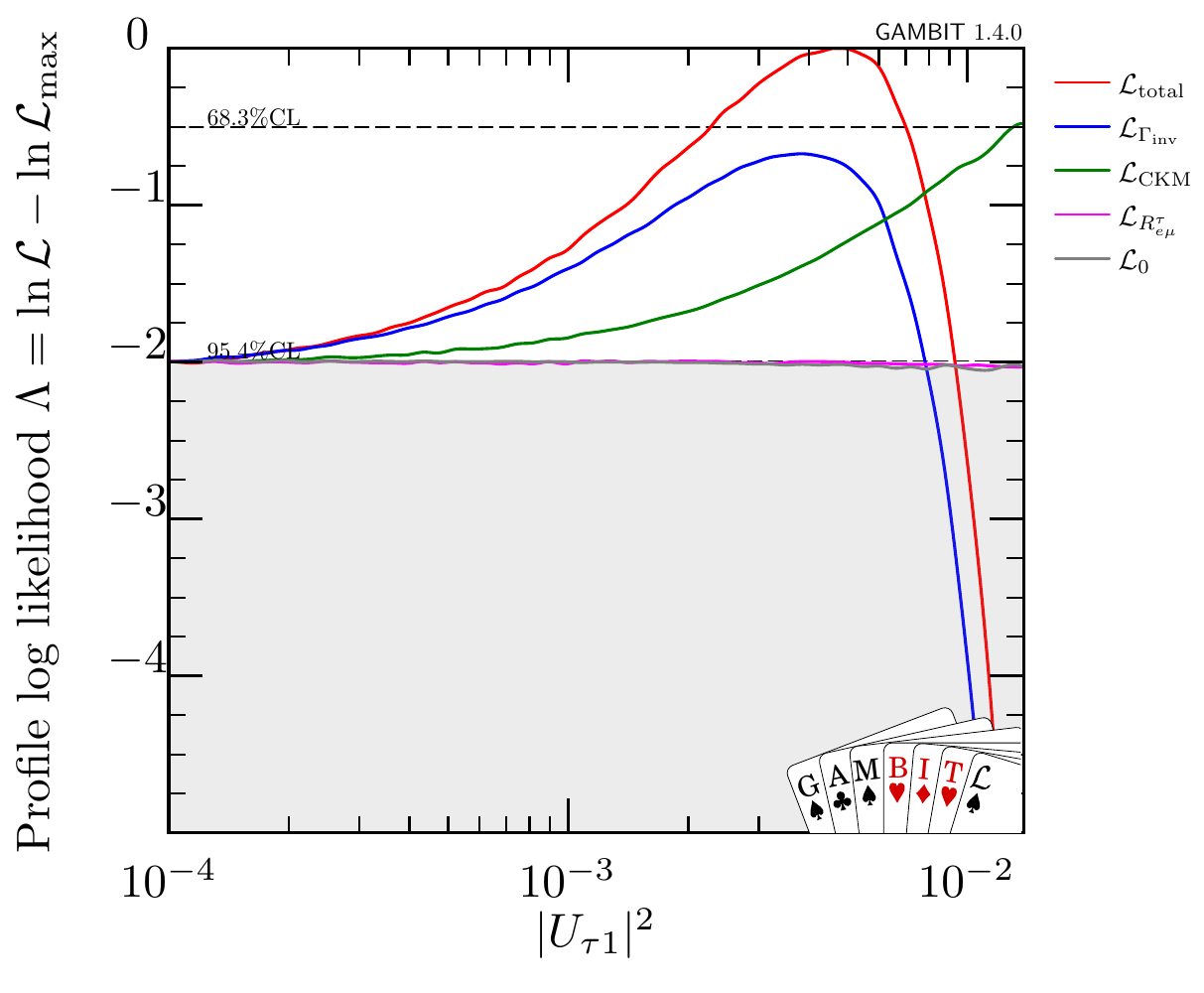}
  \caption{One-dimensional profile likelihood for $U_{\tau 1}^2$, $\mathcal{L}_{total}$, and partial likelihoods for $\Gamma_{Z}$, CKM , $R_\tau$ and combination of the rest of constraints, $\mathcal{L}_0$, in the high mass region, $M_1 > 60$ GeV, for normal (left) and inverted hierarchy (right).}
  \label{fig:Utau_lnL_highmass}
\end{figure*}
In order to study the impact of the different partial likelihoods on the total likelihood excess, we show in Figures~\ref{fig:Utau_lnL_lowmass} and \ref{fig:Utau_lnL_highmass} the partial one-dimensional likelihoods for $\Gamma_{\rm{inv}}$ (blue), CKM (green) and $R^\tau_{e\mu}$ (pink) with respect to the total likelihood (red) for $M_1 < 1$ GeV and $M_1 > 60$ GeV, respectively. All likelihoods are normalised so that they show up as a bump over the combination of all other likelihoods $\mathcal{L}_0$ (grey). These plots show that the combination of excesses from all three sources amounts to a deviation of around (high mass) or above (low mass) $2\sigma$ with respect to the background. As observed in the figures, the effect of $R_{e\mu}^\tau$ is rather negligible compared to the other two relevant likelihoods.  Even larger couplings are severely penalised by the steep drop in the $\Gamma_{\rm{inv}}$ likelihood. 

\begin{figure*}[t]
  \centering
  \includegraphics[width=0.45\linewidth]{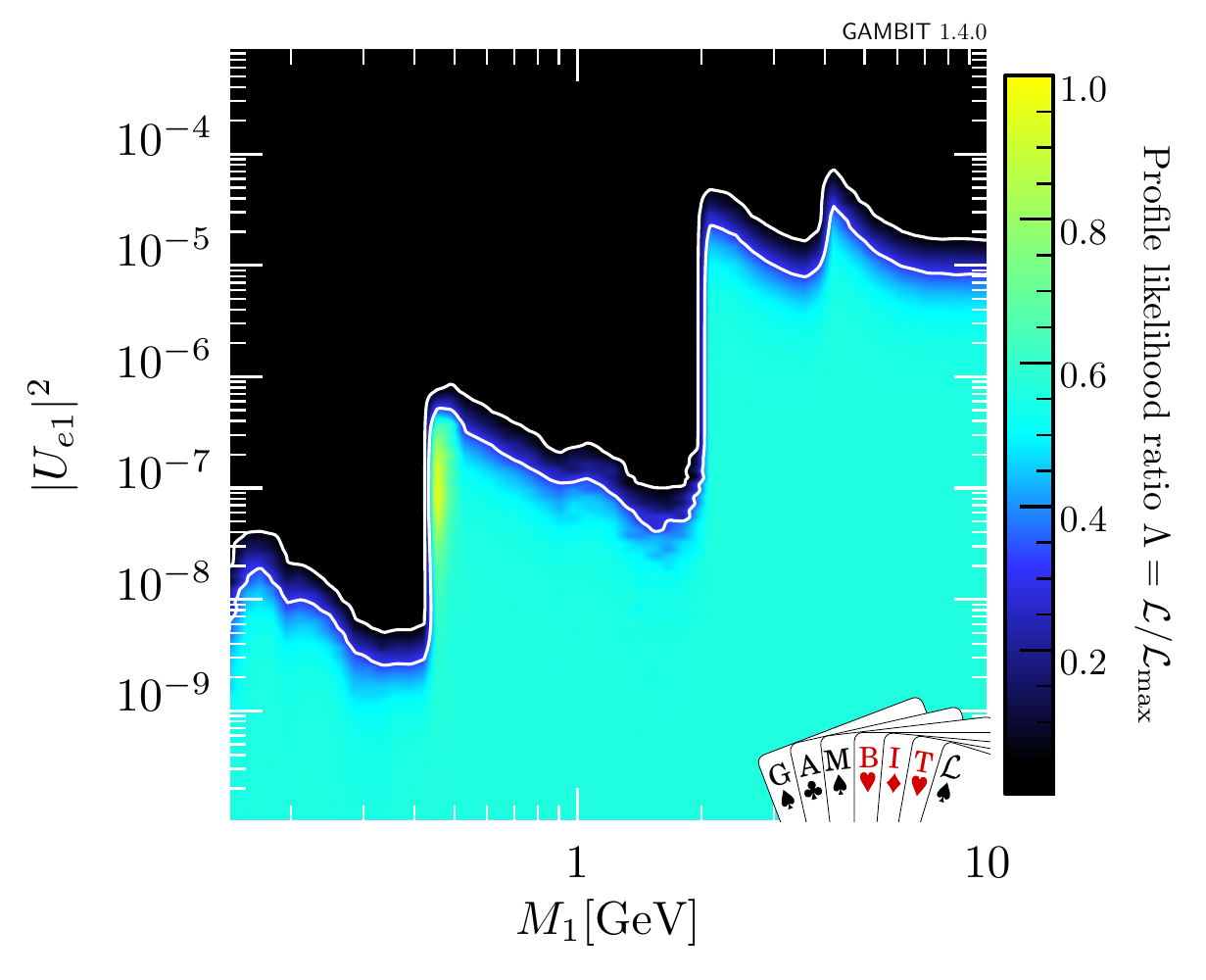}
  \includegraphics[width=0.45\linewidth]{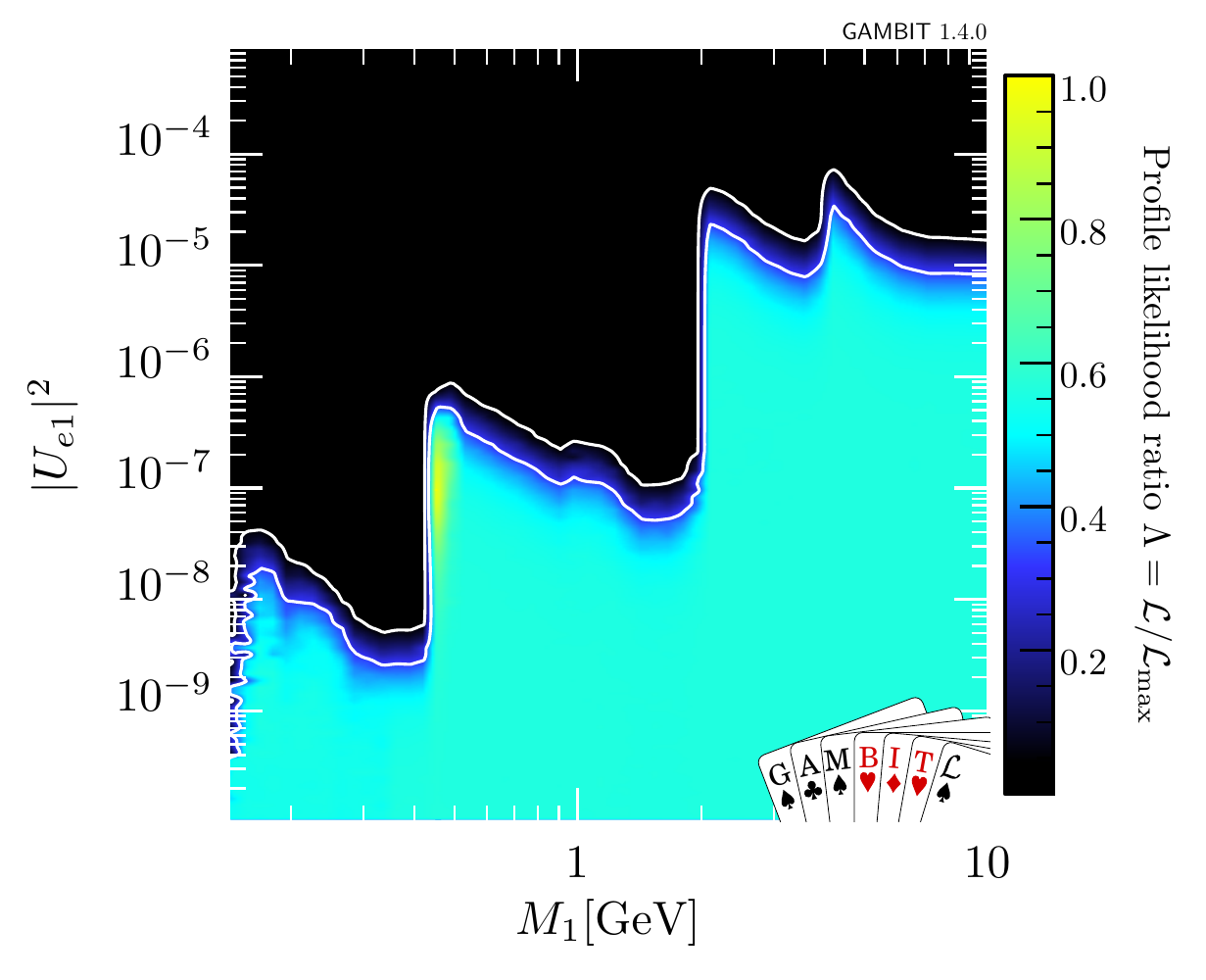}
  \caption{Profile likelihood in $M_1$ vs $U_{e 1}^2$ plane without likelihood cap showing the excesses due to the $R_K$ constraint, for normal (left) and inverted hierarchy (right).}
  \label{fig:M_Ue_RK}
\end{figure*}

\begin{figure*}[t]
  \centering
  \includegraphics[width=0.49\linewidth]{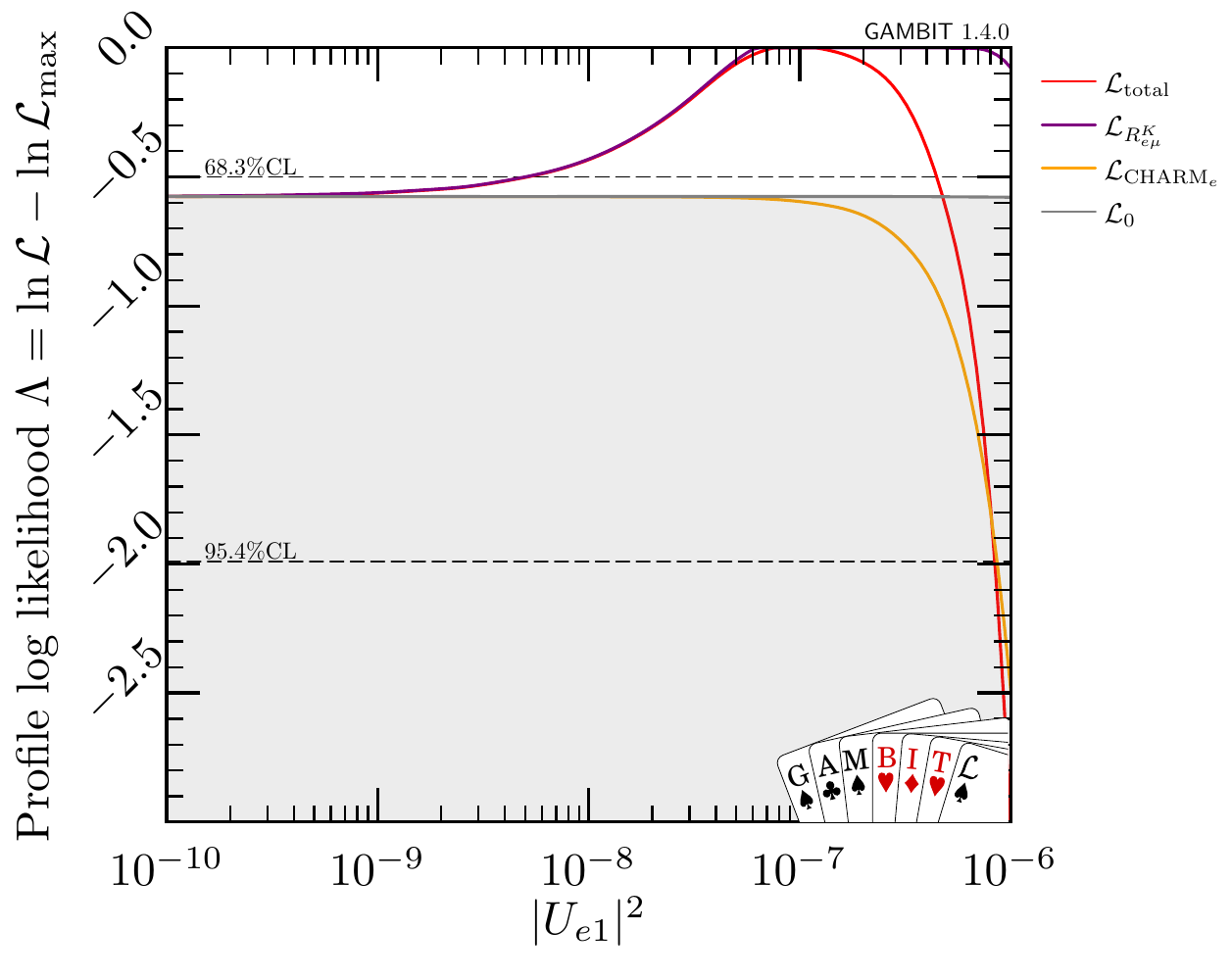}
  \includegraphics[width=0.49\linewidth]{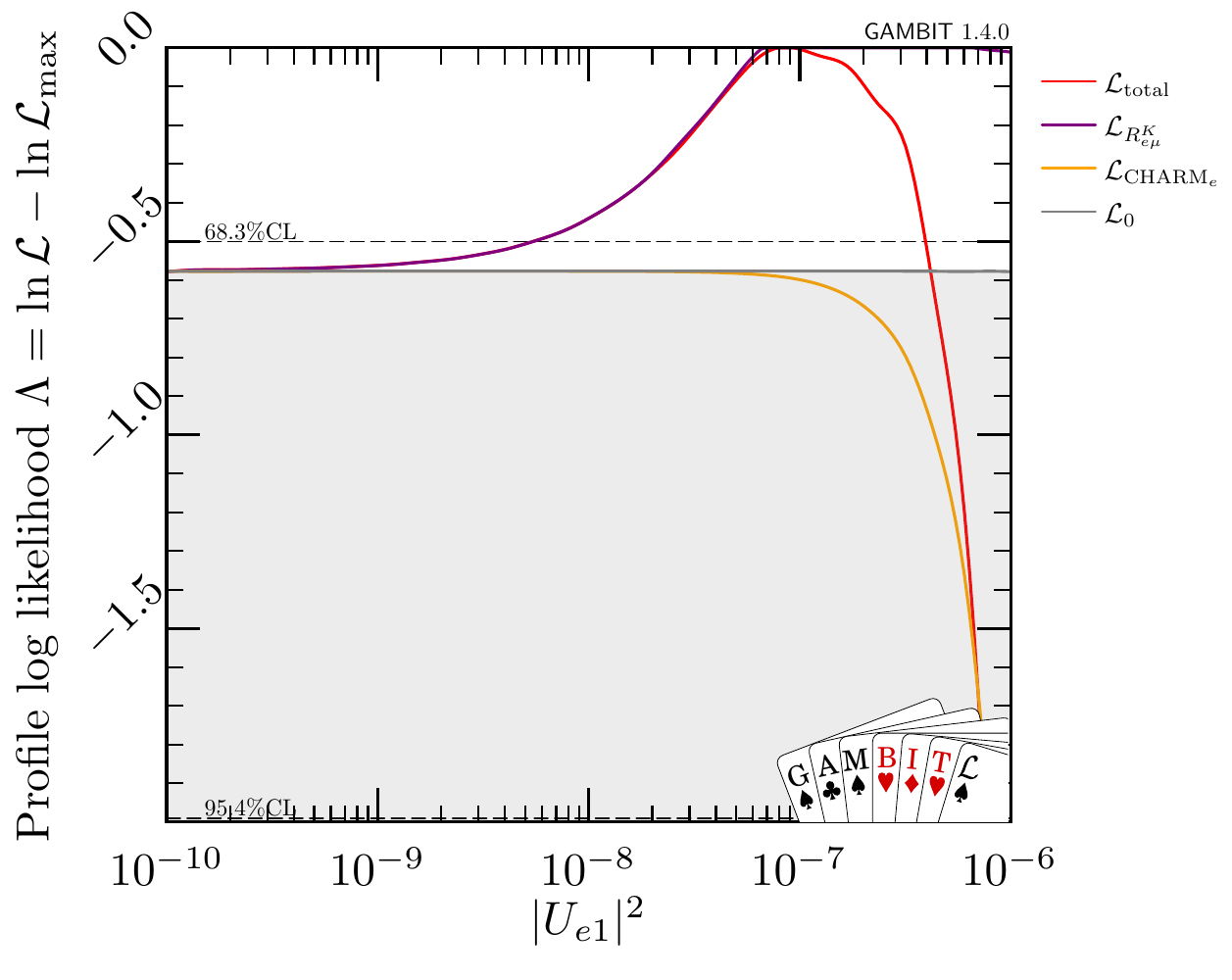}
  \caption{One-dimensional profile likelihood for $U_{e 1}^2$, $\mathcal{L}_{total}$, and partial likelihoods for $R_K$, CHARM and combination of the rest of constraints, $\mathcal{L}_0$, in the low mass region, $M_1 < 1$ GeV, for normal (left) and inverted hierarchy (right).}
  \label{fig:Ue_lnL_lowmass}
\end{figure*}

The excesses shown in Figures~\ref{fig:M_Utau_CKM_Zinv_Rtau}--\ref{fig:Utau_lnL_highmass} in $|U_{\tau 1}|^2$, for both low and high masses, are the most significant excesses arising in our three RHN scenario, but not the only ones. At masses around the $K$-meson resonance, there is an even dimmer excess in $|U_{e1}|^2$, arising from the constraint on fully leptonic decays of $K$-mesons, $R^K_{e\mu}$. As seen in Figure~\ref{fig:M_Ue_RK}, for both normal and inverted hierarchy, there is a $\sim1\sigma$ excess at $M_1 \sim 0.45$ GeV. As before, we show in Figure~\ref{fig:Ue_lnL_lowmass} the one-dimensional likelihoods for $R_{e\mu}^K$ (purple) with respect to the total likelihood (red), over the background of the combination of the rest of constraints (grey). Although the $R^K_{e\mu}$ likelihood keeps increasing for larger values of $|U_{e1}|^2$, the total likelihood drops at the limit shown in the figures due to the constraints from the CHARM experiment (orange).

\smallskip

Although the identified excesses provide interesting hints towards specific regions of the RHN parameter space, they should not be over-interpreted, since their significance remains rather small and probably consistent with statistical fluctuations. The presence of such excesses was already observed before, identified in EWPOs~\cite{Antusch:2014woa} (cf. also \cite{Akhmedov:2013hec}) and CKM constraints, and particularly in $\tau \to s$ transitions~\cite{Drewes:2015iva}.

\section{Conclusions \& Outlook}
\label{sec:concoutlook}

% CFW

We presented here the first frequentist global analysis of the extension of the Standard Model by three heavy right-handed Majorana neutrinos for a large range of their masses, from 60 MeV to 500 GeV, and for normal and inverted hierarchy of the active neutrino masses. 
As detailed in Section~\ref{subsec:MainImprovements}, our analysis improves on previous studies in numerous ways.  Most notable is the inclusion of a larger number of experimental constraints than in previous studies, such as EWPOs, all LFV decay channels, active neutrino mixing and masses, as well as many direct searches. Furthermore, we have performed a proper statistical combination of all constraints using a composite likelihood approach, and studied the overall constraints on the parameter space using robust profile likelihood methods. To this end, we have used the advanced BSM inference tool GAMBIT~\cite{gambit}, which we appropriately extended with the relevant model specifications and experimental constraints.

The results shown in Section~\ref{sec:results} cover the full studied mass range for all couplings down to $U_I^2\sim 10^{-16}$. The profile likelihood contours are consistent with the results found in previous studies. The upper limits on the heavy neutrino mixing with electron and muon flavour mostly follow the confidence levels provided by direct search experiments.
%while the mixing with the third generation is constrained by a combination of different direct and indirect searches.
In the projection of the likelihoods on the $M_I$-$U_{\alpha I}^2$ planes the interplay becomes visible only in the constraints on the third generation and for masses below a GeV.
This is qualitatively different from the model with only two heavy neutrinos ($n=2$), where combination of direct, indirect and cosmological bounds imposes stronger constraints than each of them individually on the mixing with all three SM flavours, and this interplay can rule out a considerable mass region below the kaon mass \cite{Drewes:2016jae}.
We for the first time studied the global constraints on the heavy neutrino flavour mixing pattern, which strongly depends on the mass of the lightest SM neutrino $m_{\nu_0}$.
We explicitly studied the limit of vanishing lightest neutrino masses, where we have shown that the flavour mixing pattern becomes significantly constrained for small values of $m_{\nu_0}$.
For  $m_{\nu_0} <$ 0.01 meV these constraints become independent of the precise value of $m_{\nu_0}$ in both mass hierarchies, which suggests that one heavy neutrino has effectively decoupled. In this regime we demonstrated that one can recover the results that have previously been found in the model with only two RHN in earlier works.

Furthermore, we identified a few excesses in the profile likelihood, which are due to the invisible decay width of the $Z$-boson, the CKM unitarity constraint and $R_{e\mu}^K$. Our best fit has a significance (w.r.t.~SM) slightly above $2\sigma$. Although these excesses are not significant enough to favour the $n=3$ right-handed neutrino model in favour of the SM at the moment, an improvement on the measurements of the relevant observables will increase/decrease their significance in the future.  
 Future $e^+e^-$ colliders, such as the ILC, FCC-ee or CEPC, might measure EW observables, including the $Z$ decay width, with higher precision~\cite{Fan:2014vta}
 than the current value from  LEP~\cite{Abbiendi:2000hu}. The NA62 experiment, which targets kaon decays, might be able to improve the measurments of the CKM matrix elements $V_{us}$ and $V_{ud}$, as well as the lepton universality ratio $R_{e\mu}^K$ through more precise measurement of the fully leptonic decays of kaons~\cite{Goudzovski:2010uk}.

Since the strongest constraints on the absolute value of the couplings come from direct searches, it is expected that the results obtained in this analysis will change significantly with the next generation of direct search experiments. 
An overview of projected sensitivities can e.g.~be found in Refs.~\cite{Antusch:2016ejd,Beacham:2019nyx,Alimena:2019zri}. 
Many of these searches can be performed at existing facilities, including the LHC, NA62, T2K or the DUNE near detector.
The sensitivity of the LHC will soon be upgraded 
with the
recently approved FASER experiment~\cite{Feng:2017uoz}
and other proposed dedicated detectors~\cite{Chou:2016lxi, Kling:2018wct, Gligorov:2017nwh, Curtin:2018mvb, Dercks:2018wum, Alpigiani:2018fgd, Helo:2018qej}.
In the more distant future the SHiP experiment~\cite{Alekhin:2015byh, Anelli:2015pba} can search for heavy neutrinos in the GeV mass range \cite{SHiP:2018xqw}, while future folliders such as FCC \cite{Abada:2019zxq} or CEPC \cite{CEPCStudyGroup:2018ghi} can explore larger masses.
These experimental perspectives make the study of right handed neutrinos an exciting topic for the years to come.
Additional motivation for such searches comes from cosmology because the baryon asymmetry of the universe can be explained by low scale leptogenesis for all experimentally allowed values of the mixing angles in the model considered here if the heavy neutrino masses lie below the electroweak scale \cite{Akhmedov:1998qx,Abada:2018oly}.
If any heavy neutral leptons are found in experiments then our results for their properties, such as the flavour mixing pattern as a function of light neutrino parameters, provide a powerful test to assess whether these particles are responsible for the generation of light neutrino masses and/or the baryon asymmetry of the universe \cite{Chun:2017spz}, and to distinguish the model with three heavy neutrinos considered here from the model with two heavy neutrinos or other extensions of the SM.

\section*{Acknowledgements}
The authors would like to thank S.~Antusch, C.~Weiland and R.~Ruiz as well as the rest of the GAMBIT Community for helpful discussions and comments. We acknowledge PRACE for awarding us access to Marconi at CINECA, Italy, and the Red Española de Supercomputación (Spain; FI2016-1-0021), for access to MareNostrum, Spain.  Part of this work was carried out on the Dutch national e-infrastructure (Lisa cluster) with the support of SURF Cooperative. The authors are also grateful for the computing grant on the Prometheus computer from PLGRID Infrastructure (Poland). The work of M.C.~is funded by the  Polish  National  Agency  for Academic Exchange under the  Bekker program. M.C. is also grateful to Foundation for Polish Science (FNP) for its support. The work of S.K.~and C.W.~is funded by NWO through the Vidi research grant 680-47-532. T.E.G is funded by the Australian Research Council (Discovery Project DP180102209). The work of J.H. is funded by the DFG Emmy Noether Grant No. HA 8555/1-1, and was earlier supported by the Labex ILP (reference ANR-10-LABX-63) part of the Idex SUPER, and received financial state aid managed by the Agence Nationale de la Recherche (ANR), as part of the programme Investissements d’avenir under the reference ANR-11-IDEX-0004-02. 

\clearpage
\setlength{\parskip}{0pt}

\appendix

\section{GAMBIT Implementation}
\label{app:gambit}

\GB\footnote{\href{http://gambit.hepforge.org}{gambit.hepforge.org}.} (the Global and Modular BSM Inference Tool)~\cite{gambit} is a global fitting software framework that allows for extensive calculations of observables and likelihoods in particle and astroparticle physics. It provides, out-of-the-box, a suite of statistical methods and parameter scanning algorithms, together with a hierarchical model database, a strong interface to external tools and a host of other utilities that make it one of the most powerful global fitting tools on the market.

In a nutshell, the fundamental building blocks of \GB are its \textbf{module functions}, which calculate all physical and mathematical quantities revelant to an analysis. Each module function provides a \textbf{capability} which, together with the return type of the function, unequivocally specifies the quantity calculated. 

These module functions are sorted in the different physics modules, according to their purpose, e.g. functions calculating dark matter relic density lie in \darkbit~\cite{DarkBit}. Since most observables and likelihoods computed for this analysis do not belong naturally to any of the existing \GB modules, we introduce a new \GB physics module, \neutrinobit, which contains all calculations relating to (active and sterile) neutrino physics. All of the module functions and capabilites described below are implemented in the new module \neutrinobit, unless otherwise stated.

\subsection{Neutrino models}

In \GB, models are defined by a set of parameters and relations to other models~\cite{gambit}. All the SM and active neutrino parameters are defined in a model called \textbf{StandardModel\_mNudiff}, a daughter model of the \GB model \textbf{StandardModel\_SLHA2}, which includes SM parameters written in the SLHA2 convention~\cite{Allanach:2008qq}. The former contains the parameters \term{mNu_light}, \term{dmNu21} and \term{dmNu3l}, which give the lightest neutrino mass and mass splittings, respectively. Other parameters in this model that are relevant for this study and are scanned over, as described in Section \ref{sec:scanning}, are \term{alpha1}, \term{alpha2}, \term{delta13}, \term{theta12}, \term{theta23} and \term{theta13}.

The right-handed neutrino sector is defined in another model, \textbf{RightHandedNeutrinos}, which contains the RHN masses $M_I$ and the real and imaginary parts of the $\omega_{ij}$ parameters in the C-I parametrisation (c.f. Section \ref{sec:CI}). These are \term{M_1}, \term{M_2}, \term{M_3}, \term{ReOm12}, \term{ReOm13}, \term{ReOm23}, \term{ImOm12}, \term{ImOm13}, \term{ImOm23}. To better explore the symmetry preserved region (Section \ref{sec:symmetries}), a differential model, inheriting from \textbf{RightHandedNeutrinos}, has been defined, \textbf{RightHandedNeutrinos\_diff}. This model swaps the parameters \term{M_2} for \term{delta_M21}, and defines a translation function from the parameters of the daughter model to the parent model as
\begin{equation}
 M_2 = M_1 + \delta M_{21}.
\end{equation}

Lastly, the parameter \term{Rorder} encodes the ordering of the matrices $\mathcal{R}^{ij}$ in Eq.~\eqref{Rorder}, which allows us to fully cover all the parameter space with the C-I parametrisation.

There is a number of useful quantities and observables that can be constructed from the neutrino parameters, and these are all implemented in \texttt{NeutrinoBit.cpp}. In the active neutrino sector we calculate the neutrino mass matrix, \cpp{m_nu}, and their mixing matrix, \cpp{UPMNS}, as well as useful quantities such as the type of hierarchy, \cpp{ordering}, the squared mass splittings, \cpp{md21}, \cpp{md31} and \cpp{md32}, and the minimal neutrino mass \cpp{min_mass}. It is worth noting that it is possible to fix the hierarchy of a scan by providing the option \yaml{ordering} to the capability \cpp{m_nu} in the configuration file. For example, in order to scan the normal hierarchy we would define in the \YAML file

\begin{lstyaml}
Rules:
  - capability: m_nu
    options:
      ordering: 1
\end{lstyaml}

The right-handed neutrino sector also contains a couple of useful capabilities, \cpp{SeesawI_Vnu}, which is the active neutrino mixing matrix in type-I seesaw, effectively \cpp{UPMNS} corrected by the presence of the right-handed neutrinos, and \cpp{SeesawI_Theta}, the active-sterile mixing matrix in type-I seesaw, currently implemented using the C-I parametrisation.

Another useful capability defined here is \cpp{Unitarity}, which is fulfilled by two module functions according to whether the model scanned is the SM or a RHN model, and checks whether the full mixing matrix is unitary. All these capabilities relating to neutrino masses and mixings and the module functions that fulfill them, along with their dependencies and options can be seen in Table \ref{tab:massesmixings}.

Lastly, \texttt{NeutrinoBit.cpp} also contains likelihoods for the active neutrinos, which are implemented following the results from the NuFit collaboration (c.f. Section \ref{sec:activeneutrino}). The capabilities associated with these are \cpp{md21_lnL}, \cpp{md3l_lnL} for the mass splittings, \cpp{deltaCP_lnL}, \cpp{theta12_lnL}, \cpp{theta23_lnL} and \cpp{theta13_lnL} for the phases and mixing angles, and \cpp{sum_mnu_lnL} for the cosmological limit on the sum of neutrino masses. All these capabilities, with their module functions and dependencies are listed in Table \ref{tab:NulnL}.

\begin{table*}[t]
  \centering
  \begin{tabular}{p{3cm}p{8cm}p{3cm}p{3cm}}
    \toprule
    \textbf{Capability} & \textbf{Function} (\textbf{Return Type}): \newline  \textbf{Brief description} & \textbf{Dependencies} & \textbf{Options}\\
    \midrule
    \cpp{ordering} &\cpp{ordering(bool)}: \newline  Calculates the hierarchy type. & None &\\
    \cpp{m_nu} &\cpp{M_nu(Eigen::Matrix3cd)}: \newline  Calculates the diagonalised LHN mass matrix. & \cpp{ordering} & \cpp{ordering(bool)}\\
    \cpp{md21} &\cpp{md21(double)}: \newline Calculates the square mass splitting $\Delta m^2_{21}$. & \cpp{m_nu} &\\
    \cpp{md31} &\cpp{md31(double)}: \newline Calculates the square mass splitting $\Delta m^2_{31}$. & \cpp{m_nu} &\\
    \cpp{md32} &\cpp{md32(double)}: \newline Calculates the square mass splitting $\Delta m^2_{32}$. & \cpp{m_nu} & \\
    \cpp{min_mass} &\cpp{min_mass(double)}: \newline Calculates the minimal neutrino mass. & \cpp{ordering, m_nu} &\\
    \cpp{UPMNS} & \cpp{UPMNS(Eigen::Matrix3cd)}: \newline  Calculates the PMNS matrix. & None &\\
    \cpp{SeesawI_Theta} & \cpp{CI_Theta(Eigen::Matrix3cd)}: \newline  Calculates the active-sterile mixing matrix in seesaw type-I using the C-I parametrisation. & \cpp{m_nu, UPMNS, SMINPUTS} & \\
     \cpp{SeesawI_Vnu} & \cpp{Vnu(Eigen::Matrix3cd)}: \newline  Calculates the active mixing matrix in seesaw type-I. & \cpp{UPMNS, SeesawI_Theta} &\\
     \cpp{Unitarity} & \cpp{Unitarity_UPMNS(bool)}: \newline Checks for unitarity in the SM neutrino mixing matrix. & \cpp{m_nu, UPMNS} &\\
     & \cpp{Unitarity_SeesawI(bool)}: \newline Checks for unitarity in the full neutrino mixing matrix in seesaw type-I. & \cpp{m_nu, SeesawI_Theta, SeesawI_Vnu} &\\
    \\
    \bottomrule
  \end{tabular}
  \caption{Capabilities and module functions implemented for active and sterile neutrino masses and mixings.}
  \label{tab:massesmixings}
\end{table*}

\begin{table*}[t]
  \centering
  \begin{tabular}{p{3cm}p{8cm}p{5cm}}
    \toprule
    \textbf{Capability} & \textbf{Function} (\textbf{Return Type}): \newline  \textbf{Brief description} & \textbf{Dependencies}\\
    \midrule
    \cpp{md21_lnL} &\cpp{md21_lnL(double)}: \newline  Computes the log-likehood for $\Delta m^2_{21}$. & \cpp{ordering, md21}\\
    \cpp{md3l_lnL} & \cpp{md3l_lnL(double)}: \newline  Computes the log-likehood for $\Delta m^2_{31}$ for normal hierarchy or $\Delta m^2_{32}$ for inverted. & \cpp{ordering, md31, md32}\\
    \cpp{deltaCP_lnL} & \cpp{deltaCP_lnL(double)}: \newline  Computes the log-likehood for $\delta_{CP}$. & \cpp{ordering, deltaCP} \\
    \cpp{theta12_lnL} & \cpp{theta12_lnL(double)}: \newline  Computes the log-likehood for $\theta_{12}$. & \cpp{ordering, theta12} \\
    \cpp{theta23_lnL} & \cpp{theta23_lnL(double)}: \newline  Computes the log-likehood for $\theta_{23}$. & \cpp{ordering, theta23} \\
    \cpp{theta13_lnL} & \cpp{theta13_lnL(double)}: \newline  Computes the log-likehood for $\theta_{13}$. & \cpp{ordering, theta13} \\
    \cpp{sum_mnu_lnL} & \cpp{sum_mnu_lnL(double)}: \newline  Computes the log-likehood for $\sum m_\nu$ & None
    \\
    \bottomrule
  \end{tabular}
  \caption{Capabilities and module functions implemented that calculate log-likelihoods for the active neutrino parameters.}
  \label{tab:NulnL}
\end{table*}

\subsection{Right-handed neutrino likelihood functions}

Every observable and likelihood described in Section \ref{sec:obslike} has an assigned capability within \GB. Most of these have been implemented in the new \GB module \neutrinobit, since they concern mostly neutrino physics. Their module functions are coded in the file \texttt{RightHandedNeutrinos.cpp}, to keep them separated from the likelihoods and observables concerning only active neutrinos in \texttt{NeutrinoBit.cpp}. The exception to this is the LFV observables and semileptonic lepton universality tests, which can be found in \flavbit~\cite{FlavBit}, implemented in \texttt{FlavBit.cpp} and the electroweak precision observables, which are coded up in \texttt{PrecisionBit.cpp} in \precisionbit~\cite{SDPBit}.

The implementation details for each specific observable are as follows:\\
\\
\noindent \textit{Electroweak precision observables}\\
\\
Mainly, the EWPO capabilities lie in the physics module \precisionbit and the associated module functions are implemented in \texttt{PrecisionBit.cpp}. These capabilites, \cpp{prec_sinW2_eff} and \cpp{mW}, can be seen in Table \ref{tab:ewpocaps} along with their module functions and dependencies. The log-likelihoods are provided by the capabilities \cpp{lnL_sinW2_eff} and \cpp{lnL_W_mass} can also be seen in the same Table. Additionally, the module \decaybit contains the capabilities for the invisible width of $Z$, which are \cpp{Z_gamma_nu} and \cpp{lnL_Z_inv}, and leptonic decays of the $W$ boson, \cpp{W_to_l_decays} and \cpp{lnL_W_decays}, all of which can be seen as well in Table \ref{tab:ewpocaps}.\\

\begin{table*}[t]
  \centering
  \begin{tabular}{p{2.5cm}p{7cm}p{6.5cm}}
    \toprule
    \textbf{Capability} & \textbf{Function} (\textbf{Return Type}): \newline  \textbf{Brief description} & \textbf{Dependencies}\\
    \midrule
    \cpp{prec_sinW2_eff} &\cpp{RHN_sinW2_eff(triplet<double>)}: \newline Calculates $s_{eff}^2$. & \cpp{SeesawI_Theta}\\
    \cpp{mW} & \cpp{RHN_mW(triplet<double>)}: \newline  Calculates $m_W$. & \cpp{sinW2, SeesawI_Theta}\\
    \cpp{Z_gamma_nu} & \cpp{Z_gamma_nu_2l(triplet<double>)}: \newline  Calculates the decay width of $Z$ to neutrinos. & \cpp{SM_spectrum, SeesawI_Theta, SeesawI_Vnu} \\
    \cpp{W_to_l_decays} & \cpp{RHN_W_to_l_decays(vector<double>)}: \newline  Calculates the decay width of the processes $W \to l\nu$. & \cpp{SMINPUTS, mw, SeesawI_Theta} \\
    \cpp{lnL_sinW2_eff} &\cpp{lnL_sinW2_eff_chi2(double)}: \newline Computes the log-likehood for $s_{eff}^2$. & \cpp{prec_sinW2_eff}\\
    \cpp{lnL_W_mass} & \cpp{lnL_W_mass_chi2(double)}: \newline  Computes the log-likehood for $m_W$ . & \cpp{mW}\\
    \cpp{lnL_Z_inv} & \cpp{lnL_Z_inv(double)}: \newline  Computes the log-likehood for $\Gamma_{inv}$. & \cpp{Z_gamma_nu} \\
    \cpp{lnL_W_decays} & \cpp{lnL_W_decays_chi2(double)}: \newline  Computes the log-likehood for $\Gamma_{W\to l\nu}$. & \cpp{W_to_l_decays, W_plus_decay_rates}\\
    \bottomrule
  \end{tabular}
  \caption{Capabilities and module functions that calculate electroweak precision observables and their likelihoods.}
  \label{tab:ewpocaps}
\end{table*}

\noindent \textit{Lepton flavour violation}\\
\\
The capabilities related to lepton flavour violation can be found in \flavbit and are \cpp{muegamma}, \cpp{tauegamma}, \cpp{taumugamma}, \cpp{mueee}, \cpp{taueee}, \cpp{taumumumu}, \cpp{taumuee}, \cpp{taueemu}, \cpp{tauemumu}, \cpp{taumumue}, \cpp{mueTi}, \cpp{mueAu} and \cpp{muePb}. Table \ref{tab:lfvcaps} shows these capabilities, the module functions that provide them, implemented in \texttt{FlavBit.cpp}, and their dependencies. The likelihoods, shown in Table \ref{tab:lfvlnL}, are collated into three capabilites, according to the type of process, \cpp{l2lgamma_lnL} for $l \to l \gamma$, \cpp{l2lll_lnL} for $l^- \to l^- l^- l^+$ and \cpp{mu2e_lnL} for $\mu - e$ conversion in nuclei. \\

\begin{table*}[t]
  \centering
  \begin{tabular}{p{3cm}p{5cm}p{8cm}}
    \toprule
    \textbf{Capability} & \textbf{Function} (\textbf{Return Type}): \newline  \textbf{Brief description} & \textbf{Dependencies}\\
    \midrule
    \cpp{muegamma} &\cpp{RHN_muegamma(double)}: \newline  Calculates $BR(\mu^-\to e^-\gamma)$. & \cpp{SMINPUTS, m_nu, SeesawI_Vnu, SeesawI_Theta, mu_minus_decay_rates}\\
    \cpp{tauegamma} & \cpp{RHN_tauegamma(double)}: \newline  Calculates $BR(\tau^-\to e^-\gamma)$.  & \cpp{SMINPUTS, m_nu, SeesawI_Vnu, SeesawI_Thet, tau_minus_decay_rates}\\
    \cpp{taumugamma} & \cpp{RHN_taumugamma(double)}: \newline  Calculates $BR(\tau^-\to \mu^-\gamma)$.  & \cpp{SMINPUTS, m_nu, SeesawI_Vnu, SeesawI_Theta, tau_minus_decay_rates}\\
    \cpp{mueee} & \cpp{RHN_mueee(double)}: \newline  Calculates $BR(\mu^-\to e^- e^- e^+)$. & \cpp{SMINPUTS, m_nu, SeesawI_Vnu, SeesawI_Theta, mu_minus_decay_rates}\\
    \cpp{taueee} & \cpp{RHN_taueee(double)}: \newline  Calculates $BR(\tau^-\to e^-e^-e^+)$.  & \cpp{SMINPUTS, m_nu, SeesawI_Vnu, SeesawI_Theta, tau_minus_decay_rates}\\
    \cpp{taumumumu} & \cpp{RHN_taumumumu(double)}: \newline  Calculates $BR(\tau^-\to \mu^-\mu^-\mu^+)$.  & \cpp{SMINPUTS, m_nu, SeesawI_Vnu, SeesawI_Theta, tau_minus_decay_rates}\\
    \cpp{taumuee} & \cpp{RHN_taumuee(double)}: \newline  Calculates $BR(\tau^-\to \mu^-e^-e^+)$.  & \cpp{SMINPUTS, m_nu, SeesawI_Vnu, SeesawI_Theta, tau_minus_decay_rates}\\
    \cpp{taueemu} & \cpp{RHN_taueemu(double)}: \newline  Calculates $BR(\tau^-\to e^-e^-\mu^+)$.  & \cpp{SMINPUTS, m_nu, SeesawI_Vnu, SeesawI_Theta, tau_minus_decay_rates}\\
    \cpp{tauemumu} & \cpp{RHN_tauemumu(double)}: \newline  Calculates $BR(\tau^-\to e^-\mu^-\mu^+)$.  & \cpp{SMINPUTS, m_nu, SeesawI_Vnu, SeesawI_Theta, tau_minus_decay_rates}\\
    \cpp{taumumue} & \cpp{RHN_taumumue(double)}: \newline  Calculates $BR(\tau^-\to \mu^-\mu^-e^+)$.  & \cpp{SMINPUTS, m_nu, SeesawI_Vnu, SeesawI_Theta, tau_minus_decay_rates}\\
    \cpp{mueTi} & \cpp{RHN_mueTi(double)}: \newline  Calculates $R(\mu - e)$ in a Ti nucleus. & \cpp{SMINPUTS, m_nu, SeesawI_Vnu, SeesawI_Theta}\\
    \cpp{mueAu} & \cpp{RHN_mueAu(double)}: \newline  Calculates $R(\mu - e)$ in a Au nucleus. & \cpp{SMINPUTS, m_nu, SeesawI_Vnu, SeesawI_Theta}\\
    \cpp{muePb} & \cpp{RHN_muePb(double)}: \newline  Calculates $R(\mu - e)$ in a Pb nucleus. & \cpp{SMINPUTS, m_nu, SeesawI_Vnu, SeesawI_Theta}\\
    \bottomrule
  \end{tabular}
  \caption{Capabilities and module functions to calculate LFV observables.}
  \label{tab:lfvcaps}
\end{table*}

\begin{table*}[t]
  \centering
  \begin{tabular}{p{3cm}p{8cm}p{5cm}}
    \toprule
    \textbf{Capability} & \textbf{Function} (\textbf{Return Type}): \newline \textbf{Brief description} & \textbf{Dependencies}\\
    \midrule
    \cpp{l2lgamma_lnL} &\cpp{l2lgamma_likelihood(double)}: \newline Computes the log-likelihood for $l^-\to l^-\gamma$. & \cpp{muegamma, tauegamma, taumugamma}\\
    \cpp{l2lll_lnL} & \cpp{l2lll_likelihood(double)}: \newline Computes the log-likelihood for $l^-\to l^-l^-l^+$. & \cpp{mueee, taueee, taumumumu, taumumue, tauemumu}\\
    \cpp{mu2e_lnL} & \cpp{mu2e_likelihood(double)}: \newline Computes the log-likelihood associated with $\mu - e$ conversion. & \cpp{mueTi, muAu, muePb}\\
    \bottomrule
  \end{tabular}
  \caption{Capabilities and module functions for the likelihoods computed for the LFV observables.}
  \label{tab:lfvlnL}
\end{table*}

\noindent \textit{Lepton universality}\\
\\
The observables and likelihoods associated with lepton universality constraints are spread between the \neutrinobit and \flavbit modules. Those involving fully leptonic decays are implemented in \texttt{RightHandedNeutrinos.cpp} and those for semi-leptonic decays of $B$ mesons are in \texttt{FlavBit.cpp}. The capabilities for leptonic decays are \cpp{R_pi}, \cpp{R_K}, \cpp{R_tau} and \cpp{R_W}, and for semi-leptonic \cpp{RK}, \cpp{RKstar_0045_11} and \cpp{RKstar_11_60}. They can be seen in Table~\ref{tab:lepunivcaps} together with the module functions that provide them and their dependencies. The capability \cpp{LUV_M} collates all semi-leptonic universality observables into the \flavbit-defined class \cpp{FlavBit::predictions_measurements_covariances}\footnote{For more details about \flavbit types, see \cite{FlavBit}.}. The capabilites that compute the likelihoods for lepton universality tests are \cpp{lnL_R_pi}, \cpp{lnL_R_K}, \cpp{lnL_R_tau} and \cpp{lnL_R_W} for leptonic decays, and \cpp{LUV_LL} for semi-leptonic, and they, the module functions and dependencies, can be seen in Table \ref{tab:lepunivlnL}. \\

\begin{table*}[t]
  \centering
  \begin{tabular}{p{3cm}p{8cm}p{5cm}}
    \toprule
    \textbf{Capability} & \textbf{Function} (\textbf{Return Type}): \newline  \textbf{Brief description} & \textbf{Dependencies}\\
    \midrule
    \cpp{R_pi} &\cpp{RHN_R_pi(double)}: \newline  Calculates the test of lepton universality $R^\pi_{e\mu}$. & \cpp{SMINPUTS, SeesawI_Theta, SeesawI_Vnu}\\
    \cpp{R_K} & \cpp{RHN_R_K(double)}: \newline   Calculates the test of lepton universality $R^K_{e\mu}$.. & \cpp{SMINPUTS, SeesawI_Theta, SeesawI_Vnu}\\
    \cpp{R_tau} & \cpp{RHN_R_tau(double)}: \newline   Calculates the test of lepton universality $R^\tau_{\mu e}$. & \cpp{SMINPUTS, SeesawI_Theta}\\
    \cpp{R_W} & \cpp{RHN_R_W(vector<double>)}: \newline  Calculates the test of lepton universality $R^W_{\alpha\beta}$. & \cpp{W_to_l_decays}\\
    \cpp{RK} & \cpp{RHN_RK(double)}: \newline  Calculates the test of lepton universality $R_K$. & \cpp{SMINPUTS, SeesawI_Theta}\\
    \cpp{RKstar_0045_11} & \cpp{RHN_RKstar_0045_11(double)}: \newline  Calculates the test of lepton universality $R_{K^*}$ for the range $0.045 < q^2 < 1.1 \text{GeV}^2$. & \cpp{SMINPUTS, SeesawI_Theta}\\
    \cpp{RKstar_11_60} & \cpp{RHN_RKstar_11_60(double)}: \newline  Calculates the test of lepton universality $R_{K^*}$ for the range $1.1 < q^2 < 6.0 \text{GeV}^2$. & \cpp{SMINPUTS, SeesawI_Theta}\\
    \bottomrule
  \end{tabular}
  \caption{Capabilities and module functions that calculate lepton universality observables.}
  \label{tab:lepunivcaps}
\end{table*}

\begin{table*}[t]
  \centering
  \begin{tabular}{p{3cm}p{8cm}p{5cm}}
    \toprule
    \textbf{Capability} & \textbf{Function} (\textbf{Return Type}): \newline  \textbf{Brief description} & \textbf{Dependencies}\\
    \midrule
    \cpp{lnL_R_pi} & \cpp{lnL_R_pi(double)}: \newline  Calculates the total log-likelihood for lepton universality tests on leptonic decays of $\pi$ mesons. & \cpp{R_pi}\\
    \cpp{lnL_R_K} & \cpp{lnL_R_K(double)}: \newline  Calculates the total log-likelihood for lepton universality tests on leptonic decays of $K$ mesons. & \cpp{R_K}\\
    \cpp{lnL_R_tau} & \cpp{lnL_R_tau(double)}: \newline  Calculates the total log-likelihood for lepton universality tests on leptonic decays of $\tau$ leptons. & \cpp{R_tau}\\
    \cpp{lnL_R_W} & \cpp{lnL_R_W(double)}: \newline  Calculates the total log-likelihood for lepton universality tests on leptonic decays of $W$ bosons. & \cpp{R_W}\\
    \cpp{LUV_M} & \cpp{LUV_measurements()}:\newline Collates all measurements of semi-leptonic tests of lepton universality in $B$ meson decays.& \cpp{RK, RKstar_0045_11, RKstar_11_60}\\
    \cpp{LUV_LL} & \cpp{lnL_lepuniv(double)}: \newline  Calculates the total log-likelihood for semi-leptonic tests of lepton universality in $B$ meson decays. & \cpp{LUV_M}\\
    \bottomrule
  \end{tabular}
  \caption{Capabilities and module functions for the likelihoods computed for lepton universality tests.}
  \label{tab:lepunivlnL}
\end{table*}

\noindent \textit{CKM unitarity}\\
\\
The \neutrinobit capability \cpp{calc_Vus}, implemented in \texttt{RightHandedNeutrinos.cpp}, calculates the value of $V_{us}$ that maximizes the likelihood for a given mixing matrix $\Theta$. The capabilities \cpp{lnLckm_Vus} and \cpp{lnLckm_Vusmin} compute the log-likelihood using $V_{us}$ as a scan parameter and as calculated by the profiling of \cpp{calc_Vus}, respectively. The capabilities, module functions and dependencies defined in \GB for the calculation of the observable and the likelihood connected to CKM unitarity are listed in Tab.~\ref{tab:ckmcaps}. \\

\begin{table*}[t]
  \centering
  \begin{tabular}{p{3cm}p{8cm}p{5cm}}
    \toprule
    \textbf{Capability} & \textbf{Function} (\textbf{Return Type}): \newline  \textbf{Brief description} & \textbf{Dependencies}\\
    \midrule
    \cpp{calc_Vus} &\cpp{calc_Vus(double)}: \newline  Calculates the profiling value of $V_{us}$ for a particular $\Theta$. & \cpp{SMINPUTS, SeesawI_Theta}\\
    \cpp{lnLckm_Vus} & \cpp{lnL_ckm_Vus(double)}: \newline Computes the total log-likelihood from CKM unitarity for a given parameter $V_{us}$ & \cpp{SMINPUTS, SeesawI_Theta}\\
    \cpp{lnLckm_Vusmin} & \cpp{lnL_ckm_Vusmin(double)}: \newline  Computes the total log-likelihood from CKM unitarity profiling over $V_{us}$. & \cpp{SMINPUTS, SeesawI_Theta, calc_Vus}\\
    \bottomrule
  \end{tabular}
  \caption{Capabilities and module functions implemented to calculate CKM unitarity and its likelihood.}
  \label{tab:ckmcaps}
\end{table*}

\noindent \textit{Neutrinoless double beta decay}\\
\\
In \neutrinobit there are two computations of the likelihood for neutrinoless double beta decay, one based on the half-life and one based on the invariant mass of the two electrons $m_{\beta\beta}$. The capabilities \cpp{Thalf_0nubb_Xe} and \cpp{Thalf_0nubb_Ge} calculate the half-life of the $0\nu\beta\beta$ process as studied with Xe and Ge detectors. Equivalently, the capabilities \cpp{mbb_0nubb_Xe} and \cpp{mbb_0nubb_Ge} compute $m_{\beta\beta}$ for the process in Xe and Ge detectors. The log-likelihoods are computed according to  the experiments: \cpp{lnL_0nubb_KamLAND_Zen} and \cpp{lnL_mbb_0nubb_KamLAND_Zen} calculate the log-likelihood for the KamLAND-Zen experiment based on the half-life and $m_{\beta\beta}$, respectively; and \cpp{lnL_0nubb_GERDA} and \cpp{lnL_mbb_0nubb_GERDA} for the GERDA experiment. Lastly, the total log-likelihood is given by the capabilities \cpp{lnL_0nubb}, constructed from the half-life, and \cpp{lnL_mbb_0nubb}, from $m_{\beta\beta}$. Tab.~\ref{tab:0nubb} shows the defined capabilities, associated module functions and dependencies related to neutrinoless double beta decay that are responsible for the calculation of observables and likelihoods.\\

\begin{table*}[t]
  \centering
  \begin{tabular}{p{4cm}p{7cm}p{5cm}}
    \toprule
    \textbf{Capability} & \textbf{Function} (\textbf{Return Type}): \newline  \textbf{Brief description} & \textbf{Dependencies}\\
    \midrule
    \cpp{Thalf_0nubb_Xe} & \cpp{RHN_Thalf_0nubb_Xe(double)}: \newline Calculates the half-life for Xe detector. & \cpp{m_nu, UPMNS, SeesawI_Theta} \\
    \cpp{Thalf_0nubb_Ge} & \cpp{RHN_Thalf_0nubb_Ge(double)}: \newline Calculates the half-life for Ge detector. & \cpp{m_nu, UPMNS, SeesawI_Theta} \\
    \cpp{mbb_0nubb_Xe} &\cpp{RHN_mbb_0nubb_Xe(double)}: \newline  Calculates $m_{\beta\beta}$ for Xe detector. & \cpp{m_nu, UPMNS, SeesawI_Theta}\\
    \cpp{mbb_0nubb_Ge} & \cpp{RHN_mbb_0nubb_Ge(double)}: \newline  Calculates $m_{\beta\beta}$ for Ge detector. & \cpp{m_nu, UPMNS, SeesawI_Theta}\\
    \cpp{lnL_0nubb_KamLAND_Zen} & \cpp{lnL_0nubb_KamLAND_Zen(double)}: \newline  Calculates KamLAND-Zen log-likelihood based on half-life. & \cpp{Thalf_0nubb_Xe}\\
    \cpp{lnL_0nubb_GERDA} & \cpp{lnL_0nubb_GERDA(double)}: \newline  Calculates GERDA log-likelihood  based on half-life. & \cpp{Thalf_0nubb_Ge}\\
    \cpp{lnL_0nubb} & \cpp{lnL_0nubb(double)}: \newline  Calculates the total log-likelihood based on half-life. & \cpp{lnL_0nubb_KamLAND_Zen, lnL_0nubb_GERDA}\\
    \cpp{lnL_mbb_0nubb_KamLAND_Zen} & \cpp{lnL_mbb_0nubb_KamLAND_Zen(double)}: \newline  Calculates KamLAND-Zen log-likelihood based on $m_{\beta\beta}$. & \cpp{mbb_0nubb_Xe}\\
    \cpp{lnL_mbb_0nubb_GERDA} & \cpp{lnL_mbb_0nubb_GERDA(double)}: \newline  Calculates GERDA log-likelihood based on $m_{\beta\beta}$. & \cpp{mbb_0nubb_Ge}\\
    \cpp{lnL_mbb_0nubb} & \cpp{lnL_mbb_0nubb(double)}: \newline  Calculates the total log-likelihood based on $m_{\beta\beta}$. & \cpp{lnL_mbb_0nubb_KamLAND_Zen, lnL_mbb_0nubb_GERDA}\\
    \bottomrule
  \end{tabular}
  \caption{Capabilities and module functions implemented to calculate neutrinoless double-beta decay observables and likelihood.}
  \label{tab:0nubb}
\end{table*}

\noindent \textit{Big Bang nucleosynthesis}\\
\\
There are a number of processes that contribute to the decay width of the right-handed neutrinos, relevant for Big Bang nucleosynthesis, and each of them is computed by a capability. These are   \cpp{Gamma_RHN2piplusl}, \cpp{Gamma_RHN2Kplusl}, \cpp{Gamma_RHN2Dplusl}, \cpp{Gamma_RHN2Dsl}, \cpp{Gamma_RHN2Bplusl}, \cpp{Gamma_RHN2Bcl}, \cpp{Gamma_RHN2pi0nu}, \cpp{Gamma_RHN2etanu},\; \cpp{Gamma_RHN2etaprimenu}, \cpp{Gamma_RHN2etacnu}, \cpp{Gamma_RHN2rhoplusl}, \cpp{Gamma_RHN2Dstarplusl}, \cpp{Gamma_RHN2Dstarsl}, \cpp{Gamma_RHN2rho0nu}, 
\cpp{Gamma_RHN2omeganu}, \cpp{Gamma_RHN2phinu}, \cpp{Gamma_RHN2Jpsinu}, \cpp{Gamma_RHN23nu}, \cpp{Gamma_RHN2llnu}, \cpp{Gamma_RHN2null}, \cpp{Gamma_RHN2nuuubar}, \cpp{Gamma_RHN2nuddbar} and \cpp{Gamma_RHN2ludbar}. The total decay width of each of the right-handed neutrinos is given by \cpp{Gamma_BBN} and the log-likehood for BBN by \cpp{lnL_bbn}. Tab.~\ref{tab:bbn} shows the capabilities, as defined in \GB, that pertain to Big Bang nucleosynthesis, and the module functions that satisfy them, along with dependencies that other module functions fulfill. The decay process considered in each function is mentioned below its name. \\

\begin{table*}[t]
  \centering
  \begin{tabular}{p{3cm}p{8cm}p{5cm}}
    \toprule
    \textbf{Capability} & \textbf{Function} (\textbf{Return Type}): \newline  \textbf{Brief description} & \textbf{Dependencies}\\
    \midrule
    \cpp{Gamma_RHN2piplusl} & \cpp{Gamma_RHN2piplusl(std::vector<double>)}: \newline  Calculates $\Gamma(N_I \to \pi^+l_{\alpha}^-)$. & \cpp{SMINPUTS, SeesawI_Theta}\\
    \cpp{Gamma_RHN2Kplusl} & \cpp{Gamma_RHN2Kplusl(std::vector<double>)}: \newline  Calculates $\Gamma(N_I \to K^+l_{\alpha}^-)$. & \cpp{SMINPUTS, SeesawI_Theta}\\
    \cpp{Gamma_RHN2Dplusl} & \cpp{Gamma_RHN2Dplusl(std::vector<double>)}: \newline  Calculates $\Gamma(N_I \to D^+l_{\alpha}^-)$. & \cpp{SMINPUTS, SeesawI_Theta}\\
    \cpp{Gamma_RHN2Dsl} & \cpp{Gamma_RHN2Dsl(std::vector<double>)}: \newline  Calculates $\Gamma(N_I \to D_sl_{\alpha}^-)$. & \cpp{SMINPUTS, SeesawI_Theta}\\
    \cpp{Gamma_RHN2Bplusl} & \cpp{Gamma_RHN2Bplusl(std::vector<double>)}: \newline  Calculates $\Gamma(N_I \to B^+l_{\alpha}^-)$. & \cpp{SMINPUTS, SeesawI_Theta}\\
    \cpp{Gamma_RHN2Bcl} & \cpp{Gamma_RHN2Bcl(std::vector<double>)}: \newline  Calculates $\Gamma(N_I \to B_cl_{\alpha}^-)$. & \cpp{SMINPUTS, SeesawI_Theta}\\
    \cpp{Gamma_RHN2pi0nu} &\cpp{Gamma_RHN2pi0nu(std::vector<double>)}: \newline  Calculates $\Gamma(N_I \to \pi^0\nu_{\alpha})$. & \cpp{SMINPUTS, SeesawI_Theta}\\
    \cpp{Gamma_RHN2etanu} & \cpp{Gamma_RHN2etanu(std::vector<double>)}: \newline  Calculates $\Gamma(N_I \to \eta\nu_{\alpha})$. & \cpp{SMINPUTS, SeesawI_Theta}\\
    \cpp{Gamma_RHN2etaprimenu} & \cpp{Gamma_RHN2etaprimenu(std::vector<double>)}: \newline 
    Calculates $\Gamma(N_I \to \eta'\nu_{\alpha})$. & \cpp{SMINPUTS, SeesawI_Theta}\\
    \cpp{Gamma_RHN2etacnu} & \cpp{Gamma_RHN2etacnu(std::vector<double>)}: \newline  Calculates $\Gamma(N_I \to \eta_c\nu_{\alpha})$. & \cpp{SMINPUTS, SeesawI_Theta}\\
    \cpp{Gamma_RHN2rhoplusl} & \cpp{Gamma_RHN2rhoplusl(std::vector<double>)}: \newline  Calculates $\Gamma(N_I \to \rho^+l_{\alpha}^-)$. & \cpp{SMINPUTS, SeesawI_Theta}\\
    \cpp{Gamma_RHN2Dstarplusl} & \cpp{Gamma_RHN2Dstarplusl(std::vector<double>)}: \newline  Calculates $\Gamma(N_I \to D^{*+}l_{\alpha}^-)$. & \cpp{SMINPUTS, SeesawI_Theta}\\
    \cpp{Gamma_RHN2Dstarsl} & \cpp{Gamma_RHN2Dstarsl(std::vector<double>)}: \newline  Calculates $\Gamma(N_I \to D^*_s l_{\alpha}^-)$. & \cpp{SMINPUTS, SeesawI_Theta}\\
    \cpp{Gamma_RHN2rho0nu} & \cpp{Gamma_RHN2rho0nu(std::vector<double>)}: \newline  Calculates $\Gamma(N_I \to \rho^0\nu_{\alpha})$. & \cpp{SMINPUTS, SeesawI_Theta}\\
    \cpp{Gamma_RHN2omeganu} & \cpp{Gamma_RHN2omeganu(std::vector<double>)}: \newline  Calculates $\Gamma(N_I \to \omega\nu_{\alpha})$. & \cpp{SMINPUTS, SeesawI_Theta}\\
    \cpp{Gamma_RHN2phinu} & \cpp{Gamma_RHN2phinu(std::vector<double>)}: \newline  Calculates $\Gamma(N_I \to \phi\nu_{\alpha})$. & \cpp{SMINPUTS, SeesawI_Theta}\\
    \cpp{Gamma_RHN2Jpsinu} & \cpp{Gamma_RHN2Jpsinu(std::vector<double>)}: \newline  Calculates $\Gamma(N_I \to J/\psi\nu_{\alpha})$. & \cpp{SMINPUTS, SeesawI_Theta}\\
    \cpp{Gamma_RHN23nu} & \cpp{Gamma_RHN23nu(std::vector<double>)}: \newline  Calculates $\Gamma(N_I \to \nu_{\alpha} \bar{\nu_{\beta}} \nu{\beta})$. & \cpp{SMINPUTS, SeesawI_Theta}\\
    \cpp{Gamma_RHN2llnu} & \cpp{Gamma_RHN2llnu(std::vector<double>)}: \newline  Calculates $\Gamma(N_I \to l_{\alpha \neq \beta}^- l_{\beta}^+ \nu_{\beta})$. & \cpp{SMINPUTS, SeesawI_Theta}\\
    \cpp{Gamma_RHN2null} & \cpp{Gamma_RHN2Kplusl(std::vector<double>)}: \newline  Calculates $\Gamma(N_I \to \nu_{\alpha} l_{\beta}^+ l_{\beta}^-)$. & \cpp{SMINPUTS, SeesawI_Theta}\\
    \cpp{Gamma_RHN2nuuubar} & \cpp{Gamma_RHN2nuuubar(std::vector<double>)}: \newline  Calculates $\Gamma(N_I \to \nu_{\alpha}q^u\bar{q^u}$. & \cpp{SMINPUTS, SeesawI_Theta}\\
    \cpp{Gamma_RHN2nuddbar} & \cpp{Gamma_RHN2nuddbar(std::vector<double>)}: \newline  Calculates $\Gamma(N_I \to \nu_{\alpha}q^d\bar{q^d}$. & \cpp{SMINPUTS, SeesawI_Theta}\\
    \cpp{Gamma_RHN2ludbar} & \cpp{Gamma_RHN2ludbar(std::vector<double>)}: \newline  Calculates $\Gamma(N_I \to l_{\alpha}q^u_{\beta}\bar{q^d_{\gamma}}$. & \cpp{SMINPUTS, SeesawI_Theta}\\
    \cpp{Gamma_BBN} & \cpp{Gamma_BBN(std::vector<double>)}: \newline  Calculates the total decay width for each RHN. & \cpp{Gamma_*}, as listed above\\
    \cpp{lnL_bbn} & \cpp{lnL_bbn(double)}: \newline  Calculates the log-likelihood. & \cpp{Gamma_BBN}\\
    \bottomrule
  \end{tabular}
  \caption{Capabilities and module functions implemented in \neutrinobit to calculate BBN observables and likelihood for sterile neutrino models.}
  \label{tab:bbn}
\end{table*}

\noindent \textit{Direct searches}\\
\\
As detailed in Section \ref{sec:direct}, the likelihoods for direct searchs depend on the active-sterile matrix elements $U_{\alpha I}$. Hence, for simplicity the capabilities \cpp{Ue1}, \cpp{Ue2}, \cpp{Ue3}, \cpp{Um1}, \cpp{Um2}, \cpp{Um3}, \cpp{Ut1}, \cpp{Ut2} and \cpp{Ut3} are implemented in \neutrinobit, as well as the phases of each of the matrix elements \cpp{Ue1_phase}, \cpp{Ue2_phase}, \cpp{Ue3_phase}, \cpp{Um1_phase}, \cpp{Um2_phase}, \cpp{Um3_phase}, \cpp{Ut1_phase},\cpp{Ut2_phase} and \cpp{Ut3_phase}. These can be seen in Table \ref{tab:ualphaI} where \cpp{I}=\cpp{1,2,3}. All the capabilities \cpp{UaI} can take a pair of options \cpp{upper_limit} and \cpp{lower_limit} to force the values within the given range. Using these quantities, the likelihoods for the different direct search experiments are calculated, and their capabilities are \cpp{lnLpienu} , \cpp{lnLps191e}, \cpp{lnLps191mu}, \cpp{lnLcharme}, \cpp{lnLcharmmu}, \cpp{lnLcharmtau}, \cpp{lnLdelphi_shortlived}, \cpp{lnLdelphi_longlived}, \cpp{lnLatlase}, \cpp{lnLatlasmu}, \cpp{lnLe949} and \cpp{lnLnutev}. The capabilities, module functions and their dependencies for all relevant direct search experiments are tabulated in Tab.~\ref{tab:direct_searches}. \\

\begin{table*}[t]
  \centering
  \begin{tabular}{p{2cm}p{7cm}p{3cm}p{4cm}}
    \toprule
    \textbf{Capability} & \textbf{Function} (\textbf{Return Type}): \newline  \textbf{Brief description} & \textbf{Dependencies} & \textbf{Options}\\\\
    \midrule
    \cpp{UeI} &\cpp{UeI(double)}: \newline  Magnitude of the matrix element $U_{e I} = |\Theta_{e I}|$. & \cpp{SeesawI_Theta} & \cpp{upper_limit(double), lower_limit(double)}\\
    \cpp{UmuI} &\cpp{UmuI(double)}: \newline  Magnitude of the matrix element $U_{\mu I} = |\Theta_{\mu I}|$. & \cpp{SeesawI_Theta} & \cpp{upper_limit(double), lower_limit(double)}\\
    \cpp{UtauI} &\cpp{UtauI(double)}: \newline  Magnitude of the matrix element $U_{\tau I} = |\Theta_{\tau I}|$. & \cpp{SeesawI_Theta} & \cpp{upper_limit(double), lower_limit(double)}\\
    \cpp{UeI_phase} & \cpp{UeI_phase(double)}: \newline  Argument of the matrix element $\Theta_{e I}$.  & \cpp{SeesawI_Theta}&\\
    \cpp{UmuI_phase} & \cpp{UmuI_phase(double)}: \newline  Argument of the matrix element $\Theta_{\mu I}$.  & \cpp{SeesawI_Theta}&\\
    \cpp{UtauI_phase} & \cpp{UtauI_phase(double)}: \newline  Argument of the matrix element $\Theta_{\tau I}$.  & \cpp{SeesawI_Theta}&\\
    \bottomrule
  \end{tabular}
  \caption{Capabilities and module functions that calculate the magnitude and argument of the matrix elements of $\Theta$ (\cpp{I}=\cpp{1,2,3}).}
  \label{tab:ualphaI}
\end{table*}

\begin{table*}[t]
  \centering
  \begin{tabular}{p{3cm}p{8cm}p{5cm}}
    \toprule
    \textbf{Capability} & \textbf{Function} (\textbf{Return Type}): \newline  \textbf{Brief description} & \textbf{Dependencies}\\
    \midrule
    \cpp{lnLpienu} &\cpp{lnL_pienu(double)}: \newline  Calculates the log-likelihood for PIENU. & \cpp{Ue1, Ue2, Ue3}\\
    \cpp{lnLps191e} & \cpp{lnL_ps191_e(double)}: \newline  Calculates the log-likelihood for PS-191 in the electron sector.  & \cpp{Ue1, Ue2, Ue3, Um1, Um2, Um3, Ut1, Ut2, Ut3}\\
    \cpp{lnLps191mu} & \cpp{lnL_ps191_mu(double)}: \newline  Calculates the log-likelihood for PS-191 in the muon sector.  & \cpp{Ue1, Ue2, Ue3, Um1, Um2, Um3, Ut1, Ut2, Ut3}\\
    \cpp{lnLcharme} & \cpp{lnL_charm_e(double)}: \newline  Calculates the log-likelihood for CHARM in the electron sector.  & \cpp{Ue1, Ue2, Ue3, Um1, Um2, Um3, Ut1, Ut2, Ut3}\\
    \cpp{lnLcharmmu} & \cpp{lnL_charm_mu(double)}: \newline  Calculates the log-likelihood for CHARM in the muon sector.  & \cpp{Ue1, Ue2, Ue3, Um1, Um2, Um3, Ut1, Ut2, Ut3}\\
    \cpp{lnLcharmtau} & \cpp{lnL_charm_tau(double)}: \newline  Calculates the log-likelihood for CHARM in the tau sector. & \cpp{Ut1, Ut2, Ut3}\\
    \cpp{lnLdelphi_shortlived} & \cpp{lnL_delphi_short_lived(double)}: \newline  Calculates the log-likelihood for DELPHI's short-lived RHN analyses.  & \cpp{Ue1, Ue2, Ue3, Um1, Um2, Um3, Ut1, Ut2, Ut3}\\
    \cpp{lnLdelphi_longlived} & \cpp{lnL_delphi_long_lived(double)}: \newline  Calculates the log-likelihood for DELPHI's long-lived RHN analyses.  & \cpp{Ue1, Ue2, Ue3, Um1, Um2, Um3, Ut1, Ut2, Ut3}\\
    \cpp{lnLatlase} & \cpp{lnL_atlas_e(double)}: \newline  Calculates the log-likelihood for ATLAS in the electron sector. & \cpp{Ue1, Ue2, Ue3}\\
    \cpp{lnLatlasmu} & \cpp{lnL_atlas_mu(double)}: \newline  Calculates the log-likelihood for ATLAS in the muon sector. & \cpp{Um1, Um2, Um3}\\
    \cpp{lnLlhce} & \cpp{lnL_lhc_e(double)}: \newline  Calculates the log-likelihood for CMS in the electron sector. & \cpp{Ue1, Ue2, Ue3}\\
    \cpp{lnLlhcmu} & \cpp{lnL_lhc_mu(double)}: \newline  Calculates the log-likelihood for CMS in the muon sector. & \cpp{Um1, Um2, Um3}\\
    \cpp{lnLe949} & \cpp{lnL_e949(double)}: \newline  Calculates the log-likelihood for E949. & \cpp{Um1, Um2, Um3}\\
    \cpp{lnLnutev} & \cpp{lnL_nutev(double)}: \newline  Calculates the log-likelihood for NuTeV. & \cpp{Um1, Um2, Um3}\\
    \bottomrule
  \end{tabular}
  \caption{Capabilities and module functions implemented in \neutrinobit to calculate direct search likelihoods for sterile neutrino models.}
  \label{tab:direct_searches}
\end{table*}

\noindent \textit{Other capabilities}\\
\\
The theoretical constraint for perturbativity of the Yukawa couplings has been implemented in \neutrinobit as well. The capability \cpp{perturbativity_lnL} calculates a step function likelihood for this constraint. Tab.~\ref{tab:pert} shows the module function that provides this capability and its dependencies.

Lastly, the artificial \textit{coupling slide} likelihood that was introduced to drive the scan towards high couplings, as described in Section~\ref{sec:scanning}, was also implemented in \neutrinobit with capability \cpp{coupling_slide}. The module function and dependencies of this capability can also be seen in Table~\ref{tab:pert}.

\begin{table*}[t]
  \centering
  \begin{tabular}{p{3cm}p{8cm}p{5cm}}
    \toprule
    \textbf{Capability} & \textbf{Function} (\textbf{Return Type}): \newline  \textbf{Brief description} & \textbf{Dependencies}\\
    \midrule
    \cpp{perturbativity_lnL} &\cpp{perturbativity_likelihood(double)}: \newline  Calculates the log-likelihood for the perturbativity of Yukawa couplings. & \cpp{SMINPUTS, SeesawI_Theta}\\
    \cpp{RHN_coupling_slide} &\cpp{coupling_slide(double)}: \newline  Calculates the log-likelihood for the coupling slide. & \cpp{SeesawI_Theta, Ut1, Ut2, Ut3}\\
    \bottomrule
  \end{tabular}
  \caption{Capabilities and module functions for perturbativity constraints and coupling slide.}
  \label{tab:pert}
\end{table*}

\clearpage

\section{Full expressions for the relevant observables}
\label{app:equations}

\subsection{Decay widths and form factors for LFV observables} \label{app:lfv}

The decay widths of LFV processes, as described in Section \ref{sec:lfv}, are
given by~\cite{Kuno:1999jp,Abada:2014kba}
  \begin{equation}
    \label{l2lgamma}
    \Gamma_{l_\alpha^- \to l_\beta^- \gamma} = \frac{\alpha_{\text{em}} m_{l_\alpha}^5}{4}\left(|K_2^L|^2 + |K_2^R|^2\right) 
  \end{equation}
  \begin{multline}
    \Gamma_{l_\alpha^- \to l_\beta^- l_\beta^- l_\beta^+} = \frac{m_{l_\alpha}^5}{512\pi^3} \left( e^4 |K_2^L|^2  \left(\frac{16}{3} \log\frac{m_{l_\alpha}}{m_{l_\beta}} - \frac{22}{3}\right) \right.\\
    +\left. \frac{1}{24}  (|A^S_{LL}|^2 + 2 |A^S_{LR}|^2) + \frac{1}{3} (2 |\hat{A}^V_{LL}|^2 + |\hat{A}^V_{LR}|^2) \right. \\
    +\left. \frac{e^2}{3} (K_2^L( A^{S*}_{RL} - 2 \hat{A}^{V*}_{RL} - 4\hat{A}^{V*}_{RR}) + h.c.)\right. \\
           -\left. \frac{1}{6} (A^S_{LR} \hat{A}^{V*}_{LR} + h.c.) \right)  + (L\leftrightarrow R)
  \end{multline}
  \begin{multline}
    \Gamma_{l_\alpha^- \to l_\beta^- l_\gamma^- l_\gamma^+} = \frac{m_{l_\alpha}^5}{512\pi^3} \left( e^4 |K_2^L|^2 \left(\frac{16}{3}\log\frac{m_{l_\alpha}}{m_{l_\gamma}} - 8\right)\right.\\
    +\left. \frac{1}{12}(|A^S_{LL}|^2 + |A^S_{LR}|^2) + \frac{1}{3} (|\hat{A}^V_{LL}|^2 + |\hat{A}^V_{LR}|^2) \right. \\
    +\left. \frac{2 e^2}{3} (K_2^L (\hat{A}^{V*}_{RL} + \hat{A}^{V*}_{RR}) + h.c.) \right) + (L\leftrightarrow R)
  \end{multline}
  \begin{multline}
    \Gamma_{l_\alpha^- \to l_\gamma^- l_\gamma^- l_\beta^+} = \frac{m_{l_\alpha}^5}{512\pi^3} \left(\frac{1}{24} (|A^S_{LL}|^2 + 2|A^S_{LR}|^2 ) \right. \\
    +\left. \frac{1}{3}(2|\hat{A}^V_{LL}|^2 + |\hat{A}^V_{LR}|^2) \right.\\
           -\left. \frac{1}{6} (A^S_{LR} \hat{A}^{V*}_{LR} + h.c.) \right) + (L\leftrightarrow R).
    \label{l2lll}
  \end{multline}
where we used $\hat{A}^V_{XY} \equiv A^V_{XY} + e^2 K_1^X$. The couplings $e$, $g_1$, $g_2$ correspond to the electromagnetic, hypercharge and weak couplings of the SM.

The form factors $K_1^X$, $K_2^X$, $A^S_{XY}$ and $A^V_{XY}$ are taken in the flavour basis where the charged lepton mass matrix is diagonal.

The dipole form factors
$K_1^X$ and $K_2^X$ are given as~\cite{Abada:2014kba}
  \begin{align}
    K_1^L &= 0 \\
    \label{formfactor1}
    K_1^R &= \frac{G_F}{4\sqrt{2}\pi^2}\sum_{a} \mathcal{U}_{\alpha a} \mathcal{U}^*_{\beta a} M\left(\frac{m_{\nu_a}^2}{m_W^2}\right)\\
    K_2^L &= \frac{G_F}{4\sqrt{2} \pi^2}\frac{m_{l_\beta}}{m_{l_\alpha}}\sum_{a} \mathcal{U}_{\alpha a} \mathcal{U}^*_{\beta a} G\left(\frac{m_{\nu_a}^2}{m_W^2}\right). \\
    K_2^R &= \frac{G_F}{4\sqrt{2} \pi^2}\sum_{a} \mathcal{U}_{\alpha a} \mathcal{U}^*_{\beta a} G\left(\frac{m_{\nu_a}^2}{m_W^2}\right)
\end{align}
The four lepton form factors $A^V_{XY}$ and $A^S_{XY}$ corresponding to the
process $l^-_\alpha \to l^-_\beta l^-_\gamma l^+_\delta$, with a vector or
scalar mediator respectively, are~\cite{Abada:2014kba}
  \begin{multline}
    A^V_{LL} = \Bigg[\frac{g_2^2}{32\pi^2} \frac{g_-\delta_{\gamma\delta} \Theta_{\alpha a} \Theta^*_{\beta a}}{2m_Z^2} \bigg(g_+x_aC_0(x_a,x_a)\\
    + g_2 c_w \left(1-2(B_0(1) + 2C_{00}(x_a,1) + x_a C_0(x_a, 1))\right) \\
  - \frac{g_+}{2}\left(1-2(B_0(x_a) - 2C_{00}(x_a,x_a) + C_0(x_a,x_a))\right)\bigg) \\
  + \frac{g_-^2(\Theta_{\alpha a}\Theta^*_{\beta a} m_{l_\alpha}^2-\Theta_{\beta a}\Theta^*_{\alpha a}m_{l_\beta}^2)\delta_{\gamma\delta}}{4m_Z^2(m_{l_\alpha}^2-m_{l_\beta}^2)} \\\times \left(1+2B_1(x_a)\right) \Bigg]_{\text{penguin}}
    + \Bigg[\frac{g_2^4}{32 \pi^2 m_W^2}\\ \Theta_{\alpha a} \bigg(\Theta_{\gamma a} \Theta^*_{\beta c} \Theta^*_{\delta c}x_a x_c \left(D_0(x_a,x_c) + (a\leftrightarrow c)\right) \\
    + 2\Theta_{\gamma c}\left(\Theta^*_{\beta a}  \Theta^*_{\delta c} (C_0(x_c,x_a)+D_0(x_c,x_a)) + (a \leftrightarrow c)  \right) \\
    - 6\Theta_{\gamma c}\left(\Theta^*_{\beta a}  \Theta^*_{\delta c} + \Theta^*_{\beta c}  \Theta^*_{\delta a}\right) D_{27}(x_a,x_c) \bigg)\Bigg]_{\text{box}}\\
  \end{multline}
  \begin{equation}
    A^V_{LR} = \frac{2g_1 s_w}{g_-} [A^V_{LL}]_{\text{penguin}}
  \end{equation}
  \begin{multline}
    A^V_{RL} = \frac{\delta_{\gamma\delta} g_2^2 g_- g_1 s_w}{64\pi^2 m_Z^2} \frac{\Theta_{\alpha a}\Theta^*_{\beta a} m_{l_\alpha}^2-\Theta_{\beta a}\Theta^*_{\alpha a}m_{l_\beta}^2}{m_{l_\alpha}^2-m_{l_\beta}^2} \\ \times\left(1+2B_1(x_a)\right)
  \end{multline}
  \begin{equation}
    A^V_{RR} = \frac{2g_1s_w}{g_-} A^V_{RL}
  \end{equation}
  \begin{multline}
    A^S_{LY} = \frac{Y^l_\gamma \delta_{\gamma\delta} g_2^2}{64\pi^2m_h^2} \frac{Y^l_\alpha \Theta_{\beta a} \Theta^*_{\alpha a} m_{l_\beta}^2 - Y^l_\beta \Theta_{\alpha a} \Theta^*_{\beta a} m_{l_\alpha}^2}{m_{l_\alpha}^2 - m_{l_\beta}^2} \\ \times \left(1+2B_1(x_a)\right)
  \end{multline}
  \begin{multline}
    A^S_{RY} = \frac{Y^l_\gamma \delta_{\gamma\delta} g_2^2}{64\pi^2m_h^2} \frac{Y^l_\alpha \Theta_{\beta a} \Theta^*_{\alpha a} - Y^l_\beta \Theta_{\alpha a} \Theta^*_{\beta a}}{m_{l_\alpha}^2 - m_{l_\beta}^2}m_{l_\alpha}m_{l_\beta}\\ \times \left(1+2B_1(x_a)\right) 
    \label{formfactor2}
  \end{multline}
where sum over $a$ and $c$ is assumed, $x_a = m_{\nu_a}^2/m_W^2$, $s_w =
\sin\theta_w$ and $ g_\pm = g_1 \sin \theta_w \pm g_2 \cos\theta_w$. 

\medskip

The $\mu -e$ conversion ratio described in section~\ref{sec:lfv}, for a general
nucleus $N_Z^{Z+N}$, can be written as
\begin{align}
 \notag R_{\mu -e} &= \frac{\alpha_{\text{em}}^3 m_\mu^5 Z_{\text{eff}}^4 F_p^2}{4 \pi^4 Z \Gamma_{\text{capt}}}\left|\sum_{\text{q=u,d,s}} \left\{\biggl(e^2 Q_q (K_1^L - K_2^R) \right.\right.\\
 \notag &- \left.\frac{1}{2}(B^V_{LL} + B^V_{LR})\right)\left(Z G_V^{(q,p)} + N G_V^{(q,n)}\right) \\
 \notag &- \left.\left.\frac{1}{2}(B^S_{LL} + B^S_{LR})\left(Z G_S^{(q,p)} + N G_S^{(q,n)}\right)\right\}\right|^2 \\
 &+ (L\leftrightarrow R).
 \label{mueconversion}
\end{align}
where $B_{XY}^K \leftrightarrow C_{XY}^K$ for up-type quarks and the numerical factors $G_K$ are given in~\cite{Kosmas:2001mv}. The nuclear form factor $F_p$, the effective atomic number $Z_{\text{eff}}$ and the capture rate $\Gamma_{\text{capt}}$ of the nucleus~\cite{Alonso:2012ji} can be seen, for the cases we are studying, $\text{Ti}_{22}^{48}$, $\text{Au}_{79}^{197}$ and $\text{Pb}_{82}^{208}$, in Table~\ref{tab:lfvnuclearparameters}.

\begin{table}[h]
  \centering
  \begin{tabular}{lrrr}
    \toprule
    $\mathbf{N_Z^{Z+N}}$ & $\mathbf{Z_{\text{eff}}}$ & $\mathbf{F_p}$ & $\mathbf{\Gamma_{\text{capt}}}(10^6 s^{-1})$\\
    \midrule
    $\text{Ti}_{22}^{48}$ & 17.6 & 0.54 & 2.59\\
    $\text{Au}_{79}^{197}$ & 33.5 & 0.16 & 13.07\\
    $\text{Pb}_{82}^{208}$ & 34.0 & 0.15 & 13.45\\
    \bottomrule
  \end{tabular}
  \caption{Effective atomic number $Z_{\text{eff}}$, nuclear form factor $F_p$ and capture rate $\Gamma_{\text{capt}}$ for the relevant nuclei.}
  \label{tab:lfvnuclearparameters}
\end{table}

The form factors $K_1^X$, $K_2^X$ are defined above and $B_{XY}^K$ and $C_{XY}^K$ are given by
    \begin{align}
      \notag B^V_{LL} &= -\frac{1}{3}\frac{g_d}{g_-} [A^V_{LL}]_{\text{penguin}} + \frac{g_2^4  \Theta_{\alpha a} \Theta^*_{\beta a} V_{\gamma c} V_{\delta c}}{16 \pi^2 m_W^2} \times \\
                      &\times \big(C_0(x_a, x^u_c) + D_0(x_a, x^u_c) - 3 D_{27}(x_a, x^u_c)\big)  \\
      B^V_{RL} &= - \frac{1}{3} \frac{g_d}{g_-}A^V_{RL} \\
      B^V_{XR} &= \frac{1}{3} A^V_{XR} \\
      B^S_{XY} &= \frac{Y^d_\gamma}{Y^l_\gamma} A^S_{XY} \\
      \notag C^V_{LL} &= -\frac{1}{3}\frac{g_u}{g_-} [A^V_{LL}]_{\text{penguin}} \\& + \frac{g_2^4 \Theta_{\alpha a} \Theta^*_{\beta a} V_{\gamma c} V_{\delta c} }{4 \pi^2m_W^2} D_{27}(x_a, x^d_c)\\
      C^V_{RL} &= - \frac{1}{3} \frac{g_u}{g_-}A^V_{RL} \\
      C^V_{XR} &= -\frac{2}{3} A^V_{XR} \\
      C^S_{XY} &= \frac{Y^u_\gamma}{Y^l_\gamma} A^S_{XY}
      \label{formfactor3}
  \end{align}
with $g_d = 3g_2 \cos\theta_w + g_1 \sin\theta_w$, $g_u = -3g_2\cos\theta_2 +
g_1\sin\theta_w$, $x^{(u,d)}_c = m_{(u,d)_c}^2/m_W^2$ and $V_{ij}$ is the CKM
matrix of quark mixing.

Lastly, the loop functions used in \eqref{formfactor1}-\eqref{formfactor3} are
defined as~\cite{Abada:2014kba}
\begin{align}
 G(x) = \frac{-7 + 33 x - 57 x^2 + 31 x^3 + 6 x^2 (1 - 3 x) \log(x)}{12  (-1 + x)^4}
\end{align}
\begin{multline}
 M(x) = \\\frac{6 x^2 (x - 3) \log(x) - (x - 1) (5 x^2 - 22 x + 5)}{9 (x - 1)^4}
\end{multline}
\begin{equation}
 B_0(x) = 0.252183 - \log x
\end{equation}
\begin{multline}
 B_1(x) = \\\frac{-1 + 4x -3x^2 + 0.504365(x-1)^2 +2x^2\log(x)}{4(x-1)^2}
\end{multline}
\begin{equation}
 C_0(x,y) = \frac{(x-y)\log(x) + (x-1)y\log(\frac{y}{x})}{(x-1)(x-y)(y-1)} 
\end{equation}
\begin{multline}
 C_{00}(x,y) = 0.438046 - \frac{(xy-x-y)\log(x)}{4(x-1)(y-1)} \\+ \frac{y^2\log(\frac{y}{x})}{4(x-y)(y-1)} 
\end{multline}
\begin{multline}
 D_0(x,y) = \frac{(xy-1)\log(x)}{(x-1)^2(y-1)^2} \\+ \frac{y\log(\frac{y}{x})}{(x-y)(y-1)^2} - \frac{1}{(x-1)(y-1)}
\end{multline}
\begin{multline}
 D_{27}(x,y) = \frac{(2xy-x-y)\log(x)}{4(x-1)^2(y-1)^2} \\+ \frac{y^2\log(\frac{y}{x})}{4(x-y)(y-1)^2} - \frac{1}{4(x-1)(y-1)}
\end{multline}

\clearpage

\subsection{Decay widths relevant for Big Bang Nucleosynthesis}
\label{app:bbndw}
The various decay widths of RHNs, relevant for the BBN as described in Sec.~\ref{sec:bbn}, are listed here. These expressions are taken from~\cite{Gorbunov:2007ak, Atre:2009rg, Canetti:2012kh, Bondarenko:2018ptm, Ballett:2019bgd}, among which there are slight differences that will be commented upon when relevant. We list here the decays for Majorana fermions, which differ by a factor of 2 with respect to the rates for Dirac fermions as shown, for instance, in ~\cite{Bondarenko:2018ptm}.

The decay width of a RHN, $N_I$, to a lepton, $l_\alpha$ and a charged pseudoscalar meson, $P^+ = \pi^+, K^+, D^+, D_s, B^+, B_c$, is~\cite{Gorbunov:2007ak, Atre:2009rg, Bondarenko:2018ptm, Ballett:2019bgd}
\begin{multline}
\Gamma_{N_I \to P^+ l_{\alpha}^- } = \frac{ G_F^2 |V_P|^2 f_P^2 M_I^3|\Theta_{{\alpha}I}|^2}{8\pi}\cdot\\\left( ( 1 - x_{l}^2 )^2 - x_P^2 ( 1 + x_{l}^2 ) \right) \lambda^{1/2}(1,x_P^2,x_{l}^2),
\end{multline}
with $f_P$ the decay constant of the meson $P^+$, which can be seen in Table~\ref{tab:mesondecayconstants}, $V_P$ the CKM matrix element corresponding to $P^+$ and 
\begin{align}
    x_{l} &= \frac{M_{l_\alpha}}{M_I}, \label{xl} \\
    x_P &= \frac{M_{P}}{M_I} \label{xP},\\
    \lambda(a,b,c) &= a^2 + b^2 + c^2 - 2ab - 2bc - 2ca \label{lambda}.
\end{align}
Similarly, the decay width of a RHN to a neutrino and a neutral pseudoscalar meson, $P^0 = \pi^0, \eta, \eta', \eta_c$, is~\cite{Gorbunov:2007ak, Bondarenko:2018ptm, Ballett:2019bgd} 
\begin{equation}
\Gamma_{N_I \to P^0 \nu_{\alpha}} = \frac{ G_F^2 f_{P}^2 M_I^3|\Theta_{{\alpha}I}|^2}{16\pi} \left( 1 -x_P^2 \right)^2,
\end{equation}
with $f_P$ the meson decay constant, Table \ref{tab:mesondecayconstants}, and $x_P$ as in \eqref{xP}. The expression from~\cite{Atre:2009rg} for this decay missed a factor of 2, which was corrected by the later work~\cite{Ballett:2019bgd}.

The decay width of a RHN to a lepton and a charged vector meson, $V = \rho^+, D^{*+}, D_s^*$, is~\cite{Atre:2009rg, Bondarenko:2018ptm, Ballett:2019bgd}
\begin{multline}
\Gamma_{N_I \to V^+ l_{\alpha}^-} = \frac{G_F^2 f_V^2 |V_{V}|^2 M_I^3|\Theta_{{\alpha}I}|^2}{8\pi}\cdot\\\left( ( 1 - x_{l}^2)^2 + x_V^2 ( 1 + x_{l}^2) -2x_V^4 ) \right)\lambda^{1/2}(1,x_V^2,x_{l}^2),
\end{multline}
where $f_V$ is the decay constant of $V^+$, in Table \ref{tab:mesondecayconstants}, $V_V$ is the CKM matrix element associated with $V^+$, $\lambda$ is defined in \eqref{lambda} and
\begin{equation}
    x_V = \frac{M_V}{M_I} \label{xV}.
\end{equation}
The results from~\cite{Bondarenko:2018ptm} use a different definition of decay constant $g_V = m_V f_V$, but the final values agree nevertheless. The value of $g_\rho$ in \cite{Gorbunov:2007ak} differs from that of \cite{Bondarenko:2018ptm} and there is a factor of $\tfrac{1}{2}$ missing as well with respect to the other works.

The last of the semileptonic decays of RHNs is the decay to a neutrino and a neutral vector meson, $V^0 = \rho^0, \omega, \phi, J/\psi$, with decay width~\cite{Bondarenko:2018ptm}
\begin{equation}
\Gamma_{N_I \to V^0 \nu_{\alpha}} = \frac{G_F^2 f_V^2 \kappa_V^2 M_I^3 |\Theta_{{\alpha}I}|^2}{16\pi} ( 1 + 2x_V^2 ) ( 1 - x_V^2)^2,
\end{equation}
with $f_V$ the meson decay constant, Table \ref{tab:mesondecayconstants}, $x_V$ as in \eqref{xV} and $\kappa_V$ a neutral current correction factor~\cite{Bondarenko:2018ptm}
\begin{align}
    \notag \kappa_{\rho^0} = 1 - 2\sin^2\theta_W, &\quad \kappa_\omega = \tfrac{4}{3}\sin^2\theta_W,\\
    \kappa_\phi = \tfrac{4}{3}\sin^2\theta_W - 1, &\quad \kappa_{J/\psi} = 1 - \tfrac{8}{3}\sin^2\theta_W \
\end{align}

The expressions from \cite{Bondarenko:2018ptm}, \cite{Atre:2009rg} and \cite{Ballett:2019bgd} agree but for a different definition of $\kappa_V$, whereas \cite{Gorbunov:2007ak} misses the $\kappa_V$ factor altogether.

\begin{table}[h]
 \centering
 \begin{tabular}{c|c||c|c}
  \toprule
  $f_{\pi^+}$ & 130.2~\cite{Rosner:2015wva}& $f_{\pi^0}$ & 130.2~\cite{Rosner:2015wva} \\
  $f_{K^+}$ & 155.7~\cite{Rosner:2015wva} & $f_{\eta}$ & 81.7~\cite{Bondarenko:2018ptm}\\ 
  $f_{D^+}$ & 212.6~\cite{Rosner:2015wva} & $f_{\eta'}$ & -94.7~\cite{Bondarenko:2018ptm}\\ 
  $f_{D_s}$ & 249.9~\cite{Rosner:2015wva} & $f_{\eta_c}$ & 237~\cite{Bondarenko:2018ptm}\\ 
  $f_{B^+}$ & 190~\cite{Rosner:2015wva} & &\\ 
  $f_{B_c}$ & 434~\cite{Colquhoun:2015oha} & &\\
  \midrule
  \midrule
  $f_{\rho^+}$ & 209~\cite{Ebert:2006hj} & $f_{\rho^0}$ & 209~\cite{Ebert:2006hj}\\
  $f_{D^{*+}}$ & 246.75~\cite{Dhiman:2017urn} & $f_{\omega}$ & 195~\cite{Atre:2009rg}\\
  $f_{D^*_s}$ & 284~\cite{Dhiman:2017urn} & $f_{\phi}$ & 229~\cite{Ebert:2006hj} \\
  & & $f_{J/\psi}$ & 418~\cite{Becirevic:2013bsa} \\
  \bottomrule       
 \end{tabular}
 \caption{Decay constants (in MeV) of pseudoscalar charged (top left), pseudoscalar neutral (top right), vector charged (bottom left) and vector neutral (bottom right) mesons.}
 \label{tab:mesondecayconstants}
\end{table}

The fully-leptonic three body decays of RHNs can be to three neutrinos, with decay width given by~\cite{Gorbunov:2007ak, Bondarenko:2018ptm}
\begin{equation}
\Gamma_{N_I \to \sum_{\alpha,\beta} \nu_{\alpha} \bar{\nu_{\beta}} \nu_{\beta}} = \frac{G_F^2 M_I^5}{96\pi^3} \sum_{\alpha} |\Theta_{{\alpha}I}|^2,
\end{equation}
and to two charged leptons and a neutrino. If the charged leptons have the same flavour, the decay width is~\cite{Gorbunov:2007ak, Bondarenko:2018ptm}
\begin{multline}
\Gamma_{N_I \to \nu_{\alpha} l_{\beta}^+ l_{\beta}^-} = \frac{G_F^2 M_I^5}{96\pi^3} |\Theta_{{\alpha}I}|^2\Bigg[ (C_1(1-\delta_{\alpha\beta})+C_3\delta_{\alpha\beta}).\\\left( (1-14x_l^2-2x_l^4-12x_l^6)\sqrt{1-4x_l^2}+12x_l^4(x_l^4-1)L(x_l) \right)\\+4(C_2(1- \delta_{\alpha\beta})+C_4\delta_{\alpha\beta})\Bigg( x_l^2(2+10x_l^2-12x_l^4)\sqrt{1-4x_l^2}\\+6x_l^4(1-2x_l^2+2x_l^4)L(x_l) \Bigg) \Bigg],
\label{rhn2null}
\end{multline}
with $x_{l}$ as in \eqref{xl} with $\alpha \leftrightarrow \beta$, the coefficients $C_i$ are 
\begin{align}
 C_1 &= \frac{1}{4}(1-4\sin^2\theta_W+8\sin^4\theta_W),\\
 C_2 &= \frac{1}{2}\sin^2\theta_W(2\sin^2\theta_W-1),\\
 C_3 &= \frac{1}{4}(1+4\sin^2\theta_W+8\sin^4\theta_W),\\
 C_4 &= \frac{1}{2}\sin^2\theta_W(2\sin^2\theta_W+1),
\end{align}
and the functions $S(x,y)$ and $L(x)$
\begin{align}
 S(x,y) &= \sqrt{(1-(x+y)^2)(1-(x-y)^2)},\label{Sfunction}\\
 L(x) &= \text{log}\Bigg[\frac{1-3x^2-(1-x^2)\sqrt{1-4x^2}}{x^2(1+\sqrt{1-4x^2})}\Bigg].\label{Lfunction}
\end{align}
If, on the other hand, the charged leptons are of different flavour, the decay width is given by~\cite{Canetti:2012kh}
\begin{multline}
\Gamma_{N_I \to l_{\alpha \neq \beta}^- l_{\beta}^+ \nu_{\beta}} = \frac{G_F^2 M_I^5}{96\pi^3} |\Theta_{{\alpha}I}|^2 \Bigg( S(x_{\alpha},x_{\beta})g(x_{\alpha},x_{\beta})\\- 12x_{\alpha}^4\text{log}\Bigg[\frac{1-S(x_{\alpha},x_{\beta})(1+x_{\alpha}^2-x_{\beta}^2)}{2x_{\alpha}^2}\\-\frac{2x_{\beta}^2+(x_{\alpha}^2-x_{\beta}^2)^2}{2x_{\alpha}^2}\Bigg]- 12x_{\beta}^4\text{log}\Bigg[\frac{1}{2x_{\beta}^2}\\-\frac{S(x_{\alpha},x_{\beta})(1-x_{\alpha}^2+x_{\beta}^2)-2x_{\alpha}^2+(x_{\alpha}^2-x_{\beta}^2)^2}{2x_{\beta}^2}\Bigg]\\+ 12x_{\alpha}^4x_{\beta}^4\text{log}\Bigg[\frac{1-S(x_{\alpha},x_{\beta})(1-x_{\alpha}^2-x_{\beta}^2)}{2x_{\alpha}^2x_{\beta}^2}\\-\frac{2x_{\alpha}^2-2x_{\beta}^2+x_{\alpha}^4+x_{\beta}^4}{2x_{\alpha}^2x_{\beta}^2}\Bigg]\Bigg)
\label{rhn2llnu}
\end{multline}
where $x_\alpha$ and $x_\beta$ are as in \eqref{xl}, $S(x,y)$ is as in \eqref{Sfunction} and %
\begin{align}
 g(x,y) &= 1-7x^2-7y^2-7x^4-7y^4+12x^2y^2\\
\notag &- 7x^2y^4-7x^4y^2+x^6+y^6,
\end{align}

The expressions for leptonic decays agree across all works, except minor differences in the functional forms of the decays to different flavour leptons between \cite{Canetti:2012kh} and \cite{Bondarenko:2018ptm}. These differences are, however, numerically negligible.

For large masses, above the hadronisation scale, the full hadronic decay width of the right-handed neutrinos is better approximated by computing their decay to free quarks, instead of the individual meson channels. Hence the decays of RHNs to quarks through the neutral current is~\cite{Bondarenko:2018ptm}
\begin{multline}
\Gamma_{N_I \to \nu_{\alpha} q \bar{q}} = \frac{3G_F^2 M_I^5}{96\pi^3} |\Theta_{{\alpha}I}|^2 \times \\\Bigg( \Big(12C_1^q x^4(x^4 - 1) + 6C_2^qx^4(1 - 2x^2 + 2x^4\Big)L(x) \\+  C_1^q \Big( (1 - 14x^2 - 2x^4 -12x^6)\sqrt{1 - 4x^2} \Big) \\+ 4C_2^q \Big( x^2(2 +  10x^2 - 12x^4)\sqrt{1 - 4x^2} \Big)\Bigg)
\label{rhn2nuqqbar}
\end{multline}
where the 3 upfront accounts for the number of colours, $x = M_q/M_I$, $L(x)$ as in Eq.\eqref{Lfunction}, and
\begin{align}
C_1^u &= \tfrac{1}{4}(1-\tfrac{8}{3}\sin^2\theta_W +\tfrac{32}{9}\sin^4\theta_W),\\
C_2^u &= \tfrac{1}{3}\sin^2\theta_W(\tfrac{4}{3}\sin^2\theta_W -1),
\end{align}
for up-type quarks and 
\begin{align}
C_1^d &= \tfrac{1}{4}(1-\tfrac{4}{3}\sin^2\theta_W +\tfrac{8}{9}\sin^4\theta_W),\\
C_2^d &= \tfrac{1}{6}\sin^2\theta_W(\tfrac{2}{3}\sin^2\theta_W -1),
\end{align}
for down-type quarks. 

And lastly, the decay of RHNs to free quarks through the charged current can be written as~\cite{Bondarenko:2018ptm}
\begin{equation}
\Gamma_{N_I \to l_{\alpha} u_n \bar{d_m}} = \frac{3G_F^2 M_I^5}{96\pi^3} |V_{nm}|^2 |\Theta_{{\alpha}I}|^2 I(x_u,x_d,x_l)
\end{equation}
where $x_i = M_i/M_I$ and the function $I(x_u,x_d,x_l)$ is
\begin{multline}
    I(x_u,x_d,x_l) = 12\int^{(1-x_u)^2}_{(x_d+x_l)^2} \frac{dx}{x}(x-x_l^2-x_d^2)(1+x_u^2-x)\\\sqrt{\lambda(x,x_l^2,x_d^2)\lambda(1,x,x_u^2)}
\end{multline}
The differences between the expressions for the decay to free quarks between \cite{Canetti:2012kh} and \cite{Bondarenko:2018ptm} mirror those of the leptonic decays. In this case, however, the differences can be substantial for decays to third generation quarks, in which case we opt to use the expressions from \cite{Bondarenko:2018ptm}.
\clearpage

\section{Distinguishing symmetry protected from tuned parameter choices}\label{FakeSymmetryPoints}
One goal of the present work is to fully understand the experimentally allowed range of parameters for heavy neutrinos, with minimal theoretical bias. To achieve this we employ the 
Casas-Ibarra parametrisation \eqref{CItheta} and adapt agnostic priors for the parameters in table \ref{tab:scanpars}. 
On the other hand it is also instrictive to understand what fraction of the parameter space can only be realised at the cost of fine tuning in the parameters. This requires to distinguish fine-tuned parameter choices from symmetry protected ones.

The Casas-Ibarra parametrisation \eqref{CItheta} is inherently motivated from ``bottom up'', and it is not easy to see directly from the values of its fundamental parameters whether they exhibit a symmetry protection.
 A full analytic exploration of all the possible solutions and their classification between symmetric and fine-tuned would be a  useful exercise, but lies outside the scope of this work. In our numerical scan we take a more pragmatic approach.
We first generate a huge amount of parameter choices by randomising the parameter values and the order of the matrices $\mathcal{R}^{ij}$ to ensure a maximal coverage of the parameter space. We then use the cuts \eqref{eq:tuningcut} to distinguish the symmetry protected points a posteriori.
This cut practically enforces the structure \eqref{SymmProtectParam} on the masses and couplings.

Using the cut \eqref{eq:tuningcut} requires some care for two reasons. First, the form \eqref{SymmProtectParam} does not capture all symmetry protected points, cf. footnote \ref{MassCommunism}. We may therefore misidentify some symmetry protected points as tuned. We find, however, that the number of such points is small.
Second, when using the Casas-Ibarra parametrisation, it is possible to generate points that mimic the form \eqref{SymmProtectParam} and hence 
pass the cut \eqref{eq:tuningcut}, but in fact exhibit a significant amount of tuning.

To illustrate the second point we work at tree level and approximate $\tilde{M}^\text{diag}\simeq M_M$, which yields
\begin{eqnarray}\label{CItree}
F\approx i U_\nu\sqrt{m_\nu^{\rm diag}}\mathcal{R}\sqrt{M_M}/v.
\end{eqnarray}
For the inverted hierarchy, it is straightforward to show that one qualitatively gets a pseudo-Dirac pair of heavy neutrinos for
\begin{eqnarray}\label{CIlimitIO}
M_1 = M_2=\M \ , \ (\omega_{12},\omega_{13},\omega_{23})=(\omega,0,0) \
\end{eqnarray} 
with $|{\rm Im}\omega|\gg 1$.
In addition we have to set $m_{\nu_0}=0$ to find the symmetry protected region, as a non-zero lightest neutrino mass is not consistent with $\epsilon_\alpha'\to0$.
In this case the upper left block of the matrix $\mathcal{R}\sqrt{M_M}$ 
in \eqref{CItheta}
is large and the third row that multiplies $m_3=m_{\nu_0}$ 
is small, thereby mimicking the structure in \eqref{SymmProtectParam}. One can therefore interpret the decoupling $F_{\alpha 3} = 0$ and
the vanishing mass of the lightest neutrino physically as results of the symmetry.

If we choose normal ordering, then choosing \eqref{CIlimitIO} still yields a structure that passes the cut \eqref{eq:tuningcut}, but it is in fact a tuned solution that just mimics this structure.
In that case $m_{\nu_0}=m_1$ multiplies the large components of $\mathcal{R}\sqrt{M_M}^{-1}$ in \eqref{CItheta}. 
Hence, the approximate symmetry makes the wrong light neutrino mass small ($m_3$ instead of $m_{\nu_0}=m_1$).
Of course one can set $m_1=0$ by hand in \eqref{CItheta}, but this choice cannot be justified by the symmetry. Though the limit \eqref{CIlimitIO} leads to a pseudo-Dirac structure amongst the $\nu_{R i}$ as predicted by the $B-\bar{L}$ symmetry, the vanishing mass of the lightest neutrino is not a result of that symmetry.

The problem is that the Casas-Ibarra parametrisation allows on to set $m_{\nu_0}=0$ by hand and gives no warning if this leads to accidental cancellations. 

Realising the symmetry requires to choose the eigenvalues of $M_M$ and the non-zero $\omega_{ij}$ 
consistently in a way that the large block in the matrix $\mathcal{R}\sqrt{M_M}^{-1}$ in \eqref{CItheta} multiplies the two non-zero light neutrino masses.
For normal ordering this is achieved with 
\begin{eqnarray}\label{CIlimitNO}
M_2 = M_3=\M \ , \ (\omega_{12},\omega_{13},\omega_{23})=(0,0,\omega),
\end{eqnarray}
again with $m_{\nu_0}=0$.
In particular, one cannot choose $\nu_{R3}$ to be the particle that decouples.
This is clearly no fundamental problem because the labels of the $\nu_{R I}$ have no physical meaning, but it means that the labelling and the order of the matrices $\mathcal{R}^{ij}$ have to be taken into consideration when applying a cut to identify symmetry protected points in the numerical data.

The situation is yet more tricky if one considers small perturbations around the choices \eqref{CIlimitNO} or \eqref{CIlimitIO}. In our numerical scan we randomise the order of the three matrices $\mathcal{R}^{ij}$ in \eqref{Rorder} to generate more points.
If one exactly takes the choice \eqref{CIlimitNO} or \eqref{CIlimitIO} for normal or inverted neutrino mass ordering, respectively, then the approximate $B-\bar{L}$ conserving limit is reproduced irrespectively of the ordering of the $\mathcal{R}^{ij}$. However, the effect that small perturbations around this limit have strongly depends on this ordering. The effect of perturbing $\mathcal{R}$ is the smallest if the matrices $\mathcal{R}^{ij}$ are ordered in a way that the one with large entries (controlled by $\omega$) directly multiplies $\sqrt{M_M^{-1}}$ in \eqref{CItree}.
For normal ordering this is the case with $\mathcal{R} = \mathcal{R}^{23} \mathcal{R}^{13} \mathcal{R}^{12}$, and for inverted ordering with $\mathcal{R} = \mathcal{R}^{12} \mathcal{R}^{13} \mathcal{R}^{23}$. This procedure was crucial to reproduce the constraints on the heavy neutrino flavour mixing pattern in the $n=2$ model found in ref.~\cite{Drewes:2018gkc}, cf. fig.~\ref{fig:Triangle_symmetric}.

It is worth noting that \eqref{CIlimitIO} and \eqref{CIlimitNO} are not the only combinations of parameters that yield the symmetry protected scenario, but rather the simplest. The non-trivial structure of the complex rotation matrix $\mathcal{R}$ yields many solutions to the required block layout described before. In fact, we will take advantage of this fact further below to recover the  $n=2$ case from the $n=3$ Lagrangian for normal ordering by taking
\begin{eqnarray}\label{CIlimitNO2}
M_1 = M_2=\M \ , \ (\omega_{12},\omega_{13},\omega_{23})=(0,\pi/2,\omega),
\end{eqnarray}
since in this work we will focus mainly on the case where $M_1$ and $M_2$ are almost degenerate.

\section{Partial likelihoods}
\label{app:partial}

The final result of a global fit shows the combined effect of all likelihoods on the parameter space of the model. It is, however, often useful to understand the effect on the individual partial likelihoods. Therefore, we show here a comprehensive set of scatter plots that show the contribution of each relevant partial likelihood in the $M_I$ vs $|U_{\alpha I}|^2$. In all figures throughout this section the colourbar measures the relative partial log likelihood for each observable with respect to the global best fit value. As we have seen before, away from the massless neutrino limit there is little difference between NH and IH, and thus we will only show the partial likelihoods for normal ordering. 

\begin{figure}[h]
  \centering
  \includegraphics[width=0.8\linewidth]{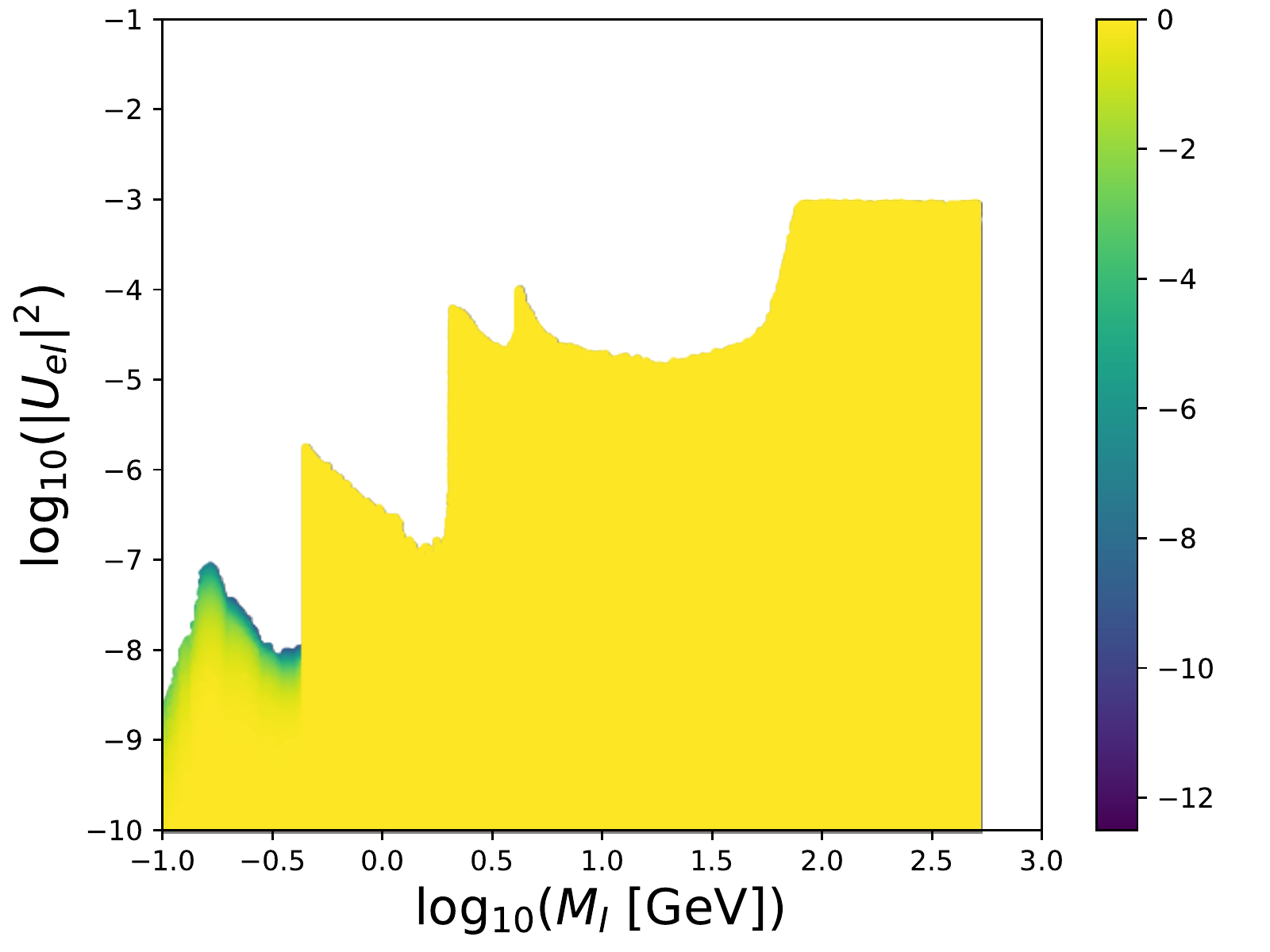}
  \caption{Partial likelihood from direct searches with PS191, $e$-channel, in the $M_I-|U_{eI}|^2$ plane.}
  \label{fig:lnL_U11_lnL_ps191_e_NH}
\end{figure}

\begin{figure}[h]
  \centering
  \includegraphics[width=0.8\linewidth]{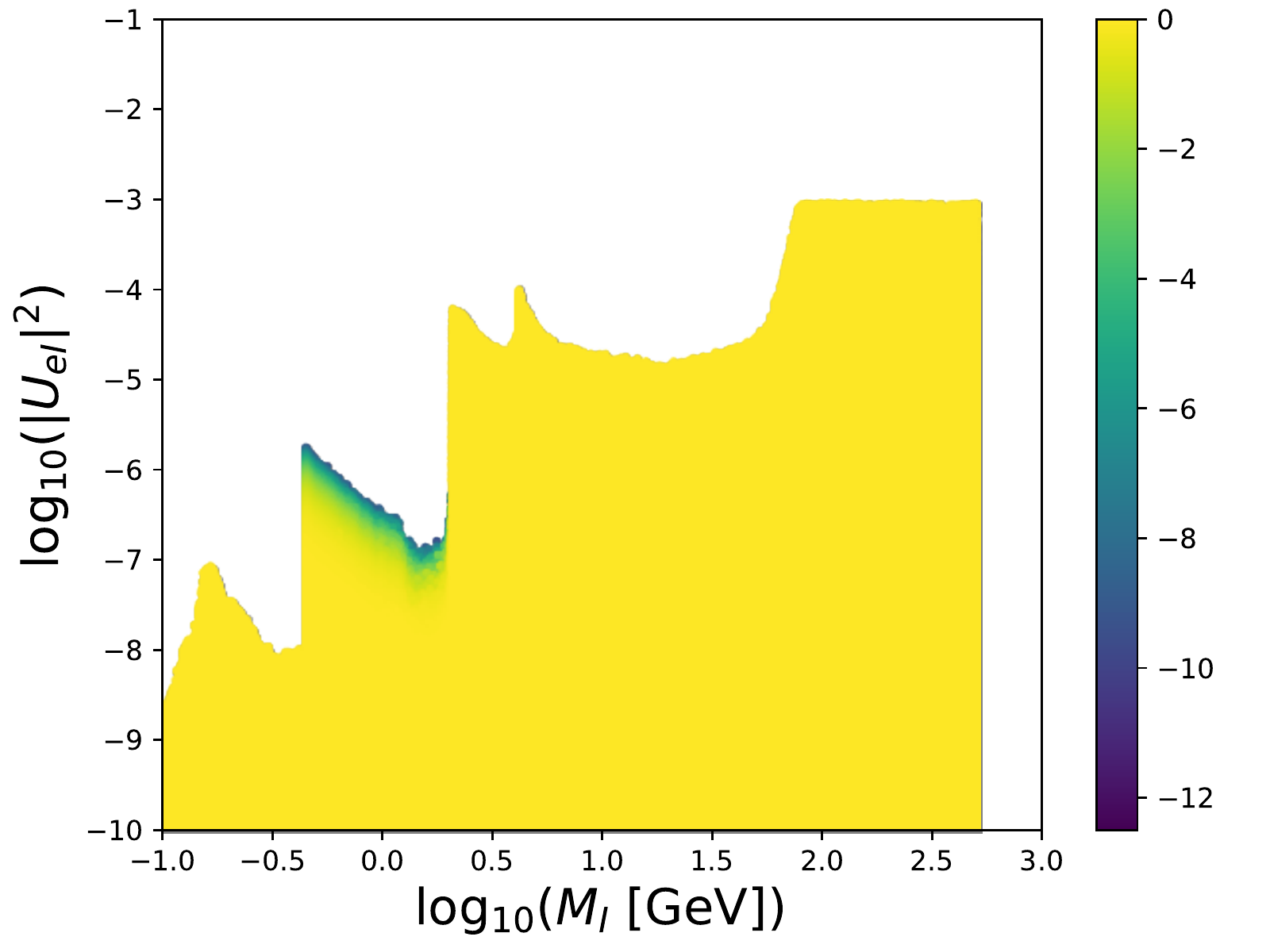}
  \caption{Partial likelihood from direct searches with CHARM, $e$-channel, in the $M_I-|U_{eI}|^2$ plane.}
  \label{fig:lnL_U11_lnL_charm_e_NH}
\end{figure}

\begin{figure}[h]
  \centering
  \includegraphics[width=0.8\linewidth]{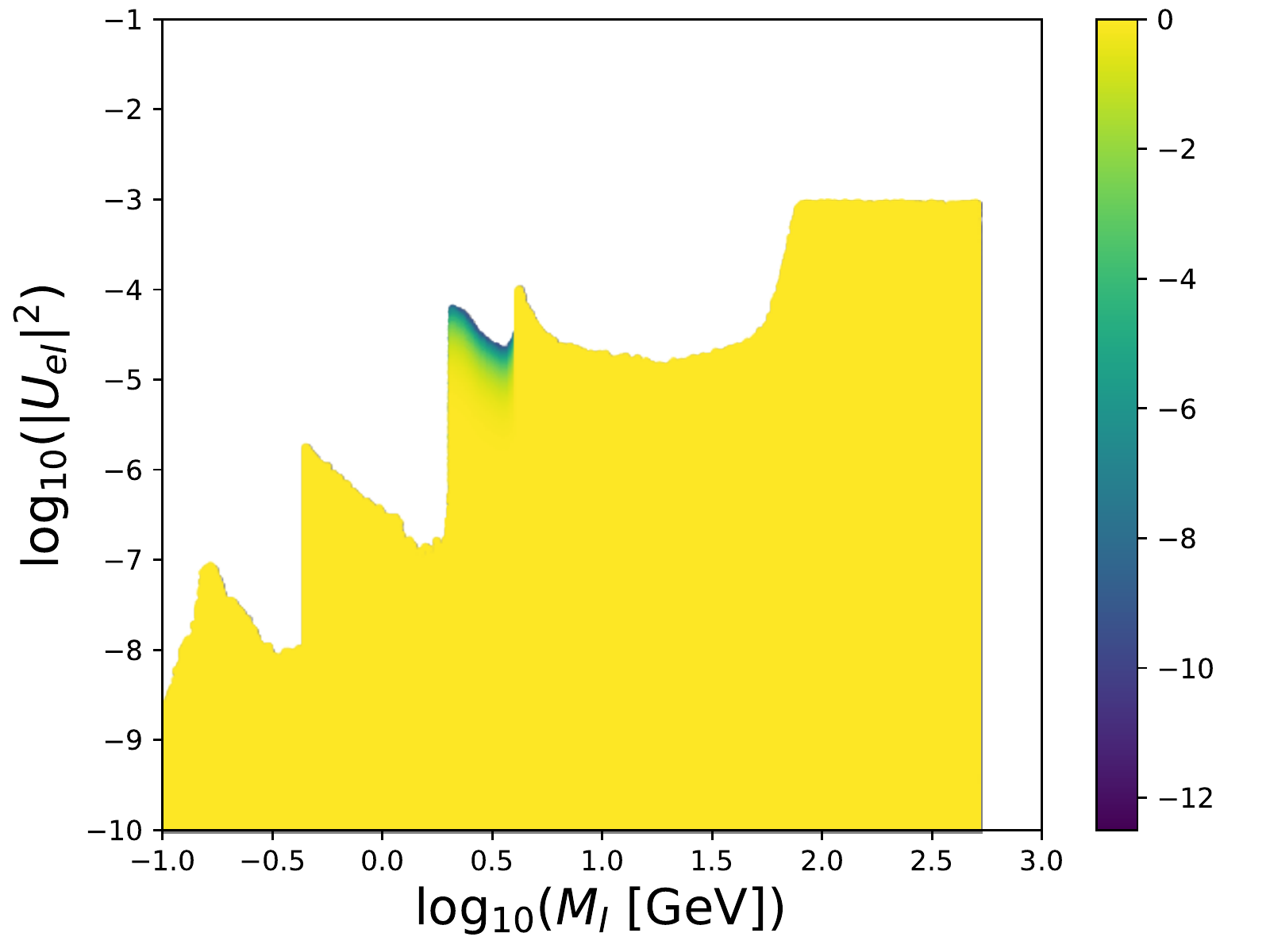}
  \caption{Partial likelihood from the long-lived particle searches with DELPHI, in the $M_I-|U_{eI}|^2$ plane.}
  \label{fig:lnL_U11_lnL_delphi_long_NH}
\end{figure}

\begin{figure}[h]
  \centering
  \includegraphics[width=0.8\linewidth]{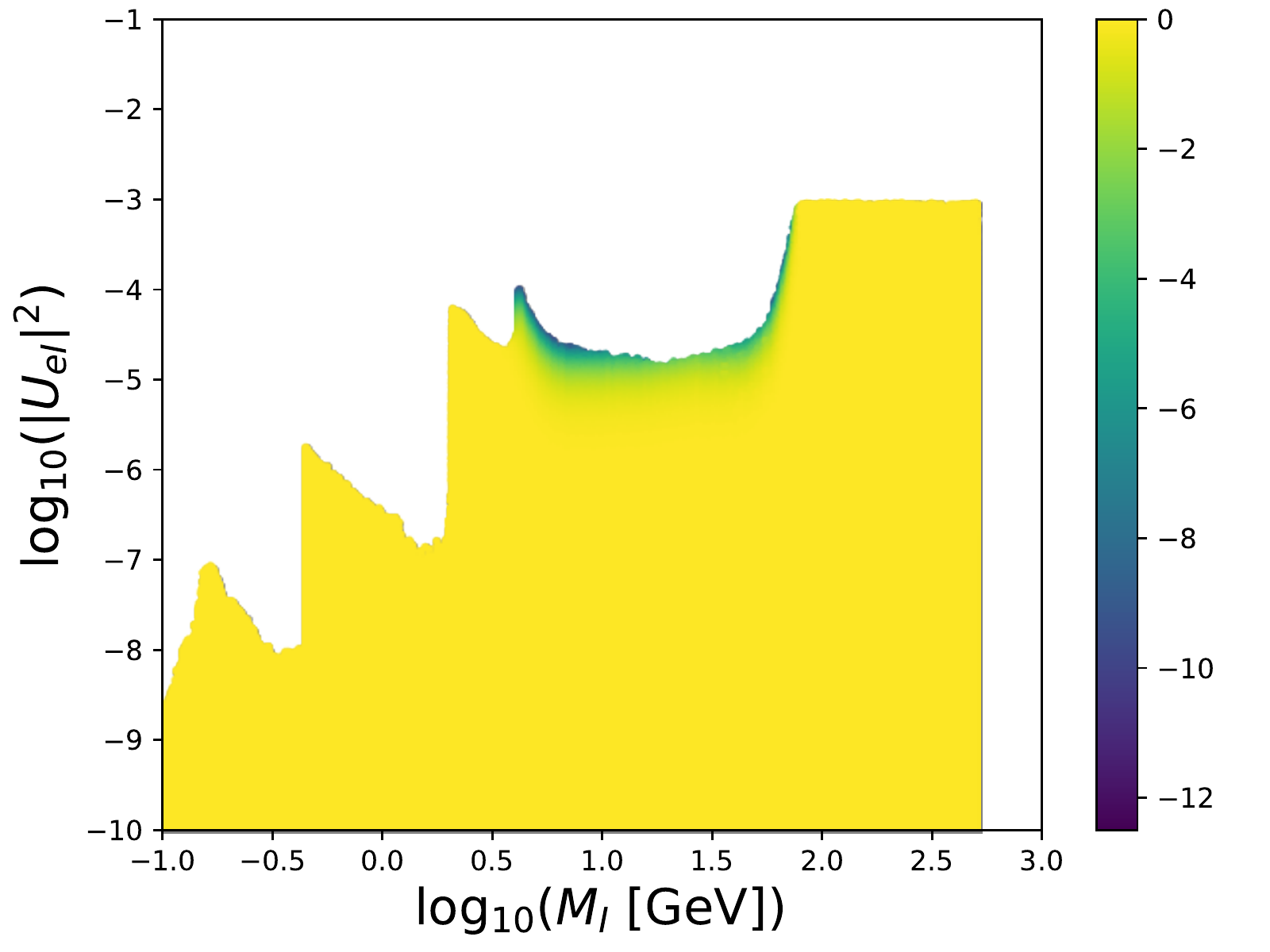}
  \caption{Partial likelihood from prompt searches with DELPHI, in the $M_I-|U_{eI}|^2$ plane.}
  \label{fig:lnL_U11_lnL_delphi_short_NH}
\end{figure}

\begin{figure}[h]
  \centering
  \includegraphics[width=0.8\linewidth]{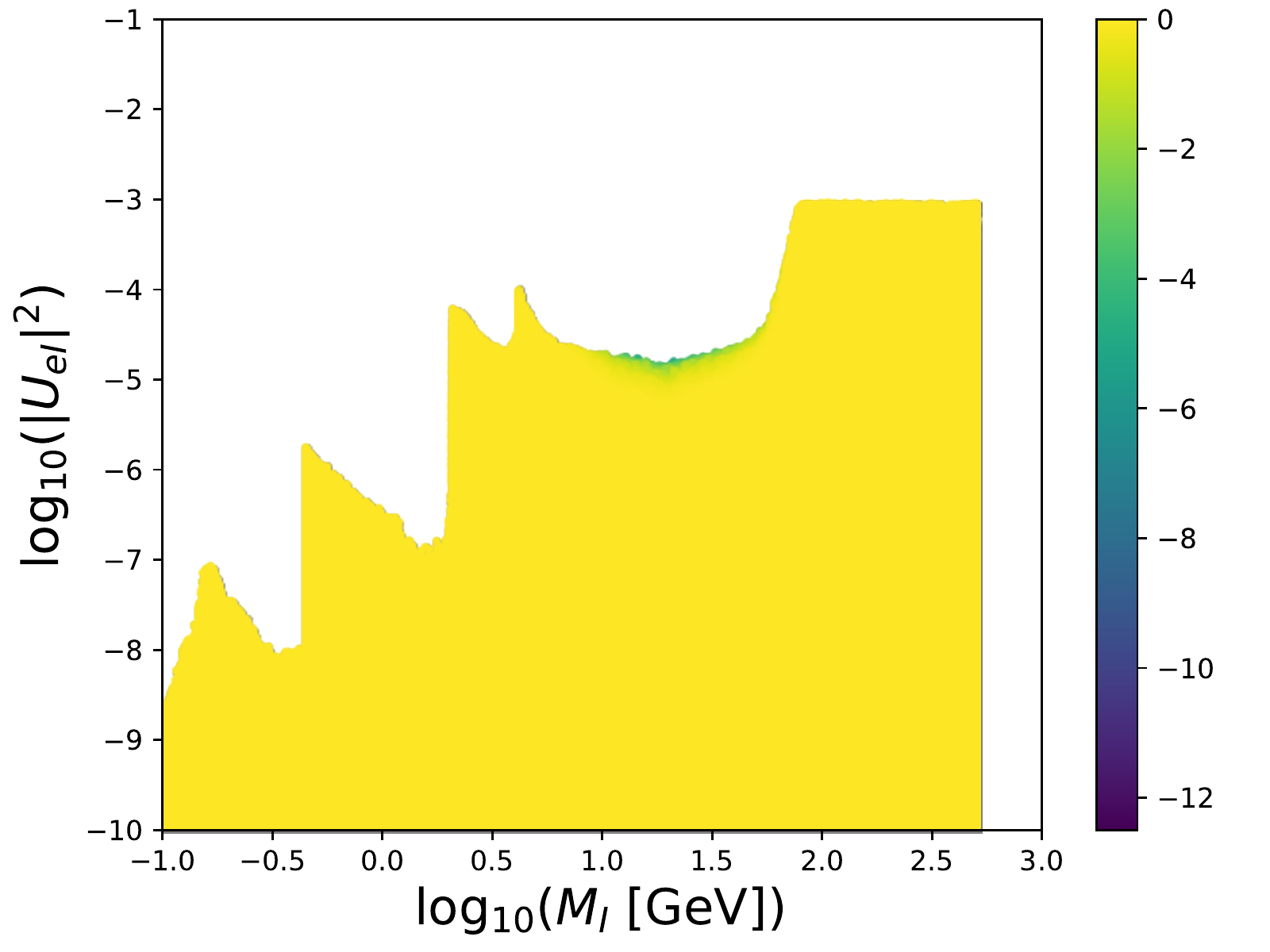}
  \caption{Partial likelihood from direct searches with CMS, $e$-channel, in the $M_I-|U_{eI}|^2$ plane.}
  \label{fig:lnL_U11_lnL_lhce_NH}
\end{figure}

\begin{figure}[h]
  \centering
  \includegraphics[width=0.8\linewidth]{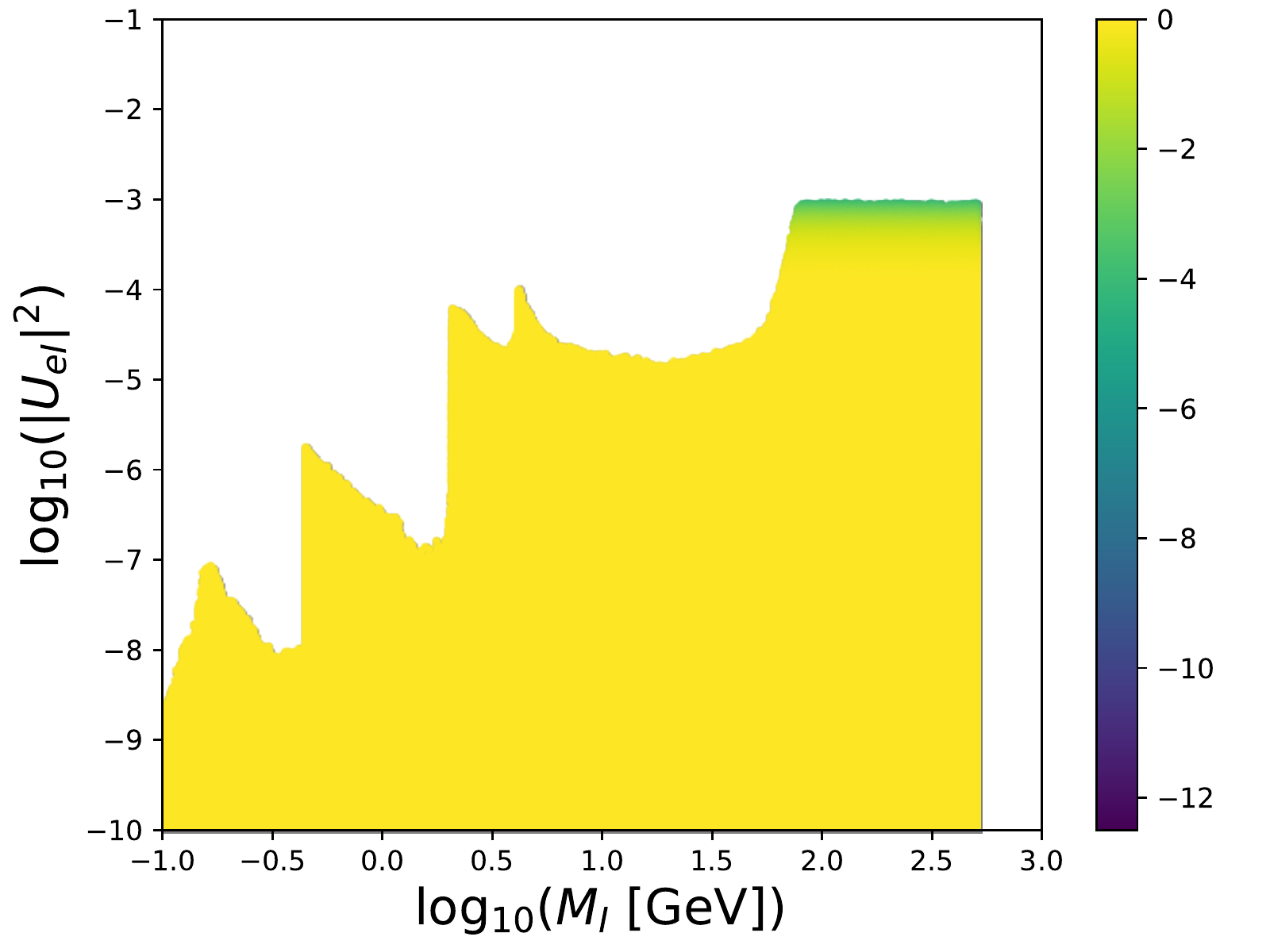}
  \caption{Partial likelihood from $\sin\theta_W$, in the $M_I-|U_{eI}|^2$ plane.}
  \label{fig:lnL_U11_lnL_sinW2_eff_NH}
\end{figure}

Figures \ref{fig:lnL_U11_lnL_ps191_e_NH} - \ref{fig:lnL_U11_lnL_sinW2_eff_NH} show the most constraining likelihoods on the $|U_{eI}|^2$ coupling. The likelihood values are normalised to the best fit value for each partial likelihood. Consistently with the results above, various direct searches constrain large values of the coupling, with PS191 dominating for $M_I \lesssim 0.45$ GeV, CHARM for $M_I \sim (0.45, 2)$ GeV, the long-lived particle search from DELPHI for $M_I \sim (2,4)$ GeV and DELPHI prompt search for $M_I \sim (4, 80)$ GeV. 

As seen in Figure~\ref{fig:lnL_U11_lnL_lhce_NH}, direct searches from CMS compete in a small mass range with DELPHI prompt searches, the statistical combination of the two setting stronger limits than each of them individually. Recent and future results from CMS and ATLAS not included in this study are expected to dominate in this range. 

Figure \ref{fig:lnL_U11_lnL_sinW2_eff_NH} shows that the larger mass range is unconstrained by direct searches, hence electroweak precision observables, in particular $\sin\theta_W$, are responsible for the upper limits in this range.

\begin{figure}[h]
  \centering
  \includegraphics[width=0.8\linewidth]{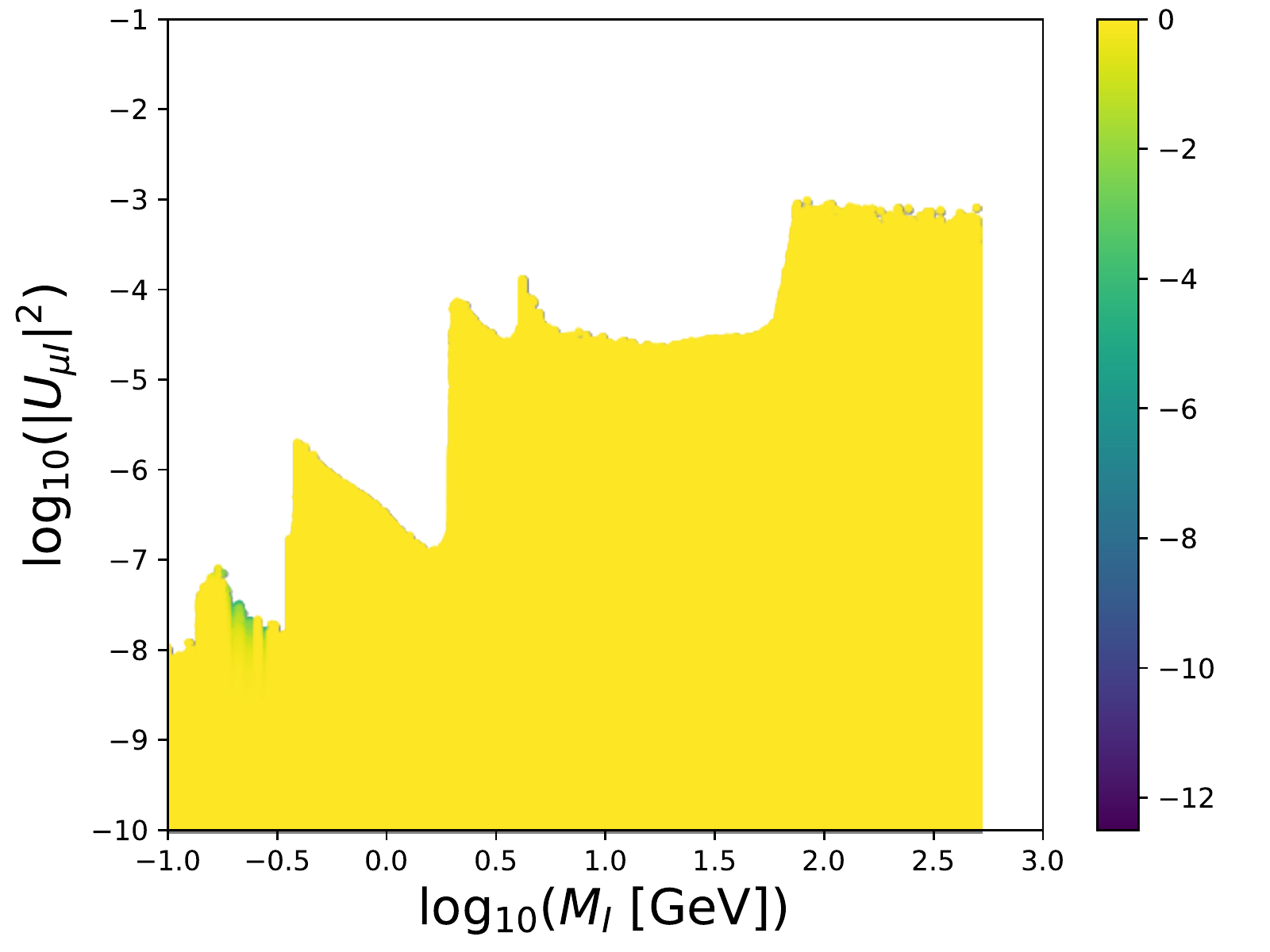}
  \caption{Partial likelihood from direct searches with E949, $\mu$-channel, in the $M_I-|U_{\mu I}|^2$ plane.}
  \label{fig:lnL_U21_lnL_e949_NH}
\end{figure}

\begin{figure}[h]
  \centering
  \includegraphics[width=0.8\linewidth]{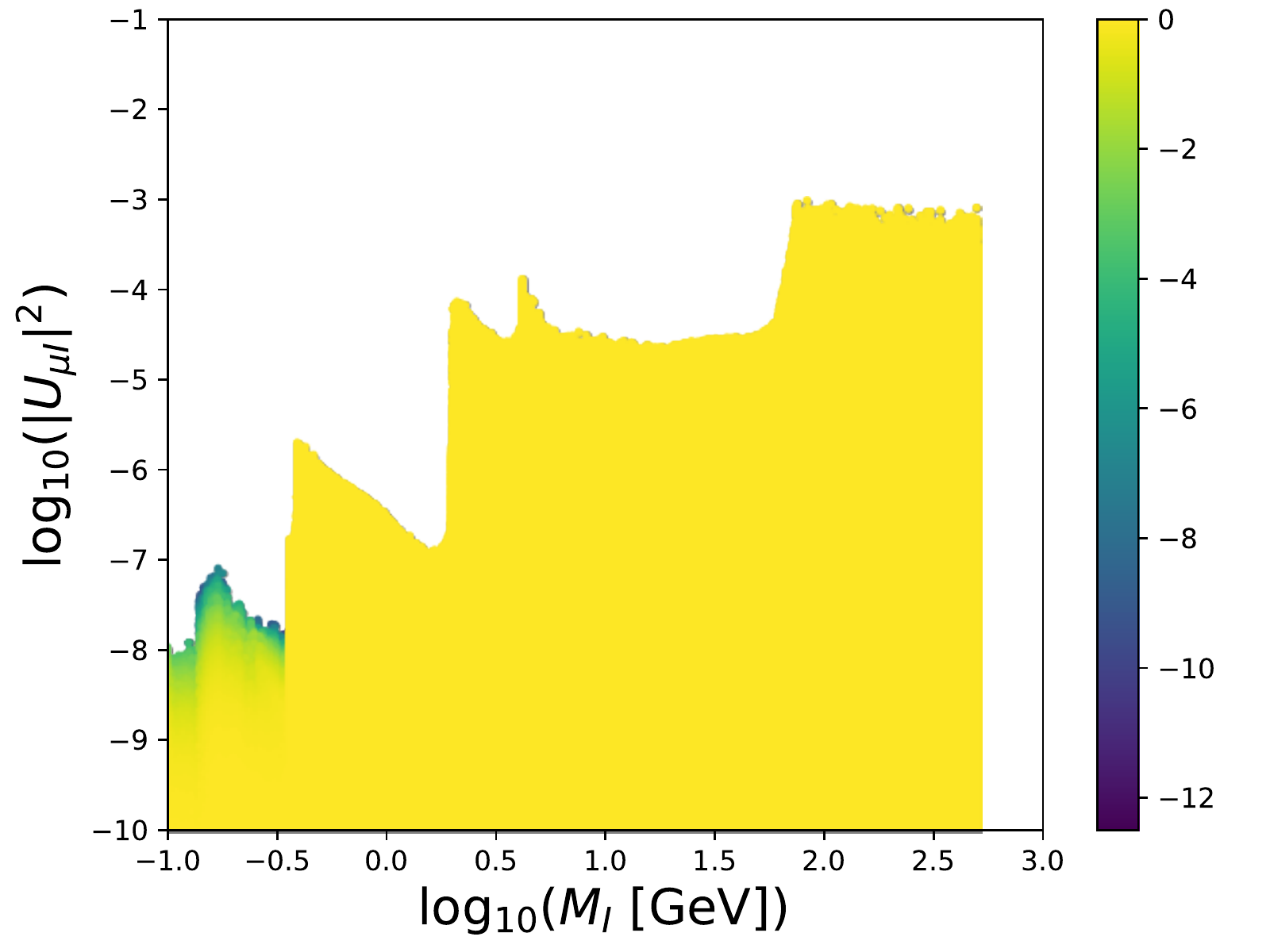}
  \caption{Partial likelihood from direct searches with PS191, $\mu$-channel, in the $M_I-|U_{\mu I}|^2$ plane.}
  \label{fig:lnL_U21_lnL_ps191_mu_NH}
\end{figure}

\begin{figure}[h]
  \centering
  \includegraphics[width=0.8\linewidth]{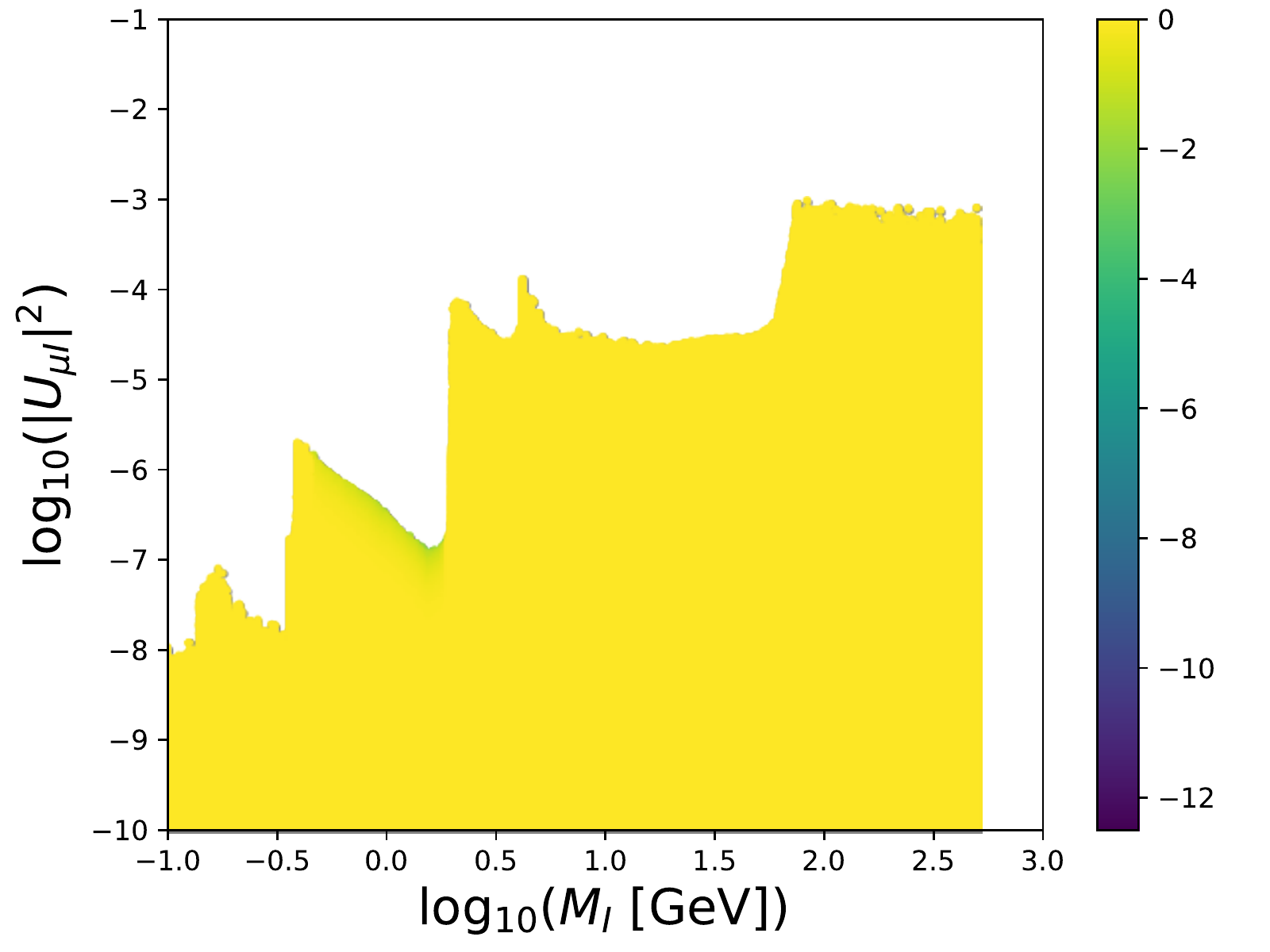}
  \caption{Partial likelihood from direct searches with CHARM, $\mu$-channel, in the $M_I-|U_{\mu I}|^2$ plane.}
  \label{fig:lnL_U21_lnL_charm_mu_NH}
\end{figure}

\begin{figure}[h]
  \centering
  \includegraphics[width=0.8\linewidth]{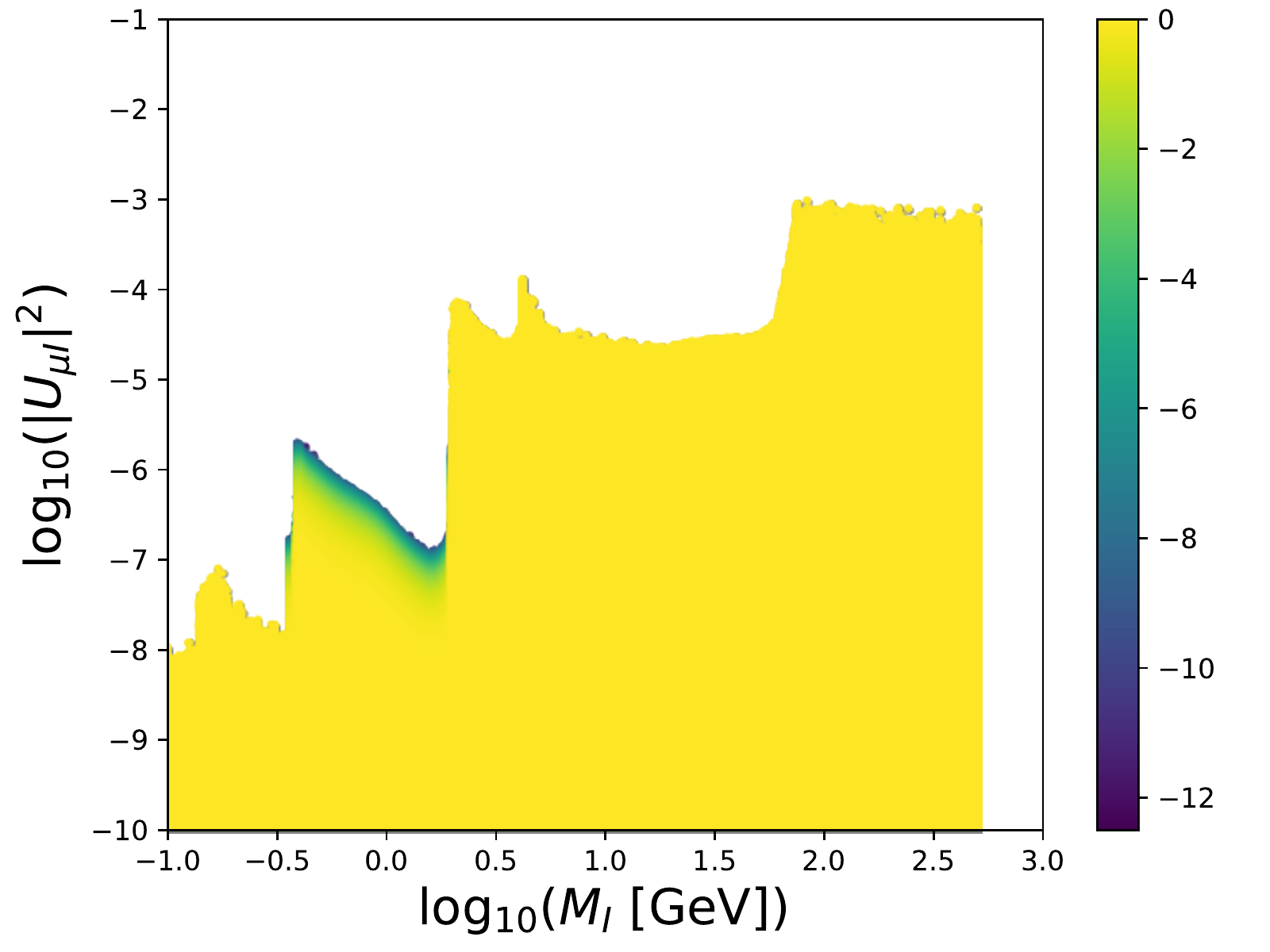}
  \caption{Partial likelihood from direct searches with NuTeV, in the $M_I-|U_{\mu I}|^2$ plane.}
  \label{fig:lnL_U21_lnL_nutev_NH}
\end{figure}

\begin{figure}[h]
  \centering
  \includegraphics[width=0.8\linewidth]{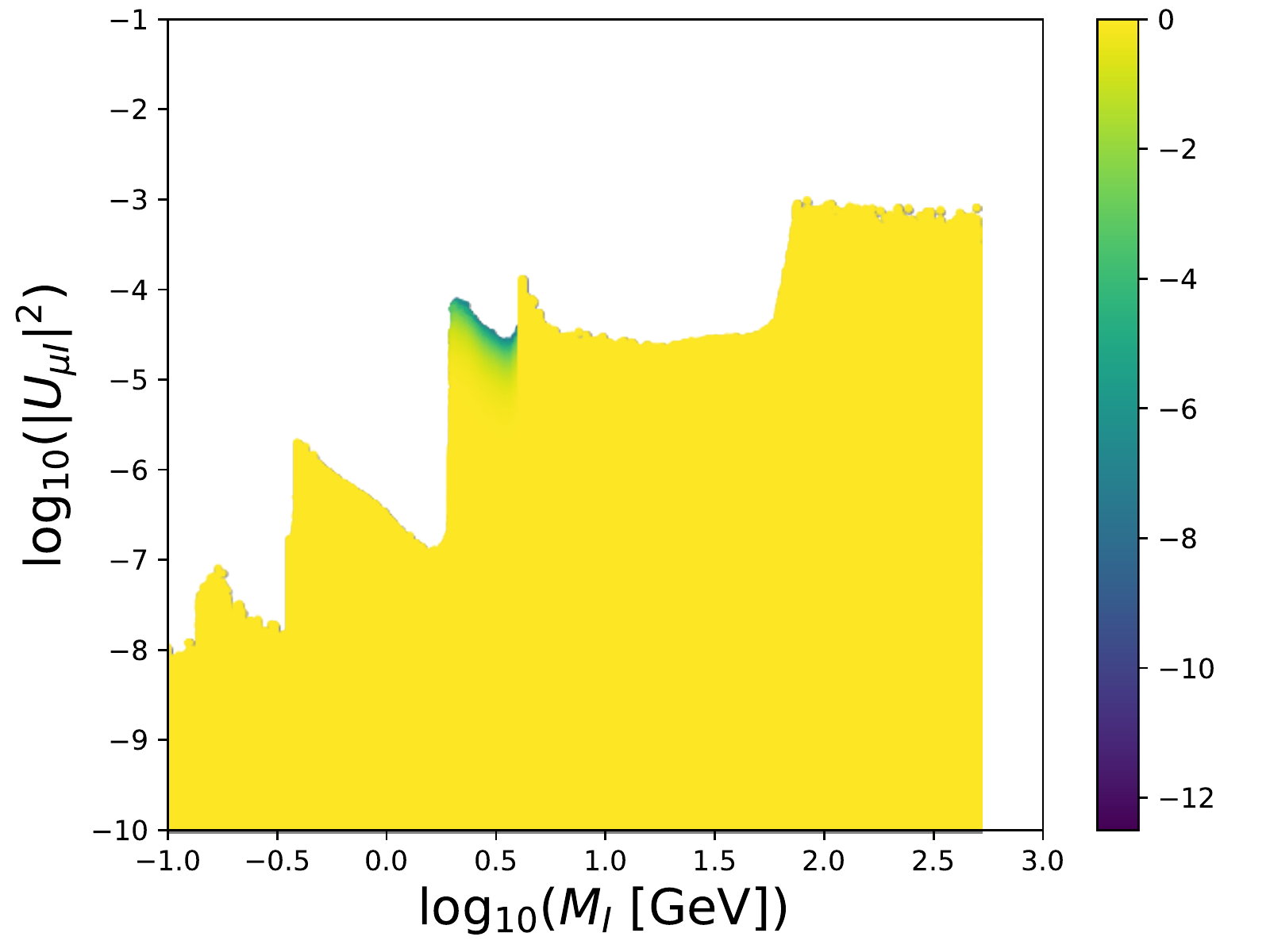}
  \caption{Partial likelihood from the long-lived particle searches with DELPHI, in the $M_I-|U_{\mu I}|^2$ plane.}
  \label{fig:lnL_U21_lnL_delphi_long_NH}
\end{figure}

\begin{figure}[h]
  \centering
  \includegraphics[width=0.8\linewidth]{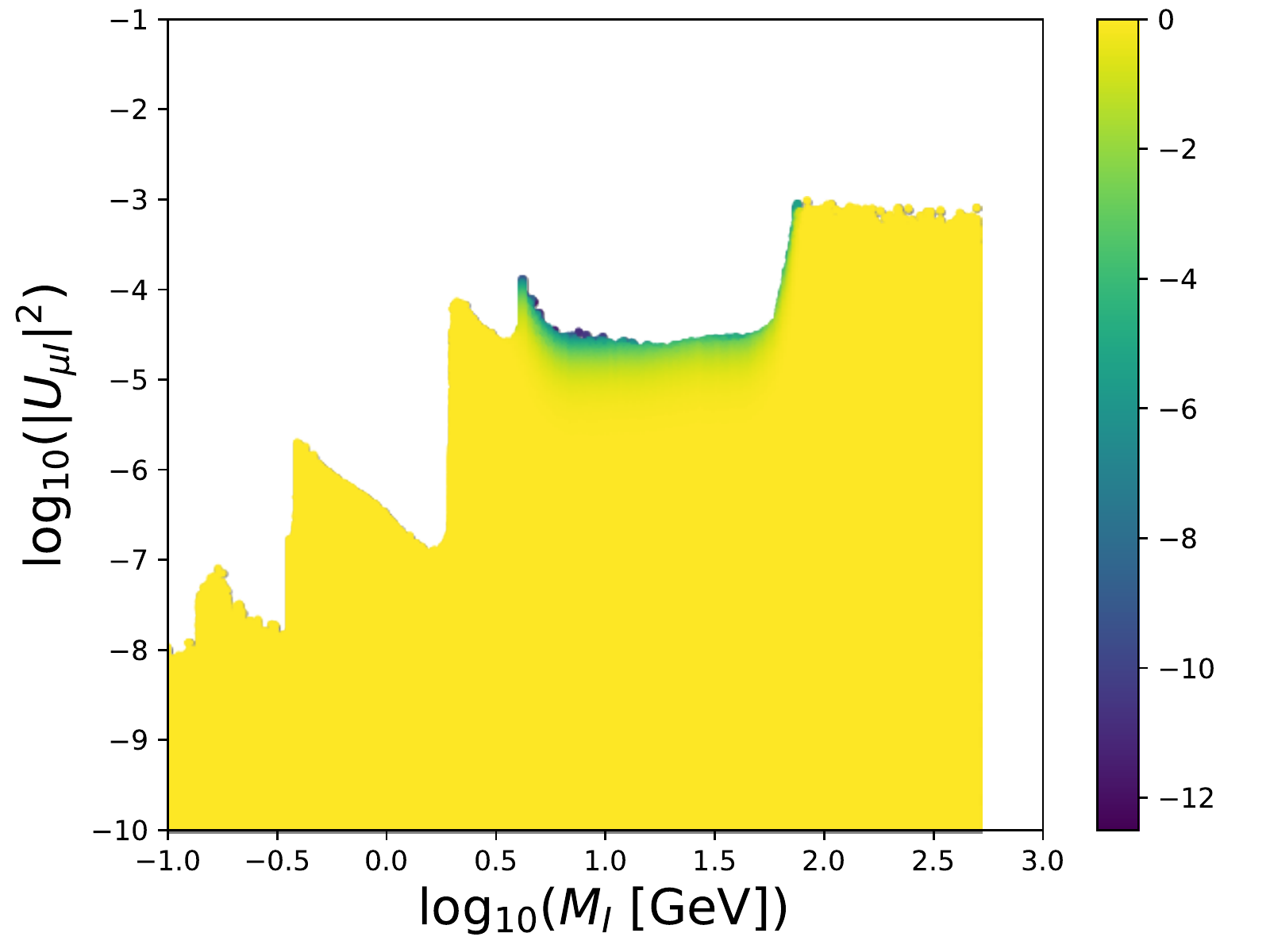}
  \caption{Partial likelihood from the prompt searches with DELPHI, in the $M_I-|U_{\mu I}|^2$ plane.}
  \label{fig:lnL_U21_lnL_delphi_short_NH}
\end{figure}

\begin{figure}[h]
  \centering
  \includegraphics[width=0.8\linewidth]{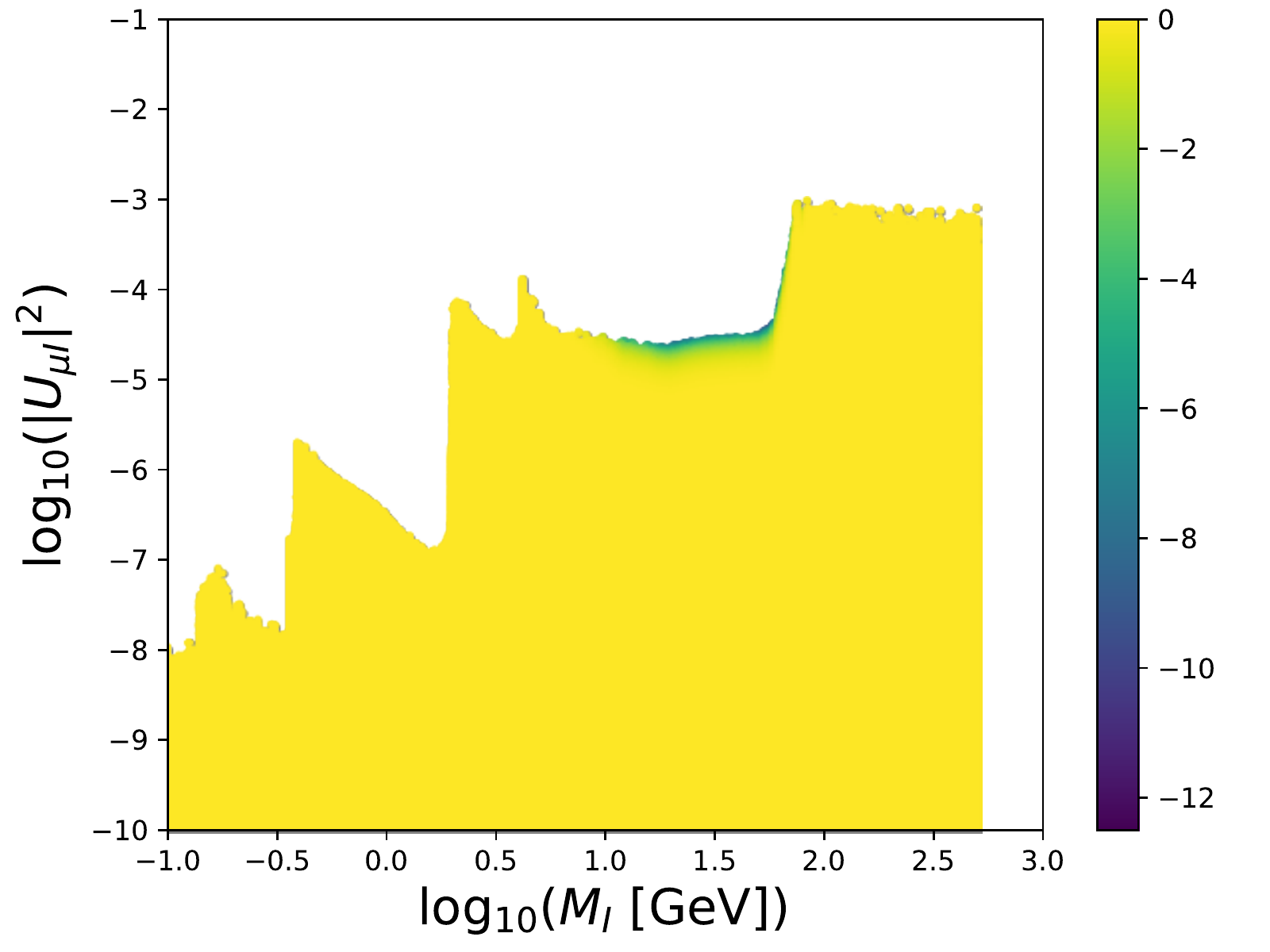}
  \caption{Partial likelihood from direct searches with CMS, $\mu$-channel, in the $M_I-|U_{\mu I}|^2$ plane.}
  \label{fig:lnL_U21_lnL_lhcmu_NH}
\end{figure}

Similar to the case above, the coupling $|U_{\mu I}|^2$ is constrained from above by several direct and precision searches. Figures \ref{fig:lnL_U21_lnL_e949_NH}-\ref{fig:lnL_U21_lnL_lhcmu_NH} show the effect of the individual likelihoods on the upper limit of $|U_{\mu I}|^2$. As opposed to the electron case, where for most mass ranges only one constraint dominated, in this case several mass ranges show competing effects from various constraints. For $M_I < 0.45$ GeV both PS191 and E949 are relevant; in the range $M_I \sim (0.45, 2)$ GeV searches at NuTeV are the most constraining, with a small contribution from the results from CHARM; the long-lived particle search from DELPHI remains unchallenged for $M_I \sim (2,4)$ GeV whereas, as before, the DELPHI prompt search competes in the range $M_I \sim (4, 80)$ GeV, with searches at CMS.

\begin{figure}[h]
  \centering
  \includegraphics[width=0.8\linewidth]{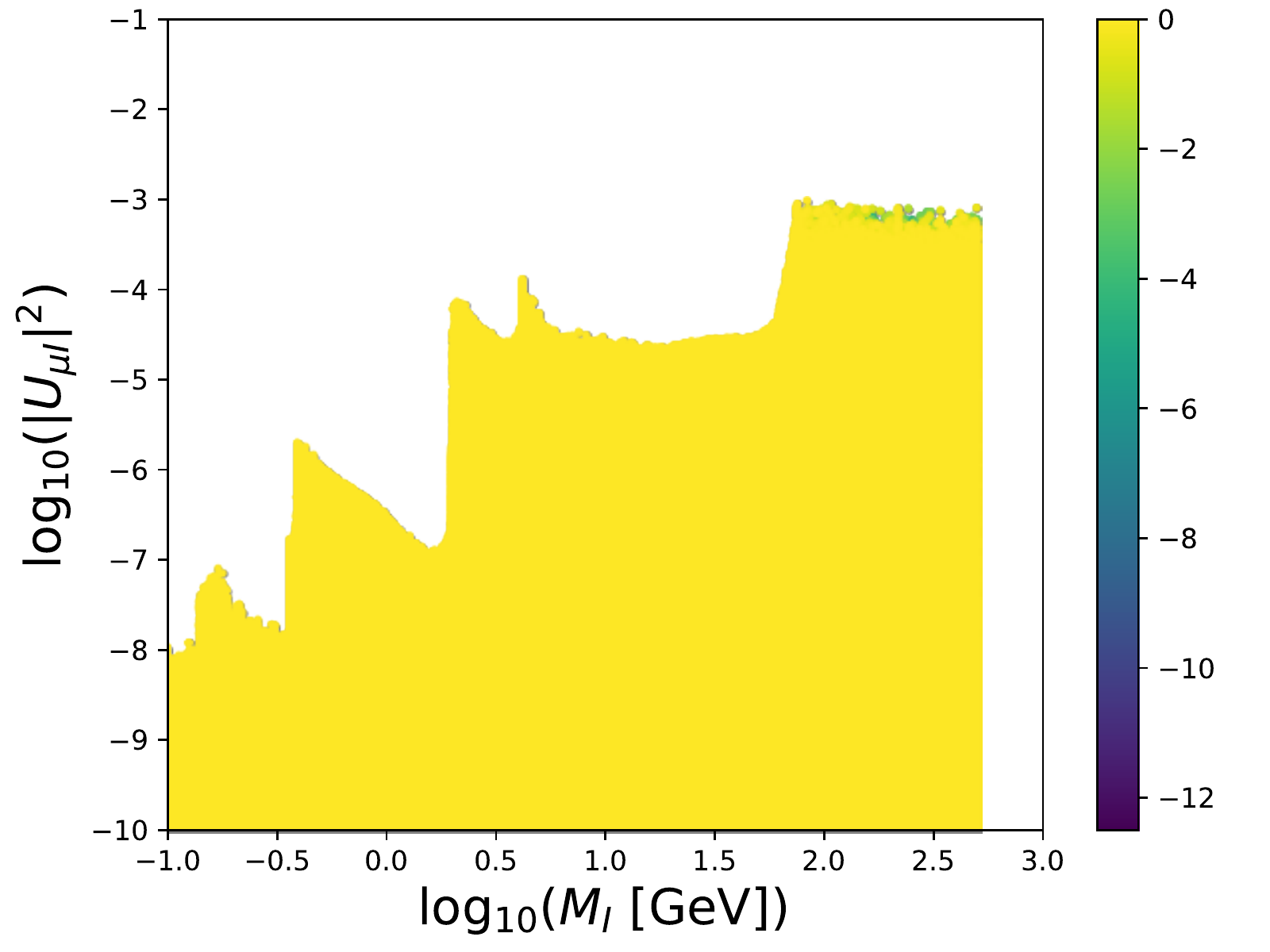}
  \caption{Partial likelihood from lepton flavour violating decays $\mu \to e\gamma$, in the $M_I-|U_{\mu I}|^2$ plane.}
  \label{fig:lnL_U21_lnL_l2lgamma_NH}
\end{figure}

\begin{figure}[h]
  \centering
  \includegraphics[width=0.8\linewidth]{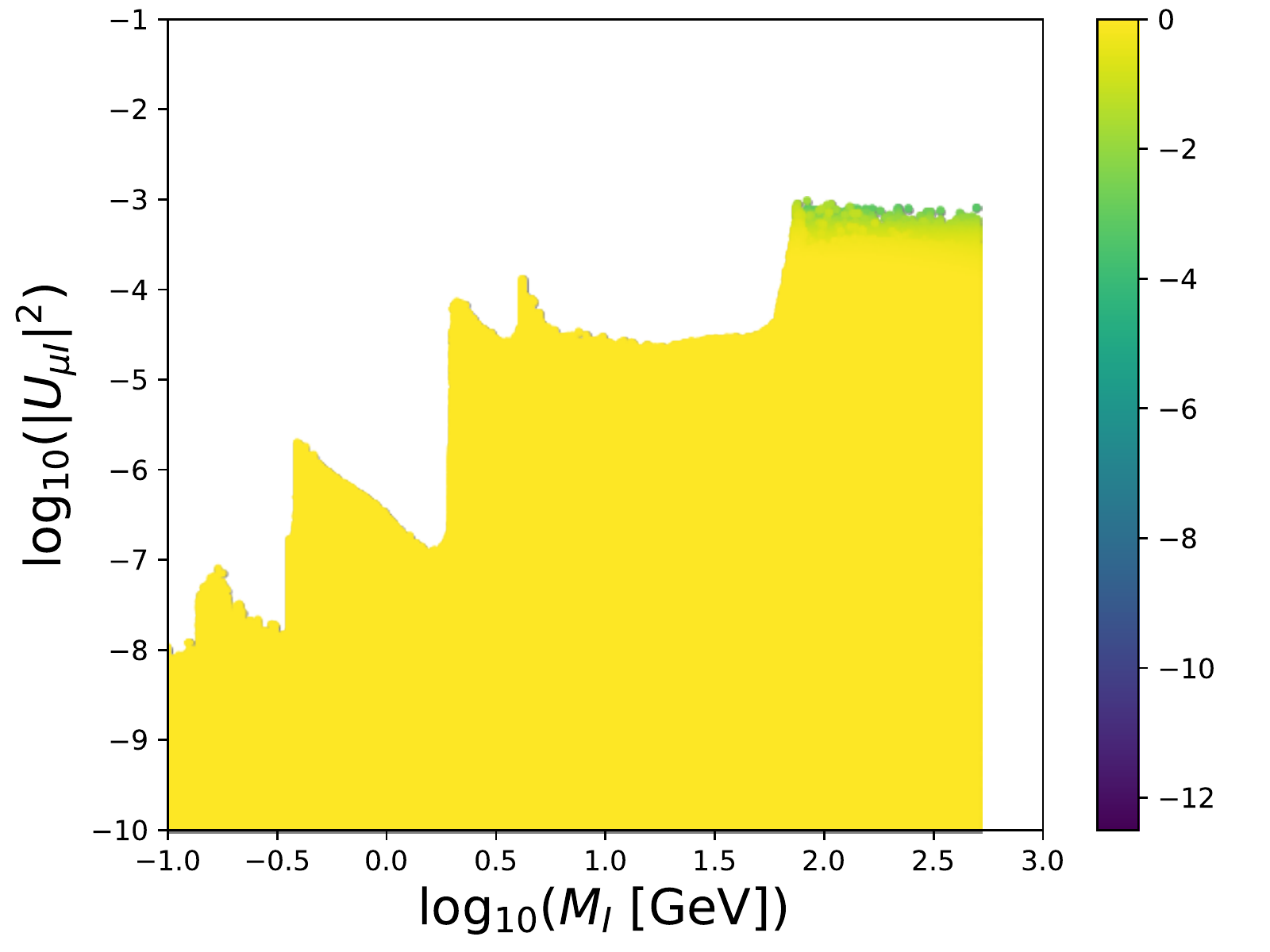}
  \caption{Partial likelihood from $\sin\theta_W$, in the $M_I-|U_{\mu I}|^2$ plane.}
  \label{fig:lnL_U21_lnL_sinW2_eff_NH}
\end{figure}

\begin{figure}[h]
  \centering
  \includegraphics[width=0.8\linewidth]{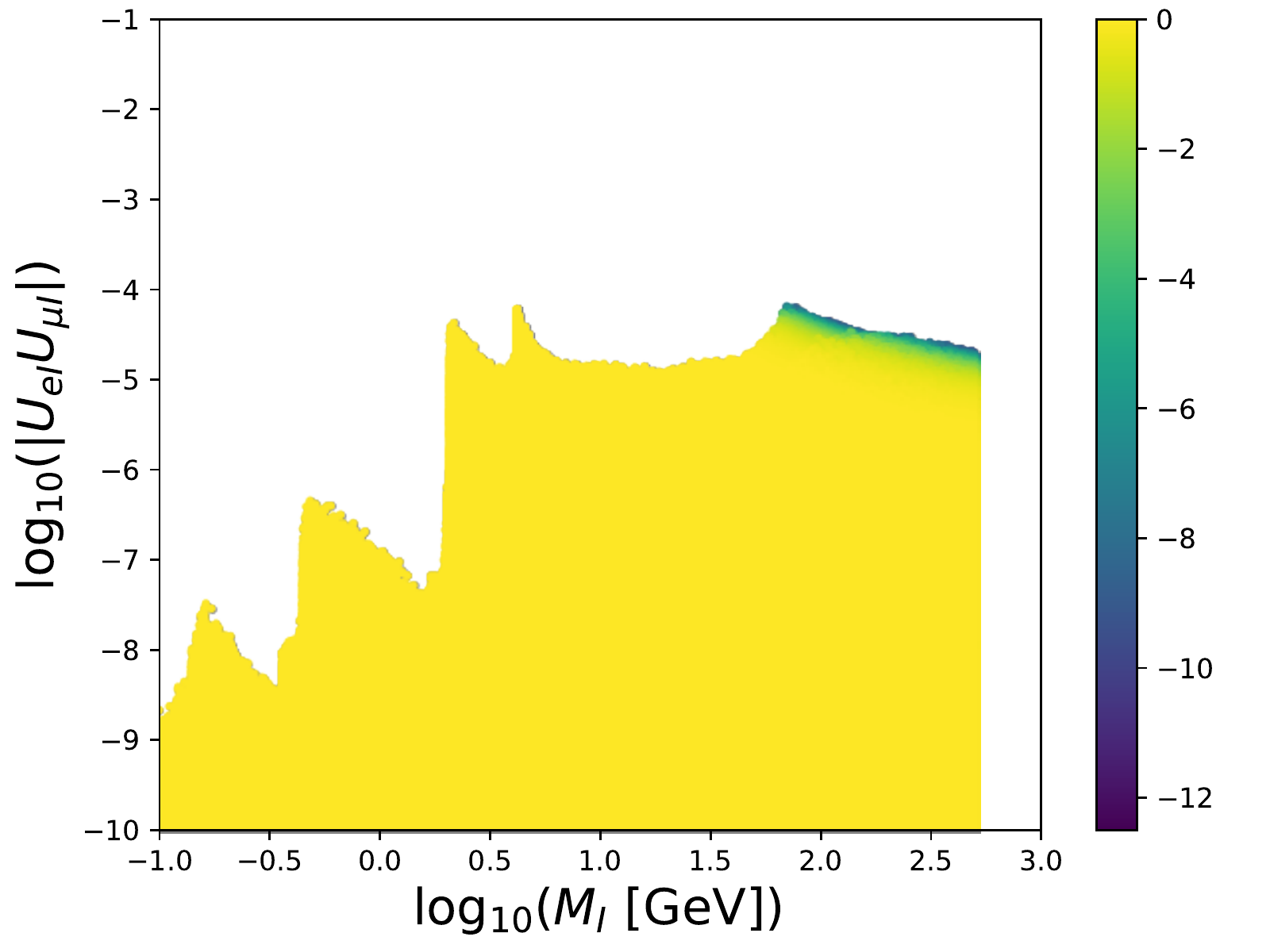}
  \caption{Partial likelihood from lepton flavour violating decays $\mu \to e\gamma$, in the $M_I-|U_{eI}U_{\mu I}|$ plane.}
  \label{fig:lnL_U121_lnL_l2lgamma_NH}
\end{figure}

\begin{figure}[h]
  \centering
  \includegraphics[width=0.8\linewidth]{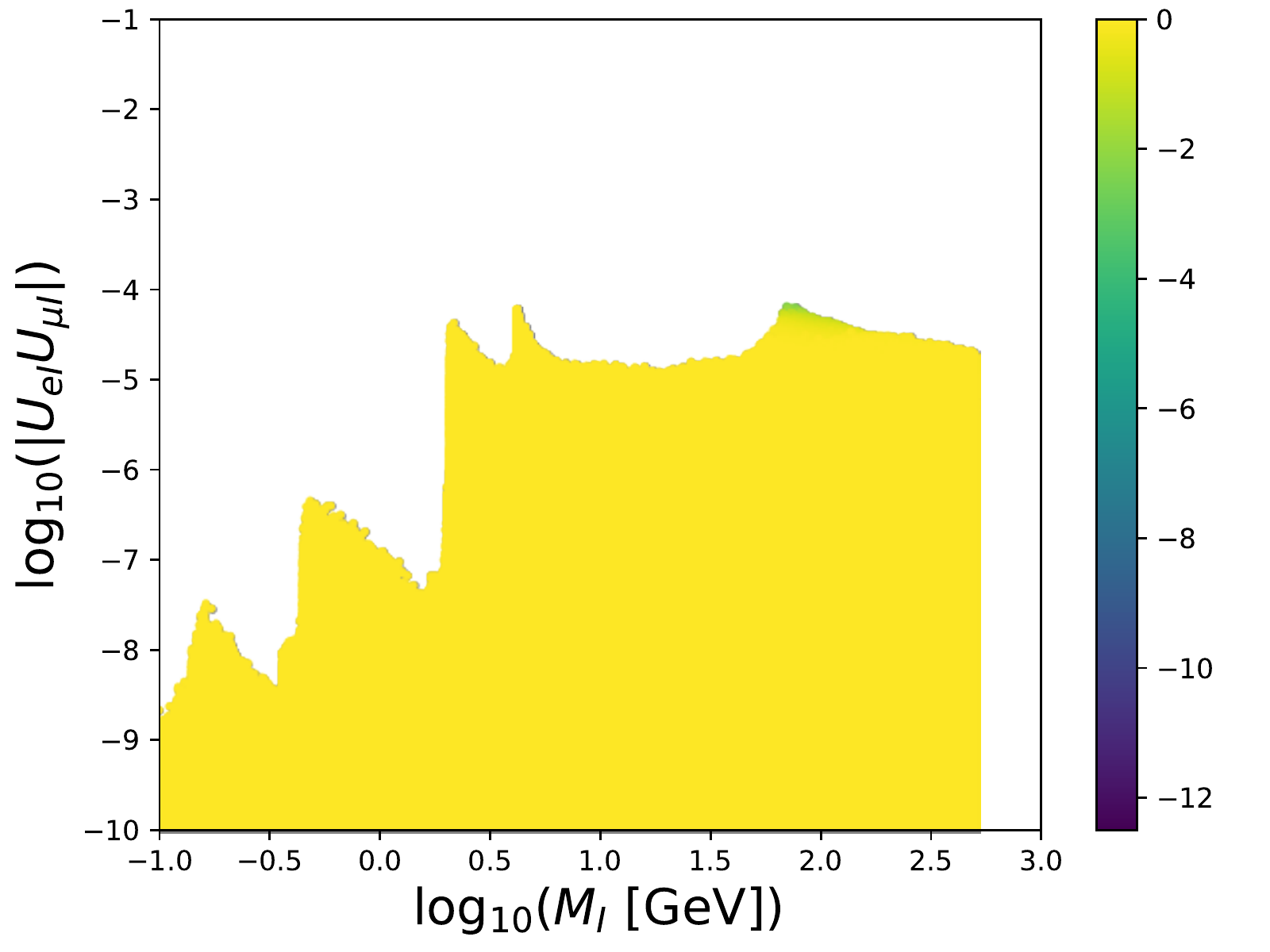}
  \caption{Partial likelihood from the lepton flavour violating $\mu -e$ conversion, in the $M_I-|U_{eI} U_{\mu I}|$ plane.}
  \label{fig:lnL_U121_lnL_mu2e_NH}
\end{figure}

Larger masses are not constrained by direct searches, but rather by a combination of precision limits. Contrary to $|U_{eI}|^2$, where only $\sin\theta_W$ dominated at large masses, upper values of $|U_{\mu I}|^2$ are also mildly constrained by lepton flavour violating decays, particularly $\mu\to e\gamma$. Hence, for $M_I \gtrsim 80$ GeV, the combination of EWPO, $\sin\theta_W$, and LFV decays, are the most constraining, as seen in Figures~\ref{fig:lnL_U21_lnL_l2lgamma_NH} and \ref{fig:lnL_U21_lnL_sinW2_eff_NH}. The effect of the LFV constraints can be better appreciated in Figs.~\ref{fig:lnL_U121_lnL_l2lgamma_NH} and \ref{fig:lnL_U121_lnL_mu2e_NH} on the combination $|U_{e I} U_{\mu I}|$ where $\mu \to e\gamma$ is the dominant constraint, supplemented slightly by $\mu -e$ conversion.

\begin{figure}[h]
  \centering
  \includegraphics[width=0.8\linewidth]{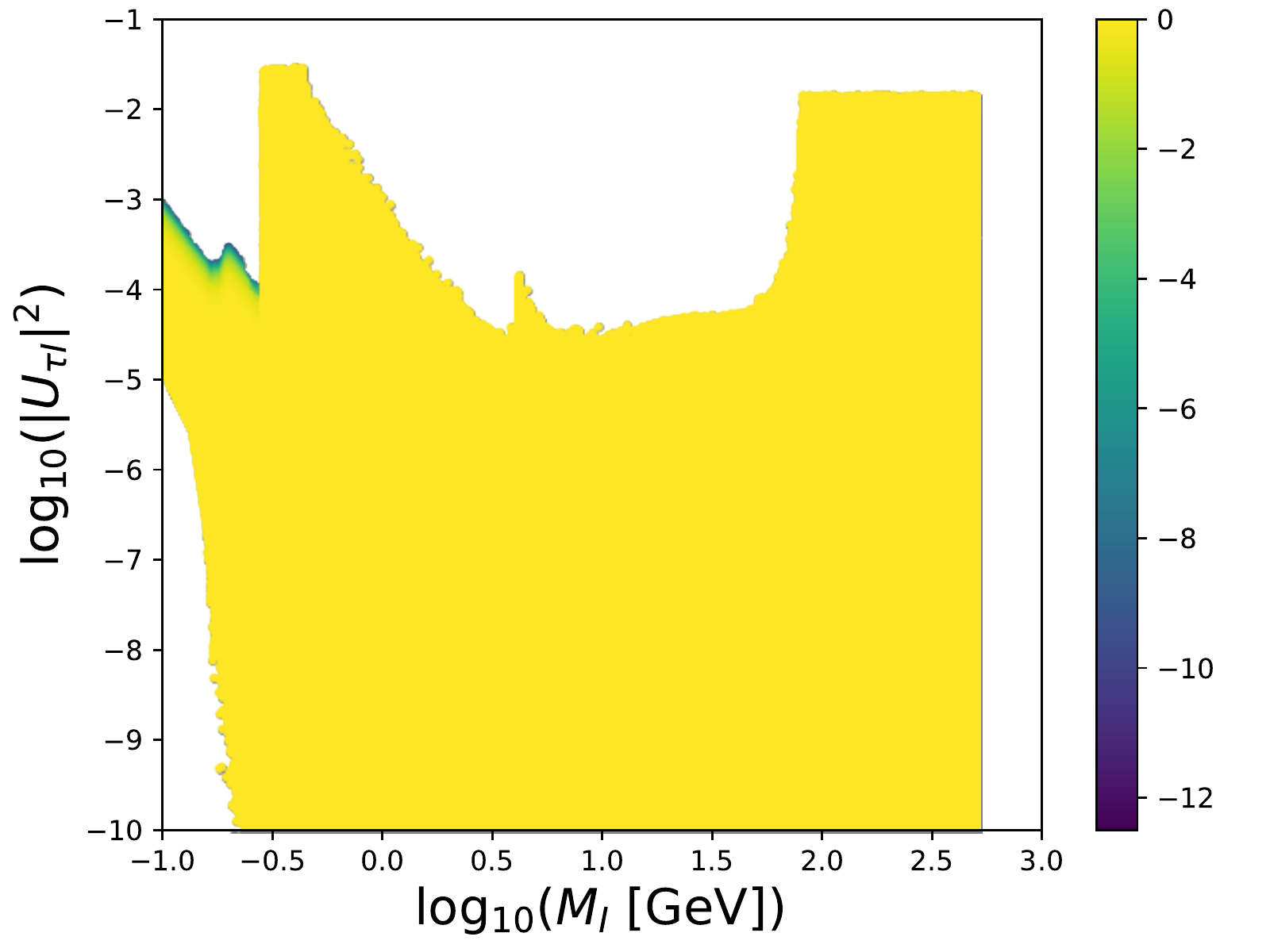}
  \caption{Partial likelihood from direct searches with CHARM, $\tau$-channel, in the $M_I-|U_{\tau I}|^2$ plane.}
  \label{fig:lnL_U31_lnL_charm_tau_NH}
\end{figure}

\begin{figure}[h]
  \centering
  \includegraphics[width=0.8\linewidth]{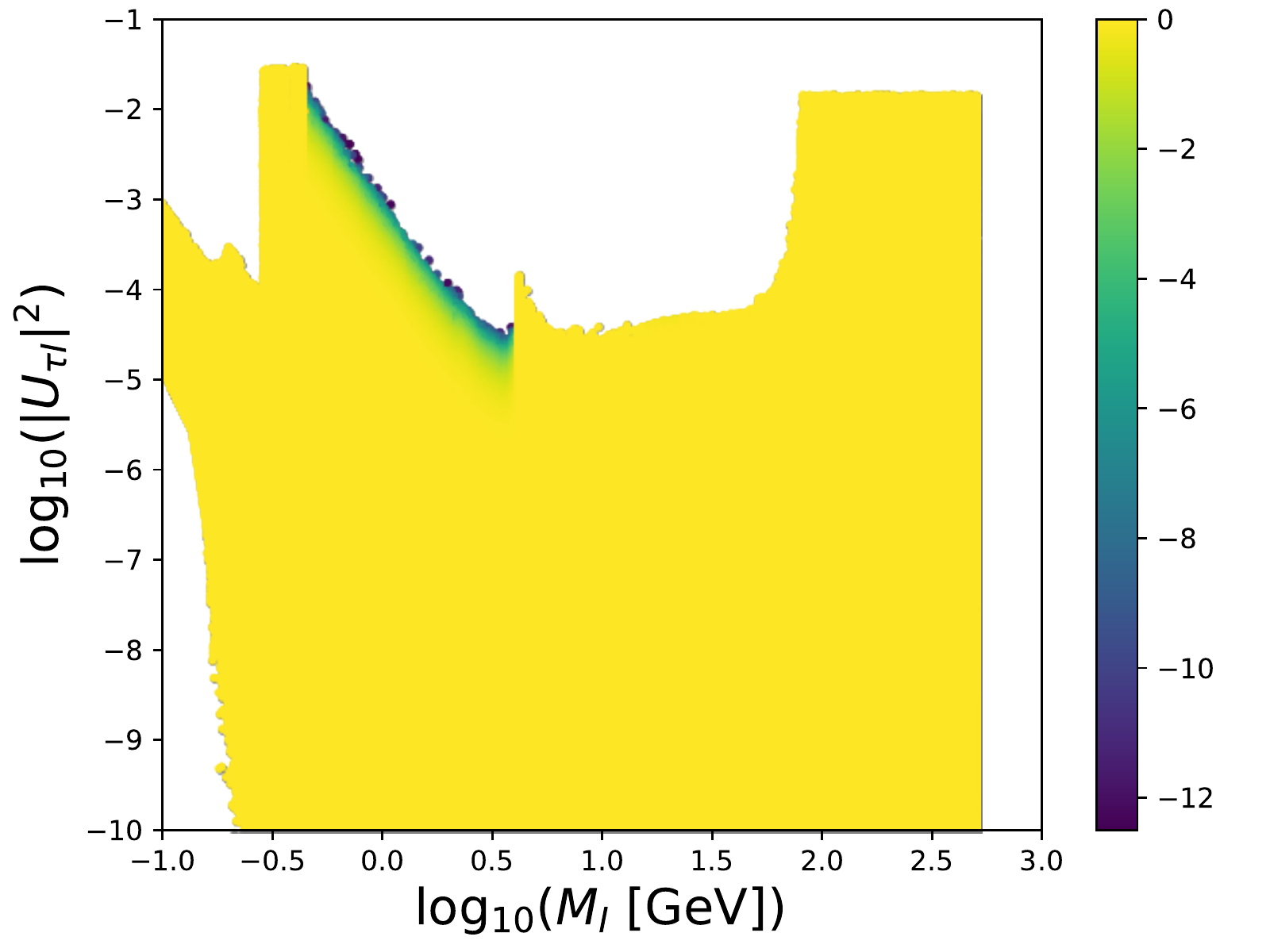}
  \caption{Partial likelihood from the long-lived particle searches with DELPHI, in the $M_I-|U_{\tau I}|^2$ plane.}
  \label{fig:lnL_U31_lnL_delphi_long_NH}
\end{figure}

\begin{figure}[h]
  \centering
  \includegraphics[width=0.8\linewidth]{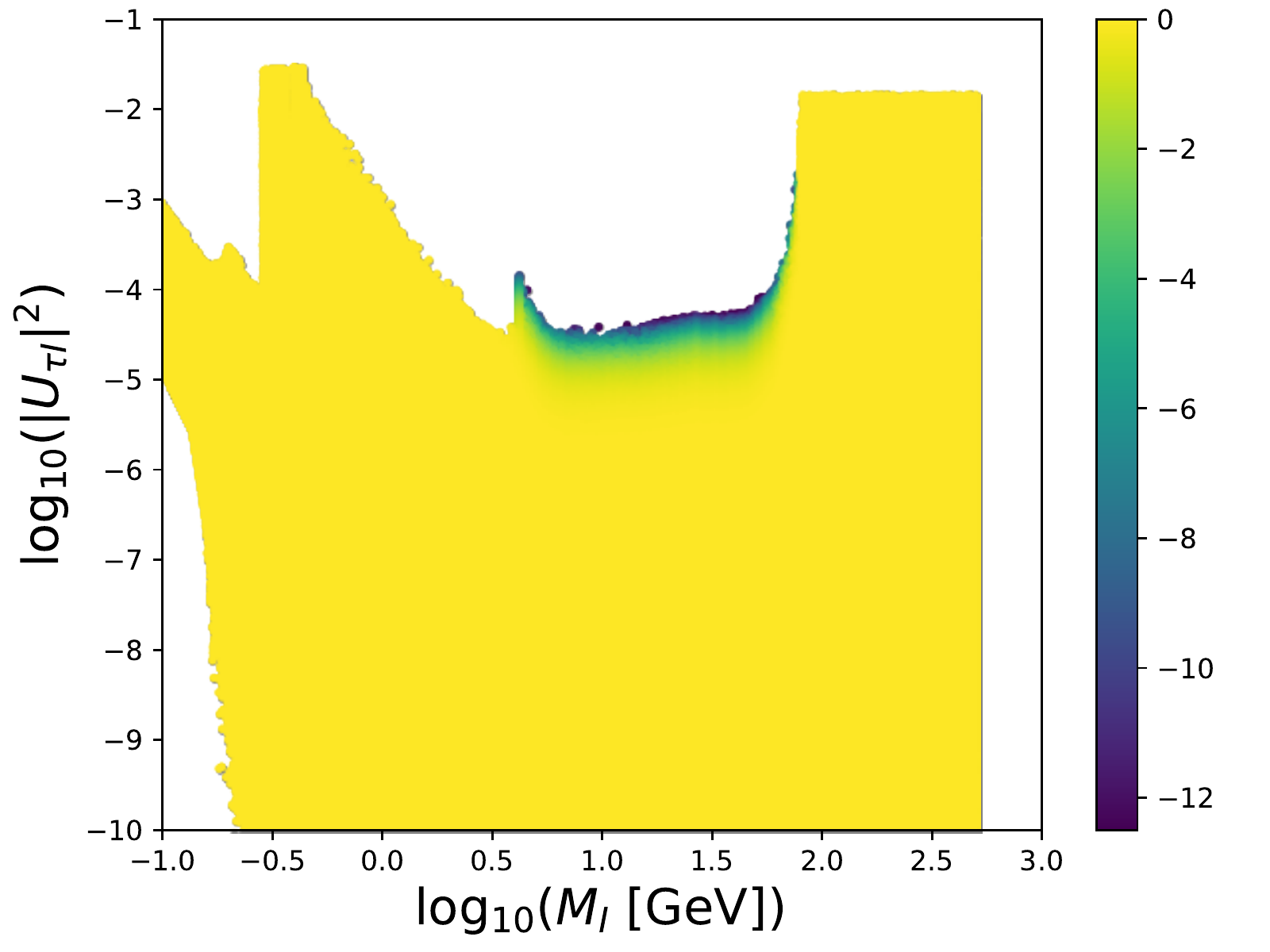}
  \caption{Partial likelihood from the prompt searches with DELPHI, on the $M_I-|U_{\tau I}|^2$ plane.}
  \label{fig:lnL_U31_lnL_delphi_short_NH}
\end{figure}

\begin{figure}[h]
  \centering
  \includegraphics[width=0.8\linewidth]{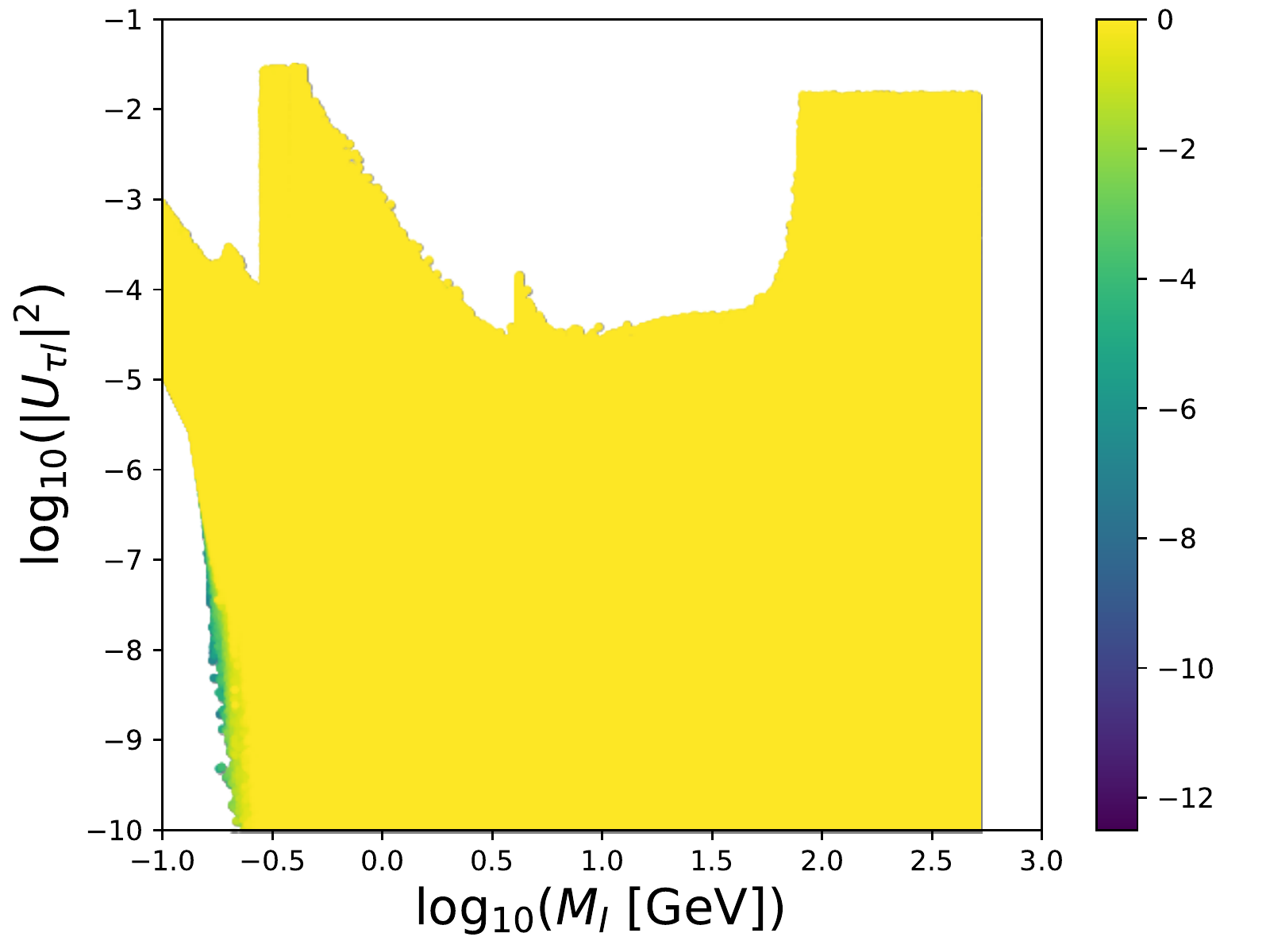}
  \caption{Partial likelihood from direct searches with PS191, $e$-channel, in the $M_I-|U_{\tau I}|^2$ plane.}
  \label{fig:lnL_U31_lnL_ps191_e_NH}
\end{figure}

The couplings of heavy neutrinos to the $\tau$ flavour, $|U_{\tau I}|^2$, are not as strongly constrained from above by direct searches. In Figures~\ref{fig:lnL_U31_lnL_charm_tau_NH}-\ref{fig:lnL_U31_lnL_delphi_long_NH}, one can see that for low masses, $M_I \lesssim 0.3$ GeV, only the direct searches from CHARM in the $\tau$ channel set an upper limit on the couplings. In the mass range $M_I \sim (0.5, 80)$ GeV, long-lived and prompt searches by DELPHI dominate. At low masses, the $|U_{\tau I}|^2$ coupling is constrained from below, as seen in Figure~\ref{fig:lnL_U31_lnL_ps191_e_NH}. This lower bound is a consequence of BBN, which sets a lower limit on the sum of couplings $|U_I|^2$, and PS191, which forces the $e$ and $\mu$ couplings to be small at low masses.

\begin{figure}[h]
  \centering
  \includegraphics[width=0.8\linewidth]{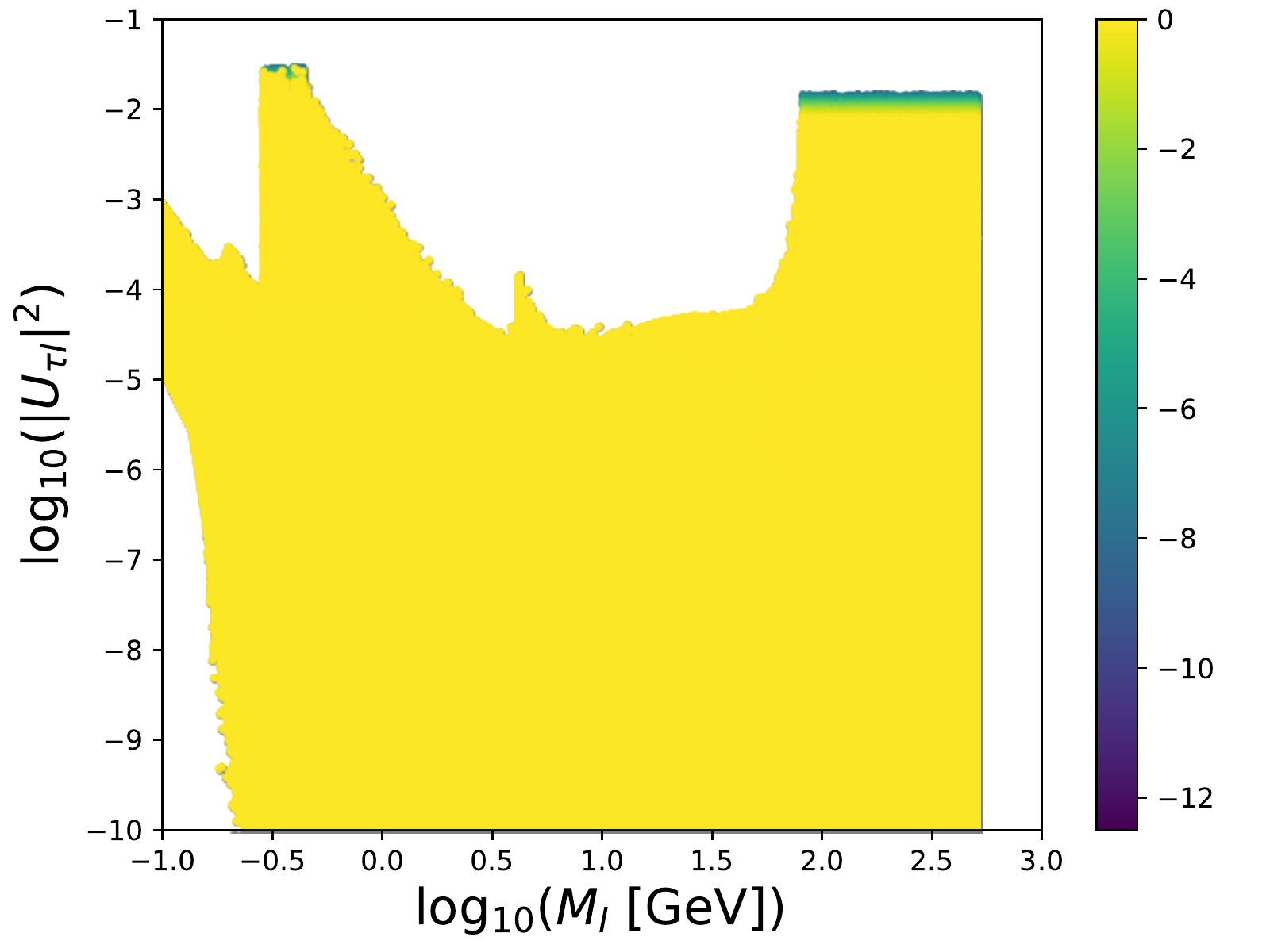}
  \caption{Partial likelihood from the invisible decay width of the $Z$-boson, in the $M_I-|U_{\tau I}|^2$ plane.}
  \label{fig:lnL_U31_lnL_Z_inv_NH}
\end{figure}

In the mass range $M_I \sim (0.3, 0.5)$ GeV, as well as for large masses $M_I \gtrsim 80$ GeV, direct searches do not constrain $|U_{\tau I}|^2$. Hence in these ranges, the strongest constraints come from the invisible decay of the $Z$ boson, as seen in Figure~\ref{fig:lnL_U31_lnL_Z_inv_NH}. This figure uses the "capped" likelihood defined previously, so the excesses in $\Gamma_{\rm inv}$ discussed in Section~\ref{sec:results} will not be visible.

\clearpage

\bibliography{R1.5}

\end{document}